\documentclass[11pt, a4paper, oneside]{report} 

\usepackage{bbm,amsthm,amsmath,amssymb}
\usepackage{wrapfig}
\usepackage[T1]{fontenc}
\usepackage{graphicx}
\usepackage{caption}
\graphicspath{{./Pictures/}} 

\newcommand{\va}[1]{\ensuremath{(\Delta#1)^2}}
\newcommand{\id}{\ensuremath{\mathbbm 1}}
\newcommand{\aver}[1]{\langle #1 \rangle}
\newcommand{\collop}[2]{\ensuremath{\sum_{n=1}^#2 #1^{(n)}}}
\newcommand{\hil}{\ensuremath{\mathcal H}}
\newcommand{\ket}[1]{\ensuremath{|#1\rangle}}
\newcommand{\ketbra}[1]{\ensuremath{| #1 \rangle \langle #1 |}}

\theoremstyle{plain}

\newtheorem{Th}{Theorem}[section]
\newcommand{\bbth}{\begin{Th}}
\newcommand{\eeth}{\end{Th}}

\newtheorem{Obsn}{Observation}[chapter]
\newcommand{\bboo}{\begin{Obsn}}
\newcommand{\eeoo}{\end{Obsn}}

\newcommand{\bbpr}{\begin{proof}}
\newcommand{\eepr}{\end{proof}}

\newtheorem{Defi}{Definition}[chapter]
\newcommand{\bbdf}{\begin{Defi}}
\newcommand{\eedf}{\end{Defi}}

\theoremstyle{definition} 

\newtheorem{Exmpl}{Example}[chapter]
\newcommand{\bbexmp}{\begin{Exmpl}}
\newcommand{\eeexmp}{\end{Exmpl}}

\theoremstyle{plain}   

\newtheorem{Princ}{Principle}
\newcommand{\bbpri}{\begin{Princ}}
\newcommand{\eepri}{\end{Princ}}

\newtheorem*{Assum}{ }
\newcommand{\bbasm}{\begin{Assum}}
\newcommand{\eeasm}{\end{Assum}}

\newtheorem*{Rmk}{Remark}
\newcommand{\bbrmk}{\begin{Rmk}}
\newcommand{\eermk}{\end{Rmk}}

\newcommand{\be}{\begin{equation}}
\newcommand{\ee}{\end{equation}}
\newcommand{\eea}{\end{eqnarray}}
\newcommand{\bea}{\begin{eqnarray}}

\newcommand{\mean}[1]{\ensuremath{\langle{#1}\rangle}}

\newcommand{\bra}[1]{\ensuremath{\langle#1|}}

\newcommand{\kommentar}[1]{}
\newcommand{\trace}{{\rm Tr}}

\newcommand{\de}{\mathrm{d}}

\newcommand{\with}{\quad \mbox{with} \quad}
\newcommand{\imply}{\ \Rightarrow \ }
\newcommand{\diag}{\mathrm{diag}}
\newcommand{\for}{\quad \mbox{for} \quad}

\usepackage{afterpage}

\newcommand\blankpage{%
    \null
    \thispagestyle{empty}%
    \newpage}

\begin{document}

\begin{titlepage}
\begin{center}

{\LARGE University of the Basque Country}\\[1.5cm] 
{\Large Doctoral Thesis}\\[0.5cm]

\vfill
{\huge \bfseries Spin Squeezing, Macrorealism and the Heisenberg uncertainty principle}\\
\vfill 
\begin{minipage}{0.4\textwidth}
\begin{flushleft} \large
\emph{Author:}\\
{Giuseppe Vitagliano} 
\end{flushleft}
\end{minipage}
\begin{minipage}{0.4\textwidth}
\begin{flushright} \large
\emph{Supervisor:} \\
{Prof. G\'eza T\'oth}  
\end{flushright}
\end{minipage}\\[3cm]
 
\textit{at the}\\[0.4cm]
Department of Theoretical Physics and History of Science\\[2cm] 
 
{\large October 2015}\\[2cm] 

\vfill
\end{center}

\end{titlepage}

\pagestyle{empty} 

\null\vfill

\textit{``I don't like it, and I'm sorry I ever had anything to do with it.''}

\begin{flushright}
Erwin Schr\"odinger
\end{flushright}

\vfill\vfill\vfill\vfill\vfill\vfill\null

\begin{flushright}
\includegraphics[width=2cm,clip]{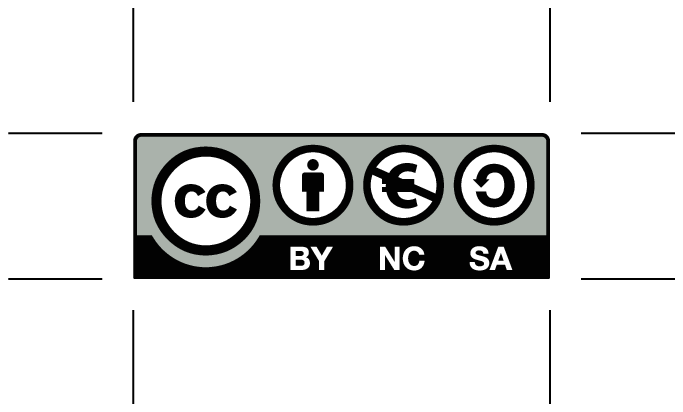}

\tiny{2015 by Giuseppe Vitagliano. This work is licensed under the Creative Commons Attribution-NonCommercial-ShareAlike International 4.0. To view a copy of this license, visit {\em http://creativecommons.org/licenses/by-nc-sa/4.0/}.}
\end{flushright}
\clearpage 

\chapter*{Abstract}

The main aim of this thesis is to investigate the foundational principles of quantum mechanics, to find methods to test them experimentally, particularly in the realm of macroscopic objects.
In other words, our study is devoted to detect effects that can be attributed to a violation of classical principles, such as 
local-realism or macroscopic-realism in macroscopic systems, e.g., large ensembles of atoms. At first,
after reviewing the basic principles and 
defining what a quantum state is, we will retrace the historical debate started around the foundations of quantum mechanics.

According to the developments that followed the celebrated result of Bell
we will argue about the difference between 
the ``truly quantum correlated'' and the classically correlated states in multipartite scenarios. 
We introduce the subject of entanglement detection and show the idea behind our specific approach, that is tied to systems composed of very many parties: detecting entanglement through uncertainty relations. 

We will focus even more specifically on the so-called spin squeezing inequalities, that
will be one of the central topics of our investigation.
We will outline the relation between spin squeezing, quantum metrology
and entanglement detection, with a particular focus on the last. 
We will derive spin squeezing criteria for the detection of entanglement and its depth that outperform past approaches, especially for unpolarized states, recently produced in experiments and object of increasing interest in the community.
Furthermore, we will extend the original definition of spin squeezed states by providing a new parameter that is thought to embrace different classes of states in a unified framework. 

Afterwards we consider a test of quantum principles in macroscopic objects originally designed by Leggett and Garg. In this case the scenario consists of a single party that is probed at different time instants and the quantum effect detected is the violation of Macrorealism (MR), due to strong correlations in time, rather than in space, between non-compatible observables.
We will look at the problems of inconclusiveness of the LG tests arising from possible explanations of the results in terms of ``clumsy'' measurements, what has been termed ``clumsiness loophole''.

We propose first a scheme to test and possibly falsify (MR) in macroscopic ensembles of cold atoms based on alternating Quantum Non-Demolition measurements and coherent, magnetically-driven collective spin rotations. Then we also propose
a way to address the clumsiness loophole by introducing and computing an invasivity quantifier to add to the original LG expression. 
We provide numerical evidence that such a clumsiness-free test is feasible under state-of-the-art realistic experimental parameters and imperfections. 

To conclude we present on one hand some preliminary results on possible further extensions of the spin squeezing framework and applications to the context of statistical physics of spin models, and also some possible future directions and applications of QND-measurement based LG tests.

\clearpage

\section*{Acknowledgements}

Since this work represents somehow the end of a research training
I would like to thank the people who contributed directly to it and 
those who contributed to create the nice research environment in which I had the 
pleasure to work. I would like to thank my supervisor, prof. G\'eza T\'oth, 
first of all for giving me this opportunity, including a research project I found very beautiful,
and for following constantly my work, teaching me and collaborating with me 
in producing the results. Also, he built during these years a very nice research group, whose members contributed very actively to my research and personal formation. I would like to thank them,
I\~nigo Urizar Lanz, Iagoba Apellaniz, Philipp Hyllus, Zoltan Zimboras and Matthias Kleinmann, for
the personal and scientific collaborations and in few words for becoming such good friends.

Then I would like to thank all the people in the departments of Theoretical Physics and Chemical Physics of the University of The Basque Country, especially the students, that represented an extension of the research group I was honoured to be part of and who formed a very  
comfortable work environment also from a personal point of view. Among them I should mention Roberto Di Candia, Simone Felicetti, Antontio Mezzacapo and Fabio Vallone, with whom I had repeated scientific and philosophical discussions and who have been an extension of my family in the day life.

During these years, I had also the possibility to meet the international scientific community, the honour and the pleasure to meet personally many great physicists, working especially in foundations of quantum mechanics and quantum information. Some of them, Costantino Budroni, Giorgio Colangelo, I\~nigo Egusquiza, Otfried G\"uhne, Morgan Mitchell, Robert Sewell, Carsten Klempt and his experimental group, Bernd L\"ucke, Jan Peise, Jan Arlt, Luis Santos, became collaborators of mine and I would like to thank them for the interchange of ideas 
and for all their work, which I benefited a lot. I would like to thank also Pasquale Calabrese for his kind hospitality and for the discussions we had during my staying in Pisa.

Finally, I should thank all the persons that contributed to my personal formation, my wife Marcella, my family and all my friends. Unfortunately there is not enough space to make a list, but I hope each of them knows how important he/she has been to me.

{\it I would like to thank the members of my panel, Antonio Ac\'in, I\~nigo Egusquiza, Otfried G\"uhne, Morgan Mitchell and Jens Siewert for carefully reading the manuscript and for their feedback, which has been very important for the completion of this work and for future developments of my research.}

\clearpage 

\chapter*{List of Publications} 

\section*{Scientific papers}

\begin{itemize}
\item  {\it C. Budroni, G. Vitagliano, G. Colangelo, R. J.
  Sewell, O, G\"uhne, G. T\'oth, M.W. Mitchell}, Quantum non-demolition measurement enables macroscopic Leggett-Garg tests, Phys. Rev. Lett. {\bf 115}, 200403 (2015) 
\item {\it B. L\"ucke, J. Peise, G. Vitagliano, J. Arlt, L. Santos, G. T\'oth, and C. Klempt}, Detecting multiparticle entanglement of Dicke states, Phys. Rev. Lett. {\bf 112}, 155304 (2014)
\item {\it G. Vitagliano, I. Apellaniz, I.L. Egusquiza, and G. T\'oth}, Spin squeezing and entanglement for arbitrary spin, Phys. Rev. A {\bf 89}, 032307 (2014)
\item {\it G. Vitagliano, P. Hyllus, I. L. Egusquiza and G. T\'oth}, Spin Squeezing Inequalities for Arbitrary Spin , Phys. Rev. Lett. {\bf 107}, 240502 (2011) 
\item {\itshape G. Vitagliano, A. Riera and J.I. Latorre}, Volume-law scaling for the entanglement entropy in spin-1/2 chains , 
New J. Phys. {\bf 12} 113049 (2010) 
\end{itemize}

\section*{Other pubblications}
\begin{itemize}
\item {\it B. L\"ucke, J. Peise, G. Vitagliano, J. Arlt, L. Santos, G. T\'oth, and C. Klempt}, Synopsis: Measuring Entanglement Among Many Particles, APS physics (2014)
\item Puntos de inter\'es: A vueltas con el entrelazamiento cu\'antico, Revista Espa\~nola de F\'isica {\bf Vol 28}, Number 2, page 31 (2014).
\end{itemize}

\clearpage 

\pagestyle{empty} 

\vfill
{\it To Bilbao, the Basque Country and the Basque People} 

\addtocontents{toc}{\vspace{2em}}

\chapter*{Introduction} 

\label{introduction} 

In parallel with the development of a well defined mathematical formalism of quantum theory \cite{dirac30,vonneumann32,bohm51}, a debate started concerning 
the striking contrast between ``spooky'' quantum phenomena (incompatibility between observables, collapse of the wave function etc.) 
and classical principles, such as the fact that outcomes of measurements 
just reveal preexisting properties of a system and can be in principle obtained 
with an arbitrarily small perturbation of the input state \cite{vonneumann32,bohreinstdeb}. 
Even more strikingly, as noted first by Einstein, Podolsky and Rosen in their seminal paper \cite{EPR1935}, 
quantum mechanics predicts effects that are in explicit tension with special relativity, 
namely with the principle that distant objects cannot instantaneously influence each other.

This debate raised the question of whether the description of a system through a \emph{quantum wave function} is complete or just emergent from a more fundamental theory, maybe impossible to discover due to practical limitations. In other words, although it was widely accepted that a quantum state is the most fundamental description of the information that an external observer can extract from a physical system, to many of the fathers of the theory, including Einstein, Schr\"odinger and others \cite{EPR1935,schrod35}, it did not provide a satisfactory description of the actual, ontic state of the system. 
This issue is still, after almost a century, very actively debated, although many steps forward have been made.

This is the first topic treated in this thesis in Chapter~\ref{Chapter1}, where we will introduce the principles and the mathematical formalism of quantum mechanics and review briefly the discussion about its foundations.

A crucial result in this respect was due to Bell in the '60s \cite{BellP1964,bell66}, that found a way to experimentally test the classical principle of \emph{local-realism} that is violated by quantum mechanics. This spooky possibility of influence at a distance was termed \emph{entanglement} by Schr\"odinger \cite{schrod35} and, after being experimentally proven in some pioneering experiments in the '80s \cite{aspect81,aspect821,aspect822}, is nowadays viewed as an important resource to be produced in experiments for different reasons. First of all it is interesting for fundamental reasons to increase the scale (in length, mass and so on) at which entanglement is detected, so to have a closer look into the \emph{quantum/classical divide}, i.e., the boundary between systems that must be described in a fully quantum mechanical formalism and systems that can be well approximated by classical physics.

In fact, although classical principles such as local-realism have been disproved in microscopic systems, it is still not clear how to resolve the tension between measured systems (that are correctly described by quantum mechanics) and measurement apparatuses, that are ultimately described in classical terms \cite{stanfmeas,ghiraldistanf}. In the formalism of quantum mechanics, in fact, this lack of a unified description is postulated as a \emph{collapse of the wave function} caused by the measurement apparatus, that is completely external to the system and not described by the theory.
This leaves an incompatibility between measurements and free evolutions that is usually referred as the \emph{measurement problem}.

On the other hand, a different line of research started with Feynmann's idea \cite{feynman82} that the most fundamental model of computation should be based on quantum mechanics. It turned out that algorithms allowing the presence of entangled states can be more efficient than ``classical'' counterparts in solving several tasks, ranging very widely from metrology, communication and computation \cite{bennett93,eckert91,shor97}. Thus, in a way it has been discovered that entanglement 
helps to enhance the efficiency in acquiring and processing information and is nowadays also a target resource to be produced for technological purposes.

Among these target entangled states a leading role is played, especially in the realm of systems composed by very many parties, by the so called \emph{squeezed states} \cite{schrodinger26,glauber63,perelomov86} that have the advantage of being relatively easy to produce and characterize. Thus, in atomic systems, \emph{spin squeezed states} \cite{Kitagawa1993Squeezed,Wineland1994Squeezed,Ma2011Quantum} are often considered important targets and have been shown to be useful for quantum information processing and quantum metrology. 

These states represent also one of the main topics of investigation of this thesis. 
In Chapter~\ref{Chapter2} we briefly review the fields of entanglement detection and spin squeezing, motivating the investigation
about these topics independently from each other and also introducing their connections.

In Chapter~\ref{Chapter3} we study deeper the connections between spin squeezing and entanglement and present our original contributions 
to the topic. We present spin squeezing criteria for the detection of entanglement and its depth in systems composed of very many parties and we show that they outperform other analogous criteria, especially in detecting unpolarized states. We mention as a practical example an experiment in which an entanglement depth of $68$ has been detected in a Dicke state based on one of our criteria \cite{Lucke2014Detecting}. We also introduce a new notion of spin squeezing parameter that generalizes past approaches in several directions and provides a compact way to
characterize the entanglement of gaussian states of multi spin-$j$ systems.

Thus, rephrasing in a resource-oriented way the discussion made above we can say that parameters connected to detection of entanglement, such as the spin squeezing parameters, represent figures of merit to certify the \emph{quantumness} of a state and in some sense also its potential usefulness for quantum information processing.

Complementarily, another way of looking at the quantum/classical divide with respect to producing and detecting macroscopically entangled states was
introduced by Leggett and Garg in 1985 \cite{LG85}, by adapting the approach of Bell to the realm of macroscopic objects. They proposed a test of the principle of macro-realism based on time correlation measurements performed on a single (possibly macroscopic) system. 
Here, the resource needed to violate LG inequalities consists in strong correlations and incompatibility between the measured observable and its time evolved.
This proposal however, had to face a fundamental problem that has been later termed \emph{clumsiness loophole} \cite{WildeFP2012,EmariNoriRPR2014} and consists in the possibility of interpreting the resulting failure of the test as coming from imperfect (clumsy) measurements, rather than from the violation of macroscopic realism per se. 
Thus, again from a resource-oriented perspective, a conclusive violation of macrorealism requires independently certified \emph{adroit} quantum measurements and witnesses their invasivity, instead of the quantumness of the input state.

Following this intuition, in Chapter~\ref{Chapter4} we present
our proposal to detect the violation of macrorealism that refines the original idea of Leggett and Garg by exploiting some features of 
Quantum Non-Demolition measurements. We propose a protocol that achieves this violation by performing several QND measurements of the collective spin of an atomic ensemble that independently evolves under the influence of an external magnetic field. Also, after reviewing in some detail the clumsiness loophole present in the original LG proposal and its experimental realizations 
made so far, we propose a scheme based on auxiliary measurement sequences that allows to close to some extent this loophole in QND measurement-based LG tests by suitably quantifying the ``clumsiness'', namely the classical noise present in the measurements.

Finally, in Chapter~\ref{Conclusions} we conclude by presenting some preliminary results on further extensions of the spin squeezing framework and suggesting future directions in the study of spin squeezing inequalities and QND measurement-based LG tests.

\chapter{Foundations of Quantum Mechanics} 

\label{Chapter1}

For completeness and in order to present a self-contained work we start in this chapter with an overview of quantum mechanics, from its basic principles and focusing on finite dimensional systems that are the ones we will be mainly interested in the next chapters. See for example  \cite{dirac30,vonneumann32,bohm51,peres_qm,Pitowsky89,parisi_qm,rovelli_qg,wolf12,bengtsson06} as some among the very many reference books on this topic. 
The point of view taken in this overview, as well as in the rest of the thesis is quantum information oriented: we will interpret quantum mechanics as the most general set of rules that one has to take into account in order to acquire knowledge about a physical system at hands. In this sense we refer to quantum theory as the most general \emph{epistemic} theory, or,  in other words, as the most general theory of information, the principles of which are confirmed by experience. 

To complete the introduction we will also discuss the problems arising when trying to interpret quantum theory as \emph{ontic}, i.e., as describing the actual state of individual natural systems.
There is an intense and still lively debate (see e.g., \cite{bell87,peres_qm,rovelli_qg,vonneumann32,bohreinstdeb,stanfkochen,stanfmeas,stanfbell,ghiraldistanf}), that started with the fathers of the theory,
on whether quantum theory can or cannot describe \emph{completely} all the properties of a system, classically assumed to exist prior to and independent of any measurement made on it. We will retrace the route along which this debate developed, by looking at the constraints imposed by various sets of classical principles and finding a fundamental incompatibility between an ontic interpretation of QM and such classical principles. This will serve us as a basis to the successive more profound discussion about possible conclusive experimental tests of (some of) the principles of quantum mechanics in macroscopic bodies.


\section{Principles of quantum mechanics}

Quantum mechanics is a mathematical framework developed to describe the observation of a physical system. Up to now quantum mechanics represents actually the most fundamental set of rules that allows to acquire knowledge and make predictions about a physical system.
It might be seen as a generalization of classical mechanics in some respects: a system can be in a state A, in a state B or in both states; the outcome of measurements are random and do not correspond to ``real'' properties of the system; there are observable properties that are incompatible, that means that they cannot be jointly known on the same state; a measurement perturbs even ``non-locally'' the state of the system; etc. 
Even more intriguingly, QM introduces a new \emph{fundamental constant} of nature $\hbar$ called the \emph{Planck constant}\footnote{This, together with the velocity of light $c$ and the gravitational constant $G$ forms a complete set, meaning that it is possible to express every physical quantity in units of (combinations of) such constants.}.
In the following we will precisely state in some details the main principles from which quantum theory follows\footnote{We are going to follow an approach close to the original axiomatic formulation proposed by von Neumann \cite{vonneumann32}.} and hopefully it will be clearer in which sense it generalizes classical physics. 

First: a system is described by a Hilbert space. When such a Hilbert space is separable (every state can be expanded in terms of a basis, that is a discrete set of independent states) the system is ``quantized'' because it can be described by a discrete set of states.

\subsection{To physical systems are associated Hilbert spaces}

\begin{figure*}[h!]
\centering
\includegraphics[width=0.7\columnwidth,clip]{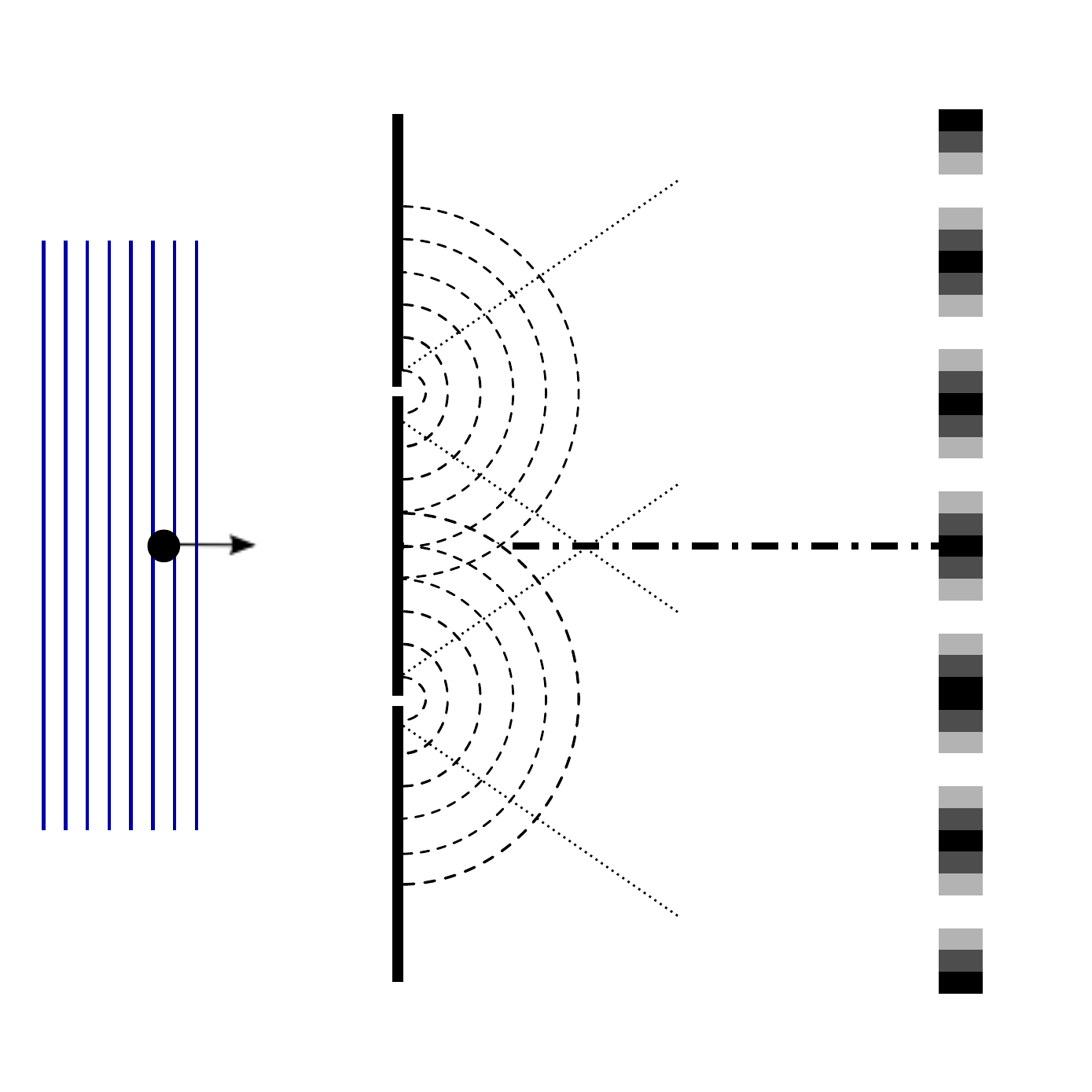}
\caption{Picture of a double-slit experiment: a particle travels through a wall with two slits. An interference pattern is seen when both slits are
open, while no interference appears when one of the slits is closed. The same behaviour is observed with incident photons, electrons or
any other particle. This shows that ``particle'' and ``wave'' are two different manifestations of a single wave/particle entity.}
\label{fig:doubsl}
\end{figure*}

To every system there is a corresponding Hilbert space $\hil_d$. Its (finite) dimension $d$ is given by the number of completely distinguishable states in which the system can be. For example a particle that can be only in two energy levels can be described by a $2$-dimensional Hilbert space $\hil_2$. 
This is a simple mathematical way to formalize the following principle, that is one of the axioms of the theory and is experimentally verified.
\bbpri{\rm \bf (Superporposition Principle).}
If a system can be in two distinguishable states, then it can be also in every linear superposition of such states.
\eepri
Note that this contrasts with the ``classical'' principle that a system (e. g. a particle) can be \emph{either} in one state or in another. According to quantum mechanics it can be also in \emph{both} states.  
Historically, an empirical verification of this principle has been the very famous \emph{double slit experiment}: a particle that interfere with itself when passes through a wall with two slits and that does not when one of the slits is closed (Fig.~\ref{fig:doubsl}).

Thus formally, if the completely distinguishable states $\ket{n}$ are a basis of the Hilbert space of the system, then a general state is any of its vectors
$\ket{\psi}=\sum_n a_n \ket{n}$. 

Afterwards, if one considers a system $\mathcal S$ as composed of multiple parties $\mathcal P_1, \dots , \mathcal P_n$ then one has to associate a Hilbert space to every party $\mathcal H_{\mathcal P_i}$ and the Hilbert space associated to the whole system will be the \emph{tensor product} 
$\mathcal H_{\mathcal S}= \bigotimes_i \mathcal H_{\mathcal P_i}$ of the parties' spaces. This is a direct consequence of the superposition principle: every party can be in every possible superposition of all its possible states. We will do a more detailed discussion about composite systems in a 
dedicated subsection.

Actually one can assume also situations in which the state of the system is not known with certainty. In that case the state is a statistical mixture of states $\{\ket{\psi_k}, p_k\}$, i.e., the system is in a certain state $\ket{\psi_k}$ with probability $p_k$. 
Then the most general quantum state, including these statistical mixtures is actually
a bounded operator acting on the Hilbert space $\rho \in \mathcal B(\hil)$, more specifically a so-called \emph{trace class operator}.
\bbdf{\bf (Density matrix).}
The most general quantum state is called \emph{density matrix}. It is defined as a bounded operator acting on the Hilbert space $\rho \in \mathcal B(\hil)$ with the following properties
\begin{equation}
\begin{aligned}
\rho &= \rho^\dagger \ , \\
\rho & \geq 0 \ , \\
\trace(\rho)&=1 \ ,
\end{aligned}
\end{equation}
and is also-called \emph{trace class operator}. A pure state is such that $\rho^2=\rho$. It is the projector $\rho=\ket{\psi} \bra{\psi}$ onto the vector $\ket{\psi}$ associated to it.
\eedf
A density matrix is a tool used to assign a probability distribution ``in a quantum way'' to every possible outcome of all the measurements that can be performed on the system.
In particular $\rho$ is needed to compute expectation values of all possible observable properties of the system.

Gleason \cite{gleason57} in his famous Theorem \ref{gletheo} proved that $\rho$ is actually the most general way to do it, as we will see in the discussion of the following principles.

The second principle generalizes classical physics to the non-ideal situation in which there exist observable properties of a system to which it is \emph{fundamentally mutually exclusive} to associate definite values independently of measurements. 
This means that perfect knowledge of the value of one of them must be associated with a large uncertainty about the value of the others. 
These are called non-compatible observables and are described by non-commuting operators acting on the systems' Hilbert space.

\begin{figure*}[h!]
\centering
\includegraphics[width=0.4\columnwidth,clip]{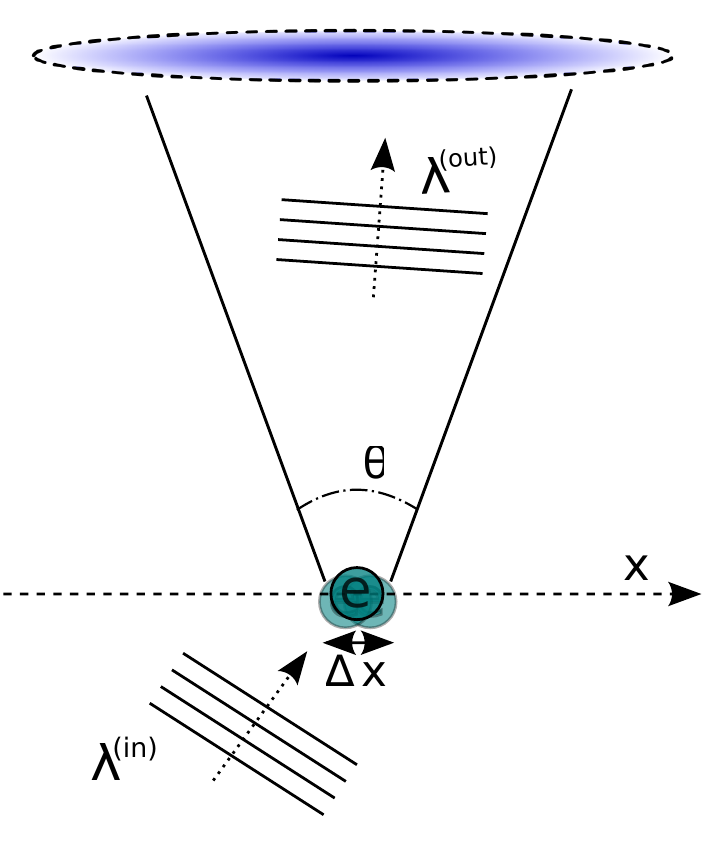}
\caption{An electron is situated near the focus of a lens and hit by photons with wavelength $\lambda^{\rm (in)}$. Photons are scattered inside the microscope deviated from the vertical by an angle less than $\frac{\theta} 2$. The momentum of the electron is consequently perturbed by the kick of the
photon. Classical optics shows that the exact electron position can be resolved only up to an uncertainty $\Delta x$ that depends on $\theta$ and $\lambda^{\rm (in)}$.
}
\label{fig:heismic}
\end{figure*}

Let us see the difference with respect to classical physics with an example. Classically one can ideally measure the position $x$ and the momentum $p$ of a particle and the couple $(x,p)$ represents the state of the point particle. According to the rules of quantum mechanics this is not anymore possible, because the position and the momentum of a particle are two non-compatible observable properties. 
In this case the state of a particle is given by either a complex wave function $\phi(x)$ that provides the probability distribution $|\phi(x)|^2$ of the position or its fourier transform $\hat{\phi}(p)$, from which the momentum probability distribution $|\hat{\phi}(p)|^2$ can be extracted. The actual values of $x$ and $p$ can be known (by hypothesis) only with a precision such that $\va{x}\va{p} \geq \frac \hbar 2$, where $\hbar$ is a new \emph{fundamental} constant.  This fact, that follows from an operational impossibility (see the Heisenberg microscope, Fig.~\ref{fig:heismic}), have several consequences that we are going to sketch briefly.

\subsection{Observables are formalized as hermitean operators}
An observable property $o$ of a system is described with a hermitean operator acting on the Hilbert space. Its eigenvalues $\lambda_i$ are the possible outcomes of a measurement of such property. It can be decomposed as a linear combination of orthogonal projectors
\begin{equation*}
 o = \sum_i \lambda_i  \pi_i \with   \pi_i   \pi_j = \delta_{ij}  \pi_i \ , \  \pi_i^\dagger =  \pi_i \ ,
\end{equation*} 
where $ \pi_i$ projects on the subspace of states that have the value $\lambda_i$ for the observable $ o$. The expectation value of $o$ on the state $\rho$ is given by
\begin{equation}
\aver{o} = \trace( o \rho) \ ,
\end{equation}
according to the general rule provided by Gleason's theorem.

\bbth\label{gletheo}{\bf (Gleason theorem).} Consider a set of projectors $\{  \pi_i \}$ on Hilbert space of at least dimension $3$ and a mapping to the real numbers $p( \pi_i)$. Then with the following hypothesis
\begin{subequations}
\begin{align}
0\leq p( \pi_i) &\leq 1 \quad \forall  \pi_i:  \pi_i^2=1 \ , \  \pi_i=  \pi_i^\dagger \ , \\
p( 0)&=0 \ \ , \ p(\id)=1 \ , \\
 \pi_i  \pi_j &=  0 \imply p( \pi_i +  \pi_j)=  p( \pi_i) + p( \pi_j )  \ , \label{eq:strongsup}
\end{align}
\end{subequations}
the most general $p( \pi_i)$ can be always written as 
\begin{equation}
p( \pi_i) = \trace( \pi_i \rho) \ ,
\end{equation}
where $\rho$ is a \emph{trace-class} operator.
\eeth

In fact, another important consequence of Gleason's theorem is exactly the fact that, under the assumption of the so-called \emph{strong superposition principle}, Eq.~(\ref{eq:strongsup}), the most general quantum state is a trace class operator. This justifies the definition of the state as a density matrix.

A complete description of a state can be given in terms of a complete set of commuting observables $\{ o_k \}$. In fact a basis of the Hilbert space is given by a set of common eigenvectors of the $\{ o_k\}$ and every density matrix can be expanded in terms of the $\{o_k\}$
\begin{equation}
\rho = \sum_k c_k  o_k \ ,
\end{equation}
where $c_k=\trace( o_k \rho)$ holds whenever the observables form a orthonormal basis $\trace( o_k  o_l)= \delta_{k,l}$. 
In particular in such cases we have 
\begin{equation}
\rho = \sum_k \aver{ o_k}  o_k \ .
\end{equation}
This formalization agrees with the following principle.
\bbpri{\rm \bf (Heisenberg uncertainty principle).}\label{pri:heisuncpri}
There exist observable properties of a system that cannot be known together with certainty on the same state. 
There is a fundamental lower bound on the joint uncertainty of two incompatible observables
that defines a fundamental constant $\hbar$.
\eepri

This assumption generalizes the classical principle that all properties of a system are compatible with each other and can be jointly known with certainty. In particular, the Heisenberg uncertainty principle holds in a two-fold way: 
\begin{enumerate}
\item it is \emph{in principle} impossible to jointly measure the values of two incompatible observables $a$ and $b$ avoiding their mutual perturbation. A relation like $\Delta a \cdot \Delta b \sim \hbar$ will always hold, where $\Delta a$ is the uncertainty in the measurement of $a$ and $\Delta b$ is an uncertainty induced by a back-action on the value of the observable $b$. 
\item then it follows as \emph{a theorem} an uncertainty relation $\Delta a \cdot \Delta b \sim \hbar$ on the preparation of a state with uncertain value of two incompatible observables $a$ and $b$. Here $\Delta a$ and $\Delta b$ are the uncertainties coming from the statistics obtained in the preparation of many copies of the state 
\end{enumerate}
This last theorem holds in many versions, the most famous of which is the following.
\bbth{\bf (Robertson-Schr\"odinger uncertainty relation).}\label{heisunctheo} 
Given two observables $ a, b$ on a state $\ket{\psi}$, their variances
$\va{ x} \equiv  \aver{\left( x- \aver{ x}\right)^2}$ obey the inequality:
\begin{equation}\label{heisenbergunc}
\va{ a}\va{ b} \geq \bigg|\dfrac{1}{2}\aver{\{ a,  b\}}-\aver{a}\aver{b}\bigg|^2+ \bigg|\dfrac{\aver{[ a,  b]}}{2i}\bigg|^2 \ ,
\end{equation}
where $[\cdot,\cdot]$ and $\{\cdot,\cdot\}$ are respectively the commutator and the anticommutator.
\eeth

Then, another way to state principle~\ref{pri:heisuncpri} is that the commutator of two observables is proportional to a fundamental constant $\hbar$ called \emph{Planck constant} 
\begin{equation}
[ a, b]\propto \hbar \ ,
\end{equation}
saying also that classical physics can be recovered in the limit $\hbar \rightarrow 0$. In classical physics all observable commute, because it is assumed that all properties of the system are compatible with each other. 

As we said, Theorem~\ref{heisunctheo} holds for every possible \emph{prepared} state. This is an uncertainty on the preparation itself. A system cannot be prepared in a state in which the values of non-commuting observables are jointly known with certainty.
On the other hand, uncertainty relations similar to Eq.~(\ref{heisenbergunc}) must hold for the measurements of two incompatible observables on an unknown state\footnote{See e.g., \cite{braginsky} for a detailed discussion on this point and also \cite{Busch2007155} for a review article about the Heisenberg principle.}. In fact precise uncertainty relations that quantify the disturbance effect of measurements have also been derived, see for example \cite{Busch13,Busch14,Busch2007155}.

Note, however, that the question of how to directly experimentally test this principle arises. The uncertainty in preparations to be tested would require to check all possible states and find a lower bound on the statistical uncertainty of two incompatible observables, while the uncertainty in joint measurements would require to span all possible measurement apparatuses. On the other hand by performing experimental tests (as e.g., those discussed in  \cite{Busch2007155}) of the disturbance effects on a single state or of the statistical uncertainty in several preparations of joint measurements of two non-compatible observables (e.g., position and momentum) one can find upper bounds to the fundamental constant $\hbar$. Thus, ideally, if the value $\hbar=0$ is found in some tests then one can conclude that the two observables were instead compatible. This is a reasoning similar to the initial trials of Einstein in disproving with ideal thought-experiments the uncertainty principle in his debate with Bohr (see Fig.~\ref{fig:einstged}). 

Apart from preparations and measurements of observables, the aim of any physical theory is to be able to predict the evolution of a physical system. In classical mechanics this is a task that is in principle possible and can be done with perfect certainty for every system during all the evolution, once a function $S$ called \emph{action}\footnote{Here we refer to the Hamilton-Jacobi formulation of the equations of motion. Of course other formulations are equivalent but for brevity we will not mention all of them.} is known. 
In other words the evolution of a system is classically completely deterministic. This is because in classical mechanics every system that is \emph{isolated}, i.e., does not interact with anything else, can be thought as \emph{closed}. Therefore the information contained in the initial state is not lost during the evolution and can be completely recovered in the final state: the evolution is a \emph{reversible} transformation of the state.

In quantum mechanics there is also a similar principle that governs the free evolution of a closed system. It is completely deterministic as well. All the information contained in the initial state can be recovered in the final state after the evolution. It is a reversible transformation of the state as well. 
A difference with respect to classical mechanics is that we have to take into account the superposition principle and this leads to the evolution to be \emph{linear} in the state.

\subsection{Free evolution is a linear operation}
A quantum state $\rho$ of a \emph{closed} system evolves through a \emph{unitary} transformation
\begin{equation}\label{unievo}
\rho \rightarrow \rho_\tau=  U(\tau) \rho U(\tau)^\dagger \ . 
\end{equation}
The parameter $\tau$ is the \emph{time} and the unitary operator can be expressed as an exponential of an (anti-)hermitean operator
\begin{equation}
U(\tau) = \exp(i  S(\tau)) \ ,
\end{equation}
where $ S(\tau)$ is a function called \emph{action} and is proportional to the fundamental constant~$\frac 1 \hbar$
\begin{equation}
S(\tau) = \frac{ S^\prime(\tau)}{\hbar} \ ,
\end{equation}
where $S^\prime(\tau)$ is an adimensional operator. This might be seen as the quantum analogue of the classical \emph{principle of least action} \cite{feynman42}. In this case, however, the state does not follow the path from $\rho$ to $\rho_\tau$ that minimizes the action. 

\begin{figure*}[h!]
\includegraphics[width=\textwidth]{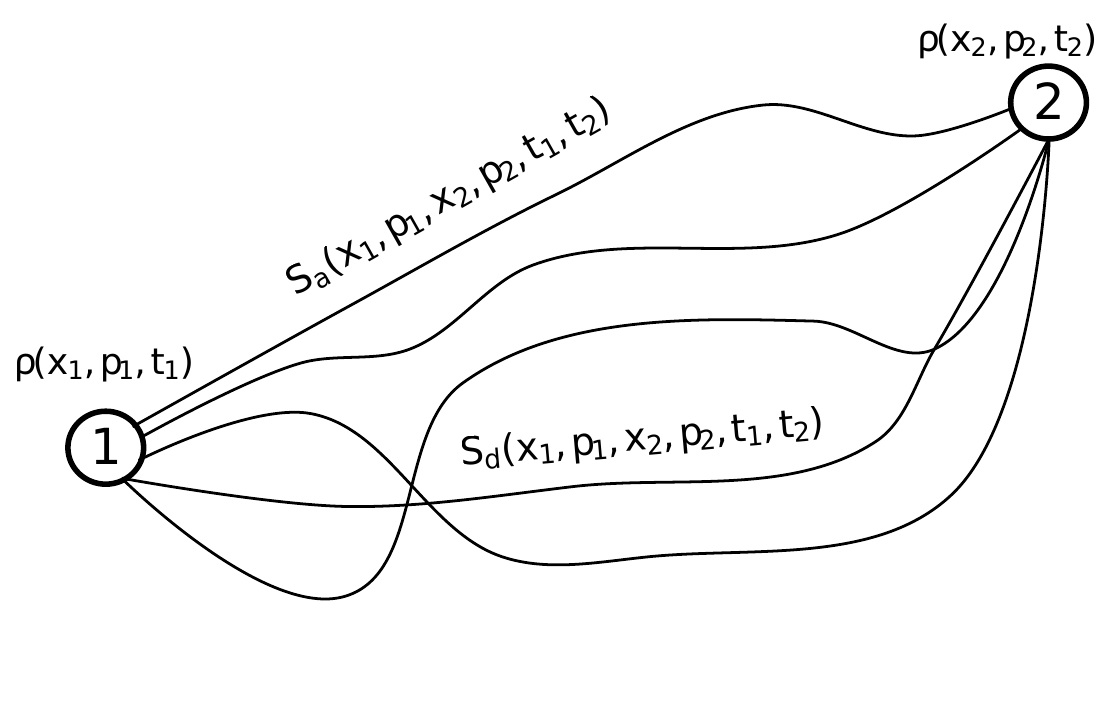}
\caption{Path integral formulation of the quantum evolution: a particle in an initial state $\rho(x_1,p_1,t_1)$ evolves to a final state $\rho(x_2,p_2,t_2)$ taking into account of all possible paths connecting the starting and end points. Every path $n$ is taken into account with equal probability, but multiplied with a phase proportional to the action $S_n(x_1,p_1,t_1,x_2,p_2,t_2)$ associated to it.}\label{fig:feypath}
\end{figure*}

The evolution takes into account all possible paths connecting the initial and final state $\rho$ and $\rho_\tau$ weighted with a phase proportional to the action of the path\footnote{This is the point of view of the {\bf path integral} formulation of quantum mechanics done by {Feynman} in his Ph.D. thesis \cite{feynman42}.}. However, the phases (i.e., actions) far away from its extremal points tend to cancel out, leading the extremal values to give the major contribution to the total \emph{integral over paths} (see Fig.~\ref{fig:feypath} for an illustration).

Eq.~(\ref{unievo}) follows from the fact that in a closed system there is no loss of information.
\bbpri{\rm \bf (Conservation of information).}
The information contained in a physical state of a \emph{closed} system must be conserved during the evolution.
\eepri
This principle immediately implies that the evolution must be \emph{reversible}. This fact, together with linearity and the requirement that the final state of the evolution has to be a physical state (i.e., a density matrix) leads to Eq.~(\ref{unievo}), namely that the evolution must be described by a unitary transformation. 

As we said this principle is valid for closed systems. 
However in general a physical system interacts with some environment that is not accessible to measurements. In this case the system is open and some information is in fact lost in the environment: the evolution is \emph{irreversible}. Therefore in the most general case of an open system the only requirement that we put to formalize the evolution is that the transformation must be \emph{linear} and that the final state must be a density matrix. More formally we define the evolution as a mapping $\mathcal{E}: \mathcal B(\hil) \rightarrow \mathcal B(\hil)$ that satisfies the following requirements
\begin{enumerate}
\item \emph{Linearity}: $\mathcal{E}(\sum_k c_k \rho_k) = \sum_k c_k \mathcal{E}(\rho_k)$ for all $\rho_k \in \mathcal B(\hil)$ and $c_k \in \mathbb C$. \\
\item \emph{Preservation of the trace}: $\trace(\mathcal{E}(\rho)) = \trace(\rho)$ for all $\rho \in \mathcal B(\hil)$. \\
\item \emph{Complete positivity}: $\mathcal{E}\otimes \id_{\rm d} \geq 0$ for all $d \in \mathbb N$.
\end{enumerate}
The requirements 2 and 3 are needed for the final state to be a density matrix. In particular \emph{complete positivity} means that the final state will be positive even if the initial state is a reduced state of a larger system for any possible size of the total system.

We can thus formalize general \emph{quantum operations}, also exploiting the following result.

\bbth\label{kratheo}{\bf (Kraus theorem).} Every linear map $\mathcal{E}: \mathcal B(\mathbb C^n) \rightarrow \mathcal B(\mathbb C^m)$ is completely positive if and only if it admits a representation of the form
\begin{equation}
\begin{aligned}\label{krausrep}
\mathcal{E}(\rho)=\sum_{k=1}^r E_k \rho E_k^\dagger \ ,
\end{aligned}
\end{equation}
where the operators $\{ E_k\}_{k=1}^r$ are called Kraus operators. The mapping $\mathcal{E}$ is also trace preserving if and only if $\sum_{k=1}^r E_k^\dagger E_k=\id$.
\eeth

Thus we can identify a quantum operation with a set of Kraus operators $\{ E_k \}_{k=1}^r$. Note that a unitary evolution is included in this framework as a trace preserving quantum operation $\{ U_k \}_{k=1}^r$ consisting of a single element, i.e., $r=1$.
Actually a general trace preserving quantum operation can even be viewed as a unitary evolution of a larger system, part of which is not accessible. This is proved in the following theorem.

\bbth\label{stitheo}{\bf ((simplified) Stinespring theorem).} Let $\mathcal{E}: \mathcal B(\mathbb C^n) \rightarrow \mathcal B(\mathbb C^m)$ be a trace preserving completely positive linear map. Then there exist a unitary $U \in \mathbb C^{n} \otimes \mathbb C^m \otimes \mathbb C^m$ and a vector $\ket{\phi}\in \mathbb C^{m}\otimes \mathbb C^m$ such that
\begin{equation}
\begin{aligned}\label{stinrep}
\mathcal{E}(\rho)=\trace_{E} \left(U \rho\otimes \ketbra{\phi} U^\dagger \right) \ ,
\end{aligned}
\end{equation}
where $\rho \in \mathcal B(\mathbb C^n)$ and the partial trace $\trace_E(\cdot)$ acts on the space of the first two tensor factors of $\mathbb C^{n} \otimes \mathbb C^m \otimes \mathbb C^{m}$.
\eeth
In particular, this last representation supports the intuition that a general quantum operation formalizes the evolution of an open system. The subsystem $E$ can be then viewed as an environment that is not accessible. Eq.~(\ref{stinrep}) is in fact also-called \emph{environmental representation} of a quantum operation.

In quantum mechanics there is another crucial difference with respect to classical mechanics: that there is also another kind of evolution that is not deterministic, but probabilistic. The concept of \emph{measurement} is introduced, representing the only way one can acquire (some of) the information contained in a state. After a measurement the state evolves in a probabilistic way and some information is lost. In this sense every system, whenever a measurement is performed, must be considered \emph{open}. 

Thus the formalization of a measurement is by itself a novelty introduced in quantum mechanics and is given as a fourth principle. This actually forces us to include open systems in the theory, because the measurement is thought to be performed by an observer that is completely external to the system.
We will give directly the most general definition, that is a generalization of the ideal measurement modelled
by von Neumann in his axioms of quantum mechanics. Contrarily to what is classically (implicitly) assumed, the results of a measurement are \emph{fundamentally random} and cannot be always predicted with certainty. They don't correspond to properties that are preexistent in the system independently of the measurement process. 
Moreover the effect of a measurement is to perturb, even strongly, the state of the system, according to the Heisenberg uncertainty principle.

\subsection{Outcomes of measurements are random}
A measurement is described by a set of operators $\{ M^2_k \}_{k=1}^r$, each corresponding to one of the possible outcomes $k\in \{1,\dots,r\}$. These operators have the properties that $M^2_k=M_k^\dagger M_k>0$ (\emph{they are positive}) and $\sum_k M_k^\dagger M_k= \id$ (\emph{they are a resolution of the identity}). They might be also non orthogonal $\trace(M^2_k M^2_l)\neq \delta_{k,l}$ and might form an overcomplete basis of the space of operators. In particular $r$ is usually larger than the dimension $d$ of the Hilbert space of the system. The set $\{ M^2_k \}$ 
is called \emph{Positive Operators Valued Measure}. 
The outcomes of the measurement $\lambda_k$ are random and to each of them is associated a probability $p_k$, that is the probability for that outcome to occur. It is given by the same formula that follows from Gleason's theorem
\begin{equation}
p_k = \trace(M^2_k \rho) \ .
\end{equation}
In particular the following principle holds.

\begin{figure*}[h!]
\includegraphics[width=\textwidth]{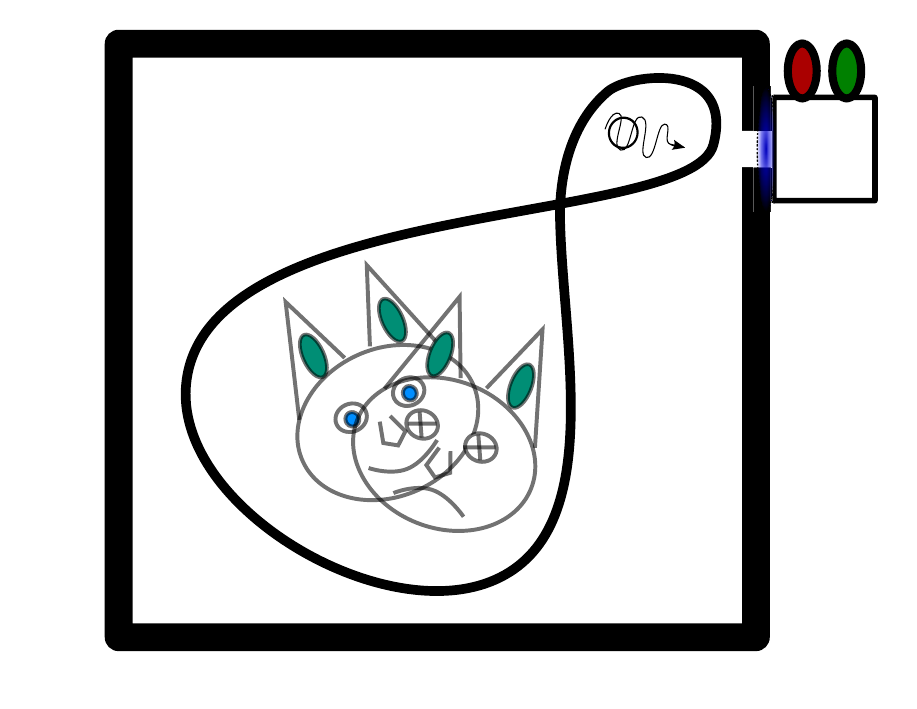}
\caption{Schr\"odinger's cat thought experiment: a box contains a macroscopic object (a cat) in an entangled state with a microscopic atom. The atom might decay with a certain probability and release a photon that can escape the box and be detected by an external measurement apparatus. Correspondingly the cat would die due to a process directly connected to the detection of the photon. Before the measurement the cat+atom state must be considered $\frac 1 {\sqrt 2} (|{\rm dead, detection} \rangle+|{\rm alive, no \ detection}\rangle)$ until the measurement is performed. Afterwards, the measurement of the photon causes a collapse of the whole cat+atom wave function and the state becomes \emph{either} $|{\rm dead, detection}\rangle$ or $|{\rm alive, no \ detection}\rangle$ with equal probability.}\label{fig:schcat}
\end{figure*}

\bbpri{\rm \bf (Collapse of the wave function).}
The outcomes of a measurement are fundamentally random. The state after a completely repeatable measurement collapses to the eigenstate corresponding to the measured outcome. 
\eepri

A particular case of a POVM is a projective measurement, that is the ideal definition given by von Neumann and to which the principle refers. A projective measurement is a POVM such that its elements are projectors pairwise orthogonal to each other $M_k M_l=\delta_{k,l} M_k$. In particular the elements of a projective measurement are exactly as many as the dimension of the Hilbert space $N_M=d$. There is also a mathematical theorem that allows to write a general POVM as a projective measurement on a space of a certain dimension bigger than $d$.
\bbth{\bf (Neumark dilation theorem).} 
Every POVM $\{ M^2_k \}$ acting on a Hilbert space of dimension $d$ can be mapped into a projective measurement $\{ E^2_k \}$, with $E_k E_l=\delta_{k,l} E^2_k$ that acts on a Hilbert space of dimension $n>d$. In other words it always exists a dimension $n>d$ such that a set $\{ E^2_k \}$ can be found.
\eeth
The state after the measurement is changed. It is mapped to $\rho \rightarrow \rho_k$ where
\begin{equation}
\rho_k = \frac{M_k \rho M_k^\dagger}{\trace(M_k \rho M_k^\dagger)} \ ,
\end{equation}
with probability $p_k = \trace(M^2_k \rho)$. Basically the state is transformed with the operator $M_k$ and then renormalized to have a unit trace.

\bbrmk
Note that we can always express the operators $M_k$ in the polar decomposition as $M_k=U_k \sqrt{M^2_k}$, where $U_k$ is a unitary (\emph{the phase}) and $\sqrt{M^2_k}$ is a positive operator (\emph{the modulus}). However the elements of the POVM are defined from just their modulus $\sqrt{M^2_k}$. There remains the ambiguity of the phase $U_k$. This ambiguity however does not affect the transformation of the state, where the operator $M_k$ itself appears. Thus, in order to describe correctly the effect of an actual measurement, one has to provide both its modulus and phase.
\eermk

The transformation of a state due to a POVM can be included in the framework of quantum operations. However we have to generalize it to include probabilistic operations as well. Thus we define a general \emph{probabilistic quantum operation} as a linear completely positive and \emph{trace non-increasing} map $\mathcal M: \mathcal B(\hil) \rightarrow \mathcal B(\hil)$. Trace non-increasing means that $\trace(\mathcal{M}(\rho)) \leq \trace(\rho)$ for all $\rho \in \mathcal B(\hil)$. As we said this is a probabilistic operation, that has a success probability given by
\begin{equation}
p=\trace(\mathcal{M}(\rho)) \ .
\end{equation}
The state after the operation has to be then normalized to unit trace, namely 
\begin{equation}
\mathcal{M}(\rho) \rightarrow \rho^\prime=\frac{\mathcal{M}(\rho)}{\trace(\mathcal{M}(\rho))} \ ,
\end{equation}
that is an operation that is \emph{nonlinear}. 
Exploiting the Kraus theorem~\ref{kratheo} we can express this operation as
\begin{equation}
\frac{\mathcal{M}(\rho)}{\trace(\mathcal{M}(\rho))} = \frac{\sum_{k=1}^r E_k \rho E_k^\dagger}{\trace(\sum_{k=1}^r E_k \rho E_k^\dagger)} \ ,
\end{equation}
where the $\{E_k\}$ are Kraus operators that are also trace non-increasing, i.e., $\sum_{k=1}^r E_k^\dagger E_k\leq \id$. We can also express a general probabilistic quantum operation in the representation provided by the Stinespring theorem~\ref{stitheo}

This last principle is the most debated and the source of a fundamental philosophical controversy of the theory that has been termed \emph{measurement problem} \cite{stanfmeas,ghiraldistanf}. 
The question arises: do the collapse of the wave function \emph{actually happen} as a physical process or is rather just an update of the information of the observer? And eventually, being this a non-linear process, how this reconciles with the linear deterministic evolution? Is the observer such an important actor in the physical processes happening in nature? These are just few of the questions raised by the theory in its modern axiomatic formulation.

\begin{figure*}[h!]
\includegraphics[width=0.45\textwidth]{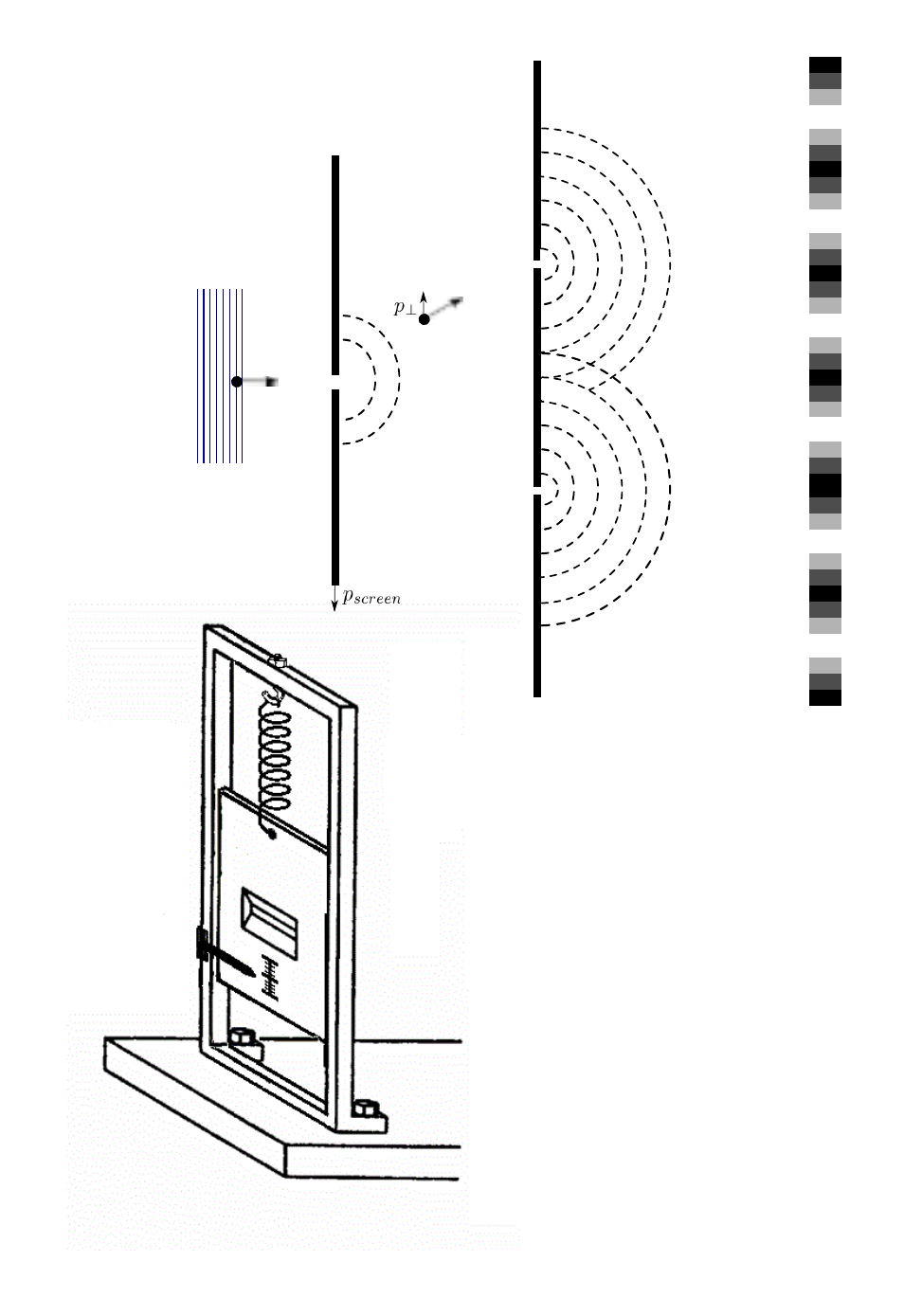}
\includegraphics[width=0.45\textwidth]{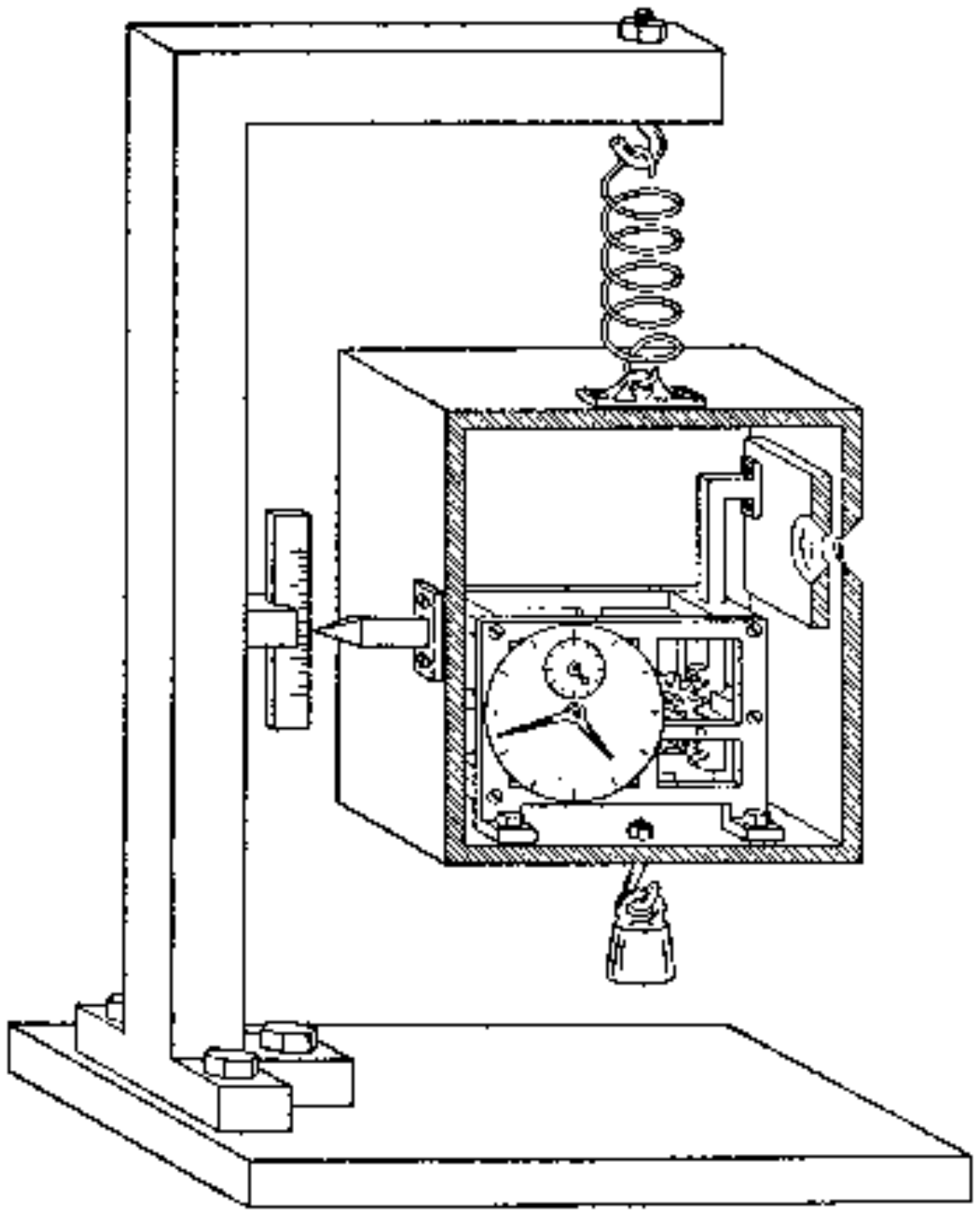}
\caption{Einstein's thought experiment of 1930 as designed by Bohr \cite{bohreinstdeb}. Einstein's box was supposed to prove the violation of the indeterminacy relation between position and momentum (left) and time and energy (right). 
(left) An additional single-slit wall is placed on the trajectory of a particle hitting a double-slit wall in order to measure its velocity. Bohr argued that every such device is still constrained to obey a fundamental position/momentum uncertainty relation.
(right) A device that was supposed to weight the mass subtracted by a photon escaping from the box during a time interval $\Delta t$ controlled by a clock. Bohr's response is that such a measurement must be made as a position measurement of the box and must have a fluctuating outcome; also, the clock position itself, being uncertain, induces a fundamental uncertainty in the time interval, due to the equivalence principle. In this way, a fundamental uncertainty relation $\Delta E \Delta t$ is restored.}\label{fig:einstged}
\end{figure*}

In fact, quantum mechanics has been since its birth a very controversial theory \cite{bohreinstdeb} because of the so radical and fundamental differences with classical mechanics. In particular many of the fathers of the theory themselves interpreted quantum mechanics as merely a set of rules that allows to predict to the maximal possible extent the results of experiments, still believing that there must be an underlying more fundamental theory with classical principles, maybe impossible to be discovered because of practical limitations. 

Taking this last point of view, the idea was that the quantum mechanical wave function was just a convenient \emph{epistemic}\footnote{This word means that the wave function was just interpreted as an object useful to acquire knowledge about a physical system, but not as a \emph{real} entity.} probabilistic description of nature.
Rather, the real value of an observable $o$ would have depended on some \emph{hidden variable} $\lambda$, distributed with a certain probability function $\Pr(\lambda)$. Then, the quantum mechanical randomness of the outcome was just due to the fundamental ignorance of the actual value of $\lambda$ and the average $\aver{o}$ had to be thought as coming from an ensemble average
\begin{equation}
\aver{o}=\int o(\lambda) \Pr(\lambda) {\rm d} \lambda \ ,
\end{equation}
analogously as in statistical mechanics.  

On the other hand, if quantum mechanics is a complete theory, then there are properties of natural systems to which it is not possible to associate univocally a value. The outcome of an experiment trying to look at such values cannot be predicted. Thus the question arises: \emph{do such properties have univocally a certain value or not}? In other words: do the \emph{ontic} state of the system have a definite value for such properties or not?

Einstein and others were thinking that even if not predictable, to every properties of a physical system must be in principle possible to associate univocally a value that is independent of any measurement. In fact, in the beginning the Einstein-Bohr debate was about the Heisenberg uncertainty principle: Einstein was trying to provide \emph{gedankenexperiments} (thought experiments) that could disprove it (see Fig.~\ref{fig:einstged}), while Bohr was arguing about the fundamental impossibility of making such a test. Einstein failed with this attempt, but still thought that it was just for practical limitations that the uncertainty principle could not be experimentally disproved.  

Nowadays, after the debate kept evolving, it is clear that the Heisenberg principle alone, or more specifically the possibility of assigning values 
to all observables independently on measurements, is something that cannot be tested experimentally on its own. Some other physical principle has to be tested jointly with it. Nevertheless, tests of classical principles can be actually made independently on a specific theory, but have to be done 
by considering correlation measurements. The most striking of such tests, the one that troubles more Einstein's viewpoint, involves correlation measurements between space-like distant particles in a composite system.

In order to explain this let us first introduce some useful tools for the analysis of composite systems in quantum mechanics.

\subsection{Operations on composite systems}

Here let us focus on systems $\mathcal S$ composed of $n$ parties $\mathcal P_1, \dots \mathcal P_n$\footnote{Here this is a purely formal subdivision, but note that in practice the subdivision of a ``system'' into ``parts'' might have different levels of arbitrariness, depending on the system itself.}.
As we said before, due to the superposition principle the Hilbert space of the system is the tensor product of the parties' Hilbert spaces, i.e.,
\begin{equation}
\mathcal H_{\mathcal S} = \bigotimes_{i=1}^n \mathcal H_{\mathcal P_i} \ .
\end{equation}
Then, to emphasize the composite structure of the system, we distinguish states and observables that act on a single party rather than on the whole system and we call them \emph{local operators}.

\bbdf{\bf (Local operators).}
An operator $O_{\rm loc}$ acting on the Hilbert space $\mathcal H_{\mathcal S}$ is called \emph{local} if it acts nontrivially only on a component $\mathcal H_{\mathcal P_i}$. In particular it can be written as
\begin{equation}
O_{\rm loc}=\id \otimes \dots \otimes o^{(i)} \otimes \dots \otimes \id \ ,
\end{equation}
for some superscript $(i)$ and some operator $o^{(i)}$ acting on $\mathcal H_{\mathcal P_i}$.
\eedf

Thus states and observables are called \emph{local} when they are relative to a single part $\mathcal P_i$. Note that this has nothing to do with locality in space; a system can be localized in space but still divided into parts, all of which share the same spatial location.

Afterwards we make another distinction within the set of operators on multipartite systems. We define the so-called \emph{separable operators} as follows.

\bbdf{\bf (Separable operators).}
An operator $O_{\rm sep}$ acting on the Hilbert space $\mathcal H_{\mathcal S}$ is called \emph{separable} if it can be written as a convex mixture of product operators
\begin{equation}\label{sepadef}
O_{\rm sep}=\sum_k p_k O_k  \ , \qquad  p_k\geq 0 \ \ , \ \sum_k p_k=1 \ ,
\end{equation}
where $\{ p_k \}$ is a probability distribution and $\{ O_k \}$ is a certain set of \emph{product operators}
\begin{equation}
O_k = o^{(1)}_k \otimes \dots \otimes o^{(n)}_k \ ,
\end{equation}
here again the superscript $(i)$ refers to the party $\mathcal P_i$. 
\eedf
Thus an operator $O_{\rm sep}$ (a state or an observable) is called separable if there exists at least one set of product operators $\{ O_k \}$ such that Eq.~(\ref{sepadef}) holds. The reasons why we are defining this special class of operators will be clearer later, but applied to density matrices this is basically the definition of a \emph{non-entangled} state. 

Given a state of a composite system $\rho\in \mathcal B(\mathcal H_{\mathcal S})$ one also wants to have access to the information available locally, i.e., on a single party $\mathcal P_i$. The way to do this is to define a quantum state for a single party $\rho_{\mathcal P_i}\in \mathcal B(\mathcal H_{\mathcal P_i})$, because as we said in the discussion of the principles this is the most general way to encode the available information of a system. 
The natural way to define a state of a single party $\rho_{\mathcal P_i}$ starting from the global state $\rho$ is to trace out all the information relative to the rest of the system. This is formally done with an operation called \emph{partial trace}, that we have already seen before. The result of this operation is what we call \emph{reduced density matrix}.

\bbdf{\bf (Reduced density matrix).} Given a state of a composite system $\rho\in \mathcal B(\mathcal H_{\mathcal S})$ we define the \emph{reduced density matrix} $\rho_{\mathcal P}\in \mathcal B(\mathcal H_{\mathcal P})$ relative to a subsystem $\mathcal P$ as the state obtained through the partial trace of $\rho$ over the rest of the system $\mathcal S / \mathcal P$
\begin{equation}
\rho_{\mathcal P}=\trace_{\mathcal S / \mathcal P} \rho \ ,
\end{equation}
where $\{ \mathcal S / \mathcal P \}$ is the set of all parties not contained in $\mathcal P$.
The reduced density matrix contains all the information of the state $\rho$ available with local measurements on $\mathcal P$.
\eedf

To conclude this section we define what the operations are that can be made locally on a subsystem
and then what the operations are that preserve the separability of a quantum state. 

\bbdf{\bf (Local operations).} A \emph{local operation} is in general a probabilistic quantum operation $\mathcal M_{\rm loc}(\rho)$ that 
can be decomposed as in the following Kraus representation
\begin{equation}\label{eq:localoperkraus}
\rho \mapsto \rho^\prime=\frac{\mathcal M_{\rm loc}(\rho)}{\trace(\mathcal M_{\rm loc}(\rho))} = \frac{\sum_{k=1}^r E_k \rho E_k^\dagger}{\trace(\sum_{k=1}^r E_k \rho E_k^\dagger)} \ ,
\end{equation}
with the set $\{ E_k \}$ formed by \emph{local operators}
\begin{equation}
E_k = \id^{(1)} \otimes \dots \otimes (E_k)^{(i)} \otimes \dots \otimes \id^{(n)} \ .
\end{equation}
\eedf
Note that almost all the possible evolutions map a separable state in general into a non-separable state. The ones that cannot do this are called \emph{separable operations}.

\bbdf{\bf (Separable operations).} A \emph{separable operation} is in general a probabilistic quantum operation $\mathcal M_{\rm sep}(\rho)$ that 
can be decomposed in as in the Kraus representation
\begin{equation}
\rho \mapsto \rho^\prime=\frac{\mathcal M_{\rm sep}(\rho)}{\trace(\mathcal M_{\rm sep}(\rho))} = \frac{\sum_{k=1}^r E_k \rho E_k^\dagger}{\trace(\sum_{k=1}^r E_k \rho E_k^\dagger)} \ ,
\end{equation}
with the set $\{ E_k \}$ formed by \emph{separable operators}
\begin{equation}
E_k = \sum_l p_l (E_k)^{(1)}_l \otimes \dots \otimes (E_k)^{(n)}_l \quad p_l \geq 0 \ , \ \sum_l p_l=1 \ .
\end{equation}
\eedf
To finish this section note that a local operation acts on the whole state $\rho$ (and not just on $\rho_{\mathcal P}$) and modifies it probabilistically, even if it is performed locally. 
This is just one of the features that is so counterintuitive that raised a very wide debate between the major authorities in the foundations of quantum mechanics since its birth until nowadays. 
This debate led until very recently to astonishing discoveries about the general assumptions that can be made on a theory that would try to compete with quantum mechanics in explaining natural phenomena. These results are what we are going to discuss in the next section.


\section{Tests of principles from correlations}\label{testcond}

In 1935, the debate about an ontic interpretation of quantum mechanics was raised again by the famous paper of \emph{Einsein, Podolsky and Rosen} (EPR) \cite{EPR1935}.
This time EPR were trying to avoid the practical limitations encountered in trying to disprove the uncertainty principle by 
looking at multipartite systems. They noted that in quantum mechanics, to quote Schr\"odinger, ``spooky'' phenomena appear when considering correlations between outcomes of some incompatible observables; phenomena that cannot have an explanation in terms of classical theories, not even with some fundamentally hidden variables. The EPR reasoning led to the theorem that either:
\begin{itemize}
\item Quantum mechanics is not a complete theory, in the sense that there must be a more fundamental theory. 
\item Keeping the local-causality principle of relativity and a strict free will assumption, then some properties of a system \emph{do not exist} until they are measured.
\end{itemize}
Note that this theorem meant a deeper and more profound difference between classical and quantum mechanics. The difference is not anymore shaded by practical limitations: if quantum mechanics is complete, then it means that the answer to the question raised before is \emph{definitely not}, there are properties of natural systems to which \emph{it is not possible} to associate any value independently of a measurement. At least this is the answer if, as Einstein did, one wants to keep the local-causality principle of relativity and a strict free will assumption. 

The answer is so definite just because, and this is the most remarkable fact,
the EPR theorem \emph{can be experimentally tested}, even if not in its original form. 
And the response of experiments up to know keep being in favour of the unrealism of quantum mechanics. This we will see later. First let us look at the original EPR argument.
 
\begin{figure*}[h!]
\includegraphics[width=\textwidth]{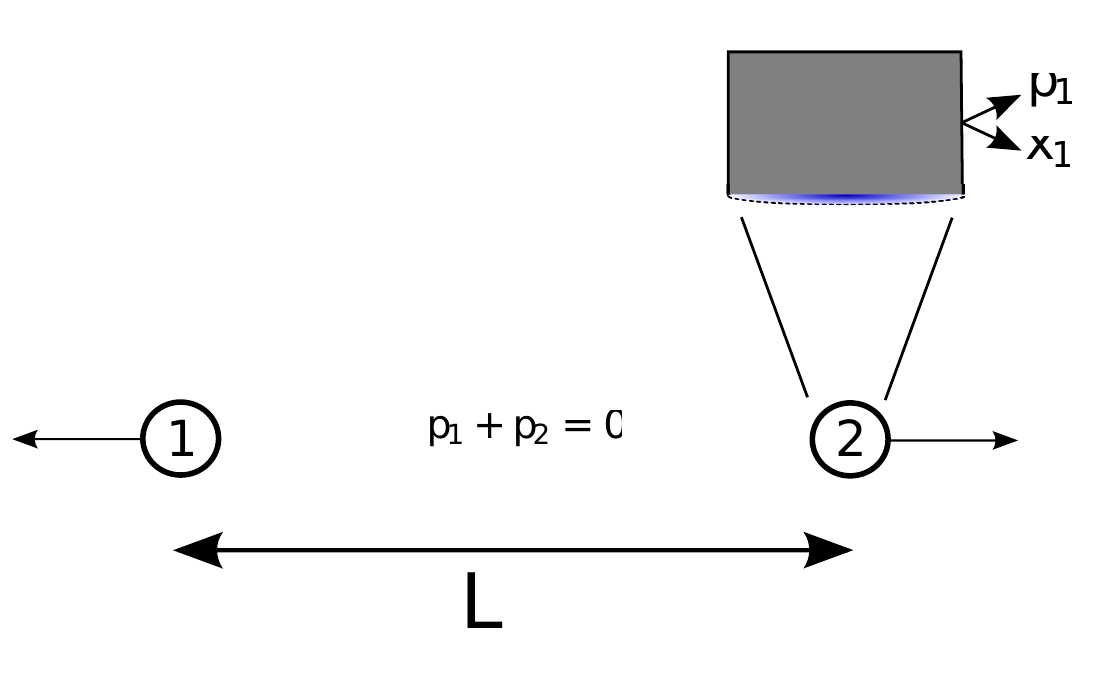}
\caption{Schematic representation of EPR thought experiment.}\label{fig:eprexp}
\end{figure*}

\paragraph{The EPR argument}
Suppose that a system of two identical particles is prepared in a state such that their relative distance is large and constant $|\vec r_1-\vec r_2|=L$ (they are space-like separated) and the total momentum is zero $\vec p_1+\vec p_2=0$ (see Fig.~\ref{fig:eprexp}). This preparation is in principle possible because the two observables say $x_1-x_2$ and $p_{x,1}+\vec p_{x,2}$ are compatible, i.e., both of them can be set to certain values with certainty on the same state. 
Correspondingly according to quantum mechanics they are in fact represented by commuting operators.

Then one can measure the value of either of the two incompatible single particle observables, say $x_1$ or $p_{x,1}$ and correspondingly deduce the value of either $x_2=L-x_1$ or $p_{x,2}=-p_{x,1}$ \emph{without interacting with particle 2}. Because of this they correspond, according to the EPR argument, to \emph{elements of reality} of the state of particle 2 that are independent of measurements and should be predictable by the theory.
On the other hand quantum mechanics cannot predict the value of \emph{both $x_2$ and $p_{x,2}$ on the same state}, because they are incompatible observables and this would be in contrast to Principle 2. Thus, conclude EPR, there are elements of reality of a state that cannot be predicted by the theory and therefore the theory is \emph{incomplete}. \qed

However, what EPR actually proved is that in principle the values of \emph{each} of the two incompatible observables $x_2$ and $p_2$ can be deduced \emph{with certainty} on the same system (i.e., particle 2) without interacting with it. Then they implicitly did the so-called \emph{realism assumption}, that is as we explained previously that the values associated to every property of a system are \emph{independent of measurements}. Thus the possibility of knowing \emph{each} of the two incompatible observables separately was sufficient to prove that \emph{both of them} should be in principle possible to know on the same state. This kind of \emph{counterfactual} reasoning is no longer permitted in quantum theory and in particular even the assumption that local measurements can be made in a way that they do not disturb the global system does not hold, being in contrast to Principle 4. 
To summarize, the EPR argument was based on the assumptions of
\bbasm{\bf (Locality or Local-Causality)}
There is no exchange of information between two systems that cannot interact. Moreover two systems can only interact locally.
\eeasm
\bbasm{\bf (Realism)}
Results of measurements are determined by properties that the system carry on prior to and independent of the measurements themselves.
\eeasm
\bbasm{\bf (Free-Will)}\footnote{We made explicit also this assumption because actually in an extreme deterministic theory it can be relaxed as well. See e.g., \cite{thooft07}.}
The choice of a measurement setting do not influence and is not influenced by the outcome of other measurements.
\eeasm
Thus the realism assumption implies that the measurement is a non-invasive action on the system, and then due to locality the results of a measurement on a system are independent of any event that is space-like separated from it, including measurements. This statement is false within quantum theory, because a measurement made on a system composed of parts causes a change of the whole system's state, even if the parts are space-like separated (there is a ``spooky action at a distance''). This can be formally understood with a simple example, that is a modern reformulation due to Bohm \cite{bohm51} of the argument of EPR itself.
\bbexmp{\rm (Maximally entangled state or EPR pair)}\label{exmp:eprpair}
Consider a system composed of two parties $\mathcal A$ and $\mathcal B$, space-like separated and in a global state
\begin{equation}\label{eprpair}
\rho= \frac{\id} 4 - \frac 1 4 \sum_{k=x,y,z} \sigma_k^{\mathcal A} \otimes \sigma_k^{\mathcal B} \ ,
\end{equation}
where $\{\sigma_x,\sigma_y,\sigma_z\}$ is a basis of the $su(2)$ algebra. This is a valid quantum state and is usually called, among other names, \emph{EPR pair}, the superscript referring to the subsystem's Hilbert space.

Now suppose that party A performs a projective measurement of the observable say $\sigma_x^{\mathcal A}$. It is a perfectly allowed measurement performed locally and with two possible outcomes $(\lambda^{\mathcal A}_{x, +}=1, \lambda^{\mathcal A}_{x, -}=-1)$ (the eigenvalues of 
$\sigma_x^{\mathcal A}$). Thus, according to the classical principle of local-realism the measurement does not disturb party B. However, assuming that the outcome is say $\lambda^{\mathcal A}_{x, +}$, then after the measurement the state is updated as
\begin{equation}\label{epr2}
\rho \mapsto \mathcal{M}_{x,+}(\rho)= \frac{\Pi_{x,+}^{\mathcal A}\otimes \id^{\mathcal B} \rho \Pi_{x,+}^{\mathcal A}\otimes \id^{\mathcal B}}{\trace(\Pi_{x,+}^{\mathcal A}\otimes \id^{\mathcal B} \rho \Pi_{x,+}^{\mathcal A}\otimes \id^{\mathcal B})} = 
\Pi_{x,+}^{\mathcal A}\otimes \Pi_{x,-}^{\mathcal B} \ ,
\end{equation}
where we performed the \emph{local} operation $\mathcal{M}_{x,+}(\rho)$ as in Eq.~(\ref{eq:localoperkraus}) with a single $E=\Pi_{x,+}^{\mathcal A}\otimes \id^{\mathcal B}$ and $\Pi_{x,\pm}=|\lambda_{x, \pm}\rangle\langle \lambda_{x, \pm}|= \frac 1 2 \left( \id\pm \sigma_x \right)$ are the projectors on the eigenstate of $\sigma_x$ with eigenvalues $\lambda_{x, \pm}$.

Now, note that the state $\rho_{x,+}^{\mathcal B}:=\trace_{\mathcal A}\left( \mathcal{M}_{x,+}(\rho) \right)$ in Eq.~(\ref{epr2}) is the projector onto the 
eigenstate of $\sigma_x^{\mathcal B}$ with eigenvalue $\lambda_{x, +}^{\mathcal B}$. Thus a measurement of $\sigma_x^{\mathcal B}$ on party B would give the result $\lambda^{\mathcal B}_{x, -}=-1$ \emph{with certainty}.

The same reasoning as before applies when measuring say $\sigma_y^{\mathcal A}$ on party A. Given an outcome, say $\lambda^{\mathcal A}_{y, +}=1$, we have $\rho \mapsto \mathcal{M}_{y,+}(\rho)=\Pi_{y,+}^{\mathcal A}\otimes \Pi_{y,-}^{\mathcal B}$ and
the outcome of $\sigma_y^{\mathcal B}$ on party B can be predicted \emph{with certainty} to be $\lambda^{\mathcal B}_{y, -}=-1$. 

Note, finally, that the observables $\sigma_x^{\mathcal B}$ and $\sigma_y^{\mathcal B}$ \emph{do not commute}, and therefore their outcomes could not be both predicted with certainty on the same state. This would be a contradiction within quantum theory that means that the theory is incomplete, according to the EPR argument.

However this is the \emph{counterfactual} reasoning that cannot be done in quantum theory. In fact the reduced initial state of party B is 
\begin{equation}\label{eq:partyBstate}
\rho^{\mathcal B}:=\trace_{\mathcal A}(\rho) = \frac \id 2 \ ,
\end{equation}
i.e., the completely mixed state. This means that a priori the outcome of every observable on party B is completely random and cannot be predicted at all. After a measurement, even if performed only locally on party A, the whole state $\rho$ changes and the reduced state of party B also changes accordingly.
\eeexmp

Thus at the end EPR gave an argument in contrast to quantum theory, but in agreement with everyday intuition based on macroscopic systems, to argue that it is an incomplete theory and this raised a debate \cite{schrod35,EPR1935,bohrEPR}, mainly still between Einstein's and Bohr's viewpoints, on the range of validity of quantum theory (i.e., microscopic vs macroscopic systems) and its interpretation. 

\bbrmk
At this point note also that, even within the realm of microscopic systems, a measurement of a composite system in an EPR pair, 
as in the example~\ref{exmp:eprpair}, raises an even bigger issue related to the measurement problem, i.e., to the problem with interpreting the collapse of the wave function as a change of the ontic state of the system. We have seen that a measurement
performed on particle A causes \emph{instantaneously} a collapse of the state of particle B. The same would be true with the parties exchanged, i.e., measuring on B would cause an immediate collapse of the state of A. Now imagine that A and B perform simultaneous measurements of two incompatible observables on their respective state: what will the output state be? According to quantum mechanics the answer would depend on the observer that is chosen as reference, but this doesn't solve completely the problem, especially if one wants to keep the principle that space-like distant observer cannot influence each other (see \cite{ghiraldistanf}).
\eermk

Within the same line of thoughts and to explain better the contradiction with classical principles, Schr\" odinger painted an extremal situation in which a microscopic state is ``entangled''\footnote{This term is meant with its linguistic meaning, and is only ``accidentally'' referring also to the quantum phenomenon of entanglement.} with a macroscopic one (see Fig.~\ref{fig:schcat} for a schematic illustration), such that apparently this last also results in a state similar to the EPR pair but with one of the parties that is a macroscopic object (the famous Schr\" odinger cat being in an undefined status of alive/dead). 

\bbrmk
Note that even if such a state would be an entangled state between a microscopic and a macroscopic system, the reduced states of the parties are not superpositions: each of the reduced density matrices is the completely mixed state (\ref{eq:partyBstate}) (i.e., the cat would be in a classical mixture of being dead and alive). 
\eermk

Soon after the EPR paper, another authority as von Neumann \cite{vonneumann32} (see also \cite{Genovese2005319,bell87}) gave an argument that claimed to prove that there could not be any classical theory behind quantum mechanics. His totally general assumption was that any reasonable classical theory designed to reproduce quantum predictions should have been written in terms of quantities $v(\cdot)$, i.e., values of observables, that are linearly related with each other
\begin{equation}\label{vnassu}
v(A)+v(B)=v(C) \ ,
\end{equation}
whenever the observables themselves are linearly related with each other, i.e., $A+B=C$.

In quantum mechanics again, one cannot assign a definite value to a couple of non-commuting observables, and this leads to a contradiction with the initial assumption. In fact consider as observables $A=\sigma_x$ and $B=\sigma_y$ for a single spin-$1/2$ particle. They are dichotomic, i.e., their values can only be $\pm1$ and thus the sum $v(\sigma_x) +v(\sigma_y)$ can only be $-2,0,2$. On the other hand, the observable given by $C=\sigma_x + \sigma_y$ can just have outcomes $\pm\sqrt{2}$ leading to a contradiction with Eq.~(\ref{vnassu}). The problem with such a proof lies in the fact that is now widely accepted that the hypothesis (\ref{vnassu}) is too restrictive\cite{Genovese2005319,bell87}. The assumption of von Neumann was later relaxed by Kochen-Specker \cite{kochen67}, that restricted Eq.~(\ref{vnassu}) to be valid only for sets of commuting observables and derived the so-called Kochen-Specker theorem that we will discuss later on. 

Historically instead, the first paper that succeeded in giving a quantitative way to test the validity of the EPR argument against experiments was written by J. S. Bell \cite{BellP1964,bell66,bell87}. His belief was, similarly as Einstein himself, that quantum mechanics was just a statistical version of a more general theory. Thus he looked for a way to test any statistical theory that respected the principles of local-causality and realism. The idea was
to write inequalities that must be satisfied by every possible local-realistic theory, based on correlations between measurements made on two parties, possibly space-like separated between each other. This in a way resembles the original EPR argument, but with the remarkable difference of being testable experimentally.

\subsection{Bell inequalities}
As we said Bell considered in total generality any theory based on classical principles that might attempt to explain and predict the probabilities associated with outcomes of experiments. Such general theories might even contain parameters that are \emph{hidden}, in the sense that they are not accessible to measurements, but still their values are independent of the measurements themselves.

More specifically in this case the realism assumption means that two parties can get correlated only \emph{prior to any measurement}. Thus
one assumes that in a realistic theory that respects also the principe of local-causality the outcomes of measurements made by distant parties might be correlated just through some variables $\lambda$ that transmits information between the parties prior to any measurement. Again, even if $\lambda$ is \emph{hidden}, i.e., not accessible to measurements, it corresponds to properties of the system that are independent of measurements. 

This reasoning leads to the definition of a \emph{local hidden variable theory} (LHV) \cite{BellP1964,bell66,bell87} (see also \cite{budronith} for a review of hidden variable theories) as the most general local-realistic model that tries to interpret some statistical data.
A simple way to define a LHV model is to consider two parties $\mathcal A$ and $\mathcal B$ and some associated measurement settings
$\{a_1,\dots,a_n \}$ and $\{b_1,\dots,b_n \}$. Then the outcomes $(x_i,y_j)$ of measurements $(a_i,b_j)$ made by the two parties can be correlated just through some hidden variable $\lambda$.\footnote{Note that it is sufficient to consider a single variable $\lambda$ that contains all the information not accessible to measurements.} such that
\begin{equation}\label{lhvdef}
\Pr(x_i,y_j)_{a_i,b_j} = \int \Pr(\lambda) \Pr(x_i|\lambda)_{a_i} \Pr(y_j|\lambda)_{b_j} {\rm d} \lambda \ , 
\end{equation}
where we called $\Pr(x_i,y_j)_{a_i,b_j}$ the probability that outcomes $(x_i,y_j)$ occur for the measurements
$(a_i,b_j)$ made by the parties, and analogously $\Pr(x_i|\lambda)_{a_i}, \Pr(y_j|\lambda)_{b_j}$, that are also conditioned on the value of $\lambda$. More generally a local hidden variables theory might be defined as follows. 

\bbdf{\bf (Local Hidden Variables theory).}\label{LHVtheory}
Consider a system composed of $n$ parties $\mathcal S=\mathcal P_1 + \dots + \mathcal P_n$ such that they cannot interact between each other. Then, a Local Hidden Variables (LHV) theory is a model that associates to every possible set of outcomes $(x_{\mathcal P_1},\dots, x_{\mathcal P_n})$ a probability distributions $\Pr(x_{\mathcal P_1},\dots, x_{\mathcal P_n})_{a_{\mathcal P_1},\dots, a_{\mathcal P_n}}$ conditioned on the choice of the measurement settings $(a_{\mathcal P_1},\dots, a_{\mathcal P_n})$ in a way such that
\begin{equation}
\Pr(x_{\mathcal P_1},\dots, x_{\mathcal P_n})_{a_{\mathcal P_1},\dots, a_{\mathcal P_n}} = \int \Pr(\lambda) \Pr(x_{\mathcal P_1}|\lambda)_{a_{\mathcal P_1}} \cdots \Pr(x_{\mathcal P_n}|\lambda)_{a_{\mathcal P_n}} {\rm d} \lambda \ ,
\end{equation}
for some \emph{hidden variable} $\lambda$ that interacts locally with each party and is not accessible to measurements.
\eedf

The idea is then to look at correlations between outcomes of different parties, as for example
\begin{equation}\label{excorr}
\aver{a_1 b_1} = \int x_1 y_1 \Pr(x_1 , y_1)_{a_1, b_1} {\rm d} x_1 {\rm d} y_1 \ ,
\end{equation}
and exploit the condition Eq.~(\ref{lhvdef}) that comes from assuming that the outcomes corresponds to pre-existing values of $a_1$ and $b_1$.

In this way Bell showed that it is possible to derive bounds on linear combinations of correlations that must hold whenever the experimental data can be explained with a LHV model. Nowadays any inequality based on statistical data that must be satisfied by every LHV model is called \emph{Bell inequality}. They are usually written for simplicity in terms of \emph{dichotomic} observables, i.e., observables that can take only $x=\pm 1$ as outcomes\footnote{Note that this is not a restriction, since given any observable one can coarse-grain the outcomes such to relabel them as $\pm 1$.}.

Although the original proposal of Bell was not experimentally realizable, many other Bell inequalities have been later derived.
In particular the most famous and the easiest to test experimentally was due to \emph{Clauser-Horne-Shimony-Holt} (CHSH) \cite{chsh69}.

\bbth{(CHSH inequality).} Consider a bipartite system $\mathcal S = \mathcal A + \mathcal B$ and a
couple of dichotomic observables $a_1,a_2$ for party $\mathcal A$ and $b_1,b_2$ for party $\mathcal B$. Then the following bound 
\begin{equation}\label{chshine}
O_{\rm CHSH}=\aver{a_1 b_1}+\aver{a_1 b_2}+\aver{a_2 b_1}-\aver{a_2 b_2} \leq 2 \ ,
\end{equation}
holds for every possible \emph{local hidden variables theory}, where
\begin{equation}\label{excorr}
\aver{a_i b_j} = \int x_i y_j \Pr(x_i , y_j)_{a_i, b_j} {\rm d} x_i {\rm d} y_j \ ,
\end{equation}
are the bipartite correlations between the outcomes of $(a_i,b_j)$.
\eeth

\bbpr
Let us compute the expression $O_{\rm CHSH}$. Defining $f_{a_i}(\lambda)=\Pr(1|\lambda)_{a_i}-\Pr(-1|\lambda)_{a_i}$ we have
\begin{equation*}
\begin{gathered}
O_{\rm CHSH} = \int \Pr(\lambda) \left[ f_{a_1}(\lambda) f_{b_1}(\lambda)+f_{a_1}(\lambda) f_{b_2}(\lambda)+f_{a_2}(\lambda) f_{b_1}(\lambda) -f_{a_2}(\lambda) f_{b_2}(\lambda) \right]{\rm d} \lambda \ ,
\end{gathered}
\end{equation*}
and thus we can bound it as
\begin{equation}
O_{\rm CHSH} \leq \int \Pr(\lambda) \max_{\lambda} F_{\rm CHSH}(\lambda) {\rm d} \lambda = 2 \int \Pr(\lambda) {\rm d} \lambda = 2 \ ,
\end{equation}
where we called $F_{\rm CHSH}(\lambda):= f_{a_1}(\lambda) f_{b_1}(\lambda)+f_{a_1}(\lambda) f_{b_2}(\lambda)+f_{a_2}(\lambda) f_{b_1}(\lambda) -f_{a_2}(\lambda) f_{b_2}(\lambda)$.
\eepr

Thus we have proven an inequality that must be satisfied by all LHV models. Then the statement of Bell's theorem is that there are quantum states from which one can produce data that violate Eq.~(\ref{chshine}) and thus
cannot be explained with a LHV model. One of such quantum states is precisely the EPR pair of the previous example. Then considering observables at each party that are non-compatible between each other (like $x$ and $p$ in the EPR argument) one can in principle observe a violation of Eq.~(\ref{chshine}).
This means that there is no LHV model that can reproduce the same predictions of quantum mechanics and thus definitely either quantum mechanics is incomplete or one of the assumptions of local-causality and realism does not hold. 

In the following we enunciate the Bell theorem and prove it using CHSH inequality. This is not exactly the original proof made by Bell himself, that derived a slightly different inequality. The formulation that we present is however closer to what has been experimentally employed afterwards.

\bbth{(Bell theorem).} No \emph{Local Hidden Variables} theory can reproduce all the predictions of quantum mechanics.
\eeth

\bbpr
Let us consider a bipartite system $\mathcal S=\mathcal A + \mathcal B$. Then let us consider the EPR pair $\rho_{\rm epr}$ of Eq.~(\ref{eprpair}) and the two couples of observables $(a_1,a_2)=(\sigma_z^{\mathcal A},\sigma_x^{\mathcal A})$ and $(b_1,b_2)=\left(\tfrac{\sigma_z^{\mathcal B}+\sigma_x^{\mathcal B}}{\sqrt 2},\tfrac{\sigma_z^{\mathcal B}-\sigma_x^{\mathcal B}}{\sqrt 2}\right)$. Then we have 
\begin{equation}
O_{\rm CHSH} = \trace(\rho_{\rm epr} o_{\rm CHSH}) = 2\sqrt 2 > 2 \ ,
\end{equation}
where we defined the operator $o_{\rm CHSH} = a_1 \otimes b_1 + a_1 \otimes b_2 + a_2 \otimes b_1 - a_2 \otimes b_2$. Thus there exists a quantum state and two pairs of observables such that the expression $O_{\rm CHSH}$ violates the CHSH inequality and therefore cannot be explained with any LHV model. 
\eepr

\paragraph{Experimental ``Bell'' tests of local-realism}

\begin{figure*}[h!]
\includegraphics[width=\textwidth]{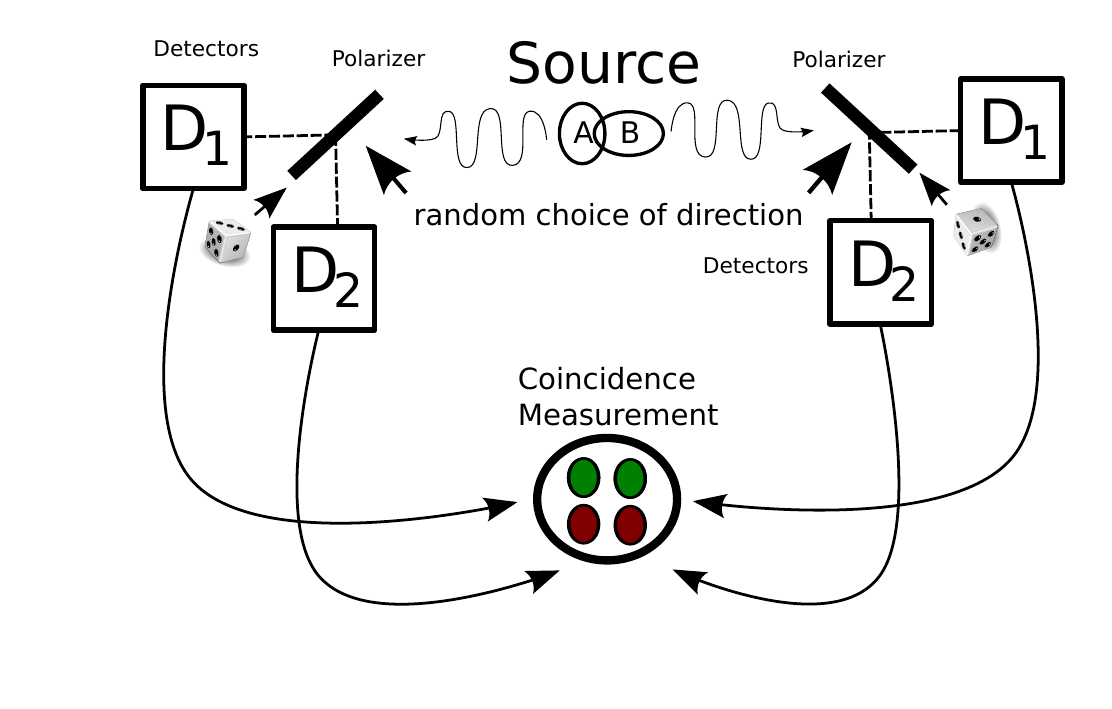}
\caption{Scheme of an experimental Bell test. A source produces pairs of entangled particles (photons), sent in opposite directions. Each photon encounters a polarizer with one randomly chosen orientation, corresponding to one observable between $a_1$ and $a_2$ (respectively $b_1,b_2$).
In the end, correlations $\aver{a_i b_j}$ are measured through a coincidence detector.}
\label{fig:belltest}
\end{figure*}

As we said, Bell's inequality, and in particular the one derived by CHSH \cite{chsh69} opened the way to possible experimental tests of local-realism, in the form of the existence of some LHV theory behind quantum mechanical predictions. However practically such tests have been and still are very difficult to perform in a doubtless way. The first test was made by Freedman and Clauser in 1972 \cite{freedmanclauser}, just few years after CHSH inequality was derived. In this test, as in most of other successive tests \cite{Klyachko03,GiustinaN2013,aspect821,aspect81,aspect822} (see also \cite{Genovese2005319} for a review), the system consisted of
pairs of photons, with an entangled polarization state analogue to Eq.~(\ref{eprpair}) (see Fig.~\ref{fig:belltest} for a scheme of the type of experiment). The result was a violation of a variant of the CHSH inequality by six standard deviations. 

However, in this first experiment \cite{freedmanclauser} an additional assumption had to be introduced, due to the lack of an efficient method to detect the photon pairs. That was the so-called \emph{fair sampling assumption}, that means that the sample detected is a fair sample of the whole set of states produced. The introduction of this additional assumption is also referred as the \emph{detection loophole} in the experiment. Its solution would consist basically of increasing the fraction of detected pairs up to around $66\%$, being at the time of the first experiment only less than $1\%$. Nowadays this loophole has been closed in experiments with other systems, as for example trapped ions \cite{RoweN2001}, and in a photonic experiment \cite{GiustinaN2013} just quite recently.

This was not the only problem. There is another important loophole, called the \emph{locality loophole}, that consists in assuring that the measurements are performed in space-like separated events. This loophole was closed later in the famous photonic experiments conducted by Aspect in the '80s \cite{aspect822}, 
but never together with the detection loophole. Thus a completely satisfactory test is still missing, thought it is strongly believed that it will be performed in the very near future \cite{Genovese2005319}.

This because, although not conclusive, almost all of the several tests performed not only violated local-realism by itself, but even more remarkably were in total agreement with the quantum mechanical predictions. Moreover each of the two main loopholes has been closed separately in photonic experiments and various groups are working to close both of them in the same experiment soon.

\subsection{Kochen-Specker theorem and contextuality inequalities}

Another milestone on the debate about the objective reality behind quantum predictions was set by Kochen and Specker \cite{kochen67} soon after Bell's theorem. Their approach was along the same line of thoughts as the old von Neumann ``no go'' theorem, trying to somehow fix its weakness. The goal was still to explore the possibility of explaining the randomness of outcomes in quantum mechanics as coming from an underlying classical definite value, possibly depending on some hidden variables. Their approach is on the one hand more general than Bell's, and thus in a sense could simply follow from Bell's argument itself. On the other hand Kochen-Specker's approach gives different insights on what the peculiar \emph{logical} differences of quantum theory are as compared to classical mechanics and does not need to restrict the analysis to multipartite scenarios. 

The principle that they wanted to test is called \emph{non-contextuality}. It states that the actual value of an observable cannot depend on the values of other observables measured jointly, provided that alltogether they form a set of \emph{compatible} observables; or in other words that the value of an observable does not depend on the \emph{context} in which it is measured\footnote{Here is the improvement with respect to von Neumann's approach. In von Neumann's theorem the assumption was implicitly that the value of observables do not depend on the value of other observables measured jointly. This assumption was in fact too strong and the theorem can be falsified by simply choosing two incompatible observables and does not show any incompleteness of QM. Nevertheless KS proved that the statement of von Neumann's theorem holds even if the observables are restricted to be compatible, and this is a much stronger requirement. By requiring this, KS showed the incompatibility between non-contextuality, realism and QM.}. In the framework of hidden variable theories one can give a definition of \emph{Non-Contextual Hidden Variables} theory (see e.g., \cite{budronith} and references therein), analogous to Def.~\ref{LHVtheory} but with the following assumptions
\bbasm{\bf (Non-contextuality)}
The value of an observable is independent of the measurement context, i.e., it is independent of which set of compatible observables is measured jointly with it. 
\eeasm
\bbasm{\bf (Realism)}
Results of measurements are determined by properties that the system carry on prior to and independent of the measurements themselves.
\eeasm
\bbasm{\bf (Free-Will)}
The choice of a measurement setting do not influence and is not influenced by the outcome of other measurements.
\eeasm

Note that both of the first two assumptions do not hold in quantum theory and thus one can try to find contradictions between NCHV theories and quantum predictions. In practice Kochen and Specker were following a slightly different approach, the same followed by Gleason \cite{gleason57} to prove his theorem (Th.~\ref{gletheo}), namely to represent results of measurements as logical yes/no answers and formalize them using projectors, and in particular commuting projectors to represent results of compatible measurements. They then proved the following theorem, of which we give just an idea of the proof in a simplified way, following an argument that Peres \cite{Peres1990107,peres91} and Mermin \cite{mermin90} gave later.

\bbth{(Kochen-Specker theorem).} No \emph{Non-Contextual Hidden Variables} (NCHV) theory can reproduce all the predictions of quantum mechanics in a system with at least $3$ possible distinguishable states.
\eeth

\bbpr To prove the theorem we will consider instead a system that can be in $4$ distinguishable states, so that the proof can be simplified a lot. Thus, consider a $4$ dimensional Hilbert space $\mathcal H$ and the following set of dichotomic observables, named Peres-Mermin square
\begin{align}
&O_{1,1}=\sigma_z \otimes \id \ ,  &O_{1,2}= \id \otimes \sigma_z & \ , &O_{1,3}=\sigma_z \otimes \sigma_z \ , & \\
&O_{2,1}=\id \otimes \sigma_x \ ,  &O_{2,2}= \sigma_x \otimes \id & \ , &O_{2,3}=\sigma_x \otimes \sigma_x \ , & \\
&O_{3,1}=\sigma_z \otimes \sigma_x \ ,  &O_{3,2}= \sigma_x \otimes \sigma_z & \ , &O_{3,3}=\sigma_y \otimes \sigma_y \ , & 
\end{align}
that is such that all the observables in the same row commute with each other, as well as the observables in the same column. Thus each row and each column form a set of compatible observables, i.e., a particular context. However all these contexts but the last column are such that their product gives the identity, e.g., $O_{1,1} \cdot O_{1,2} \cdot O_{1,3}=\id$. Instead, the last column is such that $O_{1,3} \cdot O_{2,3} \cdot O_{3,3}=-\id$. Thus in every contextual theory the values $v(\cdot)$ associated to such observables must be such that $v(O_{1,3}) \cdot v(O_{2,3}) \cdot v(O_{3,3})=-1$ and their product gives $+1$ for all other contexts. Thus in particular if we consider the contexts corresponding to the three rows then the values have to satisfy
\begin{equation}
( v(O_{1,1}) v(O_{1,2}) v(O_{1,3}) ) \cdot (v(O_{2,1}) v(O_{2,2}) v(O_{2,3}) ) \cdot (v(O_{3,1}) v(O_{3,2}) v(O_{3,3}) ) = 1 \ ,
\end{equation}
while if we consider the context corresponding to the columns we obtain
\begin{equation}
( v(O_{1,1}) v(O_{2,1}) v(O_{3,1}) ) \cdot (v(O_{1,2}) v(O_{2,2}) v(O_{3,2}) ) \cdot (v(O_{1,3}) v(O_{2,3}) v(O_{3,3}) ) = -1 \ ,
\end{equation}
that is a contradiction, since the two expressions are exactly the same.
\eepr

\bbrmk
Note that in the above theorem nothing is assumed about the composition of the system, that might be composed of simply a single party, while in Bell's theorem the multipartite scenario played a fundamental role since locality was one of the assumptions. Here moreover the minimal dimension of the system required by the theorem to hold is $3$ and not $4$\footnote{We actually gave a proof that works for a $4$ dimensional Hilbert space. However in the original proof Kochen-Specker gave an explicit set of observables that led to a contradiction even in a $3$ dimensional space, but the set is much more complicated since it is formed by 117 observables.}. 
Note also that in the proof nothing was assumed about the state of the system. As a result, contextuality conditions can be derived that are violated by \emph{any quantum state}. These last are called \emph{state-independent contextuality conditions}.
\eermk

The statement is that if one wants to believe in both quantum mechanics and the objective reality behind the measurement results, then one has to assume that the outcomes of a measurement depend on the context in which it is performed. This is a more general statement than Bell's theorem in the sense that the different (couples of) local settings in a Bell scenario (e.g., $(a_1,b_1),(a_1,b_2)\dots$) correspond to a particular case of different contexts.

It is worth to note that this KS theorem can be expressed also in a way analogous to Bell's theorem, i.e., writing an inequality that must holds in every NCHV theory and that is violated by some quantum states. An example is given by the following inequality, derived by 
\emph{Klyachko, Can, Binicio\u{g}lu and Shumovsky} (KCBS) \cite{Klyachko03}

\bbth{(KBCS inequality).} Consider a single system $\mathcal S$ and the following couples of compatible observables 
$(a_0,a_1)$, $(a_1,a_2)$, $(a_2,a_3)$, $(a_3,a_4)$, $(a_4,a_0)$. Then the inequality
\begin{equation}\label{kbcsine}
\aver{a_0 a_1}+\aver{a_1 a_2}+\aver{a_2 a_3}+\aver{a_3 a_4}+\aver{a_4 a_0} \geq -3 \ ,
\end{equation}
holds for every possible \emph{non-contextual hidden variables theory}, where
\begin{equation}\label{noncontcorr}
\aver{a_i a_j} = \int x_i y_j \Pr(x_i , y_j)_{a_i, a_j} {\rm d} x_i {\rm d} y_j \ ,
\end{equation}
are the bipartite correlations between the outcomes of $(a_i,a_j)$.
\eeth 

We don't give the explicit proof here, but the bound on the right hand side is simply obtained by trying all possible non-contextual assigment to the expression on the left hand side. 

Summarizing, the theorems that we have shown above restricts very much the kind of theories that could explain experimental results in agreement with quantum theory. In particular, if one wants to believe to quantum mechanical predictions and to the objective reality behind results of measurements (and to the free will assumption as well), then one is forced to think that 
\begin{enumerate}
\item Results of measurements depend on the context in which they are obtained.
\item Different contexts provide different results even in the case in which the measurements are not causally connected according to special relativity.
\end{enumerate}

Thus, again, there are pillar principles of classical physics that can be experimentally tested independently on the specific theory and result in explicit conflict with quantum mechanics. However so far every statement is designed to be experimentally confirmed just for microscopic systems, in which somehow the classical intuition can be given up with less effort. But then later a similar result, putting serious constraints to classical models trying to agree with quantum mechanics, has been derived explicitly considering macroscopic systems. This is what is going to be discussed next.

\subsection{Leggett-Garg inequalities}\label{sec:LGineq}

One of the main criticism made by \emph{realists}, i.e., the ones who believe that nature must have some objective reality
and that measurement are just a tool to reveal preexisting properties of a system, to the actual formulation of quantum mechanics is that it is implicitly assumed that there is a boundary between macroscopic and microscopic systems. The former are the measurement devices and the latter the physical systems. However this boundary is not fixed and can in principle be pushed forward indefinitely. For example in this regard Bell \cite{bell87} said
\begin{quote}
A possibility is that we find exactly where the boundary lies. More plausible to me is that we will find that there is no boundary. 
\end{quote}
It is therefore natural to think that quantum phenomena appear also at macroscopic scales, though more difficult to detect. 
Again, this was in fact the concern that for example Schr\"odinger wanted to express with his famous story involving the cat. With Bell-type tests however, it would not have been straightforwardly possible to detect a truly quantum (i.e., non classical) effect at macroscopic scales because of the insurmountable experimental challenges that this task would have implied.

Thus later, in 1985, Leggett and Garg \cite{LG85} tried to adapt Bell's approach to the framework of macroscopic systems and discriminate between our classical intuition and quantum mechanics. To do this they focused on correlations between measurements of a single (macroscopic) observable made on a single (macroscopic) system, but at different instants in time.

In particular to formalize our classical intuition on the behaviour of macroscopic systems they introduced the set of theories that can be called \emph{Macrorealist Hidden Variables} theories and are based on the following assumptions (see \cite{EmariNoriRPR2014})
\bbasm{\bf (Macroscopic Realism)}
It is possible at all times to assign a definite value to an observable with two or more macroscopically distinct outcomes available to it on a (macroscopic) system.
\eeasm
\bbasm{\bf (Non-Invasive Measurability)}
It is possible, in principle, to determine the value of an observable on macroscopic systems causing an arbitrarily small disturbance.
\eeasm
\bbasm{\bf (Induction)}
The outcome of a measurement on a system cannot be affected by what will or will not be measured on it later.
\eeasm

\begin{figure*}[h!]
\centering
\includegraphics[width=0.7\textwidth]{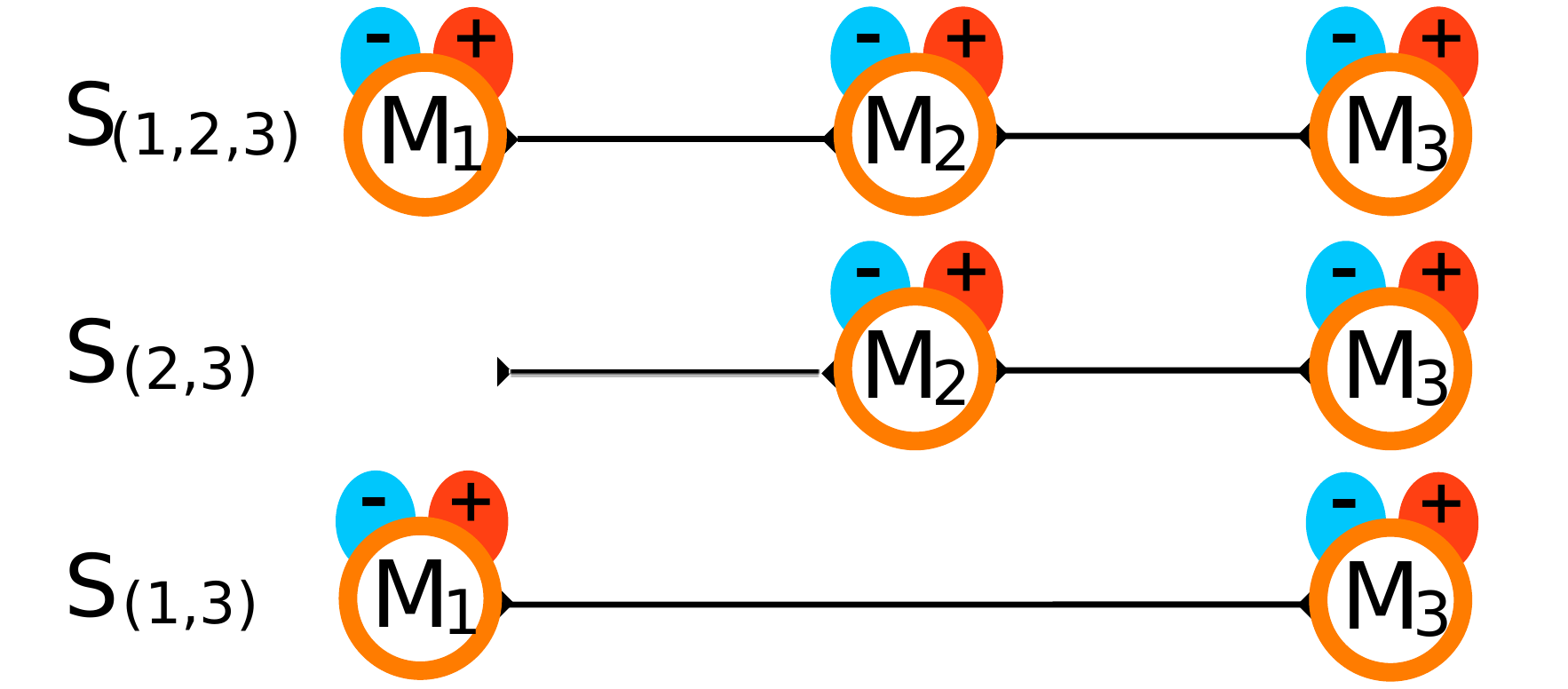}
\caption{Schematic representation of the Leggett-Garg test. Sequences of measurements $M_i$ of 
an observable $Q$ are performed giving results $x_i=\pm 1$. 
Macrorealism assumes non-invasive measurements, with the consequence that correlations, e.g., $C_{13} = \mean{Q_1 Q_3}$, are equal, independently of which sequence was performed to obtain them. $S_{\rm (1,2,3)}$ and $S_{\rm (1,3)}$, which differ by the presence or absence of $M_2$, give the same $C_{13}$ in macrorealism but not in quantum mechanics.}\label{fig:3seqschemes}
\end{figure*}

Then they assumed that one can measure the correlations 
$\aver{Q(t_i)Q(t_j)}$ of a dichotomic macroscopic variable $Q$ at two different instants of time $t_i$ and $t_j$ and, by considering the combination of three pairs of correlations for three instants $(t_1,t_2,t_3)$, they derived an inequality that must be satisfied by all MHV theories (see Fig.~\ref{fig:3seqschemes}).

\bbth{(Leggett-Garg inequality).} Consider a single system $\mathcal S$ and a macroscopic property $Q$ of such a system measured at three different instants of time $(Q(t_1),Q(t_2),Q(t_3))$. Then the inequality
\begin{equation}\label{lgine}
\aver{Q(t_1)Q(t_2)}+\aver{Q(t_2)Q(t_3)}-\aver{Q(t_1)Q(t_3)}\leq 1 \ ,
\end{equation}
holds for every possible \emph{macrorealist hidden variables theory}, where
\begin{equation}\label{timecorr}
\aver{Q(t_i)Q(t_j)}=\sum_{x_i, x_j=\pm 1} x_i x_j \Pr(x_i,x_j)_Q = \sum_{x_i, x_j=\pm 1} x_i x_j \Pr(x_j|x_i)_{Q(t_j)} \Pr(x_i)_{Q(t_i)}
\end{equation}
are the two-time correlations between the outcomes of $Q$.
\eeth 

\bbpr
The three assumptions of a MHV theory together imply that there exists a single probability distribution for the three outcomes (due to the MR assumption) $P=\Pr(x_1,x_2,x_3)_Q$ and (due to NIM) that the two-time correlations can be computed through the marginals over $P$. Then, using just the fact that $\sum_{x_1,x_2,x_3=\pm 1} P=1$ one obtains the bound of Eq.~(\ref{lgine}). 
\eepr

Now let us focus a little bit on the assumptions made here. The first is analogous to what we previously called simply realism, but just referring explicitly to macroscopic objects. Thus here basically we are assuming that the system is at any time in a single state from a set of several ``macroscopically distinct'' ones, simply meaning that we are looking at some macroscopic property of the system and assume that this property has a definite value at all times. 
Note, however, that this remains a rather vague assumption until further formal clarification of the meaning of the word \emph{macroscopic} is provided. We will discuss this issue briefly at the end of Ch.~\ref{ch:LG}.

The second assumption is also tied to macroscopic system, on which we expect to be able to perform measurements with arbitrarily small disturbance. This is an assumption that classically one usually can give up for a microscopic system because of just technical difficulties and here is somehow crucial just because we focus on macroscopic objects. Again in quantum theory both of these first two assumptions do not hold, since any system (even macroscopic) can be in a superposition state and every measurement is invasive since the state collapses afterwards. The third assumption might be thought as just the analogue of free will in this framework. 

Thus in quantum mechanics the LG inequality can be violated even in the simplest system, a single spin-$\frac 1 2$ particle, and there is no constraint that prevents it to be violated in macroscopic, e.g., very large spin, systems. In fact states of large spin systems violate Eq.~(\ref{lgine}) even up to its algebraic maximum \cite{BudroniPRL2014}. 

\bbrmk
Note also that in quantum mechanics there is not a univocal way to compute the time correlations Eq.~(\ref{timecorr}), since they depend on how the state is updated after the measurement, and thus on the nature of the measurements $\mathcal M_{Q}(\rho)$ themselves. The main example is given by projective measurements and is the one implicitly considered in the work of Leggett and Garg. In that case the correlations can be computed by means of L\"uders rule \cite{luders51}
\begin{equation}\label{projtimecorr}
\aver{Q(t_i)Q(t_j)}=\sum_{x_i, x_j=\pm 1} x_i x_j \trace\left( \pi_{x_j} U(t_j,t_i) \pi_{x_i} U(t_i,t_0) \rho_0 U^\dagger(t_i,t_0) \pi_{x_i} U^\dagger(t_j,t_i) \right) \ ,
\end{equation}
where $\pi_x$ is the projector onto the space of eigenstates with eigenvalue $x$, $\rho_0$ is the initial state and $U(t_j,t_i)$ is the unitary evolution operator.
\eermk
As in the previous cases also the incompatibility between quantum mechanics and MHV theories can be put as a theorem.

\bbth{(Leggett-Garg theorem).}\label{th:LGtheorem} No \emph{Macrorealist Hidden Variables} (MHV) theory can reproduce all the predictions of quantum mechanics.
\eeth 

\bbpr
As a proof we can show that Eq.~(\ref{lgine}) can be violated in a single spin-$\frac 1 2$ system. Consider then three spin-$\frac 1 2$ operators $Q(t_i)=\sum_k a_k(t_i) \sigma_k$ for the three instants $(t_1,t_2,t_3)$. 
It can be shown that for projective measurements the time correlations, computed according to Eq.~(\ref{projtimecorr}) can be expressed as the symmetrized product of the observables
\begin{equation}
\aver{Q(t_i)Q(t_j)}= \frac 1 2 \aver{\left\{ Q(t_i), Q(t_j) \right\}} \ ,
\end{equation}
where $\left\{ \cdot , \cdot \right\}$ is the anticommutator. Thus as a result we obtain 
\begin{equation}
\aver{Q(t_i)Q(t_j)}= \frac 1 2 \aver{\left\{ \sum_k a_k(t_i) \sigma_k , \sum_l a_l(t_j) \sigma_l \right\}} = \sum_k a_k(t_i) a_k(t_j) =\cos \theta_{i,j} \ ,
\end{equation}
for a certain phase $\theta_{i,j}$ depending on the instants $(t_i,t_j)$. In particular, choosing the time instants such that $\theta_{1,2}=\theta_{2,3}=\frac 1 2 \theta_{1,3}=\frac{\pi} 3$ we obtain
\begin{equation}
\aver{Q(t_1)Q(t_2)}+\aver{Q(t_2)Q(t_3)}-\aver{Q(t_1)Q(t_3)}=\frac 3 2 > 1 \ ,
\end{equation}
and the LG inequality (\ref{lgine}) is violated.
\eepr

Recently these LG tests have attracted increasing attention and several experiments have been performed, mainly still using microscopic systems \cite{GogginPNAS2011,xuSR2011,DresselPRL2011,SuzukiNJP2012,WaldherrPRL2011,GeorgePNAS2013,AthalyePRL2011,SouzaNJP2011,KneSimGau12,RobensPRX2015}, with few exceptions \cite{PalaciosNatPhot2010,groen13}, that used superconducting quantum devices, similarly to the original proposal of Leggett and Garg. All such tests, however, 
suffer also from a weakness called
\emph{clumsiness loophole} (see e.g., \cite{EmariNoriRPR2014} for a review on Leggett-Garg inequalities), consisting basically of the possibility of explaining the results with some hidden and unwanted clumsiness in the measurements, still obeying classical principles.

We will discuss in some detail this loophole in Ch.~\ref{ch:LG}.

\chapter{Entanglement detection and spin squeezing} 

\label{Chapter2} 

In the previous chapter we introduced the basic principles of quantum theory and discussed the interpretative questions raised by those principles. We have also presented results that show some fundamental differences between ``quantum'' and classical theories, detectable with some combinations of correlations. Here we are going to discuss in more details one of these fundamental characteristics of quantum theory, that evolved from being thought as a problem of the theory to one of the main target resources to be produced in current experiments. This is what was termed {\it entanglement} by Schr\"odinger in his response \cite{schrod35} to the EPR paper and consists in the non-separability of the description of certain states of a system into a composition of subsystems' states. As Schr\"odinger himself explains it, non-separability means the following
\begin{quote}
...Maximal knowledge of a total system does not necessarily include total knowledge of all its parts, not even when these are fully separated from each other and at the moment are not influencing each other at all... and this is what keeps coming back to haunt us.
\end{quote}
and has been initially thought as a trait of quantum mechanics difficult to interpret and contradictory to experience.

Nowadays instead, after the works of Bell the situation has changed and we know not only that entanglement is an experimental fact, but also that it can be exploited as a resource for practical tasks. In fact, since some pioneering works \cite{feynman82,bennett93,eckert91,shor97}, it has been shown that an initially prepared entangled state such as the EPR pair (\ref{eprpair}) can be used to improve the efficiency of protocols in communication, computation, metrology among other things. This, together with a progressive improvement of experimental techniques to accurately control physical systems has led to an increasing interest in the theoretical study of criteria designed to certify that entanglement has been actually observed (see \cite{Guhne2009Entanglement} for a review of the topic). Even more, entanglement intended as a resource has been studied in a way oriented to detect and quantify its usefulness \cite{horodeckirev}.

In this respect bi-partite systems are nowadays very well understood and there are
well established results that led to a satisfactory theory of entanglement and its quantification. We will thus focus on such systems to introduce the main concepts of this theory, that is on the other hand very rich and complicated.
In fact in multipartite scenarios, even for already three parties the situation is far more complicated. 
It has been discovered that there are very many ``kinds'' of entangled states, not comparable to each other and that the problem of detection and quantification of entanglement is in general very hard, if not unsolvable.

On the other hand, true multipartite entanglement has been proven to be important for technological applications.
Thus for multipartite systems, even few-partite, the theory is usually restricted to specific classes of systems or even to specific experimental setups, but still arouses a lot of interest.
In our case we will focus on systems composed of very many parties and study criteria designed to detect entangled states based on the knowledge of few collective quantities, which are relatively easy to measure in many experimental setups.

In particular we will study the so-called Spin Squeezing criteria, that define the class of Spin Squeezed States (SSS) \cite{Kitagawa1993Squeezed,Wineland1994Squeezed,Sorensen2001Many-particle,Ma2011Quantum}. These are on the one hand relatively easy to produce with different techniques and on the other hand can be used for technological improvements e.g., in metrology, due to the fact that they are entangled \cite{Giovannetti2011Advances}.

The structure of this chapter is the following. First in Sec.~\ref{entdet} we introduce the formal definition of entangled states and study the detection and quantification of entanglement in bipartite systems. In that case there exists basically a unique quantity that measures entanglement, at least in pure states. We will introduce in that scenario some different kinds of entanglement criteria generally employed.
Then, in Sec.~\ref{entinmulti} we will see how the situation complicates itself for multipartite systems. We will recall the main criteria usually considered and focus on criteria based on collective observables, adapt for systems composed by very many parties. Finally, in Sec.~\ref{spsquesec} we switch the attention to Spin Squeezed States and their practical usefulness, starting from their initial definition and going through all the successive developments of their study, leading to a multitude of possible definitions and applications.


\section{Entanglement in bipartite systems}\label{entdet}

In this section we will introduce the formal concept of entangled, as opposite to separable states, according to the mathematical definition usually attributed to R. Werner \cite{Werner1989Quantum}\footnote{Actually Werner himself in \cite{Werner14} explains that this definition was developed before his paper independently by himself and by H. Primas and members of his group.}. We will then present the fundamental results obtained in bipartite system, showing the main different approaches to the detection and quantification of entanglement. 


\subsection{Definition and basic concepts}

An entangled state can be intuitively defined as a state in which one can observe a violation of some Bell-like inequality. However this rather obscure intuition cannot be easily employed as it is within the quantum mechanical formalism, but needs some additional assumptions. One precise mathematical formulation has been given by introducing the definition of {\it separable states}, thought as the ``classical'' states that cannot violate any Bell inequality. The idea was that these should be states that can be produced by putting together states created locally in every subspace.
States that can be written as separable density matrices, according to Def.~\ref{sepadef} have such property and form a mathematically well defined class of quantum states. 

\bbdf{\bf (Separable and entangled states).}\label{def:separability}
A density matrix $\rho_{\rm sep}$ acting on a Hilbert space $\mathcal H_{\mathcal S}=\mathcal H_{\mathcal P_1}\otimes \dots \otimes \mathcal H_{\mathcal P_n}$ is called {\it separable} if it can be written as a convex mixture of product states
\begin{equation}\label{sepstatdef}
\rho_{\rm sep}=\sum_k p_k \rho_k \qquad  p_k\geq 0 \ , \sum_k p_k=1 \ ,
\end{equation}
where $\{ p_k \}$ is a probability distribution and $\{ \rho_k \}$ is a certain set of {\it product states}
\begin{equation}
\rho_k = \rho^{(1)}_k \otimes \dots \otimes \rho^{(n)}_k \ ,
\end{equation}
with $\rho^{(i)}_k \geq 0$, $\trace \rho^{(i)}_k = 1$ and the apex $(i)$ referring to the party $\mathcal P_i$. 

Otherwise, a density matrix $\rho_{\rm ent}$ that cannot be written in the form \ref{sepstatdef} is called {\it entangled}.
\eedf

Defined in this way separable states cannot violate any inequality derived with the same assumptions as the Bell theorem. However there are also entangled states that cannot violate any Bell-like inequality as well. Thus this definition does not entirely capture the concept of a state that obeys local-realism. In this respect, other inequivalent definitions can be given, such as \emph{non-locality}, that refers to a possible description of the state
with LHV models, or \emph{steering} \cite{Wiseman07}, that is somehow intermediate between the other two.  
Nevertheless, one can still provide the definition of non-separability as in Def.~\ref{def:separability} as a ``truly quantum feature'' of a state and also as a resource for practical tasks.

Unfortunately, although the property of the set of separable states being convex simplifies it a lot, the general problem of entanglement detection remains rather complicated and practically unsolvable. Thus one in general has to restrict the study to specific classes of states. 
The only relatively simple case is when the system is composed by just two parties. In particular for pure bipartite states the following theorem can be exploited to give a complete characterization of entanglement.

\bbth{(Schmidt Decomposition).} Let us consider a bipartite pure state $|\Phi\rangle \in \mathcal H_{\mathcal A} \otimes \mathcal H_{\mathcal B}$. Then there exists an orthonormal basis $\{ |a_i\rangle \}$ of $\mathcal H_{\mathcal A}$ and an orthonormal basis $\{ |b_i\rangle \}$ of $\mathcal H_{\mathcal B}$ such that
\begin{equation}
|\Phi\rangle = \lambda_i |a_i\rangle |b_i\rangle \ , 
\end{equation}
where $\lambda_i\geq 0$ are called {\it Schmidt coefficients}. The number of $\lambda_i$ different from zero is called {\it Schmidt rank} $R(|\Phi\rangle)$ of the state.

The same decomposition can be found for a bipartite density matrix. Thus every bipartite density matrix $\rho$ can be decomposed as
\begin{equation}
\rho=\sum_k \lambda_k o_k^{\mathcal A} \otimes o_k^{\mathcal B} \ ,
\end{equation}
with $\lambda_k\geq 0$ and for some couple of orthonormal basis $\{ o_k^{\mathcal A} \}$, $\{ o_k^{\mathcal B} \}$ of the space of operators.
\eeth

We will omit the proof, that can be found in many textbooks.
The above result immediately implies that a state with Schmidt rank equal to $1$ can be written in a product form in some basis and thus it is separable. On the other hand, states $|\Phi\rangle$ such that $R(|\Phi\rangle)>1$ must be entangled. Thus the quantity $R$ alone is sufficient to characterize entanglement in pure bipartite states.
For mixed bipartite states already the situation gets a lot more complicated. There are several useful criteria either necessary or sufficient to prove entanglement but no simple one is both necessary and sufficient. We are going to present some important examples of such criteria in what follows. The first, and most famous is due to Peres \cite{peresPPT} and is called {\it Positive Partial Transpose} (PPT) criterion. 

\bbth{(PPT criterion).}\label{theo:PPT} Consider a separable state $\rho_{\rm sep}$ of a bipartite system $\mathcal H_{\mathcal A}\otimes \mathcal H_{\mathcal B}$. Then the operator $\rho_{\rm sep}^{T_{\mathcal A}}$ obtained by transposing the indices relative to one subsystem $\mathcal H_{\mathcal A}$ is positive $\rho_{\rm sep}^{T_{\mathcal A}}\geq 0$. Thus any state $\rho$ such that $\rho^{T_{\mathcal A}} < 0$ must be entangled.
\eeth

\bbpr Directly from the definition of separable states we have
\begin{equation}
\rho_{\rm sep} = \sum_k p_k \rho_{k}^{\mathcal A} \otimes \rho_{k}^{\mathcal B} \ ,
\end{equation}
and thus
\begin{equation}\label{pptcondproof}
\rho_{\rm sep}^{T_{\mathcal A}} = \sum_k p_k (\rho_{k}^{\mathcal A})^T \otimes \rho_{k}^{\mathcal B} \geq 0 \ ,
\end{equation}
where we denoted as $(\cdot)^T$ the transposed matrix.
Eq.~(\ref{pptcondproof}) follows from the fact that by definition $\rho_{k}^{\mathcal A} \geq 0$ and $p_k\geq 0$. Note that it is equivalent to take the partial transposition with respect to party $\mathcal B$ since $\rho_{\rm sep}^{T_{\mathcal A}} \geq 0 \Leftrightarrow \rho_{\rm sep}^{T_{\mathcal B}} \geq 0$. 
\eepr

This theorem gives a necessary but not sufficient condition for separability and thus a sufficient criterion to prove entanglement. It exploited the fact that the transposition is an operation that is positive but {\it not completely positive} and thus the partial transposition might not preserve in general the positivity of the density matrix. However this cannot happen when the state is separable. In fact the PPT criterion can be also turned into a very general \emph{if and only if} statement, that
we are going to enunciate omitting the proof.

\bbth{(Positive maps criterion).} A state $\rho_{\rm sep}$ of a bipartite system $\mathcal H_{\mathcal A}\otimes \mathcal H_{\mathcal B}$ is separable {\it if and only if} 
\begin{equation}
\mathcal L \otimes \id (\rho_{\rm sep}) \geq 0
\end{equation}
holds for all positive but not completely positive operations $\mathcal L$ acting on one party.
\eeth

This provides a mapping from the separability problem to the study and classification of positive but not completely positive maps. In fact, other criteria have been derived using positive but not completely positive operations, but we are not going to present them. Rather, we present another very important criterion belonging to a different class. It is called the {\it Computable Cross Norm} or {\it Realignment} (CCNR) criterion \cite{rudolph00,chen03}.

\bbth{(CCNR criterion).} Consider a state $\rho$ of a bipartite system $\mathcal H_{\mathcal A}\otimes \mathcal H_{\mathcal B}$ in its Schmidt decomposed form $\rho=\sum_k \lambda_k o_k^{\mathcal A} \otimes o_k^{\mathcal B}$, with $\{ o_k^{\mathcal A} \}$, $\{ o_k^{\mathcal B} \}$ orthonormal basis of the space of operators of the respective parties. The singular values $\{ \lambda_k \}$ must satisfy
\begin{equation}
\sum_k \lambda_k \leq 1
\end{equation}
for all separable states. Thus if $\sum_k \lambda_k > 1$ the state $\rho$ must be entangled.
\eeth

\bbpr Consider a pure product state $\rho_{\rm sep}$
\begin{equation}
\rho_{\rm prod} = \rho^{\mathcal A} \otimes \rho^{\mathcal B} \ ,
\end{equation}
with $\trace (\rho^{\mathcal A})^2 =\trace (\rho^{\mathcal B})^2=1$. It is already in its Schmidt decomposed form and we have $\sum_k \lambda_k = 1$. Then consider a mixed separable state
\begin{equation}
\rho_{\rm sep} = \sum_k p_k \rho_k^{\mathcal A} \otimes \rho_k^{\mathcal B} \ .
\end{equation}
Since $\sum_k \lambda_k:=\| \rho \|_1$ is a norm in the space of density matrices we have that
\begin{equation}
\| \rho_{\rm sep} \|_1 \leq \sum_k p_k \| \rho_k^{\mathcal A} \otimes \rho_k^{\mathcal B} \|_1 = 1
\end{equation}
holds for all separable states. 
\eepr

This criterion, as the PPT, can be seen as a particular case of a larger class of criteria, that are based on the fact that the trace norm $\| \rho \|_1$ cannot increase under the action of a trace preserving positive map.

\bbth{(Contraction criterion).} A state $\rho_{\rm sep}$ of a bipartite system $\mathcal H_{\mathcal A}\otimes \mathcal H_{\mathcal B}$ is separable {\it if and only if} 
\begin{equation}
\|\mathcal T \otimes \id (\rho_{\rm sep})\|_1 \leq \|\rho_{\rm sep} \|_1 = 1
\end{equation}
holds for all trace preserving positive operations $\mathcal T$ acting on one party.
\eeth

In this case again, since the condition is an if and only if, we can map the separability problem into the study of trace preserving positive maps. This is also a sort of complementary class of criteria with respect to the previous one. In fact the PPT and CCNR criteria are used as complementary to each other in the sense that none of them detects all possible entangled states, but they can detect different states.

Since the problem of deciding whether a state is separable or not is very hard, an approach widely used to find a solution in many practical cases is based on numerical algorithms. In this respect it is very helpful that the set of separable states is convex. By exploiting this property a test of separability can be formulated as a convex optimization problem and in many cases can be solved efficiently with semidefinite programming. An important numerical method that does this is called {\it symmetric extensions} method, that is based on the following observation

\bboo{(Symmetric extension criterion).} For every bipartite separable state $\rho_{\rm sep} = \sum_k p_k \rho_k^{\mathcal A} \otimes \rho_k^{\mathcal B}$ there is an extension to a multipartite state
\begin{equation}
\sigma_{\rm sep} = \sum_k p_k \rho_k^{\mathcal A} \otimes \rho_k^{\mathcal B} \otimes \rho_k^{\mathcal A} \otimes \dots \otimes \rho_k^{\mathcal A}
\end{equation}
that has the following properties: (i) it is PPT with respect to each possible bipartition, (ii) has $\rho_{\rm sep}$ as reduced state with respect to the first two parties, and (iii) it is symmetric under the exchanges of first party with any other of the additional parties. 

Thus, for each $k$, any state $\rho$ for which such a symmetric extension to a $k$-partite state does not exist must be entangled. Moreover for every entangled state such symmetric extension to a $k$-partite state must not exist for some $k$. 
\eeoo

This facts results as quite powerful because it turns out that the problem of finding a symmetric extension is solvable efficiently within semidefinite programming and there are algorithms that either give a solution or prove that no solution exists. In this last case the state in proven to be entangled. Moreover in this way a hierarchy of separability criteria can be defined: if an extension to a $k$-partite state does exist, then one looks for a $k+1$-partite state. If the state is entangled then at some $k$ the symmetric extension must not exist and the state will be detected by this procedure. Thus this hierarchy is complete, in the sense that all entangled states are detected at some step. On the other hand, it requires a hard computational effort, so that in practice only the first step is usually feasible and only for small dimensional systems. 


\subsection{Entanglement witnesses}

In the criteria presented above, we have implicitly assumed that a certain state $\rho_{\rm sep}$ is completely known and looked for a method to decide whether it is separable or not. In practice, however, one has only very partial information about a state $\rho$, and this information is usually encoded in expectation values of observables, i.e., $\aver{O}=\trace (O\rho)$. Thus a usually more important practical question is to derive entanglement criteria based directly on such observable quantities. Formally one can define an {\it entanglement witness} as an observable that does the job, i.e., detects a certain state $\rho$ as entangled based just on its average value.

\bbdf{\bf (Entanglement witness).} An observable $W$ is called {\it entanglement witness} if
\begin{equation}
\begin{aligned}
\trace(W \rho_{\rm sep}) &\geq 0 \qquad \mbox{for all separable states $\rho_{\rm sep}$} \ , \\
\trace(W \rho_{\rm ent}) &< 0 \qquad \mbox{for at least one entangled state $\rho_{\rm ent}$}
\end{aligned}
\end{equation}
holds. This means that the state $\rho_{\rm ent}$ is detected as entangled by just measuring the average value of the witness W.
\eedf 

Thus an entanglement witness is some operator that detects at leas one state. On the other hand, it is also known that for every entangled state there is at least one witness that detects it \cite{Horodecki19961}. This fact can be directly connected with the correspondence between positive but not completely positive maps and separable states.

\bbth{(Completeness of witnesses).}\label{thcompwitn} For each entangled state $\rho_{\rm ent}$ there exists a witness $W$ that detects it, i.e., such that $\trace(W \rho_{\rm ent}) < 0$.
\eeth

\bbpr Consider a bipartite entangled state $\rho_{\rm ent}$. Then there exists a positive map $\mathcal L$ such that $\mathcal L \otimes \id (\rho_{\rm ent})$ has an eigenvector $|\eta \rangle$ with negative eigenvalue $\lambda_- < 0$. Thus, consider the adjoint map $\mathcal L^*$, i.e., such that $\trace(A\cdot\mathcal L(B))=\trace(\mathcal L^*(A)\cdot B)$ for all operators $A,B$. We have that 
\begin{equation}
\trace(\mathcal L^* \otimes \id(|\eta\rangle \langle \eta |) \cdot \rho) = \trace(|\eta\rangle \langle \eta | \cdot \mathcal L \otimes \id(\rho)) \geq 0  
\end{equation}
holds for all separable states due to the positivity of $\mathcal L$ and the positive maps criterion. On the other hand
\begin{equation}
\trace(\mathcal L^* \otimes \id(|\eta\rangle \langle \eta |) \cdot \rho_{\rm ent}) = \trace(|\eta\rangle \langle \eta | \cdot \mathcal L \otimes \id(\rho_{\rm ent})) < 0 \ .
\end{equation}
Thus, $W=\mathcal L^* \otimes \id(|\eta\rangle \langle \eta |)$ is a witness detecting $\rho_{\rm ent}$.
\eepr

Due to the previous theorem the problem of separability can be mapped into the problem of finding a witness for a certain entangled state. This is in practice what is done for most experiments that aim at producing a certain target $\rho_{\rm ent}$ and at proving that what they produced, although not exactly the target, is indeed an entangled state.
In a way, this task is feasible because of the convex geometry of the set of separable states. In fact
note that since the space of separable states is convex in the space of density matrices and $\trace(W\rho)$ is linear, an entanglement witness defines a hyperplane $\trace(W\rho)=0$ that cuts the space in two halves, one of which, $\trace(W\rho) \geq 0$, includes all separable states. Thus from an experimental point of view it is plausible that although the target is not reached perfectly, if the produced state is sufficiently close it will still be detected by the same witness $W$ as the target.
In this respect, it is important to find an optimal witness for the target state $\rho$, i.e.,
an operator $W_{\rm opt}$ that detects the maximal possible set of states which includes $\rho$.

\bbdf{\bf (Optimal entanglement witness).} A witness $W_1$ is defined {\it finer} than another witness $W_2$ if 
\begin{equation}
W_2=W_1+ P 
\end{equation}
holds for some positive operator $P$. This means that the states detected by $W_2$ are a subset of the states detected by $W_1$. Then, a witness $W$ is called {\it optimal} if there are no other witnesses finer than it. Furthermore, a witness $W$ is called {\it weakly optimal} if there exists a separable state $\rho$ such that $\trace(W_{\rm opt} \rho)=0$.
\eedf 

Therefore, a weakly optimal witness $W$ is such that the hyperplane $\trace(W_{\rm opt} \rho)=0$ touches the set of separable states, while it is optimal if and only if it has the additional property that the product states $|\psi_i\rangle$ such that $\langle \psi_i | W | \psi_i \rangle=0$ span the whole space of separable states. If a witness is not optimal, it can be in principle optimized with minimization algorithms, although they might be computationally very hard.

\begin{figure*}[h!]
\centering
\includegraphics[width=\columnwidth,clip]{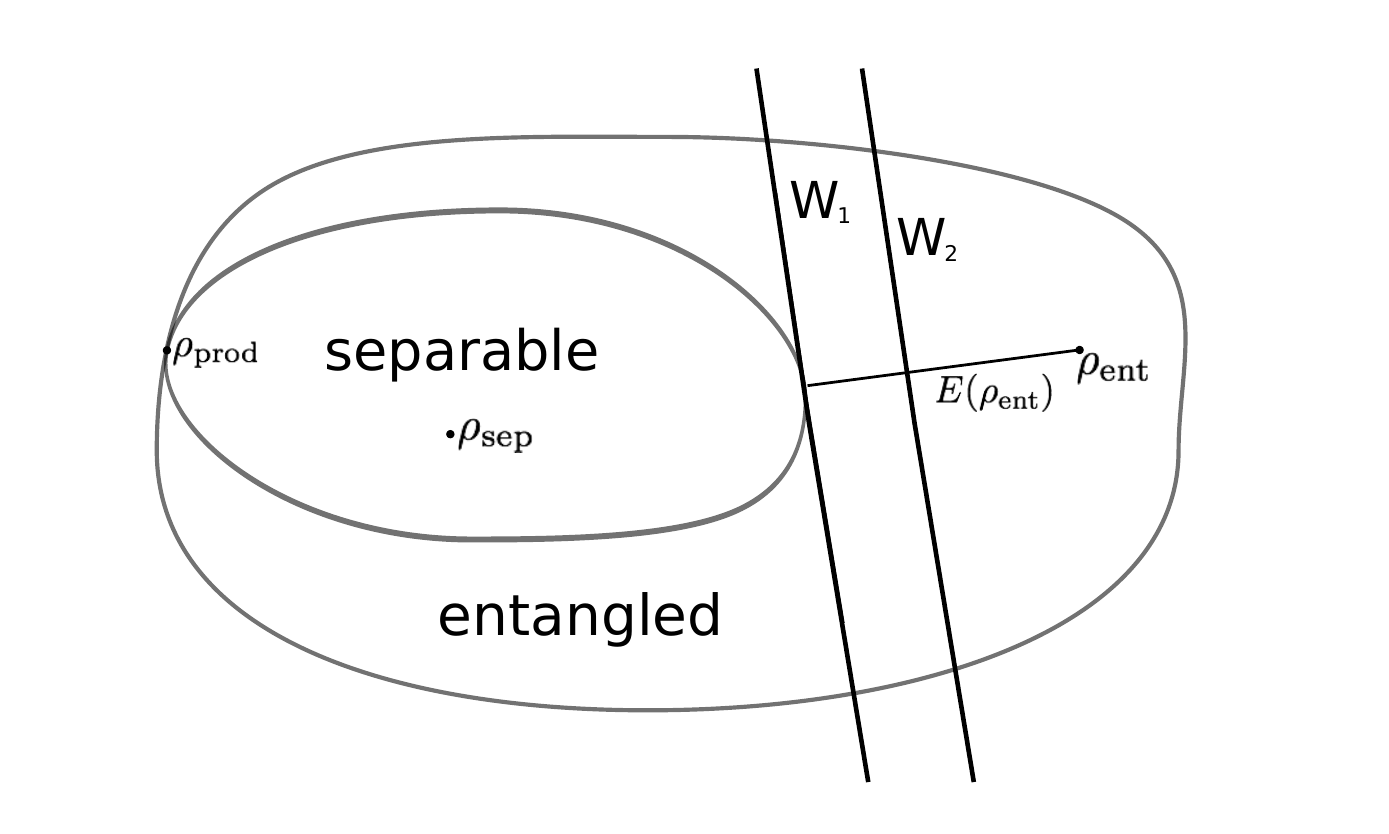}
\caption{Schematic representation of the set of separable states as a convex subset of the set of all quantum states. The boundary points are pure states and the common boundary represents pure product states. Two entanglement witnesses are represented; one of them, $W_1$ is strictly finer than $W_2$. An entanglement measure $E(\rho_{\rm sep})$ has been also represented as a distance from the state to the set of separable states.}
\label{fig:biparent}
\end{figure*}

As we have seen in Th.~\ref{thcompwitn} there is a close connection between witnesses and positive but not completely positive maps. This connection is made precise and clear by exploiting a general result mapping bipartite operators $E$ acting on $\mathcal H_\mathcal A \otimes \mathcal H_\mathcal B$ into
maps $\mathcal M: \mathcal B(\mathcal H_\mathcal A) \rightarrow \mathcal B(\mathcal H_\mathcal B)$, called {\it Choi-Jamio\l{}kowski isomorphism} \cite{depillis67,jamiolkowski72,choi82,choi75}.

\bbth{(Choi-Jamio\l{}kowski isomorphism)} The {\it Choi-Jamio\l{}kowski isomorphism} is a one to one correspondence
between bipartite operator $E$ acting on $\mathcal H_\mathcal A \otimes \mathcal H_\mathcal B$ and maps $\mathcal E$ from $\mathcal H_\mathcal A$ to $\mathcal H_\mathcal B$ given by the relations
\begin{subequations}
\begin{align}
\mathcal E(\rho)= \trace_{\mathcal A} (E \rho^T \otimes \id) \ , \label{mapchoij} \\
E= \sum_k (o_k^{\mathcal A})^T \otimes \mathcal E(o_k^{\mathcal B}) \ ,
\end{align}
\end{subequations}
where $\{ o_k^{\mathcal A} \}$, $\{ o_k^{\mathcal B} \}$ are orthonormal basis of the corresponding space of operators. 

The isomorphism has, among others the following important properties.
(i) The map $\mathcal E$ is completely positive if and only if $E\geq 0$. (ii) The map $\mathcal E$ is trace preserving if and only if $\trace_{\mathcal A}(E)=\frac{\id}{{\rm dim} \mathcal H_\mathcal B}$. (iii) The map $\mathcal E$ is positive but not completely positive if and only if $E$ is an entanglement witness.
\eeth

The previous theorem provides a very useful tool to map results obtained on bipartite operators, e.g., bipartite quantum states, to analogous for quantum operations (or quantum ``channels'') and viceversa. In particular, concerning the detection of entangled states, an interesting result is that, given an entanglement witness W, the corresponding map given by Eq.~(\ref{mapchoij}) detects all the states detected by an improved witness, namely
\begin{equation}
\tilde{W}=(F_{\mathcal A} \otimes \id) W (F_{\mathcal A}^\dagger \otimes \id) \ ,
\end{equation}
for arbitrary invertible matrices $F_{\mathcal A}$ (sometimes called {\it local filters}).


\subsection{Entanglement measures}

A further step made in the light of exploiting entanglement as a resource has been to quantify it. This idea can been pursued following an axiomatic approach that starts intuitively from the features that one expects to exploit in practice \cite{pleniovirmani,horodeckirev}. In this respect, one asks to any reasonable measure of entanglement to be a function of density matrices that satisfies a monotonicity property under maps that cannot create entanglement, namely maps that can be implemented locally and with at most a classical communication channel between the parties. In particular the simplest axiomatic definition of a measure of entanglement is called {\it entanglement monotone} \cite{vidal00}

\bbdf{\bf (Entanglement monotone).} An entanglement monotone $E(\rho)$ is a function from the space of bipartite density matrices $\mathcal H_\mathcal A \otimes \mathcal H_\mathcal B$ to the real numbers, with the following properties: (i) It vanishes on separable states, i.e., $E(\rho)=0$ iff $\rho$ is separable. (ii) It cannot increase under LOCC maps, i.e., $E(\mathcal L(\rho))\leq E(\rho)$ for all maps $\mathcal L$ belonging to the class of {\it Local Operations and Classical Communications}. In particular property (ii) also implies that (ii-a) It is invariant under local change of basis, i.e.,  $E(U_{\mathcal A}\otimes U_{\mathcal B} \rho U^\dagger_{\mathcal A}\otimes U^\dagger_{\mathcal B}))=E(\rho)$ for all unitaries $U_{\mathcal A}$, $U_{\mathcal B}$.
\eedf   

This is the minimal requirements that a reasonable entanglement measure should satisfy. However sometimes stronger properties are also required, such as for example the monotonicity under LOCC is replaced by a {\it strong monotonicity} condition. This last states that for LOCC operations that have output $\rho_k$ with a certain probability $p_k$, then the measure should not increase on average
\begin{equation}
\sum_k p_k E(\mathcal L(\rho)_k)\leq E(\rho) \ .
\end{equation}

Unfortunately this definition of entanglement monotone cannot lead to a single well defined notion of entanglement measure. This is because the space of density matrices cannot be totally ordered with respect to LOCC operations. This means that there are states that cannot be connected between each other with LOCC operations and thus cannot be compared between each other with an entanglement monotone. A quite special case in this sense is that of pure bipartite states, where at least a {\it maximally entangled state} can be univocally defined, that is precisely the EPR pair: from an EPR pair every other pure bipartite state can be reached through a LOCC. This partial order is completely lost already for mixed bipartite states. Thus, already for bipartite system, many inequivalent entanglement monotones can be found.
In many cases they are strictly related to some entanglement criteria and try to quantify how much the states violates a certain criterion.
For example, one can try to measure how negative is the partially transposed density matrix and define in this way an entanglement measure called {\it negativity}

\bbdf{\bf (Negativity).} Consider a bipartite state $\rho$ acting on $\mathcal H_{\mathcal A} \otimes \mathcal H_{\mathcal B}$. The negativity is an entanglement monotone defined as 
\begin{equation}
N(\rho)=\frac{\| \rho^{T_{\mathcal A}} \|_1-1} 2 \ ,
\end{equation}
that quantifies the violation of the PPT criterion by the state $\rho$.
\eedf  

Note that the negativity is also a convex function of the state 
\begin{equation}
N\left(\sum_k p_k \rho_k \right) \leq \sum_k p_k N(\rho_k) \ ,
\end{equation}
i.e., it cannot increase under mixing two or more states. This property is also usually employed to define entanglement measures starting from pure states through what is called {\it convex roof construction}.

\bbdf{\bf (Convex roof construction).} Given an entanglement measure defined for pure states $E_{\rm pure}(|\phi\rangle)$ and a mixed state $\rho$ we define
\begin{equation}
E(\rho)=\inf_{p_k,|\phi\rangle_k} \sum_k p_k E_{\rm pure}(|\phi\rangle_k)
\end{equation}
as a suitable entanglement measure for all states, where the infimum is taken over all the possible decompositions of $\rho$ as $\rho=\sum_k p_k |\phi\rangle_k$. With this construction $E(x)$ is basically defined as the largest convex function smaller than $E_{\rm pure}(x)$.
\eedf  

With this construction it is defined for example one of the most used measures for bipartite systems: the {\it concurrence} \cite{wootters97}

\bbdf{\bf (Concurrence).} Given a pure bipartite state $|\phi\rangle \in \mathcal H_{\mathcal A} \otimes \mathcal H_{\mathcal B}$ the concurrence is defined as
\begin{equation}
C(|\phi\rangle) = \sqrt{2(1-\trace(\rho_{\mathcal A}^2))} \ ,
\end{equation}
where $\rho_{\mathcal A}=\trace_{\mathcal B}(|\phi\rangle \langle \phi |)$ is the reduced density matrix relative to one party and the definition is independent on the choice of the party.

For mixed states the concurrence is defined via convex roof construction.
\eedf  

Another idea to define measures of entanglement is to take a distance measure from the set of separable states.

\bbdf{\bf (Distance measures).} An entanglement monotone can be defined as
\begin{equation}
E_{\rm D}(\rho)= \inf_{\sigma \ \rm separable} D(\rho,\sigma) \ ,
\end{equation}
where $D(\cdot,\cdot)$ is a distance measure in the space of operators that is also {\it monotonic} under general operations and the infimum is taken over the set of separable states.
\eedf  

Actually not all the properties of a true distance are needed. In fact a typical distance measure is defined by taking the {\it relative entropy} function 
\begin{equation}
S(\rho \| \sigma) = \trace(\rho \log \rho - \rho \log \sigma) \ ,
\end{equation}
that is not a true distance, but an entropic distance function. Note also that a distance function need only to be
monotonic under general operations in order to obtain an entanglement monotone from it.

To conclude this discussion about entanglement measures we observe that in general entanglement measures are very difficult to compute on a given state, since they are complicated non-linear functions. Because of this sometimes it is useful to derive bounds on a certain measure of entanglement based only on linear expectation values, e.g., entanglement witnesses, or some function simpler to evaluate coming from e.g., entanglement criteria. Different methods have been derived to solve this task, and we present here just one example. The idea is that measuring just the mean value of an entanglement witness $\aver{W}=w$ one wants not just to detect the state as entangled, but also to quantify its entanglement by estimating a measure. One way is to find a lower bound $E(\rho) \geq f(w)$ on some convex measure $E(\rho)$ based on its {\it Legendre transform}.

\bbth{(Legendre transform method)} Knowing an expectation value $\aver{W}=\trace(\rho W)$ of an observable $W$ on a density matrix $\rho$, an optimal lower bound on a convex function $E(\rho)$ is given by
\begin{equation}
E(\rho) \geq \sup_{\lambda} [\lambda \aver{W}- \hat{E}(\lambda W) ] \ ,
\end{equation}
where we defined the {\it Legendre transform} of $E(\rho)$ with respect to $W$ as
\begin{equation}
\hat{E}(W) := \sup_{\rho}[\trace(W\rho)-E(\rho)] \ .
\end{equation}
Note that the observable $W$ is not needed to be an entanglement witness, nor $E(\rho)$ is needed to be an entanglement monotone.
\eeth

Apart from this, other methods can be used to estimate quantitatively the entanglement of a state based only on linear expectation values or on some non-linear measurable quantities. An important example of this last case is the variance of some observables, as we are going to see in the following discussion about multipartite entanglement detection.


\section{Entanglement detection in multipartite systems}\label{entinmulti}

\begin{figure*}[h!]
\centering
\includegraphics[width=0.7\columnwidth,clip]{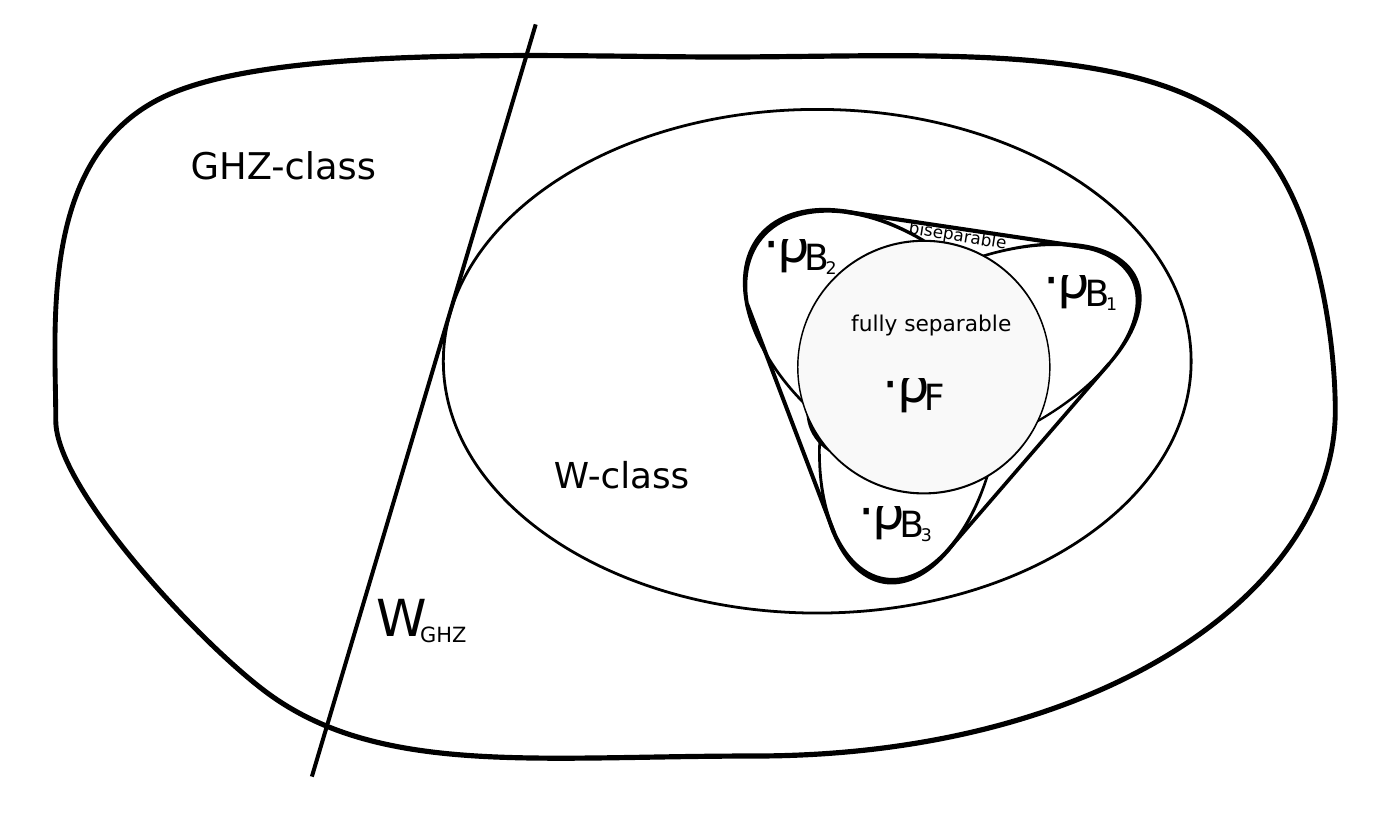}
\caption{Schematic representation of the set of tripartite states from the point of view of entanglement classification. The set of fully separable states is a subset of the set of biseparable states, that is composed by three regions, corresponding to the three different bipartitions $(A|BC,B|AC,C|AB)$. Outside the states are genuinely tripartite entangled, but still can belong to two different classes: the W-class and the GHZ-class.}
\label{fig:threent}
\end{figure*}

When the system considered is decomposed into more than two parties, then the study of quantum entanglement between all the parties becomes even much more difficult than in the bipartite case. In fact one observes from the very beginning that many different classes of entangled states can be defined. To show this consider a tripartite state $\rho$ acting on $\mathcal H_{\mathcal A}\otimes \mathcal H_{\mathcal B} \otimes \mathcal H_{\mathcal C}$. Then, the following ways to define a separable state
\begin{subequations}\label{eq:trisep}
\begin{gather}
\rho_{\rm F} =\sum_k p_k (\rho_{\mathcal A} \otimes \rho_{\mathcal{B}}\otimes \rho_{\mathcal{C}})_k \ , \label{fullysep} \\
\rho_{\rm B_1} = \sum_k p_k (\rho_{\mathcal A} \otimes \rho_{\mathcal{BC}})_k   \ , \ 
\rho_{\rm B_2} = \sum_k p_k (\rho_{\mathcal{AB}} \otimes \rho_{\mathcal{C}})_k  \ , \ 
\rho_{\rm B_3} = \sum_k p_k (\rho_{\mathcal{AC}} \otimes \rho_{\mathcal{B}})_k  \ ,  \label{bisep}
\end{gather}
\end{subequations}
with $p_k \geq 0$ and $\sum_k p_k=1$
are not equivalent. In fact, through LOCC operations, it is not possible to reach all states written as $\rho_{\rm B_i}$ starting from states of the form $\rho_{\rm F}$. The same holds for the different forms $\rho_{\rm B_i}$, for instance it is in general not possible to transform a state written as $\rho_{\rm B_1}$ into a state written as $\rho_{\rm B_2}$ through LOCC.
Therefore one must simply define different classes of entangled states: states as in Eq.~(\ref{bisep}) are called {\it biseparable}, while (\ref{fullysep}) are called {\it fully separable}. States that are not fully separable are defined as entangled and states that cannot be written in no one of the forms (\ref{eq:trisep}) are called {\it genuine tripartite entangled} (see Fig.~\ref{fig:threent}).

Furthermore, even within the genuine tripartite entangled states there are two inequivalent classes of states \cite{dur00,ghz}. For parties with $2$ dimensional Hilbert spaces (the so-called {\it qubits}) the representative states of the classes are respectively the {\it Greenberger-Horne-Zeilinger} (GHZ) state \cite{ghz}
\begin{align}\label{ghz3part}
|\Psi_{\rm GHZ}\rangle = \frac 1 {\sqrt 2} (|000\rangle + |111\rangle) \ ,
\end{align}
and the $W$-state \cite{dur00}
\begin{equation}
|\Phi_{\rm W}\rangle = \frac 1 {\sqrt 3} (|100\rangle + |010\rangle+|001\rangle) \ ,
\end{equation}
where we denoted with $|0\rangle$ and $|1\rangle$ the two basis states, as in the quantum information language \cite{keyl02}.

For systems with more parties the situation gets even more complicated, since already for $4$ parties there are infinitely many inequivalent LOCC classes, but still there remains the classification between separable, biseparable, and in general $k$-{\it separable} states

\bbdf{\bf ($k$-separable and genuine $k+1$-partite entangled states).}\label{ksepdef}
A density matrix $\rho$ of an $N$-partite system $\mathcal H_1 \otimes \dots \otimes \mathcal H_N$ is called 
$k$-{\it separable} iff it can be decomposed as
\begin{equation}\label{kseparable}
\sum_i p_i (\rho_1 \otimes \dots \otimes \rho_k)_i \ ,
\end{equation}
with $p_i \geq 0$ and $\sum_i p_i=1$ and for some partition of the system into $k$ subsystems.

A state that cannot be written in the form (\ref{kseparable}) for any partition of the system into $k$ subsystems is called {\it genuinely $(k+1)$-partite entanged}.
\eedf  

A slightly different definition is also often employed. It considers whether a state can be decomposed into a separable mixture of states involving at most $k$-parties. 

\bbdf{\bf ($k$-producible and $k+1$-entangled states).}\label{kproddef}
A density matix $\rho$ of an $N$-partite system $\mathcal H_1 \otimes \dots \otimes \mathcal H_N$ is called 
$k$-{\it producible} iff it can be decomposed as
\begin{equation}\label{kproducible}
\sum_i p_i (\rho_1 \otimes \dots \otimes \rho_M)_i \ ,
\end{equation}
with $p_i \geq 0$ and $\sum_k p_i=1$ and where the states $\{\rho_1,\dots,\rho_M\}$ are at most $k$-partite.

Any state that cannot be written in the form (\ref{kproducible}) is called $k+1$-entangled.
\eedf 

Note that although the two definitions above are different, e.g., states that are $3$-entangled are in general different from genuine $3$-partite entangled states, while states that are $N$-entangled are also genuine $N$-partite entangled and viceversa.
Given these definitions and the fact that there are infinitely many different classes of entangled states one can derive criteria that aim to detect specific classes of entangled states or specific degrees of multipartite entanglement. For example, a criterion that can rule out full separability on a state is a generalization of the CCNR criterion \cite{wocjan05,clarisse06,horodecki06}. 

\bbth{(Permutation criteria)} Consider an $N$-partite density matrix $\rho$ expanded in a certain basis
\begin{equation}
\rho=\sum_{i_1,j_1,\dots,i_N,j_N} \rho_{i_1,j_1,\dots,i_N,j_N} |i_1\rangle\langle j_1| \otimes \dots \otimes |i_N\rangle\langle j_N| \ ,
\end{equation}
and consider an arbitrary permutation of the indices $\pi(i_1,j_1,\dots,i_N,j_N)$. Then for all fully separable states 
\begin{equation}
\| \rho_{\pi(i_1,j_1,\dots,i_N,j_N)} \|_{1} \leq 1 
\end{equation}
holds.
\eeth

More generally one can find criteria based on positive but not completely positive maps, trace preserving maps or entanglement witnesses and generalize them to the multipartite case as constraints on full separability. Or one can consider for example a certain (or all) bipartition(s) and check bipartite criteria. Moreover also the correspondence between positive maps and entanglement witnesses can be exploited and generalized for the multipartite case. 
Even more specifically for many experiments it is interesting to derive criteria designed to detect entanglement in the vicinity of a given target state. This is done easier by considering entanglement witnesses optimized to detect such target states. An example can be given by a witness for GHZ-like states. 

\bbth{(GHZ entanglement witness)} Consider a tripartite system $\mathcal H_{\mathcal P_1} \otimes \mathcal H_{\mathcal P_2} \otimes \mathcal H_{\mathcal P_3}$. The operator
\begin{equation}\label{witghz3}
W_{\rm GHZ} = \frac 3 4 \id - |\Psi_{\rm GHZ}\rangle \langle \Psi_{\rm GHZ}| \ ,
\end{equation}
where $|\Psi_{\rm GHZ}\rangle \langle \Psi_{\rm GHZ}|$ is the projector onto the $3$-partite GHZ state (\ref{ghz3part}) is an optimal entanglement witness that detects states close to the GHZ state. 
\eeth

Analogously as Eq.~(\ref{witghz3}) other witnesses can be constructed to detect for instance genuine tripartite states or states in the vicinity of a given target $|\psi\rangle$. Furthermore, there can be defined measures of entanglement for a multipartite scenario. As usual a straightforward way is to consider bipartite measures and either compute them for some bipartitions or generalized them for a multipartite state. On the other hand there are also measures specifically defined to a multipartite setting. An important example is the so-called {\it three-tangle} \cite{coffman00}, that is a measure of entanglement for $3$-partite states.

\bbdf{\bf (three-tangle).} For any pure $3$-partite state $|\phi\rangle$ in $\mathcal H_{1} \otimes \mathcal H_{2} \otimes \mathcal H_{3}$ the three-tangle $\tau$ is defined as
\begin{equation}
\tau(|\phi\rangle\langle \phi|)=C^2(|\phi\rangle_{1|23}) - C^2(\rho_{12}) - C^2(\rho_{13}) \ ,
\end{equation}
where $C^2(\cdot)$ is the square of the concurrence, $\rho_{ij}$ are the bipartite reduced states relative to parties $i$,$j$ and $|\phi\rangle_{1|23}$ is the state considered as $1|23$ bipartite, i.e., a vector of the space $\mathcal H_{1} \otimes \mathcal H_{2,3}$, $\mathcal H_{2,3}=\mathcal H_{2} \otimes \mathcal H_{3}$. For general density matrices $\rho \in \mathcal B(\mathcal H_{1} \otimes \mathcal H_{2} \otimes \mathcal H_{3})$ the three-tangle is defined via the convex roof construction.
\eedf 

This is a definition based on the concurrence, that tries to generalize it to the multiparty setting. Other measures can be given to quantify the degree of multipartite entanglement, as in Defs.~\ref{ksepdef},\ref{kproddef}. For example one can take a distance measure from the state to the set of $k$-separable states, as in the definition of the {\it geometric entanglement} \cite{wei03}

\bbdf{\bf (Geometric entanglement).} For pure multipartite states $|\psi\rangle$ the geometric entanglement is defined as 
\begin{equation}
E^{\rm (k)}_G(|\psi\rangle) = 1 - \sup_{|\phi\rangle \in k-\mbox{\rm sep}} |\langle \phi | \psi\rangle|^2 \ ,
\end{equation}
where the supremum over $|\phi\rangle$ is taken over the set of $k$-separable states.

For general density matrices the geometric entanglement is defined through the convex roof construction. 
\eedf 

After briefly reviewing some different methods of entanglement detection in multipartite systems, we now focus on a method to write entanglement criteria based on the violation of Local Uncertainty Relations (LURs) in finite-dimensional systems. 
Especially when applied to detect entanglement in systems of very many particles
it reveals itself as a simple but very powerful method because allows to derive separability inequalities that are relatively easy to check experimentally.
This idea was introduced in \cite{hofman03} and was developed in several ways \cite{devicente05,gittsovich08,giovannetti04,Guhne2004Characterizing,guhnemechler,guhnelew,guhnecova},
among which there is also the derivation of some generalized spin squeezing inequalities \cite{tothpra04,He2011Planar}. 

We will follow Ref.~\cite{hofman03} 
and write a simple uncertainty relation that can be violated only by entangled states. The idea has some analogies with the original EPR-Bell approach and can be understood considering pairs of non-commuting single particle observables $(a_1,a_2)$, $(b_1,b_2)$ in a bipartite system $\mathcal H_{\mathcal A}\otimes \mathcal H_{\mathcal B}$. Since the $a_i$, as well as the $b_i$ do not commute with each other, they don't have a common eigenstate and their uncertainties cannot be both zero. However in the joint system, the uncertainties of $M_i=a_i\otimes \id + \id \otimes b_i$ can be both zero, but this must be associated to entanglement in the state. Thus an entanglement criterion can be derived based on the sum of variances as follows.

\bbth{(Entanglement detection with LURs)}\label{entdetwithlurth} Consider a bipartite system $\mathcal H_{\mathcal A}\otimes \mathcal H_{\mathcal B}$ and two pairs of observables $(a_1,a_2)$ acting on $\mathcal H_{\mathcal A}$, $(b_1,b_2)$ acting on $\mathcal H_{\mathcal B}$ such that 
\begin{equation}\label{uncab}
\va{a_1} + \va{a_2} \geq U_a \ , \qquad \va{b_1} + \va{b_2} \geq U_b \ ,
\end{equation}
holds. Then, considering the joint observables $M_i=a_i\otimes \id + \id \otimes b_i$, the inequality
\begin{equation}\label{entcritlur1}
\va{M_1}_\rho + \va{M_2}_\rho \geq U_a + U_b 
\end{equation}
must hold for all separable states $\rho$. A violation of Eq.~(\ref{entcritlur1}) implies entanglement.
\eeth

\bbpr Consider first a product state $\rho_{\rm p}=\rho_{\mathcal A}\otimes \rho_{\mathcal B}$. We have that 
\begin{equation}
\va{M_i}_{\rho_{\rm p}} = \va{a_i}_{\rho_{\mathcal A}} + \va{b_i}_{\rho_{\mathcal B}} \ ,
\end{equation}
and thus due to Eq.~(\ref{uncab})
\begin{equation}
\va{M_1}_{\rho_{\rm p}} + \va{M_2}_{\rho_{\rm p}} \geq U_a + U_b  \ .
\end{equation}
Then, we exploit the fact that the variance is a concave function of density matrices, i.e., $\va{A}_{\rho} \geq \sum_k p_k \va{A}_{\rho_k}$ for $\rho=\sum_k p_k \rho_k$ with $p_k\geq 0$, $\sum_k p_k=1$ and all observables $A$. Thus we have
\begin{equation}
\va{M_1}_{\rho} + \va{M_2}_{\rho} \geq \sum_k p_k [\va{M_1}_{(\rho_{\rm p})_k} + \va{M_2}_{(\rho_{\rm p})_k}] \geq U_a + U_b
\end{equation}
for all separable states $\rho=\sum_k p_k (\rho_{\rm p})_k$, $p_k \geq 0$, $\sum_k p_k=1$.
\eepr

Note that there is the advantage that the bounds $U_a, U_b$ in Eq.~(\ref{uncab}) need to be computed on pure states only, since the variance is a concave function of density matrices. Although in general they could still be very difficult to find, and in practice in many cases have to be found numerically, there are also interesting cases in which $U_a, U_b$ are straightforwardly computed analytically. In fact consider the three pauli matrices $\{\sigma_k \}_{k=x,y,z}$ on a single spin-$1/2$ particle system (qubit). They have to satisfy the following LUR
\begin{equation}
\sum_k \va{\sigma_k} \geq 2 \ .
\end{equation}
Thus in this case, defining the joint observables $\Sigma_k=\sigma_k \otimes \id + \id \otimes \sigma_k$, the following entanglement criterion 
\begin{equation}\label{sigmaunccrit}
\sum_k \va{\Sigma_k}_\rho \geq 4 \ ,
\end{equation}
is easily derived as in the previous theorem \ref{entdetwithlurth}. Moreover it can be easily seen that Eq.~(\ref{sigmaunccrit}) gives a {\it non-linear improvement} over the optimal entanglement witness
\begin{equation}
W=\id \otimes \id + \sum_k \sigma_k \otimes \sigma_k \ .
\end{equation}
Hence we can see that LURs are a tool to derive non-linear entanglement criteria that can detect more entangled states requiring the same measurements as linear witnesses. Furthermore, such entanglement criteria can be used also to provide experimentally measurable bounds on entanglement monotones.

\subsection{Entanglement detection with collective observables} 

Many experimental systems aiming at producing entangled states are composed of very many particles, that moreover cannot be individually addresses, either because they are indistinguishable or for a practical impossibility. The easiest and almost only quantities that can be measured in such experiments are collective observables of the ensemble. For example, in an ensemble of $N$ spin-$j$ atoms the quantities that can be measured are the collective spin operators $J_k=\sum_{n=1}^N j_k^{(n)}$.
Thus, for those situations it is important to derive entanglement criteria based on expectation values of (possibly few) collective quantities, such as $\aver{J_k}$ and $(\Delta J_k)^2$. 

For this task the previously mentioned criteria based on uncertainty relations are one of the main tools. In fact let us consider two collective operators
$A_1=\collop{a_1}{N}$ and $A_2=\collop{a_2}{N}$. Then the following theorem provides an uncertainty relation that can be violated only by entangled states and is based on few collective measurements.

\bbrmk
Note that having the possibility to measure collective quantities might already imply that the parties are not space-like separated, and thus the states detected with these methods are not violating classical principles such as local-realism. Also, for systems of identical particles the formal definition of non-separability raises interpretative problems due to the forbidden possibility to address individually the subsystems. Nevertheless we refer here to entanglement intended as a resource for quantum information processing, as a figure of merit to certify the non-classicality of the state in an information-theoretic perspective.
\eermk

\bbth\label{uncssi} 
Every fully separable state $\rho_{\rm sep}$ of $N$ particles must satisfy
\begin{equation}
\va{A_1}_{\rho_{\rm sep}} + \va{A_2}_{\rho_{\rm sep}} \geq NU \ ,
\end{equation}
where the constant $U$ is such that the LUR $\va{a_1} + \va{a_2} \geq U$ holds for all single particle states.
Thus, every state $\rho$ such that 
\begin{equation}
\va{A}_{\rho} + \va{B}_{\rho} < NU 
\end{equation}
must be entangled.
\eeth

\bbpr
Let us consider a pure product state $\rho_{\rm prod}= \bigotimes_{n=1}^{(N)} \rho^{(n)}$.
Since $A_1=\collop{a}{N}$ and the variance is additive for a product state we have $\va{A_1}_{\rho_{\rm prod}}=\sum_n \va{a_1}_{\rho^{(n)}}$. The same holds, for $\va{A_2}$, namely $\va{A_2}_{\rho_{\rm prod}}=\sum_n \va{a_2}_{\rho^{(n)}}$. Thus, exploiting the LUR we have that for a product state 
\begin{equation}
\va{A_1}_{\rho_{\rm prod}}+ \va{A_2}_{\rho_{\rm prod}} \geq  \sum_n U = NU 
\end{equation}
holds. Finally the bound can be extended also to mixed separable states
\begin{equation}
\rho_{\rm sep}= \sum_k p_k \rho_{k, \rm prod}  \quad p_k > 0 \ \ , \ \sum_k p_k=1 
\end{equation}
due to the concavity of the variance $\va{A}_{\rho_{\rm sep}} \geq \sum_k p_k \va{A}\rho_{k, \rm prod}$.
\eepr
Theorem \ref{uncssi} can be straightforwardly extended to an arbitrary number of non-commuting observables $\{A_1, \dots ,A_M\}$, namely
\begin{equation}\label{Mobsssi}
\sum_{k=1}^M \va{A_k}_{\rho_{\rm sep}} \geq NU \ ,
\end{equation}
holds for all separable states, provided that a local uncertainty relation of the form $\sum_{k=1}^M a_k \geq U$ holds for the single particle operators.

As the main practical example for our purposes we mention the following \cite{tothpra04}

\bbth{($SU(2)$ invariant Spin Squeezing Inequality.)}\label{su2invcrit} 
Every separable state $\rho_{\rm sep}$ of $N$ spin-$j$ particles must satisfy
\begin{equation}\label{su2invssi}
\va{J_x}_{\rho_{\rm sep}}+\va{J_y}_{\rho_{\rm sep}}+\va{J_z}_{\rho_{\rm sep}} \geq Nj \ ,
\end{equation}
where $J_k=\sum_{n=1}^N j_k^{(n)}$ are the collective spin components and $(j_x^{(n)})^2+(j_y^{(n)})^2+(j_z^{(n)})^2=j(j+1)$ for all particles $(n)$.

Every state $\rho$ such that 
\begin{equation}
\va{J_x}_{\rho}+\va{J_y}_{\rho}+\va{J_z}_{\rho} < Nj
\end{equation}
must be entangled.
\eeth

\bbpr
Let us consider three orthogonal spin directions $j_x$,  $j_y$ and $j_z$. Since $\sum_k  j_k^2 = j(j+1) \id$ and $\sum_k \aver{j_k}^2 \leq j^2$ we have that 
\begin{equation}
\va{j_x}+\va{j_y}+\va{j_z} \geq j 
\end{equation}
must hold for every single particle state. Thus Eq.~(\ref{su2invssi}) follows directly from Th.~\ref{uncssi}, generalized for 3 observables, as in Eq.~(\ref{Mobsssi}). 
\eepr

We called it Spin Squeezing Inequality because it detects states such that one or more collective spin variances $\va{J_k}$ are {\it squeezed}, with respect to the bound given by the right-hand side of Eq.~(\ref{su2invssi}). In the next section we will more properly define spin squeezed states as they were introduced in the literature and give some generalized definitions in the next chapter. 

The original spin squeezing parameter itself $\xi^2:=\frac{N(\Delta J_z)^2}{\aver{J_x}^2+\aver{J_y}^2}$ is an expression that provides an entanglement criterion with collective observables based on LURs. In fact S\o rensen {\it et al.} \cite{Sorensen2001Many-particle} proved the following

\bbth{(Original Spin Squeezing entanglement criterion.)}
Every separable state $\rho_{\rm sep}$of $N$ spin-$\frac 1 2$ particles must satisfy
\begin{equation}\label{oriSSIent}
\xi^2(\rho_{\rm sep}):=\frac{N(\Delta J_z)_{\rho_{\rm sep}}^2}{\aver{J_x}_{\rho_{\rm sep}}^2+\aver{J_y}_{\rho_{\rm sep}}^2} \geq 1 \ ,
\end{equation}
where $J_k=\frac 1 2 \sum_{n=1}^N \sigma_k^{(n)}$ are the collective spin components.

Every state $\rho$ such that 
\begin{equation}
\xi^2(\rho) < 1
\end{equation}
must be entangled and is also-called {\it Spin Squeezed State} (SSS).
\eeth

\bbpr Let us consider the expression $N(\Delta J_z)^2-\aver{J_x}^2-\aver{J_y}^2$. For product states $\rho_{\rm prod}= \bigotimes_{n=1}^{(N)} \rho^{(n)}$ we have that
\begin{equation}
N(\Delta J_z)^2-\aver{J_x}^2-\aver{J_y}^2 \geq N \sum_n \left(\aver{(j_z^{(n)})^2}-\frac 1 4 \right) = 0
\end{equation}
follows from the Cauchy-Schwarz inequality and the LUR $\va{j_x}+\va{j_y}+\va{j_z} \geq \frac 1 2$. Then, since the left hand side is a concave function of density matrices, the bound on the right-hand side also holds for all separable states and Eq.~(\ref{oriSSIent}) follows.
\eepr

So far the number of particles in the system has been considered fixed, and thus in particular it was a discrete number. However in actual experiments the total particle number might be fluctuating. Thus a natural extension to make would be to drop the fixed particle number assumption and consider states that are mixtures or even coherent superpositions of different number of particles \cite{hyllus12}.  
In principle, in fact, following the rules of quantum mechanics one should allow for coherent superpositions of states with different $N$, and consider a Hilbert space like $\mathcal H=\bigotimes_N \mathcal H_N$, where $\mathcal H_N$ is the state space of fixed-$N$ particles. 
However in practice coherences between different particle numbers are not observed and there is an ongoing debate on whether they are even in principle allowed.
Usually an assumption called {\it superselection rule} is employed, that forbids to have superpositions of states with different particle numbers.

\bbasm{\bf (Superselection rule on the particle number).}
A physical state cannot be in a coherent superposition of states with different number of particles.
\eeasm

In the context of entanglement detection with collective observables we also employ this assumption and restrict the framework to spaces like $\mathcal H=\bigoplus_N \mathcal H_N$
rather than the full Hilbert space. This also means that all the observables that we consider as physical must be in a reducible representation with respect to the particle number, i.e., they must be of the form $O=\bigoplus_{N=0}^\infty O_{N}$, where $O_{N}$ act on the fixed-$N$ particle space $\mathcal H_N$. In particular the collective spin components have the form
\begin{equation}\label{collspinvarN}
J_k = \bigoplus_{N=0}^\infty J_{k, \rm N} \ ,
\end{equation}
where $J_{k, \rm N}=\sum_{n=1}^N j_k^{(n)}$ are the fixed-$N$ particle observables.
Effectively, what we are doing is to consider all the projections $\rho \rightarrow \Pi_{\rm N} \rho \Pi_{\rm N}$ of the state $\rho$ belonging to the full $\mathcal H$ onto $\mathcal H_N$ for all $N$ and reducing the most general state to a mixture such as 
\begin{equation}\label{eq:rhomixNpar}
\rho= \sum_N Q_N \rho_{\rm N}  \quad Q_N \geq 0 \ , \quad \sum_N Q_N = 1 \ ,
\end{equation}
where $\rho_{\rm N}\propto \Pi_{\rm N} \rho \Pi_{\rm N}$ is an $N$-particle state (i.e., it belongs to $\mathcal H_N$). In this framework we extend the definitions of separable and more in general $k$-producible states to states that are $k$-producible in every fixed-$N$ subspace \cite{hyllus12}.

\bbdf{\bf ($k$-producible states of fluctuating number of parties).}\label{kproddeffluctN}
A density matix $\rho$ of a system with a fluctuating number of particles  $\mathcal H=\bigoplus_N \mathcal H_N$ is called 
$k$-{\it producible} iff the reduced state $\rho_N$ corresponding to every subspace $\mathcal H_N$ can be decomposed as in Eq.~(\ref{kproducible}).

Otherwise $\rho$ is called $(k+1)$-entangled.
\eedf 

Here it is also implicitly assumed that the state has probabilities $Q_N$ different from zero only when $N\geq k$ and usually one can also safely assume that $Q_N\neq 0$ only for $N \gg k$. 
According to this definition one can also try to extend the various criteria that we have discussed to the case of fluctuating number of parties. At least in the case of criteria based on LURs this extension can be done relatively easily by exploiting the concavity of the variance, namely 
\begin{equation}
(\Delta J_k)^2 \geq \sum_N Q_N (\Delta J_{k, \rm N})^2 \ ,
\end{equation}
as shown by Hyllus {\it et al} \cite{hyllus12} in the context of spin squeezing inequalities. We will see more details about this in the next chapter. However note that the same extension can be possibly done also for other criteria based on convex (or concave) functions.


\section{Spin Squeezing}\label{spsquesec}

In this section we will review briefly the very wide theory of Spin Squezeed States (SSS)\cite{Kitagawa1993Squeezed,Wineland1994Squeezed,Sorensen2001Many-particle,Ma2011Quantum}  and some of their potential technological applications \cite{Kitagawa1993Squeezed,Wineland1994Squeezed,Giovannetti2011Advances,tothapellaniz14}. 
These states are usually created in systems composed of very many particles: some collective spin operators $J_k$ are manipulated and entanglement can be created and detected with such observables. In particular then it will be necessary to use criteria for entanglement detection that are based on collective observables. Interestingly, such criteria are provided by the defining parameters themselves \cite{Sorensen2001Many-particle}. These last being directly connected with the maximal precision achievable in some very general metrological protocols \cite{Kitagawa1993Squeezed,Wineland1994Squeezed,Giovannetti2011Advances,tothapellaniz14}. 
This is thus a practical example in which the information-theoretic concept of entanglement can be exploited as a resource and because of this both the study of SSS and of collective observables based entanglement criteria is attracting a lot of attention, especially since a couple of decades, in which experimental manipulation of different many particle systems led to the possibility of actually creating and detecting them \cite{appel09,Esteve2008Squeezing,Fernholz2008Spin,Gross:2010aa,jo07,Julsgaard:2001aa,KoschorreckPRL2010b,Kuzmich1998Atomic,li07,LouchetChauvet10,orzel01,Riedel2010Atom-chip-based,SchleierSmith10,vengalatorre}.

\subsection{Definition of generalized coherent states}

The definition of spin squeezed states can be given with some analogy with a widely studied class of single particle states called generalized squeezed states \cite{perelomov86}. The latter are mathematically defined considering the action of a certain group of transformations on the Hilbert space of single particle states. Then, coherent states relative to a reference state are obtained through a group transformation that acts irreducibly on the Hilbert space. A complete treatment of the theory of coherent states is far beyond the scope 
of this thesis. However we give here a simple introduction in order to include the SSS into the general framework of squeezed states, as the name itself would suggest. The first concept of squeezed states has been introduced already by Schr\"odinger in 1926 \cite{schrodinger26}, the name termed later by Glauber \cite{glauber63} in 1963. Rephrased in modern terms \cite{zhang90,bengtsson06} it is referred to states that are obtained from a certain {\it vacuum} state $|\emptyset \rangle$ of an infinitely dimensional hilbert space $\mathcal H_{\infty}$ through the action of the so-called {\it Heisenberg-Weyl} group. The latter is defined by exponentiating the Lie algebra generated by three operators $a,a^\dagger,\id$ with the following commutation relations
\begin{equation}\label{Heiweylalg}
[a,a^\dagger]= \id \ , \qquad [a,\id]=[a^\dagger,\id]=0 \ ,
\end{equation}
defined by means of the relation $a|\emptyset\rangle=0$. Thus, starting from the vacuum (or reference) state $|\emptyset\rangle$ we define the {\it coherent states} as follows.

\bbdf{\bf (Coherent States).} 
A coherent state $|z\rangle$ can be defined as an eigenstate of the annihilation operator $a$ with eigenvalue $z$, namely
\begin{equation}
a|z\rangle=z|z\rangle \ ,
\end{equation}
where $z$ is a complex number. It can be obtained from the vacuum state $|\emptyset \rangle$, i.e., the state such that $a|\emptyset \rangle=0$, through the transformation
\begin{equation}
|z\rangle=\exp(za^\dagger-z^* a) |\emptyset \rangle \ ,
\end{equation}
that is a transformation belonging to the group generated by the algebra (\ref{Heiweylalg}).
\eedf 

Now let us briefly list some important properties of the coherent states just defined. First of all they form an overcomplete basis of the Hilbert space. In fact the set of projectors onto all coherent states is a resolution of unity
\begin{equation}
\frac 1 \pi \int dz_1 dz_2 |z\rangle\langle z|  = \id \ ,
\end{equation}
where $z=z_1+iz_2$. This in particular means that the set of projectors $\{ |z\rangle\langle z| \}$ can be viewed as a POVM (see definition in the previous chapter). The coherent states also saturate the Heisenberg uncertainty relation for the operators $(x,p)=(\frac{a+a^\dagger}{\sqrt 2},\frac{a-a^\dagger}{i\sqrt 2})$, namely
\begin{equation}\label{satuheis}
(\Delta x)^2(\Delta p)^2 = \frac{|\aver{[x,p]}|^2} 4 = \frac 1 4 \ , 
\end{equation}
with the additional property that $(\Delta x)^2=(\Delta p)^2=\frac 1 2$. 

Starting from (\ref{satuheis}) a wider class of states can be defined, namely states satisfying Eq.~(\ref{satuheis}) but such that the two variances might not be equal between each other. These states are in fact called {\it squeezed}, in the sense that one variance is squeezed as compared to a coherent state.

\bbdf{\bf (Squeezed States).} 
A squeezed state $|z, \eta \rangle$ is defined as the result of applying a squeezing operator
\begin{equation}\label{squeezingoperator}
S(\eta)=\exp\left( \frac 1 2 \eta^* a^2 - \frac 1 2 \eta (a^\dagger)^2\right)
\end{equation}
to a coherent state $|z\rangle$, namely
\begin{equation}\label{squeezedstate}
|z, \eta \rangle = S(\eta)|z\rangle \ .
\end{equation}
Squeezed states saturate the Heisenberg uncertainty relation $(\Delta x)^2(\Delta p)^2 = \frac 1 4$ and are in general such that $(\Delta x)^2\neq(\Delta p)^2$.
\eedf 

In analogy with the previous, there can be constructed generalized coherent states based on the action of other groups of transformations \cite{perelomov86}. In particular we are interested in coherent states constructed with the action of $SU(2)$ rotations on a finite dimensional Hilbert space, namely the space of a spin-$j$ particle.
In this case the algebra that generates the group is $3$-dimensional, the basis $(J_z,J_+,J_-)$ satisfying the commutation relations
\begin{equation}\label{su2commu}
[J_z,J_\pm]=\pm J_\pm \ \ , \qquad [J_+,J_-]= J_z \ .
\end{equation}
We take as reference a state $| j,j\rangle$ such that $J_+| j,j\rangle=0$ and define the $SU(2)$ coherent states by applying to it a
{\it Wigner rotation matrix} 
\begin{equation}\label{wigrotmat}
\mathcal D(\theta,\phi)= \exp(\zeta J_+ - \zeta^* J_-) = \exp(\eta J_+) \exp\left[\ln(1+|\eta|^2) J_z\right] \exp(-\eta J_+) \ ,
\end{equation}
where $\zeta=-\frac{\theta} 2 \exp(-i\phi)$, and $\eta= -\tan \frac{\theta} 2 \exp(-i\phi)$ are complex numbers bringing the dependence on two angles $(\theta, \phi)$. Here the analogy of the role played by $J_\pm$ and previously by the creation annihilation operators $a,a^\dagger$ is clear. Furthermore, operators analogous to $(x,p)$ are defined as $(J_x,J_y)=(\frac{J_++J_-}{\sqrt 2},\frac{J_+-J_-^\dagger}{i\sqrt 2})$ satisfying, together with $J_z$, the usual spin components commutation relations $[J_a,J_b]=i\varepsilon_{abc} J_c$. 

In terms of the spin operators the rotation matrix is the following explicit function of the angles
\begin{equation}
\mathcal D(\theta,\phi)=\exp\left[i\theta(\sin \phi J_x - \cos \phi J_y) \right] \ .
\end{equation}

Thus we define the {\it $SU(2)$-coherent states} as follows.

\bbdf{\bf ($SU(2)$-coherent states or Coherent Spin States (CSS)).} 
An $SU(2)$-coherent state $|\theta, \phi \rangle$ is defined as the result of applying a rotation matrix $\mathcal D(\theta,\phi)$ to the completely $z$ polarized state $| j,j\rangle$
\begin{equation}
|\theta, \phi \rangle = \mathcal D(\theta,\phi)| j,j\rangle \ ,
\end{equation}
where $\mathcal D(\theta,\phi)$ is given in Eq.~(\ref{wigrotmat}) and the state $| j,m\rangle$ is written in terms of the eigenvalues of the Casimir operator $J_x^2+J_y^2+J_z^2=j(j+1)\id$ and $J_z| j,m\rangle=m| j,m\rangle$.
The state $|\theta, \phi \rangle$ is completely polarized along the direction $\hat{n}_0=(\sin \theta \cos \phi,\sin \theta \sin \phi, \cos \theta)$, i.e., $J_{\hat{n}_0}|\theta, \phi \rangle=j|\theta, \phi \rangle$.

\begin{figure*}[h!]
\centering
\includegraphics[width=0.8\textwidth,clip]{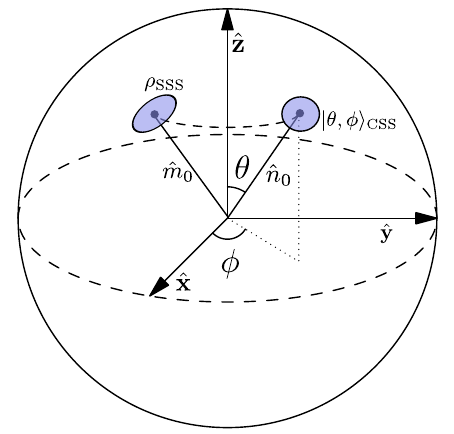}
\caption{Picture of a spin-$J$ Bloch sphere and (i) a Coherent Spin State, completely polarized along a direction $\hat n_0$ with circular uncertainty region in an orthogonal plane and (ii) a Spin Squeezed State, completely polarized along $\hat m_0$ with elliptical uncertainty region, squeezed in a direction orthogonal to $\hat m_0$.}\label{fig:blochsphere}
\end{figure*}

They form an overcomplete basis of a spin-$j$ particle Hilbert space, since
\begin{equation}
\frac{2j+1}{4\pi} \int d\Omega |\theta, \phi\rangle\langle \theta, \phi|  = \id \ ,
\end{equation}
is a resolution of the identity satisfied by the set of projectors $\{|\theta, \phi\rangle\langle \theta, \phi| \}$.
They saturate the uncertainty relation
\begin{equation}\label{spinunc}
(\Delta J_a)^2(\Delta J_b)^2=\frac{|\aver{[J_a,J_b]}|^2} 4 = \frac{\aver{J_c}^2} 4 \ ,
\end{equation}
for a certain choice of the axis $\hat{a},\hat{b},\hat{c}$, namely the one such that $\langle \theta, \phi|J_c|\theta, \phi \rangle=j$, i.e., $\hat{c}=\hat{n}_0$.
They also saturate the $SU(2)$ invariant uncertainty relation
\begin{equation}\label{eq:su2invheisunc}
(\Delta J_x)^2+(\Delta J_y)^2+(\Delta J_z)^2=j \ ,
\end{equation}
independently on the direction of polarization. For $j=\frac 1 2$ all the pure states are CSS and  define what is called {\it Bloch sphere}, namely a 2-dimensional sphere in which
every pure state can be identified with a point corresponding to the MSD $\hat{n}_0$ (see Fig.~(\ref{fig:blochsphere})).
\eedf 

As a further analogy we can see that in the limit $j\rightarrow \infty$ the $su(2)$ algebra becomes completely analogous to the Heisenberg-Weyl's. This relation can be made more precise with a mapping from a single bosonic mode operators $(a^\dagger,a)$, to the three spin direction operators. This is called {\it Holstein-Primakoff transformation} \cite{holstein40}.

\bboo{(Holstein-Primakoff transformation).}\label{obs:holprimtransf}
Let us consider the creation/annihilation operators of a single bosonic mode $(a^\dagger,a)$ and an integer number $j$. Then, the following mapping
\begin{equation}
\begin{aligned}\label{holprimtrans}
J_{+} &:=  \sqrt{2j-a^\dagger a} a \ , \\
J_- &:=  a^\dagger \sqrt{2j-a^\dagger a} \ , \\
J_z &:= j - a^\dagger a
\end{aligned}
\end{equation}
defines three operators $J_+, J_-, J_z$ that obey the $su(2)$ commutation relations (\ref{su2commu}). Eq.~(\ref{holprimtrans}) is called Holstein-Primakoff transformation and is
well defined whenever $a^\dagger a \leq 2j$. 
\eeoo

Expanding Eq.~(\ref{holprimtrans}) in a sort of Taylor series, whenever $a^\dagger a$ can be considered ``small'' as compared to $j$, we have that $J_z\simeq j$ can be considered constant and consequently
\begin{equation}\label{Heiweylalglimj}
[J_+,J_-]\simeq 2j \id \ , \qquad [J_+,J_z]= [J_+,J_z]\simeq 0 \ ,
\end{equation}
hold in that limit, reproducing the commutation relations of the Heisenberg-Weyl algebra.

We have defined in close analogy coherent states coming from the Heisenberg-Weyl and the $SU(2)$ groups. 
All of these have the property of saturating a Heisenberg uncertainty relation.
However for the purpose of defining squeezed states based on such uncertainty relation there is one main difference. In this last case the right-hand side of Eq.~(\ref{spinunc}) is not constant, but depends on the state itself. 

In the following we introduce the first definition of spin squeezing proposed and its connection with entanglement and quantum enhanced metrology. 

\subsection{Original definition of Spin Squeezed States, metrology and entanglement}

A first definition of spin squeezing was given by Kitagawa and Ueda \cite{Kitagawa1993Squeezed} considering that $|\theta, \phi \rangle$ 
is completely polarized along some direction $\hat{n}_0$. They defined the parameter
\begin{equation}\label{kituepar}
\xi^2_{\rm S}=\frac{2 \min_{\hat{n}_{\bot}}(\Delta J_{\hat{n}_{\bot}})^2}{|\aver{J_{\hat{n}_0}}|} = \frac{2 \min_{\hat{n}_{\bot}}(\Delta J_{\hat{n}_{\bot}})^2}{j} \ ,
\end{equation}
where $\hat{n}_{\bot}$ is a direction orthogonal to the {\it mean spin direction} $\hat{n}_0$. Thus, all pure CSS must satisfy $\xi^2_{\rm S}=1$ and therefore a state can be defined {\it Spin Squeezed} whenever $\xi^2_{\rm S}<1$. 

Later, the same definition was considered by Wineland {\it et al.} \cite{Wineland1994Squeezed} and proved to be 
connected to an improved sensitivity of the state under rotations. In fact let us consider a state such that the mean spin direction is $\hat{z}$ and assume to measure the operator $J_x$ after a rotation of $\varphi$ about the $\hat{y}$ axis. We have
\begin{equation}
J_x^{\rm (out)}=\exp(i\varphi J_y) J_x \exp(-i\varphi J_y) \ ,
\end{equation}
and from the error propagation formula $\Delta \varphi = \frac{\Delta f(\varphi)}{|\partial f(\varphi)/\partial \varphi|}$ we can compute 
the uncertainty in the estimation as 
\begin{equation}\label{eq:naivprec}
\Delta \varphi = \frac{\Delta J_x^{\rm (out)}}{|\partial \aver{J_x^{\rm (out)}}/\partial \varphi|}= \frac{\Delta J_x^{\rm (out)}}{|\cos \varphi \aver{J_z}|}
\end{equation}
obtaining $\Delta \varphi \simeq \frac{\Delta J_x}{|\aver{J_z}|}$ for small rotations $\varphi \sim 0$, since in that case $\Delta J_x^{\rm (out)} \sim \Delta J_x$. Thus for a CSS we have $(\Delta \varphi)_{\rm CSS}^2 = \frac{1}{2j}$ and therefore Eq.~(\ref{kituepar}) can be expressed as
$\xi^2=\frac{(\Delta \varphi)^2}{(\Delta \varphi)_{\rm CSS}^2}$, meaning that a SSS is more sensitive to small rotations around an axis perpendicular to the mean spin direction than a CSS.
Again, note that a single particle with $j=\frac 1 2$ cannot be spin squeezed.

Furthermore, noting that a spin-$j$ particle can be always seen as a collective symmetric state of $N$ spin-$\frac 1 2$ particles, one can extend the same definition (\ref{kituepar}) to states composed of many spin-$\frac 1 2$ particles. Thus, considering a spin-$j$ particle as a symmetric state of $N$ spin-$\frac 1 2$ particles we define a collective spin squeezed state based on a parameter reformulated in a slightly more general way $\xi^2_{\rm N}=\frac{N \min_{\hat{n}_{\bot}}(\Delta J_{\hat{n}_{\bot}})^2}{|\aver{J_{\hat{n}_0}}|^2}$, such that 
the same definition can be used for all collective states of $N$ qubits, even if not in the symmetric subspace. In that case the spin quantum number $j$ is not fixed, but we have $0\leq j \leq \frac N 2$. Moreover, as we have seen in the previous section (cf. Eq.~\ref{oriSSIent}) S\o rensen {\it et al.} \cite{Sorensen2001Many-particle} proved that all separable states of $N$ spin-$\frac 1 2$ particles must satisfy $\xi^2_{\rm N^\prime}\geq 1$ for a similar parameter $\xi^2_{\rm N^\prime}$. Therefore with such a definition, the spin squeezing parameter is an entanglement criterion based on collective observables. Thus exploiting these connections between Heisenberg uncertainty relations, metrology and entanglement we give the following definition of {\it Spin Squeezed States}, that is valid also for collective states of $N$ spin-$\frac 1 2$ particles (see also Fig.~\ref{fig:blochsphere} for a picture).

\bbdf{\bf (Spin Squeezed States (SSS)).} 
A state $\rho$ is defined as Spin Squeezed in the $x$ direction through the parameter 
\begin{equation}\label{ssoripar}
\xi^2(\rho)= \frac{2j_{\rm max}(\Delta J_{x})_\rho^2}{\aver{J_{y}}_\rho^2+\aver{J_{z}}_\rho^2} \ ,
\end{equation}
where $j$ is the value of the total spin, $j_{\rm max}=\frac N 2$ for an ensemble of $N$ spin-$\frac 1 2$ particles
and the mean spin direction is in the $\hat{z}$ axis.
The state $\rho$ is called Spin Squeezed whenever $\xi^2(\rho)<1$. It is such that 
$(\Delta J_{x})_\rho < \frac{|\aver{J_z}_\rho|}{2}=\frac j 2\leq \frac N 4$. The quantity $\frac{|\aver{J_z}_\rho|}{2}$ is called {\it Standard Quantum Limit}.
Every Spin Squeezed state of $N$ spin-$\frac 1 2$ particles must be entangled and is more sensitive to rotations around the $\hat{y}$ axis with respect to a CSS.
\eedf 

From the Heisenberg uncertainty relation (\ref{spinunc}) we can see that the maximal possible degree of spin squeezing achievable in a completely polarized state $|\aver{J_z}|=j_{\rm max}$ is $\xi^2(\rho) \geq \frac 1 {2j_{\rm max}} \rightarrow 0$ in the limit $j\rightarrow \infty$, as in the case of bosonic squeezing.

Thus, SSS are entangled states that can be used to enhance the 
estimation of a phase $\varphi$ acquired in an unitary dynamics 
\begin{equation}\label{unievvarphi}
\rho_\varphi=\exp(-iH\varphi)\rho\exp(iH\varphi) \ ,
\end{equation}
in which $H=J_y$ and one constructs an estimator $\hat{\varphi}$ based on measurements of $J_x$.
Speaking more generally, there exists a fundamental bound on the accuracy $(\Delta \varphi)^{-2}$ achievable in an estimation
of $\varphi$ based on an unitary evolution (\ref{unievvarphi}). It is called {\it Cram\'er-Rao} bound (see e.g., \cite{Giovannetti2011Advances,tothapellaniz14}).

\bbth{(Cram\'er-Rao bound and Quantum Fisher Information)}
The precision achievable for the estimation of a parameter $\varphi$ governing a unitary dymanics (\ref{unievvarphi}) is bounded by 
\begin{equation}\label{cramerrao}
(\Delta \varphi)^{-2} \leq m F_{\rm Q}[\rho,H] \ ,
\end{equation}
where $m$ is the number of repetitions of the experiment and 
\begin{equation}
F_{\rm Q}[\rho,H] = \trace(\rho_\varphi L_\varphi^2) \ ,
\end{equation}
with $L_\varphi$ defined by $2\partial_\varphi \rho_\varphi = \rho_\varphi L_\varphi+ L_\varphi \rho_\varphi$ 
is called {\it Quantum Fisher Information}. The QFI is a convex function that for pure states is proportional 
to the variance of the generator $H$ of the dynamics $F_{\rm Q}[|\psi\rangle \langle \psi|,H] = 4 (\Delta H)^2_{|\psi\rangle}$.
For general mixed states it assumes the form
\begin{equation}
F_{\rm Q}[\rho,H] = 2\sum_{k,l} \frac{(\lambda_k-\lambda_l)^2}{\lambda_k+\lambda_l} |\langle k | H |l\rangle|^2 \quad
\mbox{with} \quad \rho=\sum_k \lambda_k |k\rangle \langle k|
\end{equation}
and can be proved to be the convex roof of $(\Delta H)^2$ \cite{tothpra13}. Moreover there exists always an optimal measurement such that
the bound (\ref{cramerrao}) can be reached.
\eeth

As a result then, Eq.~(\ref{cramerrao}) provides the ultimate bound on the precision achievable in measuring the phase $\varphi$ obtained after a rotation generated by $J_y$, no matter which measurement we perform. In particular, measuring $J_x$ 
in an ensemble of $N$ qubits we have
seen that the accuracy is related to spin squeezing as $(\Delta \varphi)^{2} = \frac{\xi^2}{N}$. The value of the denominator $(\Delta \varphi)^2_{\rm CSS}=\frac 1 {N}$, that is the precision achievable with a CSS of $N$ qubits, is usually called {\it shot-noise limit}. On the other hand, the maximal possible precision scales as $(\Delta \varphi)^2_{\rm Heis}=\frac 1 {N^2}$ and is called {\it Heisenberg limit}, corresponding to the maximal possible degree of squeezing, namely $\xi^2=\frac 1 N$.
Thus we have that the relation
\begin{equation}
\xi^2 \geq \frac{N}{F_{\rm Q}[\rho,J_y]}
\end{equation}
holds between spin squeezing and the optimal precision achievable in estimating $\varphi$. 
To conclude it is also interesting to mention that the inequality
\begin{equation}
\chi^2:=\frac{F_{\rm Q}[\rho,J_y]}{N} \leq k 
\end{equation}
holds for all $k$-producible states, whenever $N$ is divisible by $k$. In particular then the relation $\chi^2 < 1$, i.e., in a sense the metrological usefulness of the state itself provides 
an entanglement criterion more stringent than $\xi^2 <1$, although more difficult to verify.

\subsection{Generation of Spin Squeezed States}\label{sec:geneses}

Here let us briefly review what are the possibilities to generate a SSS from a CSS {\it deterministically} with a unitary evolution (see \cite{Ma2011Quantum} for a review). Picturing it in the Bloch sphere this operation corresponds to ``squeeze'' the uncertainty region of the state along a certain direction orthogonal to $\hat n_0$ at expenses of increasing it in the third remaining direction (see Fig.~\ref{fig:blochsphere}) resulting in an uncertainty region with elliptical shape. This is due to the constraint imposed by the Heisenberg uncertainty relations (\ref{spinunc}).

This task has some analogies with producing a squeezed bosonic state from a coherent one, that is done as in Eq.~(\ref{squeezedstate}) by employing the operator $S(\eta)$ (\ref{squeezingoperator}), i.e., an evolution $S=\exp(-iHt)$ with the following Hamiltonian
\begin{equation}\label{squeezham}
H=i(g(a^\dagger)^2-g^*a^2) \ ,
\end{equation}
where the coupling $g$ is related to $\eta$ through $\eta=-2|g|t$.
In fact the original proposal \cite{Kitagawa1993Squeezed} was to generate spin squeezing in a direction $\hat{x}$ by a relatively simple evolution, governed by the so-called {\it one-axis twisting} Hamiltonian
\begin{equation}\label{oneaxistw}
H_1=\chi J_x^2 \ ,
\end{equation}
which is quadratic in the spin operator. Thus, through Eq.~(\ref{oneaxistw}), a state initially $\hat{z}$-polarized $|\Phi\rangle=|j,j\rangle_{\rm z}$ evolves as
\begin{equation}
|\Phi\rangle(t)=\exp(-i\theta J_x^2) |j,j\rangle_{\rm z} \ ,
\end{equation}
with $\theta=\chi t$. It has been shown, then, that in the case $j\gg 1$ and for $|\theta| \ll 1$ the spin squeezing parameter scales as
$\xi^2 \sim j^{-2/3}$, and therefore the state becomes spin squeezed.
An improvement over $H_1$ can be obtained by adding an external control field $B$ in the polarization direction \cite{law01}, so that the Hamiltonian 
becomes
\begin{equation}\label{oneaxfield}
H_{1,\rm field}=\chi J_x^2 + B J_z \ .
\end{equation}
It has been numerically shown that evolving a CSS with $H_{1,\rm field}$ leads to spin squeezing during a wider time interval as compared to $\theta$.
Apart from this simple one-axis twisting Hamiltonian, the maximal degree of spin squeezing $\xi^2=\frac 1 N$ can be in principle generated with a {\it two-axis twisting} Hamiltonian \cite{Kitagawa1993Squeezed}
\begin{equation}\label{twoaxestw}
H_2=-i \frac{\chi} 2 (J_+^2 - J_-^2) \ ,
\end{equation}
that is completely analogous to (\ref{squeezham}). These Hamiltonians, especially (\ref{oneaxfield}) can be practically employed in different experimental setups, mainly consisting of ensembles of very many cold atoms manipulated through interactions with light fields (see \cite{Ma2011Quantum} and references therein). 

\paragraph{Spin Squeezing through Quantum Non Demolition measurements.}
In systems consisting of atomic clouds, also a different technique can be used to generate spin squeezing: performing Quantum Non-Demolition Measurements of the atomic collective spin through probes with polarized light fields. This method differs substantially due to the fact that here spin squeezing is not produced deterministically through a unitary evolution, but it is obtained, still starting from a CSS, through a {\it probabilistic quantum operation}. Therefore the resulting spin squeezing depends on random outcomes of measurements of the light field polarization. Because of this the result of this operation is sometimes called {\it conditional spin squeezing}. Here we shall review briefly what is the theoretical idea behind this method (see e.g., \cite{KoschorreckPhD,Ma2011Quantum}), also because it allows us to understand the main features of QND measurements, that we will see also in following chapters, thought in the different context of Legget-Garg-like tests. More theoretical details about QND measurements, especially in the context of atomic ensemble system are given in Appendix~\ref{sec:gaussstatqnd}. Now let us just mention few important properties of an ideal QND measurement, that basically consists in measuring indirectly a certain observable $O$ of a system, by transferring the information to a {\it meter} observable $M$ and without perturbing its value.
A QND measurement is performed through a Hamiltonian interaction $H_{\rm I}$ between the {\it target observable} $O$ (e.g., collective spin component $J_z$ of the atoms) and a {\it meter} $M$ (e.g., the polarization $S_y$ of pulses of light fields) such that: (i) the value of $O$ remains constant of motion during the evolution, i.e., $[H_{\rm I},O]=0$; (ii) the value of $M$ is perturbed in relation to the information acquired about the value of $O$, and in particular $[H_{\rm I},M]\neq 0$. 

In particular, in the cases that we are considering, the simplest QND interaction is 
\begin{equation}
H_{\rm I} = gS_zJ_z \ ,
\end{equation}
where $J_z$ is the collective spin component of the atoms (i.e., the system), $S_z=\frac{a_L^\dagger a_L - a_R^\dagger a_R}{2}$ is a Stokes component of the light field and the meter observable is $S_y=\frac{a_L^\dagger a_R - a_R^\dagger a_L}{2i}$ (that in fact does not commute with $H_{\rm I}$).
In order to show how spin squeezing is generated in this case let us consider an initial state in which both the light and the atoms are in a collective $\hat{x}$-polarized CSS, namely
\begin{equation}\label{cssinqnd}
|\Phi(0)\rangle = |S,S_x\rangle_{\rm x} |J,J_x\rangle_{\rm x} \ ,
\end{equation}
where $S$ and $J$ are the integer numbers such that $S_x^2+S_y^2+S_z^2=S(S+1)$ and $J_x^2+J_y^2+J_z^2=J(J+1)$ respectively. 
This initial state can be also conveniently expressed as a product joint probability distribution for the atomic and light states, given by the expansion coefficients of $|J,J_x\rangle_{\rm x}$ in the $|J,J_z\rangle_{\rm z}$ basis
\begin{equation}\label{eq:cssbinprob}
P_{\rm J}(J_z) := \bigg| \langle J,J_z |J,J_x\rangle_{\rm x} \bigg|^2 = \frac 1 {2^{N_A}} \binom{N_A}{\frac{N_A} 2 + J_z} \stackrel{N_A\rightarrow \infty}{\simeq}
\frac 1 {\sqrt{\pi N_A/2}} \exp\left( - \frac{2J_z^2}{N_A} \right) \ ,
\end{equation}
and $|S,S_x\rangle_{\rm x}$ in the $|S,S_y\rangle_{\rm y}$ basis
\begin{equation}
P_{\rm S}(S_y) := \bigg| \langle S,S_y |S,S_x\rangle_{\rm x} \bigg|^2 = \frac 1 {2^{N_L}} \binom{N_L}{\frac{N_L} 2 + S_y} \stackrel{N_L\rightarrow \infty}{\simeq}
\frac 1 {\sqrt{\pi N_L/2}} \exp\left( - \frac{2S_y^2}{N_L} \right) \ ,
\end{equation}
where $N_A$ and $N_L$ are respectively the number of atoms in the ensemble and photons in each probe and are assumed to be very large.
In fact, the state (\ref{cssinqnd}) can be expressed as
\begin{equation}
|\Phi(0)\rangle = \sum_{J_z=-J}^J \sum_{S_y=-S}^S \sqrt{P_{\rm J}(J_z) P_{\rm S}(S_y)} |S,S_y\rangle_{\rm y} |J,J_z\rangle_{\rm z} \ ,
\end{equation}
and the fact that it is a CSS is translated in having binomial probability distributions $P_{\rm J}(J_z) P_{\rm S}(S_y)$, that become {\it gaussian} for large number of particles.

After turning on the interaction for a small time $\tau$, the operator $S_y$ has evolved to (at first order in $\kappa=g\tau$)
\begin{equation}
S_y(\tau)=S_y(0) + \kappa J_z(0)S_x(0) \ ,
\end{equation}
which means that the average value is shifted to
\begin{equation}
\aver{S_y(\tau)}=\aver{S_y(0)} + \kappa \aver{J_z(0)S_x(0)} = \kappa S \aver{J_z(0)} \ ,
\end{equation}
and the conditional probability $\Pr(|S,S_y\rangle_{\rm y} | J_z= J_z^{\rm (out)})$ of obtaining the output state $|S,S_y\rangle_{\rm y}$ after measuring $J_z$ and obtaining $J_z^{\rm (out)}$ is
\begin{equation}
\Pr(|S,S_y\rangle_{\rm y} | J_z= J_z^{\rm (out)})=\frac 1 {\sqrt{\pi N_L/2}} \exp\left( - \frac{2(S_y-\kappa S J_z^{\rm (out)})^2}{N_L} \right) \ ,
\end{equation}
i.e., it has an average shifted by a quantity $\mu_{\rm S_y}=\kappa S J_z^{\rm (out)}$. Correspondingly, due to Bayes theorem, we can obtain the conditional probability for the output atomic state
\begin{equation}
\Pr(|S,S_y\rangle_{\rm y} | J_z= J_z^{\rm (out)}) P_{\rm J}(J_z^{\rm (out)}) = \Pr(|J,J_z\rangle_{\rm z} | S_y= S_y^{\rm (out)}) P_{\rm S}(S_y^{\rm (out)}) \ ,
\end{equation}
that results to
\begin{equation}
\Pr(|J,J_z\rangle_{\rm z} | S_y= S_y^{\rm (out)})=\frac 1 {\sqrt{\pi \xi^2 N_A/2}} \exp\left( - \frac{2(J_z-\kappa \xi^2 J S_y^{\rm (out)})^2}{\xi^2N_A} \right) \ ,
\end{equation}
with $\xi^2=\frac 1 {1+\zeta^2}$ being the spin squeezing parameter and $\zeta^2=SJ\kappa^2$. Thus the output state will be spin squeezed whenever $\zeta^2>0$, i.e., basically there is interaction for a certain non-zero time $\tau$. Note that the value of $\xi^2$ is given deterministically, i.e., independently on the outcome $S_y^{\rm (out)}$, while on the other hand
the average $\aver{J_z}$ has been shifted by a quantity proportional to the outcome. Then, in order to obtain back the original state with one squeezed variance it is needed a feedback scheme that restores $\aver{J_z}=0$ and puts back the mean spin into its initial direction.

Spin squeezing produced with this technique has been recently achieved experimentally \cite{Kuzmich1998Atomic,appel09,Julsgaard:2001aa,Koschorreck10,KoschorreckPRL2010b,LouchetChauvet10}.
Finally, before concluding this chapter it is worth to mention that squeezing can be also transferred from light to atoms, i.e., from quadrature squeezing of the light $(x,p)$ operators to spin squeezing of the collective $(J_x,J_y)$ operators. 
This idea is also intriguing for a quantum information perspective because it allows to think about protocols that transfer information from light to atomic ensembles and store it in the latter systems, that thus would function as {\it quantum memories}, while the former being information carriers.

\afterpage{\blankpage} 
\chapter{Generalized Spin Squeezing} 

\label{Chapter3} 

In the previous chapter we have introduced the concept of Spin Squeezed states, first as opposite to coherent $su(2)$ states in a single particle framework and then extended
to a general multiparticle setting. 
In that last case, Spin Squeezing has been shown to be connected with both entanglement and potential technological developments. Here, in the course of the chapter, we shall extend further this definition in several directions, in a way especially oriented to explore deeper the connections with entanglement. 

It will be our first main original work contained in this thesis.
First, also to motivate it further, let us observe that in the case of $su(2)$ other definition of squeezed states can be given when more than a single spin component has a squeezed variance. 

\begin{figure*}[h!]
\centering
\includegraphics[width=0.8\columnwidth]{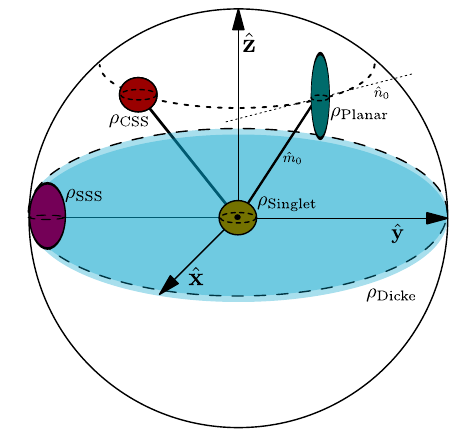}
\caption{Some examples of generalized spin squeezed states depicted in a spin-$J$ Bloch sphere. (i) $\rho_{\rm CSS}$ is a mixed state completely polarized and close to a CSS, that has three variances of the order of $(\Delta J_k)^2\simeq \frac J 2$, (ii) $\rho_{\rm SSS}$ is completely polarized $|\aver{J_y}|\simeq J$ and has a single squeezed variance, (iii) $\rho_{\rm Planar}$, a \emph{planar squeezed state}, almost completely polarized and has two squeezed variances, (iv) $\rho_{\rm Singlet}$, a \emph{macroscopic singlet state}, with all three variances below a SQL, and (v) the \emph{unpolarized Dicke state} $\rho_{\rm Dicke}$, with a tiny planar uncertainty region $(\Delta J_z)^2 \simeq 0$ and $(\Delta J_x)^2=(\Delta J_y)^2\simeq J^2$.
}\label{fig:blochother}
\end{figure*}

\paragraph{Generalized Spin Squeezing: Planar Squeezed and Singlet states.}
Loosely speaking a Spin Squeezed state as we defined it in the previous chapter can be pictured in a spin-$j$ Bloch sphere as a state completely polarized in a $\hat{m}_0$ direction and occupying a region of uncertainty which has a shape that is squeezed in an orthogonal direction. In Fig.~\ref{fig:blochsphere} it has been depicted in an ideal case of an eigenstate of $J_{m_0}$, while in Fig.~\ref{fig:blochother} it is depicted in a more realistic situation in which one variance is squeezed, while the other two are increased with respect to a CSS due to the Heisenberg uncertainty principle.

Very naively, then, a Spin Squeezed State can be thought as a quantum-enhanced clock arm which can resolve a phase accumulated during a rotation about the $\hat{z}$ direction (setting the axes as in Fig.~\ref{fig:blochother}) more accurately than a ``classical'' one, which occupies a spherical region of uncertainty in the Bloch sphere (see $\rho_{\rm CSS}$ in Fig.~\ref{fig:blochother}). 

This picture has some analogies with the quadrature squeezing of bosonic modes, and in fact reduces to it in the $j \rightarrow \infty$ limit in some sense. 
However, while in the bosonic case the squeezing of the variance of a quadrature is reflected in a back-action on the conjugate one independently of other quantities, here the back-action on $(\Delta J_z)$ due to the squeezing of $(\Delta J_x)$ depends on the average polarization in the $\hat y$ direction $|\aver{J_y}|$. In other words, the Standard Quantum Limit (SQL) on the squeezing of one variance with respect to another, depends on the average polarization in the direction orthogonal to their plane. On the other hand, we can try to provide a definition of SQL that is independent of the state and in particular of its polarization. 

In this sense it can be worth to consider a different uncertainty relation as a figure of merit to define squeezing in two directions, namely a constraint on the sum of the two variances, that cannot be zero whenever the two observables are incompatible. This was precisely the approach followed by He \emph{et al.} in \cite{He2011Planar} in defining \emph{planar squeezing}. They studied a single spin-$j$ particle uncertainty relation and found that 
\begin{equation}
(\Delta J_{x})+(\Delta J_{y}) \geq C_j \ ,
\end{equation}
where the constant on the right hand side depends on $j$ and scales as $C_{j} \sim j^{\frac 2 3}$ for $j\gg 1$. 

They also studied the states that saturate the inequality and defined them as
\emph{planar squeezed states}. They verified that the so-defined planar squeezed states are almost completely polarized in the plane of squeezing for $j\gg 1$, i.e., they satisfy $\aver{J_x}^2+\aver{J_y}^2 \sim j^2$, and therefore a natural choice of Standard Quantum Limit as $(\Delta J_k)^2_{\rm SQL}=\frac{\sqrt{\aver{J_x}^2+\aver{J_y}^2}} 2$ immediately follows. This choice can also be justified further again by the fact that it is related to the precision achievable in the measurement of a phase accumulated in a unitary rotation around the $\hat{z}$ direction. 

In fact, He \emph{et al.} also noted that actually
Eq.~(\ref{eq:naivprec}), namely the uncertainty in a unitary phase estimation process is given by
\begin{equation}
\Delta \varphi = \frac{\Delta J_x^{\rm (out)}}{|\partial \aver{J_x^{\rm (out)}}/\partial \varphi|}= \frac{\sqrt{(\Delta J_x^{\rm (in)})^2\cos^2 \varphi+ (\Delta J_y^{\rm (in)})^2\sin^2 \varphi }}{|\aver{J_y^{\rm (in)}} \cos \varphi - \aver{J_x^{\rm (in)}} \sin \varphi |}
\end{equation}
for a general phase $\varphi$ and for a unitary dynamics given by $U(\varphi)=\exp(-i\varphi J_z)$. Thus, especially away from $\varphi \sim 0$ the estimation can be made more precise using input states $\rho$ that minimize a parameter like $\zeta_{\rm P}^2(\rho)= \frac{(\Delta J_{x})_\rho^2+(\Delta J_{y})_\rho^2}{\sqrt{\aver{J_x}_\rho^2+\aver{J_y}_\rho^2}}$, thought as a comparison between the input state and a CSS completely polarized in the plane of squeezing. Then, motivated also by this usefulness, they defined planar squeezed states already for the general multiparticle scenario. 

\bbdf{\bf (Planar Squeezed States (PSS)).} 
A state $\rho$ is defined as Planar Squeezed in the $x$ and $y$ directions through the parameter 
\begin{equation}\label{ssoriplan}
\xi_{\rm P}^2(\rho)= \frac{j_{\rm max} \left((\Delta J_{x})_\rho^2+(\Delta J_{y})_\rho^2\right)}{\aver{J_x}_\rho^2+\aver{J_y}_\rho^2} \ ,
\end{equation}
where $j_{\rm max}=Nj$ for an ensemble of $N$ spin-$j$ particles.
The state $\rho$ is called Planar Squeezed whenever $\xi_{\rm P}^2(\rho)<1$ holds together with $\frac{2j_{\rm max}(\Delta J_x)_\rho^2}{\aver{J_x}_\rho^2+\aver{J_y}_\rho^2}<1$ and $\frac{2j_{\rm max}(\Delta J_y)_\rho^2}{\aver{J_x}_\rho^2+\aver{J_y}_\rho^2}<1$. The SQL is then chosen to be $(\Delta J_k)^2_{\rm SQL}=\frac{\aver{J_x}_\rho^2+\aver{J_y}_\rho^2}{2j_{\rm max}}$ for a single variance and $[(\Delta J_x)^2+(\Delta J_y)^2]_{\rm SQL}=\frac{\aver{J_x}_\rho^2+\aver{J_y}_\rho^2}{j_{\rm max}}$ for the pair of variances, and the state must surpass this limit in both two variances.
The optimal PSS, i.e., states that minimise $\xi_{\rm P}^2$ are such that $\aver{J_x}^2+\aver{J_y}^2 \sim j_{\rm max}^2$ and $(\Delta J_{x})^2+(\Delta J_{y})^2 \sim j_{\rm max}^{\frac 2 3}$ for $j_{\rm max}\gg 1$.
\eedf 

Here we have provided directly a definition of planar squeezing that holds also for states of $N$ spin-$j$ particles and is a figure of merit for the usefulness of the state in a specific metrological task. In the Bloch sphere picture planar squeezed states are almost completely polarized in a certain direction $\hat{m}_0$ and have a cigar-shaped uncertainty region, that is tiny in the $(\hat{m}_0,\hat{n}_0)$ directions and large in the remaining orthogonal one (see $\rho_{\rm Planar}$ in Fig.~\ref{fig:blochother}). 
As compared to original spin squeezed states they have two variances that scale as $(\Delta J_{x})^2 \sim (\Delta J_{y})^2 \propto j_{\rm max}^{\frac 2 3}$, while the spin squeezed states are such that both variances scale as $(\Delta J_{x})^2 \sim (\Delta J_{y})^2 \propto j_{\rm max}$, even if one of them $(\Delta J_{x})^2 =\frac{\xi^2} 2 j_{\rm max}$ might be squeezed as compared to the other
$(\Delta J_{y})^2 =\frac{j_{\rm max}}{2\xi^2}$ by a factor of $(\xi^2)^2$.

However with this definition there is no clear relation between planar squeezing of multiparticle states and their entanglement. An entanglement criterion related to planar squeezing has then been provided in the same work of He \emph{et al.} \cite{He2011Planar} and is another important example of entanglement criterion with collective observables based on LURs

\bboo{(Planar squeezing entanglement criterion).}
For every separable state $\rho_{\rm sep}$ of $N$ spin-$j$ particles 
\begin{equation}\label{eq:plansqentcrit}
(\Delta J_{x})_{\rho_{\rm sep}}^2+(\Delta J_{y})_{\rho_{\rm sep}}^2 \geq NC_j \ ,
\end{equation}
must hold where $C_j$ is the constant bound of the LUR
\begin{equation}
(\Delta j_{x})^2+(\Delta j_{y})^2 \geq C_j \ ,
\end{equation}
and can be computed numerically for single spin-$j$ particle states. Every state $\rho$ such that $(\Delta J_{x})_{\rho}^2+(\Delta J_{y})_{\rho}^2 < NC_j$ must be entangled.
\eeoo

Since planar squeezed states are such that for large $N\gg 1$ they satisfy $(\Delta J_{x})_{\rho}^2+(\Delta J_{y})_{\rho}^2 \sim (Nj)^\frac 2 3$ they can be detected as entangled by the violation of Eq.~(\ref{eq:plansqentcrit}). On the other hand, the relation between planar squeezing and entanglement is not clarified, since the two parameters $\xi_{\rm P}^2$ and $\xi_{\rm P, ent}^2:=\frac{(\Delta J_{x})_{\rho}^2+(\Delta J_{y})_{\rho}^2}{NC_j}$ do not coincide. For example in the case of multiqubit systems, since $C_{\frac 1 2}=\frac 1 4$, we have that $\xi_{\rm P}^2<\frac 1 2$ is a signal of entanglement coming from Eq.~(\ref{eq:plansqentcrit}), that is then an additional stronger requirement with respect to just planar squeezing, signalled by simply $\xi_{\rm P}^2<1$.
Moreover, the criterion (\ref{eq:plansqentcrit}) itself is not tied to detecting planar squeezed states, since it is maximally violated by the so called \emph{singlet states}, i.e., states such that $(\Delta J_n)^2=0$ in all directions $\hat{n}$. In Fig.~\ref{fig:blochother} the ideal singlet corresponds to the origin of axes, while $\rho_{\rm singlet}$ represents a noisy state close to the singlet. These last actually can be thought as three variance spin squeezed states.

To go one step further in fact, one can think to define states with all the three spin variances below a certain SQL and thus define three variance SSS. However in this case again there is a problem in defining the SQL, enhanced by the fact that the LUR $(\Delta J_{x})^2+(\Delta J_{y})^2+(\Delta J_{z})^2\geq j$ cannot be violated and is saturated by every pure single spin-$j$ state. Thus, at least for single particle states such definition cannot be given, since all pure states would be three-variance CSS. This problem can be solved in the multipartite setting, exploiting the fact that entangled states can violate the inequality $(\Delta J_{x})^2+(\Delta J_{y})^2+(\Delta J_{z})^2\geq Nj$ \cite{tothpra04} (see Th.~(\ref{su2invcrit})). 

Thus one can define a SQL for three collective variances of a system of $N$ spin-$j$ particles as $[(\Delta J_{x})^2+(\Delta J_{y})^2+(\Delta J_{z})^2]_{\rm SQL}=Nj$ and correspondingly a spin squeezing parameter
\begin{equation}\label{eq:singletparini}
\xi_{\rm T, ent}^2:=\frac{(\Delta J_{x})_\rho^2+(\Delta J_{y})_\rho^2+(\Delta J_{z})_\rho^2}{Nj} \ ,
\end{equation}
that detects states close to \emph{macroscopic singlet states} and proves their entanglement. The parameter (\ref{eq:singletparini}) has been introduced in \cite{tothmitchell10} and the detected states, that in the Bloch sphere picture would correspond to a single point in the center with a small spherical uncertainty region, have been proposed as useful resources for gradient magnetometry \cite{urizar13} (see also Fig.~\ref{fig:blochother}).

Thus, we have seen that other definitions of spin squeezing can be thought and that they can provide figures of merit for quantum metrology and for entanglement detection, although possibly not related to each other. The states detected by those different parameters can still be produced relatively easily in many experimental settings, cold and ultra cold atomic ensembles among others, and can be exploited for technological purposes. 
Furthermore, from the point of view of entanglement detection, such states can be easily proven to be entangled with criteria involving just few collective measurable quantities.

However, there is also an ambiguity in these definition of squeezing with respect to entanglement. The definitions are in fact given in the single particle setting and then straightforwardly extended to multipartite systems. Then, it is clear that the same state with spin quantum number $J$ can be interpreted in many different ways: it can be seen as a single spin-$J$ particle, or as a system of $2J$ spin-$\frac 1 2$ particles in a symmetric states and so on. The spin squeezing parameters cannot distinguish very well the different interpretations and the same state can be thought as an entangled state of $2J$ qubits or as an unentangled single particle state. The only exception to this ambiguity is the parameter (\ref{eq:singletparini}), that is meaningful only for multipartite systems.

In the following we are going to focus deeper into this issue and develop some definitions of spin squeezing that, as Eq.~(\ref{eq:singletparini}), are intrinsecally related to interparticle entanglement in systems of general spin-$j$ particles. 


\section{Optimal spin squeezing inequalities for entanglement detection}

\subsection{A complete set of spin squeezing inequalities}\label{ssec:completessi12}

As we have seen, from the point of view of entanglement detection SSS can be seen as forming a set of detectable entangled states that can be actually produced and are interesting for technologically oriented purposes. These are basically entangled states of $N$ particles that can be detected through the measurement of just collective spin averages $\aver{J_k}$ and variances $(\Delta J_k)^2$. 

On the other hand, in this sense one can look for generalization of SSS in many directions. A natural question that arises in this context is then: given as measured data 
only the set of three first $\aver{\vec J}=(\aver{J_x},\aver{J_y},\aver{J_z})$ and second moments $\aver{\vec K}=(\aver{J^2_x},\aver{J^2_y},\aver{J^2_z})$, what is the maximal set of entangled states that can be detected? This question has been answered by T\'oth \emph{et al.} \cite{tothPRL07,tothPRA09} by finding a closed set of few inequalities that detect all entangled states that can be detected based on $(\aver{\vec J},\aver{\vec K})$. As we are going to see, these inequalities can be viewed as generalization of the original spin squeezing parameter in several respects.

\bboo{(Complete set of Spin Squeezing Inequalities).} The following set of inequalities
\begin{subequations}\label{completessiqubits}
\begin{align}
\aver{J_x^2} + \aver{J_x^2} + \aver{J_x^2} &\leq \frac{N(N+2)} 4 \label{symmine} \ , \\
(\Delta J_x)^2 + (\Delta J_y)^2 + (\Delta J_z)^2 &\geq \frac{N} 2 \label{qubitsingin} \ ,\\
\aver{J_k^2} + \aver{J_l^2} -\frac{N} 2 &\leq (N-1)(\Delta J_m)^2 \label{eq:qubitorissss} \ ,\\
(N-1)\left[(\Delta J_k)^2 +(\Delta J_l)^2\right] &\geq \aver{J_m^2}+\frac{N(N-2)} 4 \label{eq:ourqubitplanar} \ ,
\end{align}
\end{subequations}
where $k,l,m$ are three arbitrary orthogonal directions must hold for all separable states of $N$ spin-$\frac 1 2$ particles.

Moreover the set (\ref{completessiqubits}) is complete, in the sense that in the large $N$ limit it detects all possible entangled states, based on the vectors of first
 $\aver{\vec J}=(\aver{J_x},\aver{J_y},\aver{J_z})$ and second moments $\aver{\vec K}=(\aver{J^2_x},\aver{J^2_y},\aver{J^2_z})$.
\eeoo

We will give the proof later, by proving a generalization of Eq.~(\ref{completessiqubits}) to ensembles of spin-$j$ particles and its completeness in the same sense. Here let us just note that every CSS saturates all the inequalities in the set (\ref{completessiqubits}). In fact CSS states are completely polarized in a certain direction, say $\aver{\vec J}=(\frac N 2,0,0)$, while having
$\aver{\vec K}=(\frac{N^2} 4,\frac{N} 4,\frac{N} 4)$. Thus, in some sense the violation of any of Eqs.~(\ref{completessiqubits}) defines a set of states analogous to SSS, since they will have some variances $(\Delta J_k)^2$ that are squeezed as compared to CSS. Let us look explicitly at some example states 
detected by such generalized spin squeezing inequalities. 

\bbexmp{(Original Spin Squeezed States).} As we will recall also later, one inequality in the set, namely Eq.~(\ref{eq:qubitorissss}) detects all the states detected by the spin squeezing parameter (\ref{ssoripar}). In fact states $\rho$ of a large number of particles $N\gg1$ such that $\xi^2(\rho)<1$ also violate Eq.~(\ref{eq:qubitorissss}).
\eeexmp
An other important class of states that are spin squeezed according to (\ref{completessiqubits}) are \emph{symmetric Dicke states}

\bbexmp{(Symmetric Dicke States).} Eq.~(\ref{qubitsingin}) is maximally violated by the unpolarized \emph{symmetric Dicke state}. This is a permutationally symmetric eigenstate of $J_z$ such that $\aver{\vec J}=0$ and $\aver{\vec K}=(\frac{N(N+2)}{8},\frac{N(N+2)}{8},0)$. It belongs to the class of symmetric Dicke states defined as
\begin{equation}
|D_N^{(m)}\rangle_z=\binom{N}{m}^{-\frac 1 2} \sum_k \mathcal P_k(|1/2 \rangle_z^{\otimes m} |-1/2\rangle_z^{\otimes N-m}) \ ,
\end{equation}
where the sum is over all the possible permutations of the particles such that $m$ of them have $j_z=\frac 1 2$ and $N-m$ have $j_z=-\frac 1 2$. The state $|D_N^{(m)}\rangle_z$ is an eigenstate of $J_z$ with eigenvalue $\aver{J_z}=-\frac N 2+m$. The unpolarised Dicke state is obtained for $m=\frac N 2$ and can reach the Heisenberg limit $(\Delta \theta)^{-2}\sim N^2$ in some parameter estimation task, similarly as optimal SSS and the GHZ state. All of them, except for the two extremal cases $m=0$ and $m=N$ which are product states, are truly $N$-partite entangled.
\eeexmp

Symmetric states close to $|D_N^{(\frac N 2)}\rangle$ have been produced among other systems in Bose-Einstein condensates \cite{Lucke2011Twin,Lucke2014Detecting}. In the Bloch sphere picture they correspond to the $(\hat x, \hat y)$ circular section in the equator: the width due to uncertainty in the $\hat z$ direction is vey small, while it is very big in the two orthogonal directions (see $\rho_{\rm Dicke}$ in Fig.~\ref{fig:blochother}).

\bbexmp{(Macroscopic Singlet States).} Eq.~(\ref{qubitsingin}) is maximally violated by macroscopic singlet states. These are the states for which $\aver{\vec J}=\aver{\vec K}=0$ and thus minimise the left hand side of Eq.~(\ref{qubitsingin}). Macroscopic singlet states $\rho_{\rm Singlet}$ are invariant under collective rotations, i.e., $e^{-iJ_n \theta}\rho_{\rm Singlet}e^{iJ_n \theta}=\rho_{\rm Singlet}$ for all directions $\hat{n}$, and such that $(\Delta J_n)^2=0$ in all directions $\hat{n}$. They have been proposed as useful metrological resources for gradient magnetometry. 
\eeexmp

Macroscopic singlet states can be obtained as ground states of spin chains, as well as in atomic ensembles, e.g., through QND measurements \cite{tothmitchell10}.

\bbexmp{(Planar Squeezed States).} Planar Squeezed States, as defined by (\ref{ssoriplan}) are also detected as entangled by one inequality in the set, namely Eq.~(\ref{eq:ourqubitplanar}).
\eeexmp

PSS can be realized experimentally in cold \cite{puentes13} and ultra cold atomic ensembles~\cite{He2011Planar}.

Thus, in analogy with the original spin squeezing parameter, for each inequality of (\ref{completessiqubits}) we can define a Standard Quantum Limit and a corresponding generalized spin squeezing parameter. Also here there is a problem in the definition of the SQL, that is not univocal and can be arbitrarily rescaled. 
We put here some convenient choices, that also assure that the defined parameters are positive.

In particular, apart from Eq.~(\ref{symmine}) that cannot be violated by any quantum state, we can define
\begin{subequations}\label{genssparam}
\begin{align}
\xi^2_{\rm singlet}&:=\frac{(\Delta J_x)^2 + (\Delta J_y)^2 + (\Delta J_z)^2}{\frac N 2} \label{qubitsingpar} \ ,\\
\xi^2_{\rm Dicke}&:=\frac{(N-1)(\Delta J_m)^2}{\aver{J_k^2} + \aver{J_l^2} -\frac N 2}   \ ,\\
\xi^2_{\rm planar}&:=(N-1)\frac{\left[(\Delta J_k)^2 +(\Delta J_l)^2-\frac{N} 4 \right]}{\aver{J_m^2}-\frac{N} 4} \ ,
\end{align}
\end{subequations}
for collective systems of $N$ spin-$\tfrac 1 2$ particles. 
Each of Eqs.~(\ref{genssparam}) is a generalized spin squeezing parameter, such that
$\xi^2 <1$ is a proof of entanglement. In particular $\xi^2_{\rm singlet} < 1$ detects states close to macroscopic singlets, while $\xi^2_{\rm Dicke} < 1$ detects the original SSS plus states close to symmetric Dicke states. The third parameter $\xi^2_{\rm planar}$ also detects the singlet, although less efficiently than $\xi^2_{\rm singlet}$, but detects also planar squeezed states, i.e., states with a small value of the sum of two variances and a big value of the second moment in the orthogonal direction.

Thus, from an entanglement detection point of view, the set of parameters (\ref{genssparam}) represent an improvement over the original $\xi^2$, since even one single parameter in the set detects a strictly wider class of states. Furthermore, it introduces new classes of states, such as Dicke states and in general unpolarized states as spin squeezed. We will later see in more details what are the advantages and the problems of (\ref{genssparam}) and try to provide a single definition that embraces all the parameters in the set.


\subsection{Extreme spin squeezing}\label{sec:extremess}

An other natural question that arises concerning spin squeezing and in particular its importance for phase estimation protocols, is how to identify easily the SSS that reach the Heisenberg limit in the precision. These are in a sense the most squeezed spin states, since they have to minimize $(\Delta \phi)^2=\frac{\xi^2}{N}=\frac{(\Delta J_x)^2}{\aver{J_z}^2}$. Ideally, in fact an optimal SSS would have the minimal possible variance
$(\Delta J_x)_{\rm min}^2$, given the fact that it is completely polarized in the $\hat{z}$ direction, i.e., $\aver{J_z}=\frac N 2$. Interestingly it turns out that such optimal SSS are also $N$ entangled (or genuinely $N$ partite entangled), i.e., they have the highest possible depth of entanglement. 

On the other hand, an independent question that can be asked from the point of view of entanglement detection oriented study of spin squeezing is whether it is possible to distinguish higher form of multipartite entanglement with spin squeezing inequalities. In a seminal paper \cite{Sorensen2001Entanglement}, S\o rensen-M\o lmer solved both problems at the same time, developing
a method that, looking at optimal SSS, provides a family of inequalities that detect the \emph{depth} of entanglement, i.e., optimal criteria for $k$-producibility. The states detected with this method have been termed \emph{extreme spin squeezed}, because they are exactly the states that minimise $\xi^2$ for each fixed value of the polarization $\aver{J_z}$. 
Formally, the basic idea consists in first defining the functions
\begin{equation}\label{fsormol}
F_J(X)=\frac 1 J \min_{\frac 1 J \aver{j_z}=X} (\Delta j_x)^2 \ ,
\end{equation}
where $J$ is the spin quantum number and evaluating it numerically for any value of $J$. Note that $F_J(1)$ gives the minimal value of $\xi^2$ for a single spin-$J$ particle.
Then, based on Eq.~(\ref{fsormol}) a family of $k$-producibility criteria can be derived as in the following theorem.

\bbth{(Extreme Spin Squeezing inequalities).}\label{th:extSSI} The following family of inequalities
\begin{equation}\label{sormolcrit}
(\Delta J_x)^2 \geq Nj F_{kj}\left( \frac{\aver{J_z}}{Nj} \right) \ ,
\end{equation}
holds for all $k$ producible states of $N$ spin-$j$ particles, $J_l=\sum_{n=1}^N j_l^{(n)}$ being the collective spin components of the ensemble. The set of functions $F_{kj}(\cdot)$ in Eq.~(\ref{sormolcrit}) is defined as in Eq.~(\ref{fsormol}), with the identification $J=kj$, i.e.,
\begin{equation}\label{fsormolth}
F_{kj}(X)=\frac 1 {kj} \min_{\frac 1 {kj} \aver{L_z}=X} (\Delta L_x)^2 \ ,
\end{equation}
where $L_l$ are the spin components of a single particle with spin $J=kj$. These functions have the following properties: (i) $F_{J}(X)$ are convex and monotonically increasing for all $J$ and so are $F_{J}(\sqrt X)$, (ii) they are such that $F_{J}(0)=0$ for all $J$, (iii) $F_{J_1}(X) \leq F_{J_2}(X)$ holds for $J_1 \geq J_2$.

Every state $\rho$ of $N$ spin-$j$ particles that violates Eq.~(\ref{sormolcrit}) must be $(k+1)$-entangled.
\eeth

\bbpr Let us consider a $k$-producible state
\begin{equation}
\rho_{\rm k-prod}=\sum_i p_i \rho_i^{(1)}\otimes \dots \otimes \rho_i^{(n)} \ ,
\end{equation}
where the $\rho_i^{(1)}$ are states of at most $k$ particles and thus $n\geq \frac N k$. Since the variance is a concave function of density matrices and is additive on product states we have
\begin{equation}
(\Delta J_x)^2 \geq \sum_i p_i \left[(\Delta L^{(1)}_x)_{\rho_i^{(1)}}^2+\dots + (\Delta L^{(n)}_x)_{\rho_i^{(n)}}^2\right] \ ,
\end{equation}
where we introduced $L^{(n)}_m=\sum_{l=1}^{k_n} j_x^{(k_n)}$ as the $\hat{m}$ spin components of a $k_n$ particle subsystem. Note that the states $\rho_i^{(n)}$ are such that $k_n\leq k$ and that $\sum_n k_n=N$.
By employing the definition (\ref{fsormolth}) we have 
\begin{equation}
(\Delta J_x)^2 \geq \sum_i p_i \sum_n k_n j F_{k_n j}\left( \frac{\aver{L_z^{(n)}}_{i}}{k_n j} \right) \ ,
\end{equation}
and by using the property (iii) of $F_{J}(X)$ and the fact that $k_n\leq k$ 
\begin{equation}\label{secondbounddjx}
(\Delta J_x)^2 \geq \sum_i p_i \sum_n k_n j F_{k j}\left( \frac{\aver{L_z^{(n)}}_{i}}{k_n j} \right) \ .
\end{equation}
Then, by using property (i) of $F_{J}(X)$ (i.e., its convexity) and applying Jensen inequality in the form $\sum_n a_n f(x_n) \geq \sum_l a_l \cdot f\left( \frac{\sum_n a_n x_n}{\sum_l a_l} \right)$ with $a_n=k_n j$ and $x_n=\tfrac{\aver{L_z^{(n)}}_{i}}{k_n j}$ we have for the right hand side of Eq.~(\ref{secondbounddjx})
\begin{equation}
\sum_i p_i \sum_n k_n j F_{k j}\left( \frac{\aver{L_z^{(n)}}_{i}}{k_n j} \right) \geq \sum_i p_i  Nj F_{k j}\left( \frac 1 {Nj} \sum_n k_n j \frac{\aver{L_z^{(n)}}_{i}}{k_n j} \right) =
\sum_i p_i  Nj F_{k j}\left( \frac{\aver{J_z}_{i}}{Nj} \right) \ ,
\end{equation}
where we used that $\sum_n k_n j=Nj$ and $\sum_n \aver{L_z^{(n)}}_{i}=\aver{J_z}_{i}$. Finally, by applying again Jensen inequality with $a_i=p_i$ and $x_i=\frac{\aver{J_z}_{i}}{Nj}$ the statement follows.
\eepr

The proof that we have just given is similar to the original proof done by S\o rensen-M\o lmer \cite{Sorensen2001Entanglement}, but is also inspired by the proof given by Hyllus \emph{et al.} in \cite{hyllus12} concerning some aspects that were missing in the original proof itself.
Moreover, strictly speaking the properties (i-iii) of $F_{J}(X)$ are not formally proven, but can be explicitly seen by evaluating the functions numerically for different values of $J$.

With this theorem, S\o rensen-M\o lmer also introduced the depth of entanglement as a figure of merit from the point of view of metrological usefulness of SSS. In fact from the measurement of $(\Delta J_x)^2$ and $\aver{J_z}$ the depth of entanglement can be inferred, certifying hierarchically how close the state is to an optimal SSS.
These extreme spin squeezed states can be obtained as ground states of the hamiltonian
\begin{equation}\label{hspinsq}
H=J_x^2+\mu J_z \ ,
\end{equation}
where the parameter $\mu\neq 0$ can be also thought as a Lagrange multiplier that can be scanned in order to solve the optimization problem Eq.~(\ref{sormolcrit}) leading to $F_J(X)$. 

In fact, it can be shown that the functions $F_{J}(X)$ needed to evaluate the criteria can be evaluated numerically by looking at ground states of the hamiltonian in Eq.~(\ref{hspinsq})
for different values of the parameter $\mu$ \emph{in a single spin-$J$ particle space} and evaluating $((\Delta J_x)^2, \aver{J_z})=(F_{J}(X),X)$ on such states. By scanning over a wide range of $\mu$ and eventually compute the convex roof, one then obtains the continuous convex function $F_{J}(X)$.  
Note that the dimension of the space in which the numerical optimization has to be carried out is $d=2J+1=k+1$, that scales linearly with the entanglement depth. A brute force optimization of $(\Delta J_x)^2$ on $k$-producible states would have had to be performed on a roughly $\sim 2^k$ dimensional space instead and would have been hopeless for already $k\sim 10$. On the other hand, Eq.~(\ref{sormolcrit}) can be evaluated easily for $k$ of the order of several hundreds and provides a tight criterion, in the sense that the boundary on the right hand side is reached by $k$ producible states. 

Finally note also that in the two limits $\mu \rightarrow 0$ and $\mu \rightarrow \pm \infty$ the ground state of (\ref{hspinsq}) for $k$ particle collective spin operators is a symmetric Dicke state $|D_N^{(\frac N 2)}\rangle_x$ and a completely polarized state $|J,\mp J\rangle_z$ respectively. 


\section{Generalized Spin Squeezing and entanglement for multi spin-$j$ systems}\label{sec:genSSspinj}

The idea of our work \cite{vitagliano11,vitagliano14} on entanglement detection-oriented spin squeezing inequalities has been to try to generalize the framework to systems composed of spin-$j$ particles, starting from the well established results of T\'oth \emph{et al.}, Eqs.~(\ref{completessiqubits}) on spin-$\frac 1 2$ particle systems. The goal was to derive a small, closed set of spin squeezing inequalities valid for composite systems of spin-$j$ particles, with the same completeness properties of Eqs.~(\ref{completessiqubits}). 

In particular, in this way we provide a complete construction that generalizes and extends the concept of spin squeezing to composite systems of higher spin particles and to a wider class of states, including e.g., Dicke states, singlet states and general unpolarized states. 

This construction and the spin squeezing parameters so defined can be measured in present and future experiments involving, e.g., higher spin atomic systems in which entangled states are aimed to be produced for technological purposes, like quantum enhanced metrology, quantum computation and general quantum information tasks.
In fact these ideas have been already applied to some recent experiments in photons, cold atomic ensembles and Bose-Einstein condensates.

This idea can be pursued in a two fold way: one can either derive a full set of generalized SSIs directly as entanglement criteria coming from variance based single particle Local Uncertainty Relations, or one can find a mapping from SSIs that are valid for multipartite spin-$\frac 1 2$ systems to analogous SSIs valid for multipartite spin-$j$ systems for a general value of $j$. We will see that both methods lead to the same complete set of inequalities and in a sense give complementary insights on the nature of the constraints used to derive them.

\subsection{A complete set of multipartite SSIs for arbitrary spin}\label{subs:completessij}

First of all let us observe how the complete set of SSIs comes from a variance based LUR and that this relation can lead to more general results. The following observations thus, not only
prove Eqs.~(\ref{completessiqubits}) being entanglement criteria and forming a complete set, 
but also extend the same results to multipartite spin-$j$ systems. They also show explicitly the connections between spin squeezing inequalities, single particle LURs and the set of generalized coherent states.

\bboo{(Spin Squeezing Inequalities from LURs).}\label{obs:completessispinj}
The following set of generalized Spin Squeezing Inequalities (SSIs)
\begin{subequations}\label{completessiqubitsj}
\begin{align}
\aver{J_x^2} + \aver{J_x^2} + \aver{J_x^2} &\leq Nj(Nj+1) \label{symminej} \ , \\
(\Delta J_x)^2 + (\Delta J_y)^2 + (\Delta J_z)^2 &\geq Nj \label{qubitsinginj} \ ,\\
\aver{J_k^2} + \aver{J_l^2}  -Nj(Nj+1) &\leq N(\Delta J_m)^2 - N\sum_n \aver{(j_m^{(n)})^2} \ , \label{eq:dickeineqj} \\
(N-1)\left[(\Delta J_k)^2 +(\Delta J_l)^2\right] &\geq \aver{J_m^2}- N\sum_n \aver{(j_m^{(n)})^2} +N(N-1)j \ ,
\end{align}
\end{subequations}
where $k,l,m$ are three arbitrary orthogonal directions must hold for all separable states of $N$ particles, where $J_k=\sum_n j_k^{(n)}$ are the collective spin components. Eqs.~(\ref{completessiqubitsj}) hold whenever there is a constraint of the form $\sum_{k=x,y,z} \aver{j_k}^2 \leq j^2$ for the single particle expectations or
equivalently whenever the LUR
\begin{equation}\label{3spinslur}
(\Delta j_x)^2+(\Delta j_y)^2+(\Delta j_z)^2 \geq j \ ,
\end{equation}
holds on the single particle variances. For $j=\frac 1 2$ Eqs.~(\ref{completessiqubitsj}) reduces to Eqs.~(\ref{completessiqubits}).
\eeoo

\bbpr Let us consider a pure product state $\rho = \bigotimes_n \rho^{(n)}$ and let us define some modified second moments $\aver{\tilde{J}_k^2}=\aver{J_k^2}-\sum_n \aver{(j_k^{(n)})^2}$ and the corresponding modified variances $(\tilde{\Delta} J_k)^2=\aver{\tilde{J}_k^2}-\aver{J_k}^2$. Using the additivity of the variance on product states we have that $(\tilde{\Delta} J_k)^2_\rho=-\sum_n \aver{(j_k^{(n)})^2}_{\rho^{(n)}}$, while $\aver{\tilde{J}_k^2}_\rho=\aver{J_k}_\rho^2-\sum_n \aver{(j_k^{(n)})^2}_{\rho^{(n)}}$. Now let us consider the following expression
\begin{equation}
(N-1)\sum_{l \in I} (\tilde{\Delta} J_l)^2 - \sum_{l \notin I}\aver{\tilde{J}_l^2} \ ,
\end{equation}
where $I$ is any subset of indices of $\{x,y,z\}$ including $I=\emptyset$. We have, on product states
\begin{equation}\label{compacttildessis}
(N-1)\sum_{l \in I} (\tilde{\Delta} J_l)_\rho^2 - \sum_{l \notin I}\aver{\tilde{J}_l^2}_\rho \geq -\sum_n (N-1) \sum_{l=x,y,z} \aver{(j_l^{(n)})^2} \geq -N(N-1)j^2 \ ,
\end{equation}
where we have used the facts that $\aver{J_l}^2 \leq N \sum_n \aver{(j_l^{(n)})^2}$ following from the Cauchy-Schwarz inequality and the constraint $\sum_{k=x,y,z} \aver{j_k}^2 \leq j^2$ on the single particle averages. Moreover, since the left hand side of (\ref{compacttildessis}) is concave in the state, the bound on the right hand side also holds for general mixtures of product states, i.e., on all separable states.
Eqs.~(\ref{completessiqubitsj}) follow from Eqs.~(\ref{compacttildessis}) from the fact that $\sum_{l=x,y,z}\aver{J_l^2}=\sum_{l=x,y,z}\aver{\tilde{J}_l^2}+Nj(j+1)$, which is a consequence of the operator identity
\begin{equation}\label{casimirid}
j_x^2+j_y^2+j_z^2 = j(j+1) \id \ .
\end{equation}
From Eq.~(\ref{casimirid}) it also follows that $\sum_{k=x,y,z} \aver{j_k}^2 \leq j^2$ is equivalent to Eq.~(\ref{3spinslur}).
\eepr

In the proof of Eq.~(\ref{completessiqubitsj}) we have introduced some quantities that are modified version of the second moments of collective spin components, namely
\begin{equation}
\begin{aligned}\label{eq:modsecmomdef}
\aver{\tilde{J}_k^2}&=\aver{J_k^2}-\sum_n \aver{(j_k^{(n)})^2} \ , \\
(\tilde{\Delta} J_k)^2&=\aver{\tilde{J}_k^2}-\aver{J_k}^2 \ ,
\end{aligned}
\end{equation}
that are obtained basically subtracting the average local second moment from the corresponding collective second moment. 

In this way we are defining a quantity that takes account of the squeezing of the collective variance coming from just interparticle, rather than intra-particle entanglement (or spin squeezing) by subtracting the average single particle spin squeezing.  Note that in the case of $j=\frac 1 2$ the modified second moments are obtained from the true second moments by just subtracting a constant, and in fact since a single spin-$\frac 1 2$ particle cannot be spin squeezed there cannot be intraparticle squeezing in that case.

\begin{figure*}
\includegraphics[width=0.45 \textwidth]{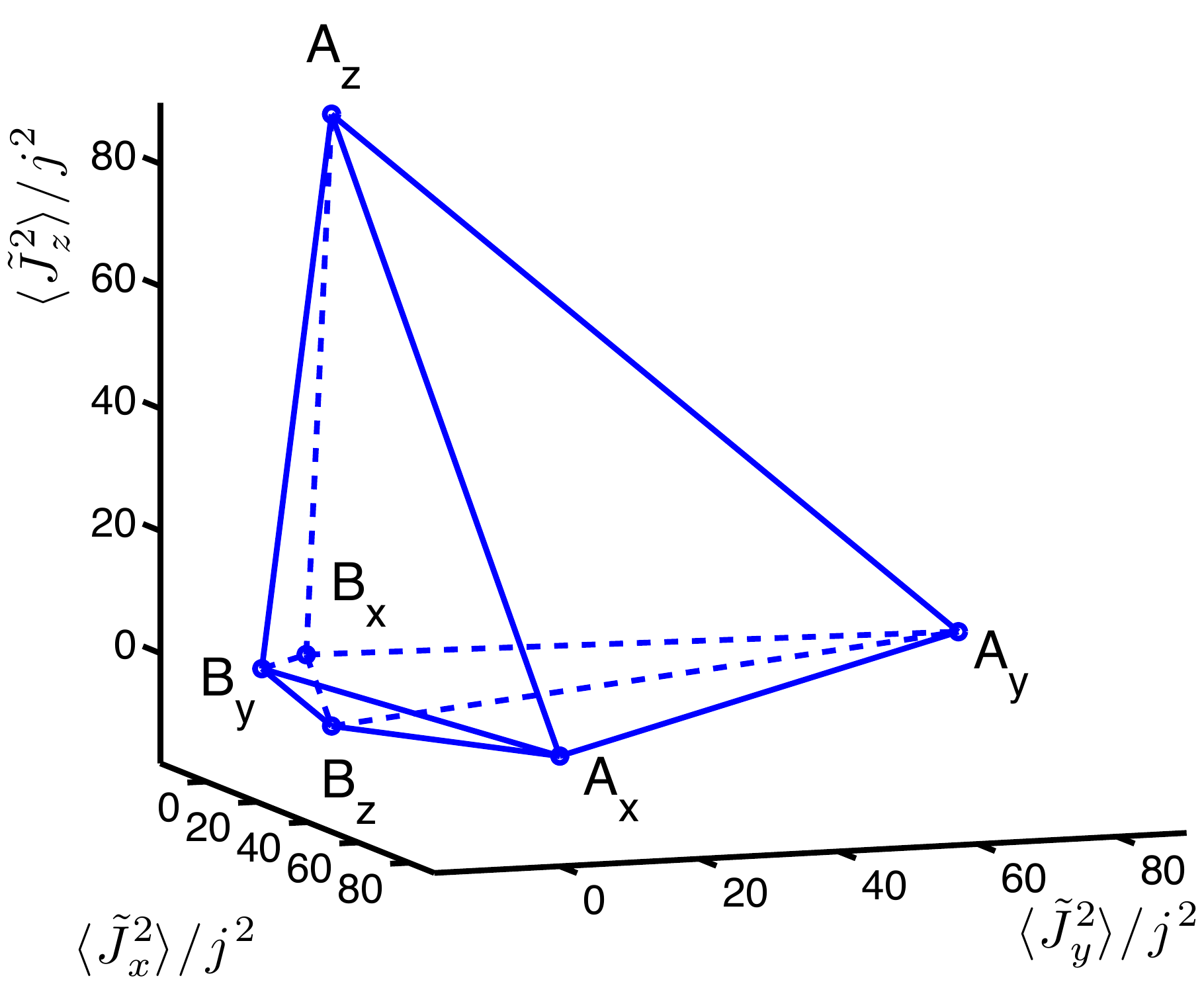}
\includegraphics[width=0.45 \textwidth]{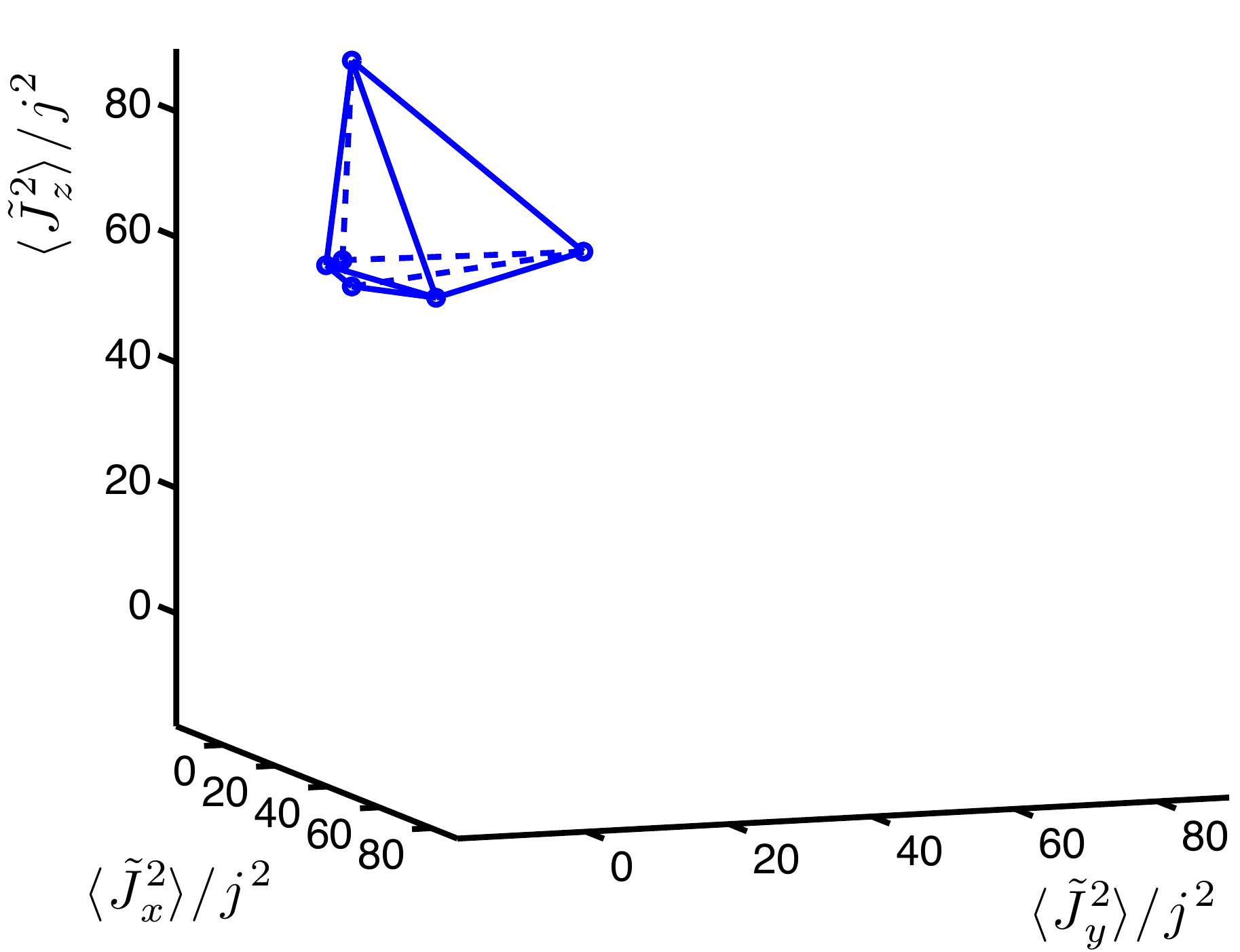}
\caption{(left) The polytope of separable states corresponding to
Eqs.~(\ref{completessiqubitsj})  for $N=10$ spin-$j$ particles and for $\vec{J}=0.$ (right)
The same polytope for $\vec{J}=(0,0,8)j.$ Note that this polytope is
a subset of the polytope in (left). For the coordinates of the points $A_l$  and $ B_l$  see Eq.~(\ref{eq:axbxcoord}).}
\end{figure*}

Afterwards, we show that the set (\ref{completessiqubitsj}) provides a complete entanglement-based definition of spin squeezing, in the sense that all possible multipartite entangled SSS are detected by one of Eqs.~(\ref{completessiqubitsj}).

\bboo{(The set Eqs.~(\ref{completessiqubitsj}) is complete and defines generalized CSS).}\label{obs:completeness}
For every value of the mean spin vector $\aver{\vec J}$ the set Eqs.~(\ref{completessiqubitsj}) defines a polytope in the space of $\aver{\vec{\tilde K}}=(\aver{\tilde{J}_x^2},\aver{\tilde{J}_y^2},\aver{\tilde{J}_z^2})$, the vertices of which have coordinates
\begin{equation}
\begin{aligned}\label{eq:axbxcoord}
A_x&=\left(N(N-1)j^2-\kappa (\aver{J_y}^2+\aver{J_z}^2), \kappa \aver{J_y}^2, \kappa \aver{J_z}^2\right) \ , \\
B_x&=\left(\aver{J_x}^2+\frac{\aver{J_y}^2+\aver{J_z}^2} N, \kappa \aver{J_y}^2, \kappa \aver{J_z}^2\right) \ , 
\end{aligned}
\end{equation}
with $\kappa=\frac{N-1} N$ and including some other analogously defined points $(A_y,B_y)$, $(A_z,B_z)$. Each of the points $(A_k,B_k)$ can be attained by a separable state in the $N\rightarrow \infty$ limit. In particular, separable states corresponding to the vertices are combinations of CSS.
\eeoo

\bbpr Let us consider a system of $N$ spin-$j$ particles in the subspace of states with a given value of $\aver{\vec J}=(\aver{J_x},\aver{J_y},\aver{J_z})$ and define the quantity $c_x$ such that $c_x^2 N^2j^2 = N^2j^2-\aver{J_y}^2-\aver{J_z}^2$.
Then, a separable state corresponding to $A_x$ is given by
\begin{equation}
\rho_{\rm A_x}=p \rho_+^{\otimes N} + (1-p) \rho_-^{\otimes N} \ ,
\end{equation}
where $\rho_{\pm}$ are the states such that $\aver{\vec j}_{\rho_{\pm}}=(\pm j c_x, \frac{\aver{J_y}} N, \frac{\aver{J_y}} N)$. This can be seen by direct computation of the corresponding coordinates $\aver{\vec{\tilde K}}_{\rho_{\rm A_x}}=(\aver{\tilde{J}_x^2}_{\rho_{\rm A_x}},\aver{\tilde{J}_y^2}_{\rho_{\rm A_x}},\aver{\tilde{J}_z^2}_{\rho_{\rm A_x}})$. Analogous states corresponding to $A_y$ and $A_z$ can be obtained with the same construction and all of them exist for every fixed value of $\aver{\vec J}$, corresponding to a different value of $p$.
Afterwards, if $M=Np$ is an integer, we can also construct separable states corresponding to $B_x$ as
\begin{equation}
\rho_{\rm B_x}=\rho_+^{\otimes M} \otimes \rho_-^{\otimes (N-M)} \ ,
\end{equation}
and analogously for $B_y$ and $B_z$. Finally, if $M$ is not an integer consider the separable state
\begin{equation}
\rho^\prime_{\rm B_x,\epsilon}= (1-\epsilon)\rho_+^{\otimes M^\prime} \otimes \rho_-^{\otimes (N-M^\prime)}
+\epsilon \rho_+^{\otimes M^\prime+1} \otimes \rho_-^{\otimes (N-M^\prime-1)} \ ,
\end{equation}
where $M^\prime=M-\epsilon$ is the largest integer smaller than $M$. This state has the same coordinates $\aver{\vec{\tilde K}}$ as $B_x$ except for $\aver{\tilde{J}_x^2}$, which differs 
from a quantity $\delta_{\aver{\tilde{J}_x^2}} = 4j^2c_x^2 \epsilon (1-\epsilon) \leq j^2$. Thus in the limit $N\rightarrow \infty$ the state $\rho^\prime_{\rm B_x,\epsilon}$ corresponds to the vertex
$B_x$. Analogously we can find states close to $B_y$ and $B_z$ that reach such points for $N\rightarrow \infty$.
\eepr

Thus, Obs.~\ref{obs:completessispinj} provides a closed set of SSIs that are separability criteria for systems of spin-$j$ particles and Obs.~\ref{obs:completeness}
proves that such set gives a complete figure of merit for entangled multi spin-$j$ squeezed states. In particular, in the case of spin-$\frac 1 2$ particle systems the results reduce to what derived by T\'oth \emph{et al.} in \cite{tothPRL07,tothPRA09} and mentioned in Sec.~\ref{ssec:completessi12}. Conversely, the same statements could be obtained through a mapping from quantities computed in the spin $j=\frac 1 2$ space to quantities defined for general spin $j$.

\bboo{(Mapping from spin-$\frac 1 2$ to higher spins entanglement criteria).}\label{obs:mapping12j}
Let us consider a necessary separability condition for $N$ spin-$\frac 1 2$ particles, written in terms of $\aver{\vec J}$ and $\aver{\vec{\tilde K}}$ as
\begin{equation}\label{eq:mapinifunc}
f(\{ \aver{J_l} \},\{ \aver{\tilde{J}_l^2}\}) \geq {\rm const.} \ .
\end{equation}
Then the inequality obtained from Eq.~(\ref{eq:mapinifunc}) by the substitutions
\begin{equation}\label{eq:12jmapp}
\aver{J_l} \rightarrow \frac 1 {2j} \aver{J_l} \ , \qquad \aver{\tilde{J}_l^2} \rightarrow \frac 1 {4j^2} \aver{\tilde{J}_l^2} \ ,
\end{equation}
is a necessary separability condition for $N$ spin-$j$ particles.
\eeoo

\bbpr Let us consider a product state of $N$ spin-$j$ particles $\rho_{\rm prod, j}=\bigotimes_{n=1}^N \rho_{\rm j}^{(n)}$ and define the quantities $r_l^{(n)}=\frac 1 j \aver{j_l^{(n)}}_{\rho_{\rm j}}$. 
Then we have
\begin{equation}
\frac{\aver{J_l}_{\rho_{\rm prod, j}}}{2j} = \frac 1 2 \sum_n r_l^{(n)} \ , \qquad \frac{\aver{\tilde{J}_l^2}_{\rho_{\rm prod, j}}}{4j^2} = \frac 1 4 \sum_{n\neq m} r_l^{(n)} r_l^{(m)} \ ,
\end{equation}
and the LUR constraint for the single particle spin-$j$ states $\rho_{\rm j}^{(n)}$ becomes equivalent to
\begin{equation}
0\leq \sum_l (r_l^{(n)})^2 \leq 1 \ ,
\end{equation}
independently on $j$. Thus for product states the set of allowed values of $( \frac 1 {2j} \aver{J_l} , \frac 1 {4j^2} \aver{\tilde{J}_l^2})$ is independent on $j$, since they just depend on 
the quantities $r_l^{(n)}$. The same holds also for general separable states, since they are just mixtures of product states. Thus the range of allowed values of a function 
$f(\{\frac 1 {2j} \aver{J_l}\} , \{\frac 1 {4j^2} \aver{\tilde{J}_l^2}\})$ on separable states is independent on $j$ and this proves the claim.
\eepr

As we claimed before, with the mapping (\ref{eq:12jmapp}) we can directly generalize Eqs.~(\ref{completessiqubits}) to higher spin particle systems and obtain precisely 
Eqs.~(\ref{completessiqubitsj}). In particular this mapping provides a normalization that allows to define spin squeezing due to interparticle entanglement independently on the value of the spin quantum number $j$. In the following we argue that actually a complete definition unifying all kinds of spin squeezing due to interparticle entanglement can be given by a single parameter.

\subsection{A unique, optimal Spin Squeezing parameter}

Let us start here a discussion on possible definitions of entanglement-based spin squeezing parameters. As discussed before, even for single spin-$j$ particles it is not straightforward to give a unique figure of merit that defines spin squeezing starting just from Heisenberg uncertainty relations, analogously as the case of bosonic squeezing.
Moreover, since for $su(2)$ there are three mutually conjugate observables, one can also ask: can we define SSS of more than just one variance? We have already seen that in fact this is the case, leading to the definition of 
planar squeezed states and singlet states. 

Furthermore, in a multiparticle scenario the situation gets even more complex. Different figures of merit for spin squeezing can be associated to either, e. g., entanglement, or metrological usefulness for a specific task, but they are not equivalent to each other. 

Now, a further question that might arise is: can we include different types of SSS in a unified framework? From a fundamental point of view it would be important to define a Standard Quantum Limit that sets univocally the quantum/classical 
border for a multi spin-$j$ state and a unified distinction between Squeezed/Coherent states can help in this sense.
Furthermore, since all spin squeezed states have been proposed to improve metrological tasks, this might help also for a unified characterization of metrologically useful states. The main obstacle to this is how to define a SQL independent from 
the properties of the specific state and 
that unifies all the different kinds of spin squeezed states.

From the point of view of multipartite entanglement we have seen that all these different SSS arise naturally as points exterior 
to a polytope of separable combinations of CSS. Thus a natural definition of SQL is provided by the polytope itself, being the boundary between CSS and (generalized) SSS and a quantification of the degree of squeezing can be given as a distance from the boundary. Thus, following our approach entanglement oriented, we are able to give a unified definition of spin squeezing as figure of merit for interparticle entanglement in multi spin-$j$ systems. The parameter that we provide solves in this way the practical problem of including states with different polarizations and states with more than a single squeezed variance in the same framework and in this sense it will be a generalization of the original spin squeezing parameter.

Naively, as a preliminary step, from each of Eqs.~(\ref{completessiqubitsj}) (apart Eq.~(\ref{symminej}) that cannot be violated by any quantum state) we can define a different parameter by, e.g., dividing the left hand side by the right hand side of the inequality. The practical problem is that the denominator is not univocal (one can always add the same quantity to both sides of the inequality) and must be always positive. A choice is for example
\begin{subequations}\label{genssparamj}
\begin{align}
\xi^2_{\rm Dicke, j}&:=(N-1)\frac{(\tilde{\Delta} J_x)^2 +Nj^2}{\aver{\tilde{J}_y^2}+\aver{\tilde{J}_z^2}} \ , \label{spinjdickepar} \\
\xi^2_{\rm planar, j}&:=(N-1)\frac{(\tilde{\Delta} J_x)^2+(\tilde{\Delta} J_y)^2+2Nj^2}{\aver{\tilde{J}_z^2}+N(N-1)j^2} \ , \\
\xi^2_{\rm singlet, j}&:=\frac{(\Delta J_x)^2 + (\Delta J_y)^2 + (\Delta J_z)^2}{Nj} \label{spinjsingpar} \ ,
\end{align}
\end{subequations}
that reduces to (\ref{genssparam}) for $j=\frac 1 2$. Thus, all separable states obey $\xi_{\rm X,j}^2 \geq 1$ for all the three parameters and any entangled spin squeezed state must be detected by one of the three because of the completeness of the set. Note that the parameters are well defined only when the denominator is positive. However it can be shown that whenever the denominator of each parameter is negative, then the corresponding inequality cannot detect the state as entangled. Thus the definitions are consistent, though valid under some restrictions. 

As a comparison we can apply the mapping (\ref{eq:12jmapp}) to the original spin squeezing parameter (\ref{ssoripar}) and define a generalized version that detects entangled states also for spin-$j$ particle systems with $j>\frac 1 2$. Note, in fact that Eq.~(\ref{ssoripar}) is not a valid entanglement criterion for higher spin systems, since even a single spin-$1$ state can have $\xi^2<1$. 

\bboo{(Original SS entanglement criterion mapped to higher spin systems).}
Exploiting the mapping in Eq.~(\ref{eq:12jmapp}) we can map the original spin squeezing parameter 
\begin{equation}
\xi^2:=\frac{N(\Delta J_x)^2}{\aver{J_y}^2+\aver{J_z}^2}
\end{equation}
of a system of $N$ spin-$\frac 1 2$ particles, to
\begin{equation}\label{eq:mapporisspar}
\xi^2_{\rm ent, j}:=N\frac{(\tilde{\Delta} J_x)^2+Nj^2}{\aver{J_y}^2+\aver{J_z}^2} \ ,
\end{equation}
that is a generalized spin squeezing parameter such that $\xi^2_{\rm ent, j} \geq 1$ for all separable states of $N$ spin-$j$ particles, even with $j>\frac 1 2$.
Thus every state $\rho$ of $N$ spin-$j$ particles such that $\xi^2_{\rm ent, j}(\rho) < 1$ must be entangled.
\eeoo

Now let us compare $\xi^2_{\rm ent, j}$ with $\xi^2_{\rm Dicke, j}$. The following observation, that has been proven in \cite{vitagliano14}, shows that $\xi^2_{\rm Dicke, j}$ gives an improvement over 
$\xi^2_{\rm ent, j}$ from the point of view of entanglement detection.

\bboo{($\xi^2_{\rm Dicke, j}$ is strictly finer than $\xi^2_{\rm ent, j}$).}\label{obs:compara}
The following statements hold when $N\gg 1$: (i)  $\xi^2_{\rm Dicke, j}(\rho)<\xi^2_{\rm ent, j}(\rho)$ for all states $\rho$ such that $\xi^2_{\rm ent, j}(\rho)<1$, (ii)  $\xi^2_{\rm Dicke, j}\left((1-p_{\rm n})\rho + p_{\rm n} \frac{\id}{(2j+1)^N}\right)\simeq (1-p_{\rm n}) \cdot \xi^2_{\rm ent, j}\left((1-p_{\rm n})\rho + p_{\rm n} \frac{\id}{(2j+1)^N}\right)$ for all completely polarized states $\rho$ and $0\leq p_{\rm n}\leq 1$.
\eeoo

Thus, from Obs.~\ref{obs:compara} it follows that in the large $N$ limit a single parameter in the set (\ref{genssparamj}) is strictly finer and also more tolerant to white noise in detecting usual SSS with respect to $\xi^2_{\rm ent, j}$. Furthermore (\ref{spinjdickepar}) can detect unpolarized states while (\ref{eq:mapporisspar}) cannot, since we have $\xi^2_{\rm ent, j}(\rho) \rightarrow \infty$ when $\rho$ is unpolarized.

More in general it can be also shown that every state detected by $\xi^2_{\rm ent, j}$ is also detected by either (\ref{spinjdickepar}) or (\ref{spinjsingpar}). Thus, once more we have proven that the set (\ref{genssparamj}) is a generalization of the original definition of spin squeezing that includes a wider class of entangled states. 

From Obs.~\ref{obs:completeness} it also follows that from the point of view of entanglement detection the definition of spin squeezing based on (\ref{genssparamj}) cannot be further extended. 
Thus, it makes sense to look for a single parameter that unifies the whole set. In the next we present such a single parameter, written in an explicit rotationally invariant form so to avoid a definition based on the prior knowledge of some privileged Mean Spin Direction. The idea is that one can look directly at the optimal directions, possibly more than just one, in which $\rho$ is eventually spin squeezed. 

At first let us express the set of inequalities (\ref{completessiqubitsj}) in a form that is explicitly invariant under orthogonal changes of reference axes.

\bboo{(Rotationally invariant form of Eqs.~(\ref{completessiqubitsj})).}
The set (\ref{completessiqubitsj}) can be compactly written as\footnote{Note that in \cite{vitagliano14} we gave a different definition of the matrix $Q$. Here we used Eq.~(\ref{eq:qmatdef}) because it allows to write the set (\ref{completessiqubitsj}) in a single compact way, namely (\ref{eq:su2invine})}
\begin{equation}\label{eq:su2invine}
(N-1)\trace(\Gamma) - \sum_{k=1}^I \lambda^{\downarrow}_k(\mathfrak X) -N(N-1)j\geq 0 \ ,
\end{equation}
where we defined the matrices 
\begin{align}\label{eq:corrmatrices}
C_{kl}&:=\frac 1 2 \aver{J_k J_l + J_l J_k} \ , \\
\Gamma_{kl}&:=C_{kl} -\aver{J_k}\aver{J_l}  \ , \label{eq:gammamatdef} \\
Q_{kl}&:=\frac 1 N \sum_n \left( \frac 1 2 \aver{j_k^{(n)} j_l^{(n)}+ j_l^{(n)}j_k^{(n)}} \right) \ , \label{eq:qmatdef} \\
\mathfrak X &:= (N-1) \Gamma + C - N^2 Q \ , 
\end{align}
and $\sum_{k=1}^I \lambda^{\downarrow}_k(\mathfrak X)$ is the sum of the largest $I$ eigenvalues of $\mathfrak X$ in decreasing order. Eq.~(\ref{eq:su2invine}) is invariant under
orthogonal changes of reference axes $\hat{n}\rightarrow O\hat{n}$.
\eeoo

\bbpr Let us start considering a single expression for the set (\ref{completessiqubitsj}), namely
\begin{equation}
(N-1)\sum_{l \in I} (\tilde{\Delta} J_l)^2 - \sum_{l \notin I}\aver{\tilde{J}_l^2} \geq -N(N-1)j^2 \ ,
\end{equation}
and add on both sides the quantity $(N-1)\sum_{l \notin I} (\tilde{\Delta} J_l)^2$. We obtain
\begin{equation}\label{eqpr:su2invin}
(N-1)\sum_{l=x,y,z} (\tilde{\Delta} J_l)^2 +N(N-1)j^2 \geq (N-1)\sum_{l \notin I} (\tilde{\Delta} J_l)^2 + \sum_{l \notin I}\aver{\tilde{J}_l^2} \ ,
\end{equation}
where we have also rearranged the terms. On the other hand we can express a diagonal element of $\mathfrak X$ as
\begin{equation}
\mathfrak X_{ll}= (N-1)\sum_{l \notin I} (\tilde{\Delta} J_l)^2 + \sum_{l \notin I}\aver{\tilde{J}_l^2} \ ,
\end{equation}
i.e., precisely as the expression in the right hand side of Eq.~(\ref{eqpr:su2invin}). Thus, since the left hand side is $(N-1)\sum_{l=x,y,z} (\tilde{\Delta} J_l)^2+N(N-1)j^2=(N-1)\trace(\Gamma)-N(N-1)j$ we finally have
\begin{equation}\label{eq:derssparameter}
(N-1)\trace(\Gamma) - \sum_{l \notin I} \mathfrak X_{ll} -N(N-1)j \geq 0 \ ,
\end{equation}
where $I$ is a general subset of indices. By taking the optimal diagonal elements $\mathfrak X_{ll}$, namely, since the right hand side is a lower bound its largest eigenvalues, we proved that Eq.~(\ref{eq:su2invine}) is equivalent to the compact version of Eq.~(\ref{completessiqubitsj}). To complete the proof we also show that it is invariant under orthogonal change of coordinates. In fact performing an orthogonal change of reference system $\hat{n}\rightarrow O\hat{n}$ is reflected in an orthogonal quadratic transformation of the matrices (\ref{eq:corrmatrices}), e.g., $C \rightarrow O C O^T$ and Eq.~(\ref{eq:su2invine}) is invariant under such transformations. \eepr

Next, starting from Eq.~(\ref{eq:su2invine}) we define a single spin squeezing parameter by just optimizing over the subset of directions $I$. Basically this optimization process provides the directions in which the state can have collective variances squeezed with respect to the SQL defined as the boundary of the polytope discussed in Obs.~\ref{obs:completeness} and the parameter provides a figure of merit that defines the degree of spin squeezing in such directions (to be published in the near future \cite{vitaglianototh2015}).

\bbdf{\bf (Generalized Spin Squeezed States).} \label{def:genssstates}
Let us consider an $N$ spin-$j$ particle state $\rho$ and define an \emph{optimal spin squeezing parameter} as
\begin{equation}\label{eq:genssparameter}
\xi^2_{\rm G}(\rho):=\frac{\trace(\Gamma_\rho) - \sum_{k=1}^I \lambda^{>0}_k(\mathfrak Z_\rho)}{Nj} \ ,
\end{equation}
where $I$ is the set of directions corresponding to the positive eigenvalues of $\mathfrak Z_\rho = \frac 1 {N-1} \mathfrak X_\rho$, that we called $\lambda^{>0}_k(\mathfrak Z_\rho)$ and $[\trace(\Gamma_\rho) - \sum_{k=1}^I \lambda^{>0}_k(\mathfrak Z_\rho)]_{\rm SQL}=Nj$ is a Standard Quantum Limit. The directions $\hat{k}$ such that
\begin{equation}
\lambda_k(\mathfrak Z_\rho)=(\Delta J_k)^2_\rho + \frac 1 {N-1} \aver{J_k^2}_\rho - \frac{N}{N-1} \sum_n \aver{(j_k^{(n)})^2}_\rho > 0 \ ,
\end{equation}
correspond to the directions in which $\rho$ \emph{cannot be} spin squeezed. 

We define an $N$ spin-$j$ particle state $\rho$ as generalized spin squeezed in the $\hat{n}\notin I$ directions whenever
\begin{equation}
\xi^2_{\rm G}(\rho) < 1 \ , 
\end{equation}
such that it is detected as entangled based on $\aver{\vec J}$ and $\aver{\vec{\tilde{K}}}$.
\eedf

The above definition has been obtained from Eq.~(\ref{eq:derssparameter}) by dividing by $N-1$, reordering the terms as 
$\trace(\Gamma) - \frac 1 {N-1} \sum_{l \notin I} \mathfrak X_{ll} \geq Nj$ and then defining
\begin{equation}
\xi^2_{\rm G}=\min_I \frac{\trace(\Gamma) - \frac 1 {N-1} \sum_{l \notin I} \mathfrak X_{ll}}{Nj} \ ,
\end{equation}
i.e., as a ratio between the left hand side and the right hand side, optimized over the set of directions $I$.

Exploiting the results given previously, then, it is straightforward to see that all usual CSS $\rho_{\rm CSS}$ are such that $\xi^2_{\rm G}(\rho_{\rm CSS})=1$
and that $\xi^2_{\rm G}<1$ detects all possible entangled states based on $\aver{\vec J}$ and $\aver{\vec{\tilde{K}}}$, including a set of states that goes beyond the original definition of SSS. As in Eq.~(\ref{eq:genssparameter}), the generalized spin squeezing parameter is well defined, since the denominator is always positive and provides a natural universal definition of SQL based on interparticle entanglement. 

\bbexmp{(Values of $\xi^2_{\rm G}$ on example useful states).} As an example we can easily compute the parameter $\xi^2_{\rm G}$ on some important states 
\begin{itemize}
\item for the singlet state $\rho_{\rm sing}$ it reaches the minimal value $\xi^2_{\rm G}(\rho_{\rm sing})=0$, independently of $j$. In fact the singlet is such that $\lambda_k(\mathfrak Z_{\rho_{\rm sing}})$ are all negative and the state is squeezed in all three directions.
\item  for the unpolarized Dicke state $\rho_{\rm Dicke}$ we have  $\xi^2_{\rm G}(\rho_{\rm Dicke})=\frac{Nj-1}{2Nj-1}$. This is because there are two positive eigenvalues $\lambda_x(\mathfrak Z_{\rho_{\rm Dicke}})=\lambda_y(\mathfrak Z_{\rho_{\rm Dicke}})=\frac{N^3j^3}{2jN-1}$ and just one negative $\lambda_z(\mathfrak Z_{\rho_{\rm Dicke}})=-\frac{N^2j^2}{2jN-1}$. Thus the Dicke state is spin squeezed in just one direction $\hat z$ and in the limit $N\gg 1$ it reaches $\xi^2_{\rm G}(\rho_{\rm Dicke})\rightarrow \frac 1 2$ independently of $j$.
\item A state $\rho_{\rm pol}$ of $N$ spin-$\frac 1 2$ particles that is completely polarized in a direction $\hat z$, $\aver{J_z}\simeq \frac N 2$ is such that $\lambda_z(\mathfrak Z_{\rho_{\rm pol}})\simeq (\Delta J_z)^2>0$. Assuming also that it has a squeezed variance along $\hat x$, i.e., $(\Delta J_x)^2 =\xi^2 \frac{N}{4}$, where $\xi^2<1$ is the value of the original spin squeezing parameter (\ref{ssoripar}), and correspondingly $(\Delta J_y)^2 = \frac{N}{4\xi^2}$ we have $\lambda_y(\mathfrak Z_{\rho_{\rm pol}})=\frac{N^2}{4(N-1)} \left( \xi^{-2}-1\right) >0$, while $\lambda_x(\mathfrak Z_{\rho_{\rm pol}})=-\frac{N^2}{4(N-1)} \left(1- \xi^{2}\right) <0$. The value of the generalized spin squeezing parameter is then $\xi^2_{\rm G}(\rho_{\rm pol})\simeq \frac 1 2\left(1+\xi^2 \right)<1$ and the state is spin squeezed along $\hat x$ and detected as entangled, as it should.
\item Planar squeezed states $\rho_{\rm planar}$ saturate the uncertainty relation $(\Delta J_x)^2 + (\Delta J_y)^2 \geq C_J$ for a single spin-$J$ particle. Equivalently, such states can be viewed as permutationally symmetric states of $N=2J$ spin-$\frac 1 2$ particles, almost completely polarized in the $\hat x$ direction, $\aver{J_x} \simeq \frac N 2$ and with the following variances: $(\Delta J_x)^2 \simeq \frac 1 8 N^{\frac 2 3}$, $(\Delta J_y)^2 \simeq \frac 1 4 N^{\frac 2 3}$, $(\Delta J_z)^2 \simeq \frac 1 4 N^{\frac 4 3}$. According to the definition (\ref{eq:genssparameter}), however, they are spin squeezed only in the $\hat y$ direction since for such states $\lambda_y(\mathfrak Z)<0$, while $\lambda_x(\mathfrak Z)$ and $\lambda_z(\mathfrak Z)$ are positive. The value of the complete spin squeezing parameter is $\xi^2_{\rm G}(\rho_{\rm planar})\simeq \frac 1 2\left(\frac N {N-1}+N^{-\frac 1 3} \right)\rightarrow \frac 1 2$.
\end{itemize}
\eeexmp

Table~\ref{tb:table1}
summarizes the properties of the spin squeezed states described in the previous example. It is interesting to note that planar squeezed states are squeezed only in one $\hat y$ direction according to $\xi_G^2$ and thus are similar to the original spin squeezed states and to Dicke state in this sense.
This fact can be clarified by looking at our definition of spin squeezing and providing an interpretation in terms of the scaling of the collective variances with the number of particles $N$. In fact, according to Def.~\ref{def:genssstates} a necessary condition for a state to be spin squeezed in a certain direction $\hat k$ is 
\begin{equation}
\lambda_k(\mathfrak Z_\rho)=(\Delta J_k)^2_\rho + \frac 1 {N-1} \aver{J_k^2}_\rho - \frac{N}{N-1} \sum_n \aver{(j_k^{(n)})^2}_\rho < 0 \ ,
\end{equation}
which means that, referring us for clarity to the case of spin-$\frac 1 2$ particle systems, both the variance $(\Delta J_k)^2$ and a normalized second moment $\frac 1 {N-1} \aver{J_k^2}$ must scale slower than $\frac N 4$. In the case of planar squeezed states, although both variances $(\Delta J_x)^2$ and $(\Delta J_y)^2$ scale as $N^{\frac 2 3}$, only one of the second moments, $\frac 1 {N-1}\aver{J_y^2}$ scales as $N^{-\frac 1 3}$. This is due to the fact that the state is almost completely polarized in the $\hat x$ direction  and thus $\frac 1 {N-1}\aver{J_x^2}\propto N$. On the contrary, according to our definition, a state spin squeezed in two directions would be such that e.g., $\lambda_x(\mathfrak Z) <0$ and $\lambda_y(\mathfrak Z) <0$ hold together with $\xi_G^2<1$.

\begin{table}[h!]
\caption{Collective correlation matrices (\ref{eq:corrmatrices}) and complete spin squeezing parameter (\ref{eq:genssparameter})
for interesting quantum states.}\label{tb:table1}
\begin{tabular*}{\textwidth}{@{\extracolsep{\fill}\vspace{0.03cm}}ll}
 \hline \hline \vspace{-0.2cm}\\
Multipartite Singlet states 
& $C=\diag(0,0,0)$ \\
 & $\Gamma=\diag(0,0,0)$ \\
\hspace{1.5cm} $\xi^2_{\rm G}(\rho_{\rm sing})=0$   & $Q=\diag(\frac{j(j+1)} 3,\frac{j(j+1)} 3,\frac{j(j+1)} 3)$ \\
  & $\vec{\lambda}(\mathfrak Z)=(-\frac{j(j+1)}{3}\frac{N^2}{N-1},-\frac{j(j+1)}{3}\frac{N^2}{N-1},-\frac{j(j+1)}{3}\frac{N^2}{N-1})$  \vspace{0.2cm}\\
 \hline  \vspace{-0.2cm}\\
Planar Squeezed states 
& $C\simeq\diag(\frac 1 4 N^2,\frac 1 4 N^{\frac 2 3},\frac 1 4 N^{\frac 4 3} )$ \\
 & $\Gamma\simeq \diag(\frac 1 8 N^{\frac 2 3},\frac 1 4 N^{\frac 2 3},\frac 1 4 N^{\frac 4 3})$ \\
\hspace{1.5cm} $\xi^2_{\rm G}(\rho_{\rm planar})\simeq \frac 1 2\left(\frac N {N-1}+N^{-\frac 1 3} \right)$  & $Q=\diag(\frac 1 4,\frac 1 4,\frac 1 4)$ \\
  & $\vec{\lambda}(\mathfrak Z) \simeq (\frac{1}{8}N^{\frac 2 3},-\frac{N^2(1-N^{-\frac 1 3})}{4(N-1)},\frac{N^2(N^{\frac 1 3}-1 )}{4(N-1)})$  \vspace{0.2cm}\\
  \hline  \vspace{-0.2cm}\\
Original Spin Squeezed states
& $C\simeq \diag(\frac{\xi^2} 4 N,\frac 1 {4\xi^2}N, \frac 1 4 N^2+\frac N 4)$ \\
 & $\Gamma\simeq \diag(\frac{\xi^2} 4 N,\frac 1 {4\xi^2}N,\frac N 4)$ \\
\hspace{1.5cm} $\xi^2_{\rm G}(\rho_{\rm pol})\simeq \frac 1 2\left(1+\xi^2 \right)<1$   & $Q=\diag(\frac 1 4,\frac 1 4,\frac 1 4)$ \\
  & $\vec{\lambda}(\mathfrak Z) \simeq (-\frac{N^2(1-\xi^2 )}{4(N-1)},\frac{N^2(\xi^{-2} -1 )}{4(N-1)},\frac N 4)$  \vspace{0.2cm}\\
 \hline  \vspace{-0.2cm}\\
 Unpolarized Dicke states
& $C= \diag(\frac{Nj(Nj+1)} 2,\frac{Nj(Nj+1)} 2, 0)$ \\
 & $\Gamma=\diag(\frac{Nj(Nj+1)} 2,\frac{Nj(Nj+1)} 2, 0)$ \\
\hspace{1.5cm} $\xi^2_{\rm G}(|Nj,0\rangle\langle Nj,0|)=  \frac{Nj-1}{2Nj-1}$   & $Q=\diag(\frac{j(j+1)} 2 - \frac{(N-1)j^2}{4jN-2},\frac{j(j+1)} 2 - \frac{(N-1)j^2}{4jN-2},\frac{(N-1)j^2}{2jN-1})$ \\
  & $\vec{\lambda}(\mathfrak Z) = (\frac{N^3j^3}{2jN-1},\frac{N^3j^3}{2jN-1},-\frac{N^2j^2}{2jN-1})$  \vspace{0.2cm}\\
\hline \hline
\end{tabular*}
\label{default}
\end{table}

Thus, to resume, here we have introduced $\xi^2_{\rm G}(\rho)$ in Eq.~(\ref{eq:genssparameter}) as a generalized notion of Spin Squeezing parameter that embraces in a single quantity the complete set of SSIs of Eq.~(\ref{completessiqubitsj}). It is a definition that generalizes the concept of spin squeezing to states with more than a single squeezed variance
and to a multipartite scenario for general spin-$j$ systems by distinguishing interparticle from intraparticle spin squeezing. This allows to include a wider class of states in the framework of spin squeezed states, and in particular many important unpolarized states that can be also studied for quantum enhanced technological purposes.

\subsection{Extreme Spin Squeezing near Dicke states}\label{sec:extrdicke}

A further question that can be asked is whether it is possible to detect also the depth of entanglement with generalized spin squeezing inequalities, as in the approach of S\o rensen-M\o lmer described in Sec.~\ref{sec:extremess}. In order to do this we can try to consider the complete set of generalized SSIs (\ref{completessiqubitsj}) and find a way to derive $k$-producibility inequalities from the same expressions. However in this respect 
numerical studies suggest that only one of the inequalities, namely (\ref{eq:dickeineqj}) can be possibly generalized to a $k$-producibility condition, since the others might be maximally violated already by $2$-entangled states, such as singlet states. 

In \cite{Lucke2014Detecting,vitaglianototh2015} we proceeded analogously as  S\o rensen-M\o lmer and consider a similar optimization problem, namely in our case find the minimum value achievable for a single collective variance $(\Delta J_x)^2$ for every fixed value of the sum of the orthogonal second moments $\aver{J_y^2 + J_z^2}$ in all $k$ producible states. As in the previous works described before in this Sec.~\ref{sec:genSSspinj}, the idea is to substitute the mean spin length orthogonal to $\aver{J_x}$ with the sum of two second moments, i.e., $\aver{J_y}^2+\aver{J_z}^2\rightarrow \aver{J_y^2 + J_z^2}$.

In the case of spin-$\frac 1 2$ particle systems we found as a result a family of $k$ separability condition very similar to Eq.~(\ref{sormolcrit}), written in terms of the same function (\ref{fsormol}) but with the quantity $\sqrt{\aver{J_y^2 + J_z^2}-\frac N 2 (\frac k 2 + 1)}$ in substitution of the mean spin length $\sqrt{\aver{J_y}^2+\aver{J_z}^2}$ (see supplementary material of \cite{Lucke2014Detecting} and \cite{vitaglianototh2015})

\bboo{(Improved extreme Spin Squeezing inequalities).} The following inequality
\begin{equation}\label{eq:oursormolcrit}
(\Delta J_x)^2 \geq \frac{N} 2 F_{\frac k 2}\left( \frac{\sqrt{\aver{J_y^2 + J_z^2}-\frac N 2 (\frac k 2 + 1)}}{\frac N 2} \right) \ ,
\end{equation}
holds for all $k$ producible states of $N$ spin-$\frac 1 2$ particles whenever $\aver{J_y^2 + J_z^2}-\frac N 2 (\frac k 2 + 1) \geq 0$, $J_l=\sum_{n=1}^N j_l^{(n)}$ being the collective spin components of the ensemble. The set of functions $F_{\frac k 2}(\cdot)$ in Eq.~(\ref{eq:oursormolcrit}) is defined as in Eq.~(\ref{fsormol}), with the identification $J=\frac k 2$, i.e.,
\begin{equation}\label{fsormolthj}
F_{\frac k 2}(X)=\frac 2 {k} \min_{\frac 2 k \aver{L_z}=X} (\Delta L_x)^2 \ ,
\end{equation}
where $L_l$ are the spin components of a single particle with spin $J=\frac k 2$ and the $F_J$ have the following properties. (i) $F_{J}(X)$ are convex and monotonically increasing for all $J$ and so are $F_{J}(\sqrt X)$. (ii) they are such that $F_{J}(0)=0$ for all $J$. (iii) $F_{J_1}(X) \leq F_{J_2}(X)$ holds for $J_1 \geq J_2$.

Every state $\rho$ of $N$ spin-$\frac 1 2$ particles that violates Eq.~(\ref{eq:oursormolcrit}) must be $(k+1)$-entangled. Eq.~(\ref{eq:oursormolcrit}) is maximally violated by unpolarized Dicke states, that are in fact detected as $N$-entangled states. Moreover, since $F_{J}(X) \leq \frac 1 2$ states such that $(\Delta J_x)^2 \geq \frac N 4$ cannot be detected with (\ref{eq:oursormolcrit}).
\eeoo

\bbpr The proof follows a reasoning analogous to the proof of Eq.~(\ref{sormolcrit}) 
plus the fact that for pure $k$-producible states of $N$ qubits
\begin{equation}\label{eq:condextrssfoll}
\sqrt{\aver{J_y^2 + J_z^2}-\frac N 2 \left(\frac k 2 + 1\right)}  \leq \sqrt{\aver{J_y}^2+\aver{J_z}^2} = \aver{J_n}
\end{equation}
holds. Eq.~(\ref{eq:condextrssfoll}) follows from $(\Delta J_y)^2+(\Delta J_z)^2 = \sum_{l=1}^M (\Delta j_y^{(l)})^2+(\Delta j_z^{(l)})^2 \leq \sum_{l=1}^M \frac{k_l} 2 \left(\frac{k_l} 2 + 1\right)\leq \frac N 2 \left(\frac k 2 + 1\right)$, where a pure $k$ producible state 
$|\phi_{\rm k-prod}\rangle=\bigotimes_{l=1}^M |\phi^{(l)}\rangle$
has been separated in total generality in $M$ groups of $k_l$ particles, indexed by $l$ and subject to the constraints $\sum_ {l=1}^M k_l =N$ and $\max_l k_l=k$, and 
we called $\hat n$ the direction of polarization of the state in the $(\hat y, \hat z)$-plane.
Thus for pure $k$-producible states $(\Delta J_x)^2 \geq \frac N 2 F_{\frac k 2}\left( \frac{\aver{J_n}}{\frac N 2} \right)
\geq \frac N 2 F_{\frac k 2}\left( \frac{\sqrt{\aver{J_y^2 + J_z^2}-\frac N 2 \left(\frac k 2 + 1\right)}}{\frac N 2} \right)$ follows from 
properties (i)-(iii) of the $F_{J}(X)$. Then, for mixed $k$-producible states Eq.~(\ref{eq:oursormolcrit}) follows 
from the fact that $F_{J}(\sqrt X)$ are convex, i.e., $\sum_i p_i F_{J}(\sqrt{X_i}) \geq F_{J}(\sqrt{\sum_i p_i X_i})$ where $\{p_i\}$ is a probability distribution.
\eepr

As we also said in Sec.~\ref{sec:extremess}, the functions $F_{J}(X)$ and correspondingly the criteria in Eq.~(\ref{eq:oursormolcrit}) can be easily evaluated numerically for $k$ of the order of several hundreds. Moreover also in this case the bound on the right hand side is tight for $N\gg 1$, meaning that there are $k$ producible states reaching the boundary. These last are product of $k$ particle generalized spin squeezed states, namely ground states of (\ref{hspinsq}), which include Dicke states.

\begin{figure}[h!]
\centering
\includegraphics[width=0.7\columnwidth,clip]{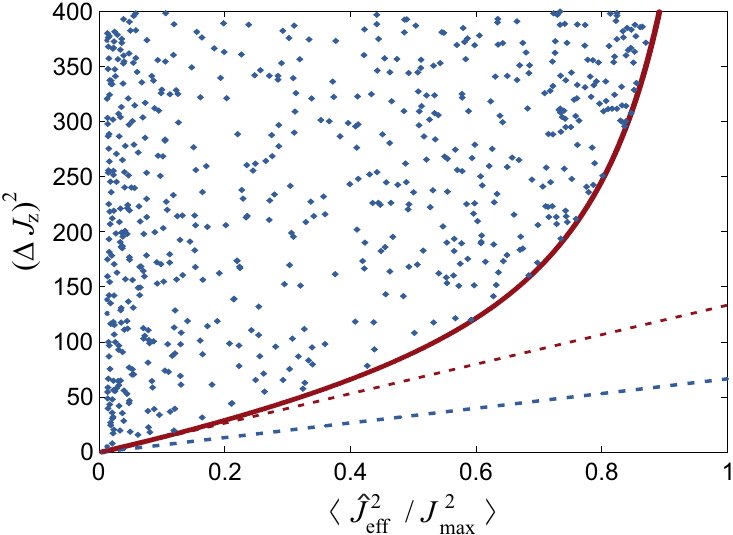}
\caption{Detection of $k$-particle entanglement based on the total spin. The red line marks the boundary for $k$-particle entangled states with $N=8000$ and $k=28$ in the $(\aver{\hat J_{\rm eff}^2/J_{\rm max}^2}, (\Delta J_z)^2)$-plane, where we defined $\hat J_{\rm eff}^2:=J_x^2+J_y^2$ and $J_{\rm max}^2:=\frac N 2$. As a cross-check, random states with $k$-particle entanglement are plotted as blue dots, filling up the allowed region. The criterion of Ref.~\cite{Duan2011Entanglement} only detects states that correspond to points below the dashed blue line. An improved linear criterion is gained from calculating a tangent to the new boundary (dashed red line).}\label{fig:kentcond}
\end{figure}

\begin{figure}[h!]
\centering
\includegraphics[width=0.7\columnwidth,clip]{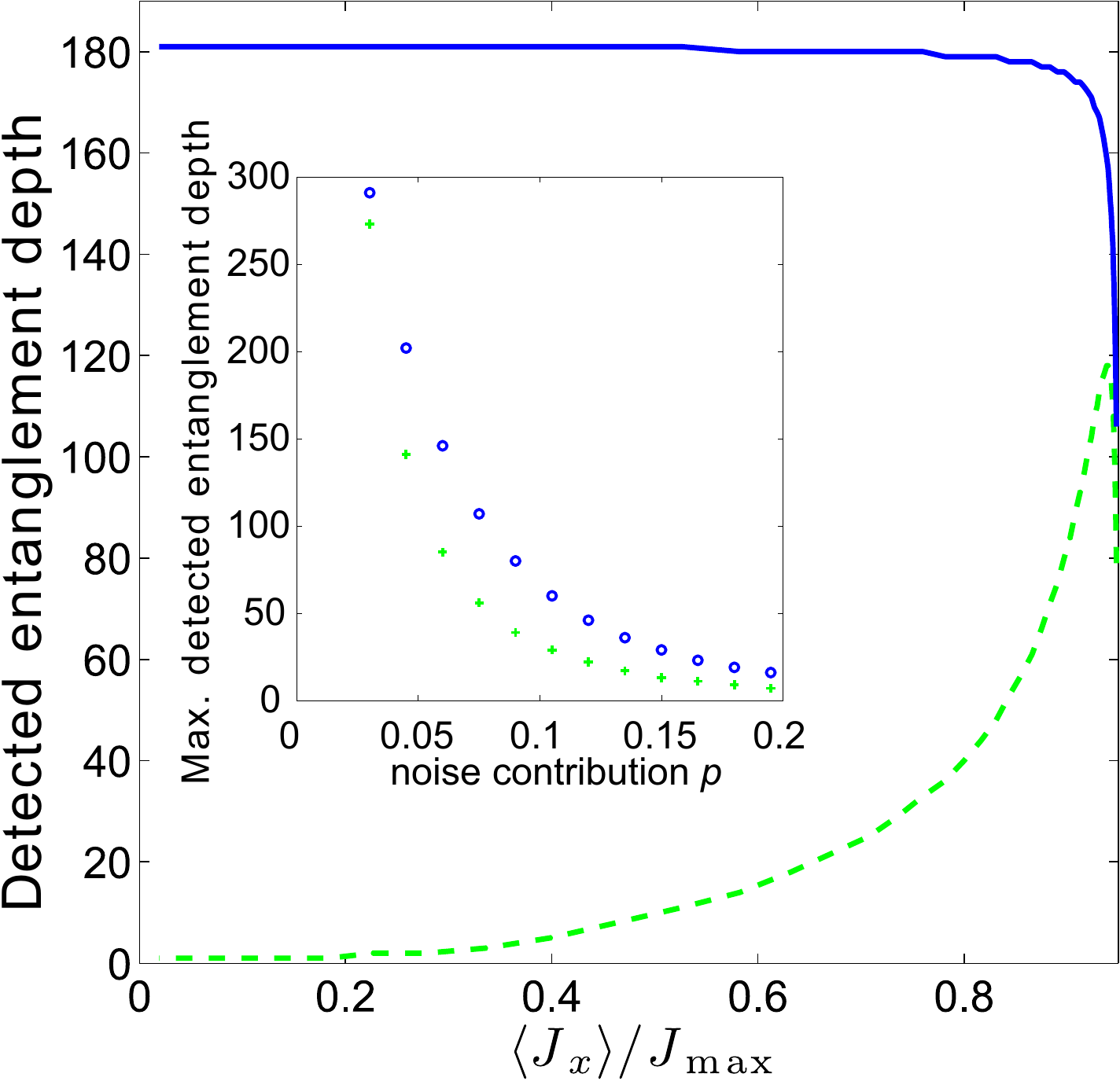}
\caption{The
graph shows the entanglement depth detected by the condition (\ref{eq:oursormolcrit}) (solid line) 
and the S\o rensen-M\o lmer condition (\ref{sormolcrit})
(dashed line) for N = 4000 spin-1 particles with additive
white noise to account for imperfections. For states that are not completely polarized, Eq.~(\ref{eq:oursormolcrit}) detects a considerably larger entanglement depth. The inset shows that the maximal detected entanglement depth depending on the noise contribution is larger for our criterion (circles) than for the S-M criterion (crosses) if some very small noise is present.}\label{fig:oursolmoncomp}
\end{figure}

The $F_{J}(X)$ can be computed analytically straightforwardly just in the cases $J=\frac 1 2$ and $J=1$, the results being
\begin{equation}
\begin{aligned}
F_{\frac 1 2}(X) &= \frac 1 2 X^2 \ , \\
F_1(X) &=\frac 1 2 (1-\sqrt{1-X^2}) \ ,
\end{aligned}
\end{equation}
and leading to the following analytical separability and $2$-producibility conditions
\begin{subequations}
\begin{align}
N (\Delta J_x)^2 &\geq  \aver{J_y^2 + J_z^2}- \frac 3 4 N  \ , \label{eq:sepcondextdic} \\
(\Delta J_x)^2 &\geq\frac N 4 -\frac 1 2 \sqrt{\frac{N^2} 4 -\aver{J_y^2 + J_z^2}+N} \ ,
\end{align}
\end{subequations}
respectively. Note that Eq.~(\ref{eq:sepcondextdic}) provides a worst bound as compared to Eq.~(\ref{eq:dickeineqj}), the difference being nevertheless negligible for large $N$.
Afterwards, it is interesting to compare the two criteria (\ref{eq:oursormolcrit}) and (\ref{sormolcrit}). Due to the monotonicity of $F_{\frac k 2}(X)$ we can then just compare the arguments of the functions. Then, whenever 
\begin{equation}
\aver{J_y^2 + J_z^2}-\frac N 2 \left(\frac k 2 + 1\right) \leq  \aver{J_y}^2+\aver{J_z}^2
\end{equation}
holds, we have that (\ref{sormolcrit}) implies (\ref{eq:oursormolcrit}) and therefore the former gives a stronger condition with respect to the latter. Vice versa, 

\bboo{(Comparison between Eq.~(\ref{eq:oursormolcrit}) and S-M inequality).}
Whenever 
\begin{equation}\label{eq:ourstonger}
\aver{J_y^2 + J_z^2}-\frac N 2 \left(\frac k 2 + 1\right) \geq  \aver{J_y}^2+\aver{J_z}^2
\end{equation}
holds, then (\ref{eq:oursormolcrit}) is stronger than the S\o rensen-M\o lmer criteria (\ref{sormolcrit}) in the sense that it detects a wider set of $k$-entangled states. 
\eeoo
In practice, a particular case in which Eq.~(\ref{eq:ourstonger}) holds, is on states close to Dicke states and in all states in which the mean polarization $\aver{J_y}^2+\aver{J_z}^2$ is small or even just slightly reduced by noise. In fact, our criteria have then a stronger white noise tolerance and outperform (\ref{sormolcrit}) when already a very small amount of noise (e.g., around $5\%$) is present. See Fig.~\ref{fig:oursolmoncomp} for a numerical comparison of the entanglement depth detected by the two criteria.

Thus, the entanglement criteria in Eq.~(\ref{eq:oursormolcrit}) extend S\o rensen-M\o lmer's criteria since they 
allow to detect more efficiently the depth of entanglement in a wider class of states that goes beyond usual SSS states and includes,
e.g., unpolarized states close to Dicke states. They can thus be thought as a generalization of Eq.~(\ref{sormolcrit}), in the same spirit as the set (\ref{completessiqubitsj}) being a generalization of the original SSI. 

Furthermore, we can compare our criteria with an other important set of conditions that are designed to detect the entanglement depth of Dicke-like states. These are linear criteria derived by L.-M Duan \cite{Duan2011Entanglement}
\bboo{(L.-M Duan's linear $k$ producibility conditions).}
The following inequality
\begin{equation}\label{eq:duancon}
N(k+2)(\Delta J_x)^2 \geq \aver{J_y^2 + J_z^2}-\frac N 4 \left(k + 2\right)
\end{equation}
holds for all $k$ producible states of $N$ spin-$\frac 1 2$ particles. Any state that violates Eq.~(\ref{eq:duancon}) is detected as $k+1$-entangled.
\eeoo
We can as well obtain easily linear criteria from Eq.~(\ref{eq:oursormolcrit}) by just taking the tangents to the curve given by $F_{\frac k 2}(X)$. Since Eq.~(\ref{eq:oursormolcrit}) is optimal, in the sense that the bound on the right hand side is tight, the tangents provide also optimal linear criteria, that outperform Eq.~(\ref{eq:duancon}). In particular this can be seen explicitly from the numerical plots in Fig.~\ref{fig:kentcond} and analytically in the $k=2$ case by computing the second derivative of $F_1(X)$. We obtain so the following linear $2$ producibility condition
\begin{equation}
2N(\Delta J_x)^2 \geq  \aver{J_y^2 + J_z^2}-N 
\end{equation}
that is finer than Duan's and improves the slope by roughly a factor 2. 

\begin{figure}[h!]
\centering
\includegraphics[width=0.7\columnwidth,clip]{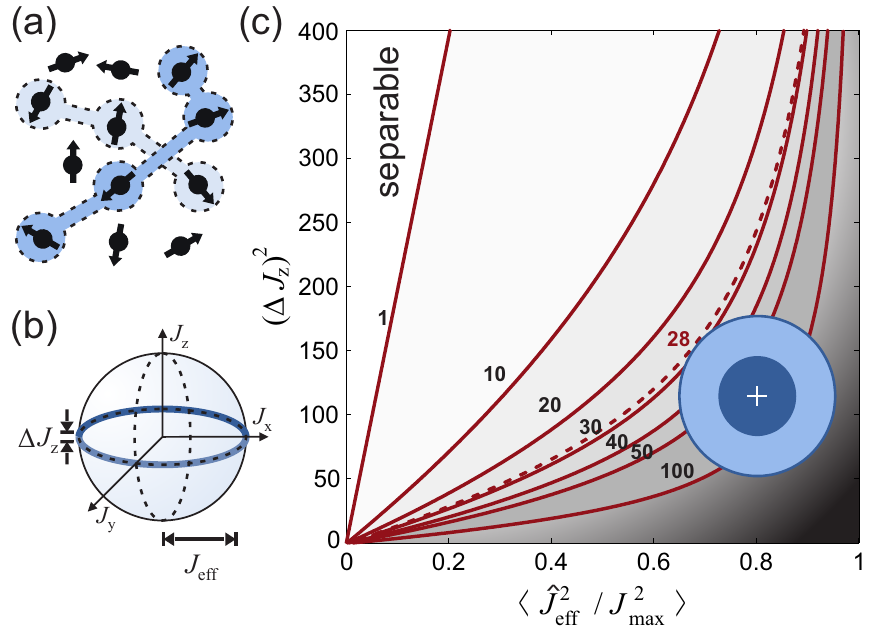}
\caption{(left) Measurement of the entanglement depth for an $8000$ atoms Dicke state. (a) The entanglement depth is given by the number of atoms in the largest non-separable subset (shaded areas). (b) Dicke states are represented by an equatorial plane: it has an ultralow width $\Delta J_z$ and a large radius $J^2_{\mathrm{eff}}=J_x^2+J_y^2$.  (c) The red lines indicate the boundaries for various entanglement depths. The experimental result is shown as blue uncertainty ellipses with one and two standard deviations, proving an entanglement depth larger than $28$ (dashed line). (right) A photo of the experiment.}\label{fig:luckeexp}
\end{figure}

Figure~\ref{fig:luckeexp}(c) shows the entanglement depth of a created Dicke-like state of $8000$ atoms \cite{Lucke2014Detecting}. The red lines present the newly derived boundaries for $k$-particle entanglement. All separable states are restricted to the far left of the diagram, as indicated by the $k=1$ line. The measured values of $(\Delta J_z)^2$ and 
 $(\aver{\hat J_{\rm eff}^2/J_{\rm max}^2}$ (where we defined $\hat J_{\rm eff}^2:=J_x^2+J_y^2$ and $J_{\rm max}^2:=\frac N 2$) are represented by uncertainty ellipses with one and two standard deviations. The center of the ellipses corresponds to an entanglement depth of $k=68$. With two standard deviations confidence, the data prove that the state had an entanglement depth larger than $k=28$.


\section{Spin squeezing for fluctuating number of particles}

The concept of spin squeezing has been introduced for a single spin-$j$ and then developed to include composite systems of many spins. In this last case the natural question that was addressed was the connection with multipartite entanglement between the spins. 

Then, we can ask the question of how to extend also the various Spin Squeezing entanglement criteria to the fluctuating $N$ case. This has been studied by Hyllus \emph{et al.} \cite{hyllus12}, who found a very general result by basically noticing that all such criteria can be generalized to the fluctuating number of particles case by simply exploiting the concavity of variance, which leads to 
\begin{equation}
(\Delta J_k)^2 \geq \sum_N Q_N (\Delta J_{k, \rm N})^2 \ ,
\end{equation}
holding for all states of the form (\ref{eq:rhomixNpar}) and basically substituting the constant value of $N$ with its average value $\aver{N}$. In this way for example, the seminal result of S\o rensen-M\o lmer can be generalized to 

\bbth{(Extreme SSI for fluctuating $N$).} The following inequality
\begin{equation}\label{sormolcritfluctN}
(\Delta J_x)^2 \geq \aver{N}j F_{kj}\left( \frac{\aver{J_z}}{\aver{N}j} \right) \ ,
\end{equation}
holds for all $k$ producible states with an average number $\aver{N}$ of spin-$j$ particles, $J_k = \bigoplus_{N=0}^\infty J_{k, \rm N}$ being the collective spin components of the ensemble. The set of functions $F_{kj}(\cdot)$ in Eq.~(\ref{sormolcritfluctN}) is defined as in Th.~(\ref{th:extSSI}). 

Every state $\rho$ of an ensemble of a fluctuating number $\aver{N}$ of spin-$j$ particles that violates Eq.~(\ref{sormolcritfluctN}) must be $k+1$-entangled.
\eeth

Analogously, the complete set of generalized SSIs of Eq.~(\ref{eq:su2invine}) can be extended to

\bboo{(Complete set of SSIs for fluctuating $N$).}
Let us consider the matrices 
\begin{align}
C_{kl}&:=\frac 1 2 \aver{J_k J_l + J_l J_k} \ , \\
\Gamma_{kl}&:=C_{kl} -\aver{J_k}\aver{J_l}  \ , \\
\mathfrak Q_{kl}&:= \frac 1 2 \aver{ N(N-1)^{-1} \sum_n \left( j_k^{(n)} j_l^{(n)}+ j_l^{(n)}j_k^{(n)} \right)} \ , \label{eq:qmatdefluctN} \\
\mathfrak Z &:= \Gamma + \frac 1 2 \aver{(N-1)^{-1}(J_k J_l + J_l J_k)} - \mathfrak Q \ , 
\end{align}
where $J_k = \bigoplus_{N=0}^\infty J_{k, \rm N}$ are the collective spin operators of an ensemble of fluctuating number of spin-$j$ particles.
 
Then, the following set of inequalities
\begin{equation}\label{eq:su2invinefluctN}
\xi^2_{G, \rm N}:=\frac{\trace(\Gamma) - \sum_{k=1}^I \lambda^{>0}_k(\mathfrak Z)}{\aver{N}j} \geq 1 \ ,
\end{equation}
holds for all separable states with an average number $\aver{N}$ of spin-$j$ particles, where $\lambda^{>0}_k(\mathfrak Z)$ are the positive eigenvalues of $\mathfrak Z$. 

Every state $\rho$ of an ensemble of a fluctuating number $\aver{N}$ of spin-$j$ particles that violates Eq.~(\ref{eq:su2invinefluctN}) must be entangled.
\eeoo

Thus we can define \cite{vitaglianototh2015} generalized SSS of an ensemble of fluctuating number of spin-$j$ particles based on the parameter $\xi^2_{G, \rm N}$, namely whenever $\xi^2_{G, \rm N}<1$ the state is called \emph{generalized Spin Squeezed} in this framework. The proof of Eq.~(\ref{eq:su2invinefluctN}) has been omitted because it is analogous to the result that we are going to show. Namely we can also extend our set of $k$ entanglement conditions (\ref{eq:oursormolcrit}) to the fluctuating $N$ case. In the following actually we prove a slightly stronger result as compared to (\ref{eq:oursormolcrit}), 
tightening the bound on the right hand side by a term of the order of $\frac 1 N$ (in preparation for a future publication \cite{vitaglianototh2015}).

\bboo{(Improved extreme SSIs for fluctuating $N$).} 
Consider an ensemble of spin-$\frac 1 2$ in a mixture $\rho=\sum_N Q_N \rho_N$ of $N$ particle states $\rho_N$ and define the collective spin operators to be in the reducible representation $J_k = \bigoplus_{N=0}^\infty J_{k, \rm N}$. Then,
the following inequality
\begin{equation}\label{eq:oursormolcritfluctN}
\aver{J_y^2 + J_z^2} \leq \aver{N} \frac{k+2} 4 + \frac{\aver{N(N-1)}} 4 F^{-2}_{\frac k 2}\left( \frac{2\aver{N(N-1)(J_{x,\rm N}-\aver{J_{x,\rm N}})^2}}{\aver{N(N-1)}} \right) \ ,
\end{equation}
must be satisfied by all $k$ producible states of an average number $\aver{N}$ of qubits whenever 
\begin{equation}
\frac{2\aver{N(N-1)(J_{x,\rm N}-\aver{J_{x,\rm N}})^2}}{\aver{N(N-1)}} \leq \frac 1 2
\end{equation}
holds. The set of functions $F_{\frac k 2}^{-1}(\cdot)$ in Eq.~(\ref{eq:oursormolcritfluctN}) is defined as the inverse of Eq.~(\ref{fsormol}) and has the following properties. (i) $F_{J}^{-1}(X)$ are concave and monotonically increasing for all $J$ and so are $F_{J}^{-2}(X)$ for $X\leq \frac 1 2$. (ii) they are such that $F_{J}(0)=0$ for all $J$. (iii) $F_{J_1}(X) \leq F_{J_2}(X)$ holds for $J_1 \geq J_2$.

Every state $\rho$ of a fluctuating number $\aver{N}$ of spin-$\frac 1 2$ particles that violates Eq.~(\ref{eq:oursormolcritfluctN}) must be $k+1$-entangled. 
\eeoo

\bbpr Let us look for the maximum of $\aver{J_y^2 + J_z^2}$ on $k$ producible states, as a function of expectations involving $J_{x, \rm N}$ and $N$. For states of the form (\ref{eq:rhomixNpar}) we have $\aver{J_y^2 + J_z^2}=\sum_N Q_N \aver{J_{y, \rm N}^2 + J_{z, \rm N}^2}$. Thus, at first let us look for bounds to $\aver{J_{y, \rm N}^2 + J_{z, \rm N}^2}$ in each fixed-$N$ subspace. In particular we look for the maximal value of $\aver{J_{y, \rm N}^2 + J_{z, \rm N}^2}$ for every fixed value of $(\Delta J_{z, \rm N})^2$, and therefore, due to the concavity of the variance, we focus on pure $k$ producible states. We have 
\begin{equation}
\aver{J_{y, \rm N}^2 + J_{z, \rm N}^2} = (\Delta J_{y, \rm N})^2+(\Delta J_{z, \rm N})^2+\aver{J_{y, \rm N}}^2 + \aver{J_{z, \rm N}}^2 \ ,
\end{equation}
which becomes 
\begin{equation}\label{eq:firbound}
\aver{J_{y, \rm N}^2 + J_{z, \rm N}^2} = \sum_{n=1}^{N_k}\left( (\Delta L_{y, \rm k_n})_n^2+(\Delta L_{z, \rm k_n})_n^2\right) +\aver{J_{y, \rm N}}^2 + \aver{J_{z, \rm N}}^2 \ ,
\end{equation}
on pure $k$ producible states $\rho_{\rm k-prod, N}=\bigotimes_{n=1}^{N_k} \rho_{k_n, n}$, where $\rho_{k_n}$ are states of $1\leq k_n \leq k$ particles, $L_{l, \rm k_n}$ are the corresponding collective spin components and $\sum_{n=1}^{N_k} k_n =N$. Now, by using Jensen inequality in the form 
\begin{equation}\label{eq:jensinefn}
-\sum_{n=1}^{N_k} k_n f^2_n \leq -\frac 1 N \left( \sum_{n=1}^{N_k} k_n f_n \right)^2 \ ,
\end{equation}
where $f_n$ are real numbers, we can bound 
\begin{equation}\label{eq:secbound}
-\sum_{n=1}^{N_k} \aver{L_{l, \rm k_n}}^2_n \leq -\sum_{n=1}^{N_k} k_n \frac{\aver{L_{l, \rm k_n}}^2_n}{k_n^2} \leq -\frac 1 N \left( \sum_{n=1}^{N_k} \aver{L_{l, \rm k_n}}_n \right)^2=-\frac 1 N \aver{J_{l, \rm N}}^2 \ ,
\end{equation}
where the first inequality comes from the fact that $k_n \geq 1$ and the second comes from Eq.~(\ref{eq:jensinefn}) with the choice $f_n=\frac{\aver{L_{l, \rm k_n}}_n}{k_n}$.
Thus from Eq.~(\ref{eq:firbound}) and Eq.~(\ref{eq:secbound}) we obtain
\begin{equation}
\aver{J_{y, \rm N}^2 + J_{z, \rm N}^2} \leq \sum_{n=1}^{N_k} \aver{L_{y, \rm k_n}^2+L_{z, \rm k_n}^2}_n +\frac{N-1} N (\aver{J_{y, \rm N}}^2 + \aver{J_{z, \rm N}}^2) \leq  
\frac{N(k+2)} 4 + \frac{N-1} N (\aver{J_{y, \rm N}}^2 + \aver{J_{z, \rm N}}^2)  \ ,
\end{equation}
where the second inequality follows from $\sum_{n=1}^{N_k} \aver{L_{y, \rm k_n}^2+L_{z, \rm k_n}^2}_n \leq \sum_{n=1}^{N_k} \frac{k_n(k_n+2)}{4} \leq N\frac{k+2} 4$. Now we employ the definition of $F_{J}(X)$ and we exploit the result of S\o rensen-M\o lmer, Eq.~(\ref{sormolcrit}) to bound 
\begin{equation}
\sqrt{\aver{J_{y, \rm N}}^2 + \aver{J_{z, \rm N}}^2} \leq \frac N 2  F_{\frac k 2}^{-1}\left( \frac{2(\Delta J_{z, \rm N})^2}{N} \right) \ ,
\end{equation}
which is valid for $k$ producible states and obtain finally
\begin{equation}
\aver{J_{y, \rm N}^2 + J_{z, \rm N}^2} \leq N\frac{k+2} 4 + \frac{N(N-1)} 4 F_{\frac k 2}^{-2}\left( \frac{2(\Delta J_{z, \rm N})^2}{N} \right) \ .
\end{equation}
To extend this result to arbitrary mixtures of $N$ particle states we simply use again Jensen inequality $\sum_N Q_N N(N-1) F_{\frac k 2}^{-1}(X_N) \leq \left( \sum_N Q_N N(N-1)\right) \cdot
F_{\frac k 2}^{-1}(\frac{\sum_N Q_N N(N-1) X_N}{\sum_N Q_N N(N-1)})$ that is valid for concave functions $F_{\frac k 2}^{-1}(X_N)$. In this way we arrive at 
\begin{equation}
\aver{J_{y}^2 + J_{z}^2} \leq \aver{N}\frac{k+2} 4 + \frac{\aver{N(N-1)}} 4 F_{\frac k 2}^{-2}\left( \frac{2\sum_N Q_N N(N-1) (J_{x,\rm N}-\aver{J_{x,\rm N}})^2}{\aver{N(N-1)}} \right) \ ,
\end{equation}
and we conclude the proof by exploiting the monotonicity of $F_{\frac k 2}^{-1}(X_N)$, which is valid for its argument being bounded by $X_N\leq \frac 1 2$. 
\eepr

In this chapter we have presented our results contained in Refs.~\cite{vitagliano11,vitagliano14,Lucke2014Detecting,vitaglianototh2015} concerning an extension of the concept of spin squeezing to the framework
of multi spin-$j$ systems. We have extended the well established results of T\'oth \emph{et al}, namely Eq.~(\ref{completessiqubits})
and the proof of its completeness, to multipartite systems of spin-$j$ particles, resulting in Eq.~(\ref{completessiqubitsj}). 
We have shown that this result can be directly obtained with a very general mapping (\ref{eq:12jmapp}) from entanglement criteria valid for spin-$\frac 1 2$ particle systems to analogous criteria valid for spin-$j$ systems.

Then, we have defined a single generalized spin squeezing parameter, $\xi^2_{\rm G}(\rho)$ in Eq.~(\ref{eq:genssparameter}),
that extends the original definition of spin squeezing in several respects: (i) it detects states with more than a single squeezed variance, (ii) it detects important unpolarized states, (iii) it detects all entangled spin states that can be detected based only on the collective correlation matrices (\ref{eq:corrmatrices}). This parameter in summary, provides a unified and complete figure of merit for spin squeezing, valid even when more than a single variance is squeezed. 

Afterwards, we also extended the well-known $k$-entanglement criteria of S\o rensen-M\o lmer, Eq.~(\ref{sormolcrit}), to a set criteria that detects the entanglement depth of a wider class of states, including unpolarized Dicke states. The resulting
criteria, Eq.~(\ref{eq:oursormolcrit}), are more efficient than the original S\o rensen-M\o lmer criteria in detecting the entanglement depth already when a very small amount of depolarizing noise is present and also outperform the linear criteria (\ref{eq:duancon}) of L.-M. Duan. They have been employed to detect a depth of entanglement of $28$ with two standard deviations confidence in an experimentally produced Dicke state of a BEC \cite{Lucke2014Detecting}, see Fig.~\ref{fig:luckeexp}.

Finally, all the above mentioned results have been extended to systems in which the number of particles is fluctuating.

\afterpage{\blankpage}
\chapter{Toward a conclusive Leggett-Garg test with QND measurements}\label{ch:LG} 

The other topic treated in this thesis concerns a foundational question behind the modern formulation of quantum mechanics: where is the boundary between the microscopic realm governed by QM and the measuring devices? Quoting Bell, the answer is (believed to be) that there is no boundary at all and that a quantum mechanical description is needed even at macroscopic scales. However this fact would conflict with classical principles, such as Realism and non-invasive Measureability, i.e., essentially with the possibility of assigning values to macroscopic properties independently of measurements. 
To make clearer the distinction between a macrorealist theory and quantum mechanics Leggett and Garg \cite{LG85} designed a rigorous test of such principles, that we described in Sec.~\ref{sec:LGineq}. We mentioned previously that the proposal, however, suffers from a fundamental {\it loophole}: non-invasive Measurability, cannot be tested independently from Macrorealism {\it per se} because of the practical impossibility of making perfectly and rigorously tested non-disturbing measurements.

In a way we can say that the two assumptions of (MR) and (NIM) can be interpreted as two faces of a single property of quantum mechanics, syntethized as the {\it Heisenberg uncertainty principle} holding also for macroscopic systems.
A failure of an LG test thus witnesses such a property, but without additional refinements it could be interpreted as a practical rather than fundamental problem of the measurements made. In the next section~\ref{sec:cluloophint} we are going to see in more details this weakness of the LG approach and some proposals made to address it as much as possible. We will mention the original idea of {\it ideal negative choice measurement} \cite{LG85} and a successive different proposal by Wilde and Mizel \cite{WildeFP2012}.

Then we discuss our two-fold original contribution to this topic \cite{budronivitagliano,vitaglianobudroni15}. On the one hand in Sec.~\ref{sec:LGatomic} we are going to adapt the original LG proposal to the realm of ensembles of cold atoms probed with {\it Quantum Non-Demolition} (QND) measurements \cite{MitchellNJP2012,KoschorreckPhD} and show that a violation of LG inequalities can be reached in a realistic setting. 
In a sense, we show then that QND measurements are an ideal tool for disproving macrorealism in a real experiment with a macroscopic system \cite{budronivitagliano}.
 
Afterwards, in Sec.~\ref{sec:invasivityquant}, we study in more details the invasivity of these QND measurements and propose a way to rigorously distinguish the ``clumsiness'' of the measurements from the unavoidable disturbance due to quantum principles. We finally show by suitably quantifying it, that this classical clumsiness is not sufficient to explain the predicted failure of the LG test in realistic cold atomic ensembles \cite{vitaglianobudroni15}.


\section{The clumsiness loophole in a LG test.}\label{sec:cluloophint}

As we said, the aim of a Leggett-Garg test is to witness a genuine macroscopic quantum effect, i.e., an 
effect that is inconsistent with a realist view at a macroscopic level. Such a viewpoint has been formalized via the joint hypothesis of (i) macroscopic realism (MR), i.e., the existence of a definite value for a macroscopic quantity at any time, and (ii) non-invasive measurability (NIM), i.e., the possibility of measuring such value with an arbitrary small perturbation of the system. Hence, to make a conclusive LG test on macroscopic systems one has to address separately the invasivity problem and show that the measurement by itself does not disturb the system from a macrorealist viewpoint. This is a very difficult and even to some extent fundamentally prohibited goal since any experimental realization of a LG test will have to face disturbances, both unwanted and fundamental (see e.g., the review article \cite{EmariNoriRPR2014}). In fact, already very soon after the publication of the LG idea, a debate started on whether it is even in principle possible to address the (NIM) assumption by itself in an actual experiment \cite{ballentine87,LGcomment87,peresLG88,LGcomment89,Elby199217,benatti95,Leggett08}. Such debate is not yet closed and this rather fundamental problem of LG tests has been termed {\it clumsiness loophole} \cite{WildeFP2012}. 

To make a comparison with Bell-type tests, the analogous loophole in that framework would be 
the {\it communication loophole}, namely the possibility that the two party can communicate between each other, violating the locality assumption by classical means, maybe through some hidden variable. However, the communication loophole can be conclusively closed by simply putting the two parts in space-like distant regions and invoke the principle of local causality of special relativity. Thus, when faced with a negative result in a Bell test performed with such shrewdness, one has to choose between renounce to realism or special relativity.
Here there is no analogous fundamental principle to call upon and one can always attribute the failure of an LG test to the modification of some hidden variable made by the measurements. On the other hand, it is also generally understood that the sole principle of realism cannot be tested \cite{EmariNoriRPR2014}. Thus in this case the best that one can do is to contrive possible explanations in terms of classical clumsiness so that they become as unacceptable as the violation of realism itself.

Leggett and Garg \cite{LG85}, as an example of such strategy proposed a measurement thought to be ideally non-disturbing at least from a macrorealist perspective, called {\it ideal negative-choice measurement}. To describe it they, considering again a macroscopic dichotomic variable $Q$, supposed that one can arrange the measurement apparatus, say a detector, such that it interacts with the system only when $Q=1$. In this way one can acquire the information of the variable having value $Q=-1$ {\it without interacting with the system}, i.e., without a click of the detector. Thus, taking only this negative results into account one can argue that the measurements were non-invasive from a macrorealist point of view and can witness just the quantum unavoidable invasivity due to the Heisenberg uncertainty principle. 

Nevertheless one can still try to explain a negative answer to an LG test involving these negative-choice measurements as coming from some hidden and unwanted \emph{clumsiness}. Some more quantitative analysis of classical disturbances 
is needed, leaving still open the clumsiness loophole \cite{WildeFP2012}. This means that in a realistic scenario one has to explicitly quantify any possible classical clumsy effects of the measurements, showing that these are not sufficient to explain the failure of the LG test.
To give more details on this point let us consider the LG inequality 
\begin{equation}\label{lgthree1}
\aver{Q_2Q_1}+\aver{Q_3Q_2}-\aver{Q_3Q_1}\leq 1 \ ,
\end{equation}
for dichotomic observables $Q_i$ and a sequence of three measurements $\mathcal S_{\rm (1,2,3)}=(\mathcal M_1 \rightarrow \mathcal M_2 \rightarrow \mathcal M_3)$ at times $t_1,t_2,t_3$ (see Fig.~\ref{fig:3seqschemes}). Clearly, with such a sequence it is not possible to violate the LGI with ideal non-disturbing measurements, 
since from these observations we obtain a joint probability distribution $\Pr(x_1,x_2,x_3)_{\rm (Q_1,Q_2,Q_3)}$ and compute the correlations $\mean{Q_i Q_j}$ from the relative marginals\footnote{Note that it would be possible to violate Eq.~(\ref{lgthree1}) with a sequence like $\mathcal S_{\rm (1,2,3)}$ if one allows a disturbance on $Q$ that could change its value and push it outside of the interval $[-1,1]$, see \cite{EmariNoriRPR2014} for more details.}.

A violation of LGI can be instead achieved by performing two different sequences of measurements, say $(\mathcal S_{\rm (1,2,3)}, \mathcal S_{\rm (1,3)})$, the second of which,
$\mathcal S_{\rm (1,3)}=(\mathcal M_1 \rightarrow \mathcal M_3)$ does not include a measurement at $t_2$. In this way we compute $\mean{Q_1Q_3}$ from a probability distribution $\Pr(x_1,x_3)_{\rm (Q_1,Q_3)}$ that is not obtained as a marginal over $\Pr(x_1,x_2,x_3)_{\rm (Q_1,Q_2,Q_3)}$. However an eventual falsification of the test coming from such a measurement scheme could be also interpreted as   
coming from the invasivity of the measurement at $t_2$.
Formally, to address this possibility we define an invasivity parameter relative to $\mathcal M_2$ \cite{WildeFP2012}
\begin{equation}\label{invpar}
\mathcal{I}(\mathcal M_2)=\int dx_1 dx_3 \left|\Pr (x_1,x_3)_{\rm (Q_1,Q_2,Q_3)}- \Pr(x_1,x_3)_{\rm (Q_1,Q_3)}\right| \ ,
\end{equation}
where $\Pr (x_1,x_3)_{\rm (Q_1,Q_2,Q_3)}=\int \Pr(x_1,x_2,x_3)_{\rm (Q_1,Q_2,Q_3)} {\rm d} x_2$ is the marginal relative to the outcomes of $Q_1$ and $Q_3$ obtained performing the sequence $\mathcal S_{\rm (1,2,3)}$ and $\Pr(x_1,x_3)_{\rm (Q_1,Q_3)}$ is the actual probability distribution coming from the sequence $\mathcal S_{\rm (1,3)}$.

The parameter $\mathcal{I}(\mathcal M_2)$ can give us the corrections to the bound for the LG inequality due to invasiveness of $\mathcal M_2$ since
\begin{equation}\label{diffcorr}
|\mean{Q_3 Q_1}_{\rm (1,2,3)}-\mean{Q_3 Q_1}_{\rm (1,3)}|\leq \mathcal{I} (\mathcal M_2) \ ,
\end{equation}
where $\mean{\ }_{\rm (1,2,3)}$ and $\mean{\ }_{\rm (1,3)}$ are the correlations computed with probabilities $\Pr (x_1,x_3)_{\rm (Q_1,Q_2,Q_3)}$ and $\Pr (x_1,x_3)_{\rm (Q_1,Q_3)}$, respectively.
Such a parameter, however, would by definition compensate any possible violation of the LGI with such protocol.

\begin{figure*}[h!]
\centering
\includegraphics[width=\textwidth,clip]{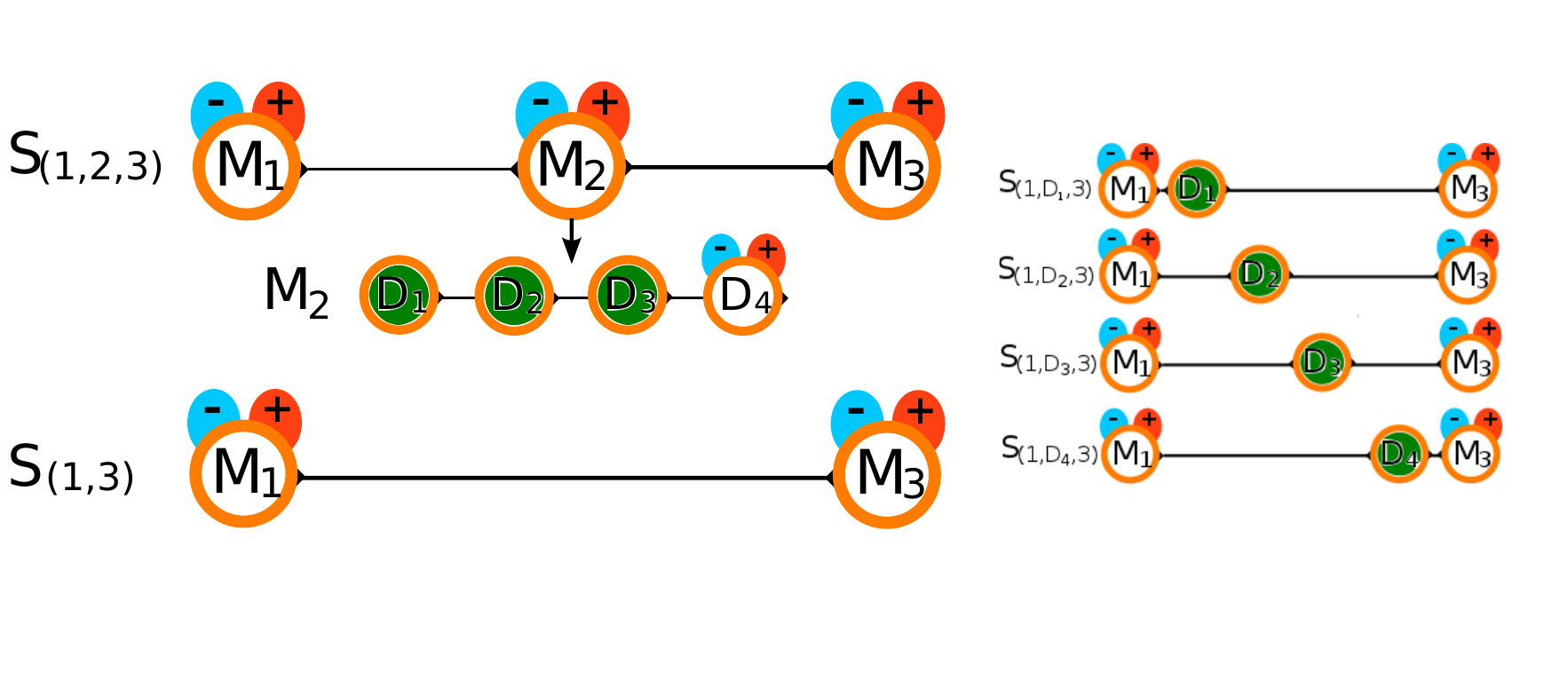}
\caption{Schematic representation of the Wilde-Mizel proposal. (left) Measurement $\mathcal M_2$ 
is decomposed in four steps $\mathcal D_i$ and the outcome is recorded only in the last step $\mathcal D_4$ giving the result $Q_2=x_2$.
(right) The invasivity of each of the $\mathcal D_i$ can be tested separately by performing auxiliary sequences $\mathcal S_{\rm (1,D_i,3)}$ and comparing them
with $\mathcal S_{\rm (1,3)}$ as in Eq.~(\ref{adroit}).}\label{fig:wmproposal}
\end{figure*}

As a possible solution to this problem, Wilde and Mizel in \cite{WildeFP2012} proposed to decompose the second measurement $\mathcal M_2$ in four measurement steps $\mathcal D_i$, and read the outcome only in the last step, as depicted in Fig.~\ref{fig:wmproposal}. 
Then, the invasiveness of each of the $\mathcal D_i$ can be tested separately by performing auxiliary sequences $\mathcal S_{\rm (1,D_i,3)}=\mathcal M_1 \rightarrow \mathcal D_i \rightarrow \mathcal M_3$ and computing the parameters 
\begin{equation}\label{adroit}
\mathcal{I}(\mathcal D_i)=\int dx_1 dx_3 |\Pr(x_1,x_3)_{\rm (Q_1,Q_i,Q_3)} - \Pr(x_1,x_3)_{\rm (Q_1,Q_3)}| \ ,
\end{equation}
where $\Pr(x_1,x_3)_{\rm (Q_1,Q_i,Q_3)}$ comes as marginal from a sequence $\mathcal S_{\rm (1,D_i,3)}$ that includes $\mathcal D_i$.

The trick is that in principle the $\mathcal D_i$ can be chosen such that ideally $\mathcal{I}(\mathcal D_i)=0$. By adding the above parameters to the original LG expression, we can substitute (NIM) with the weaker

\bbasm{\bf (Non-colluding measurements)}
The total clumsiness of a sequence of measurements is given by the sum of the contributions of every single measurement. 
\eeasm

With this assumption, an ideal sequence of adroit measurements is also adroit. In non-ideal situations in which $\mathcal{I}(\mathcal D_i)\geq 0$, we can compute the parameter (\ref{invpar}) as $\mathcal{I}(\mathcal M_2)=\sum_i \mathcal{I}_i (\mathcal D_i)$ since $\mathcal M_2$ is a sequence $\mathcal M_2=(\mathcal D_1 \rightarrow \mathcal D_2 \rightarrow \mathcal D_3\rightarrow \mathcal D_4)$ of four steps $\mathcal D_i$. 

Then, we can write a modified LG inequality as

\bboo{(Wilde and Mizel inequality).}
The following inequality
\begin{equation}\label{NCMbound}
\aver{Q_2Q_1}+\aver{Q_3Q_2}-\aver{Q_3Q_1} \leq 1 +\sum_i \mathcal{I}_i (\mathcal D_i) \ ,
\end{equation}
holds under the assumption of (i) {\it Macrorealism} (MR) and (ii) {\it Non-Colluding Measurements} (NCM).
\eeoo

A violation of Eq.~(\ref{NCMbound}) would thus prove that either or both (i) the system is not macrorealistic in strict sense, (ii) two seemingly non-invasive measurements can collude to strongly modify the macroscopic variable $Q$. 
In this way the clumsiness loophole is substituted with a weaker {\it collusion loophole}. 

As an example let us see explicitly how Wilde and Mizel inequality can be violated in a single-qubit system \cite{WildeFP2012}. 
\bbexmp 
Let us write a single-qubit observable at $6$ different times as $Q_i=\sum_k a_k(t_i) \sigma_k$ (see also the proof of Th.~\ref{th:LGtheorem}) for some time dependent coefficients $a_k(t_i)$ and $t_i \in \{t_1,\dots, t_6 \}$. The measurements of $Q_1$ and $Q_6$ correspond to $\mathcal M_1$ and $\mathcal M_3$, while the measurement $\mathcal M_2$ is decomposed in the four measurements of $Q_2, \dots, Q_5$ and the outcome is registered as the outcome of $Q_5$.
Let us without loss of generality choose the basis of observables such that $Q_6=\sigma_z$ and 
$Q_1=\cos \theta \sigma_z+\sin \theta \sigma_x$ for some phase $\theta$. Then, let us choose the time steps such that $Q_1=Q_3=-Q_5$ and $Q_2=Q_4=Q_6$. In this way we have ideally $\mathcal{I}(\mathcal D_i)=0$ and thus $\mathcal{I}(\mathcal M_2)=0$ since all the observables measured in the $\mathcal D_i$ commute either with $Q_1$ or $Q_6$ and are thus non-disturbing when performed between $\mathcal M_1$ and $\mathcal M_3$.

Then, assuming that the initial state of the qubit is the completely mixed state $\rho=\frac{\id} 2$, we have that $\aver{Q_3Q_1}=-\aver{Q_3Q_2}= \cos \theta$ and $\aver{Q_2Q_1}=-\cos^4 \theta$. Thus we have
\begin{equation}
\aver{Q_2Q_1}+\aver{Q_3Q_2}-\aver{Q_3Q_1}-\sum_i \mathcal{I}_i(\mathcal D_i)=-\cos^4 \theta-2\cos \theta \ ,
\end{equation}
that is larger than $1$ for $0.683 \pi < \theta < \pi$. 
\eeexmp

Thus, recalling also the result of Th.~\ref{th:LGtheorem}, we have seen that already a single-qubit can violate both the original LG inequality (\ref{lgthree1}) and its modified version (\ref{NCMbound}) that substitutes the assumption (NIM) with the weaker (NCM). No entanglement is needed in order to violate macrorealism, but just strong correlations and non-compatibility of the observable with its time-evolved.
Analogous violation can be achieved also with many particle states. 

In particular we are going to focus on the so called {\it Gaussian states}, which are in a sense among the many body states most similar to a single-qubit state. They can be depicted in a many particle Bloch sphere and can be described by a Gaussian probability distribution for the collective operators. See Appendix~\ref{sec:gaussstatqnd} (or e.g., \cite{KoschorreckPhD} and references therein) for details about 
Gaussian states and some Gaussian operations such as QND measurements.


\section{Leggett-Garg tests in atomic ensembles}\label{sec:LGatomic}

Here we show how QND measurements on an atomic spin ensemble can be used to test an LGI under circumstances closely resembling the original Leggett-Garg proposal and open the possibility of tightening the clumsiness loophole in a macroscopic system \cite{budronivitagliano}. Using the Gaussian state formalism reviewed in Appendix~\ref{sec:gaussstatqnd} we predict violation of LGIs for realistic experimental parameters in schemes involving at least $7$ measurements. Furthermore our calculation method allows a clear discrimination between incidental disturbances from, e.g., spontaneous scattering, and unavoidable disturbance due to quantum back-action. In fact later we also show formally that the clumsiness loophole can be tightened by suitably quantifying the contribution
to the LG expression due to classical imperfections in the measurements.
Doing so a failure of our test forces macrorealists to a position strongly resembling quantum mechanics.

Let us first mention an extension of the simplest LG inequality in the form
\begin{equation}\label{LGdef}
K_3:=\aver{Q_2Q_1}+\aver{Q_3Q_2}+\aver{Q_3Q_1}+1\geq 0
\end{equation}
to protocols with more measurements, a result of Avis {\it et al.} \cite{AvisPRA2010}. 

\bboo{(n-measurement LGIs).}
The following inequality
\begin{equation}\label{K_n}
K_n := \sum_{1\leq i<j\leq n}  \aver{Q_i Q_j} + \left\lfloor \frac{n}{2}\right \rfloor\geq 0 \ ,
\end{equation}
where $\left\lfloor k \right \rfloor$ denotes the integer part of $k$, holds for every $n$-measurement scheme under the assumption of (i) MR and (ii) NIM. Eq.~(\ref{K_n}) are optimal tests of macrorealism with $n$ measurements in time.
\eeoo

In the following we will show that Eq.~(\ref{K_n}) with at least $n=7$ measurements can be violated in realistic measurement schemes in cold atomic ensembles.
 
\subsection{Violation of LGI in atomic ensembles}

To introduce our proposal let us look at the canonical example of protocol violating Eq.~(\ref{K_n}) for a single qubit.
Let us consider a single spin-$\frac 1 2 $ particle initialized in the $| +\frac 1 2\rangle_{\rm z}$ state and evolving through the Hamiltonian
\begin{equation}
H=\frac 1 2 \omega \sigma_x \ ,
\end{equation}
and let us imagine that the observable $Q=\sigma_z$ is measured at different time instants $t_i$. The time correlators result to be $C_{i j}:= \aver{Q_i Q_j} = \cos \omega (t_i - t_j)$ and thus with equally delayed measurement $\omega (t_i-t_j):=\theta$ we have 
\begin{equation}\label{K_nviolsingqbit}
K_n = \sum_n \sum_{d=1}^{n-1} \cos (\theta \cdot d) + \left\lfloor \frac{n}{2}\right \rfloor \ ,
\end{equation}
that can become negative for suitable choices of the delay $\theta$. 

This example has many similarities with the protocol that we are going to propose in atomic ensembles. In fact we propose to measure the time correlators appearing in Eq.~(\ref{K_n})
performing QND measurements of a collective spin component $J_z$ of the ensemble externally-driven by a Hamiltonian $H= \kappa B J_x$. 

More concretely, our system is an atomic ensemble of $N_A$ spin-$1$ atoms, of which we consider the collective spin vector $\vec J$. The system is initially in a state completely $\hat x$ polarized $\aver{\vec J}=(N_A,0,0)$ (see \cite{ColangeloNJP2013} and references therein).
The probing light, i.e., the meter, consists of pulses of $N_L$ photons completely $\hat x$-polarized $\aver{\vec S^{(i)}}=(\frac{N_L} 2,0,0)$ as well.
The  system+meter can be compactly described in terms of the vector of observables, $\vec V =  \vec J \oplus  \vec S^{(1)}\oplus \dots \oplus \vec S^{(n)} $, where $n$ is the total number of light pulses, corresponding to the total number of measurements.
For our purposes, the quantum state of the system can be described by means of just the vector of averages $\mean{V}$ and the  covariance matrix $\Gamma_V$ as in the {\it Gaussian approximation} (see Appendix~\ref{sec:gaussstatqnd}).

We consider a Gaussian spin state rotating under the influence of an external magnetic field $B$ acting along the $x$ direction, i.e., with Hamiltonian $H=\kappa B J_x$, where $B$ is a classical external field\footnote{In our protocols $J_x$ can be also effectively treated as a classical variable, since the state remains always completely $\hat x$ polarized}. Calling $\theta:=\kappa B \Delta t$, the atomic variables are modified as 
in Eq.~(\ref{eq:gaussatorot}) where $O_{\rm B}(\theta)$ is a rotation about the $\hat x$ axis. The QND measurement instead modifies the atomic and light variables as in Eq.~(\ref{eq:gaussqnd}) in an ideal situation.

In a more realistic scenario, we have to take account of possible
noisy effects that might disturb our experiment.
Phenomenologically noise can be described as a reduction of the atomic polarization and a consequent modification of the atomic covariance matrix as \cite{ColangeloNJP2013}
\begin{equation}\label{loss}
\Gamma_J\mapsto \chi^2 \Gamma_J+\chi(1-\chi)\frac{N_A}{2}+(1-\chi)\frac{2}{3}N_A,
\end{equation}
where we basically described dehocerence as an assignment of a random polarization to a fraction $1-\chi$ of the $N_A$ atoms (see Eq.~(\ref{eq:gaussopgen})). Here $\chi=\exp(-\eta N_L)$ depends on the \emph{scattering rate} $\eta$, as well as on the number of incoming photons $N_L$.

\begin{figure*}[h!]
\centering
\includegraphics[width=0.8\textwidth,clip]{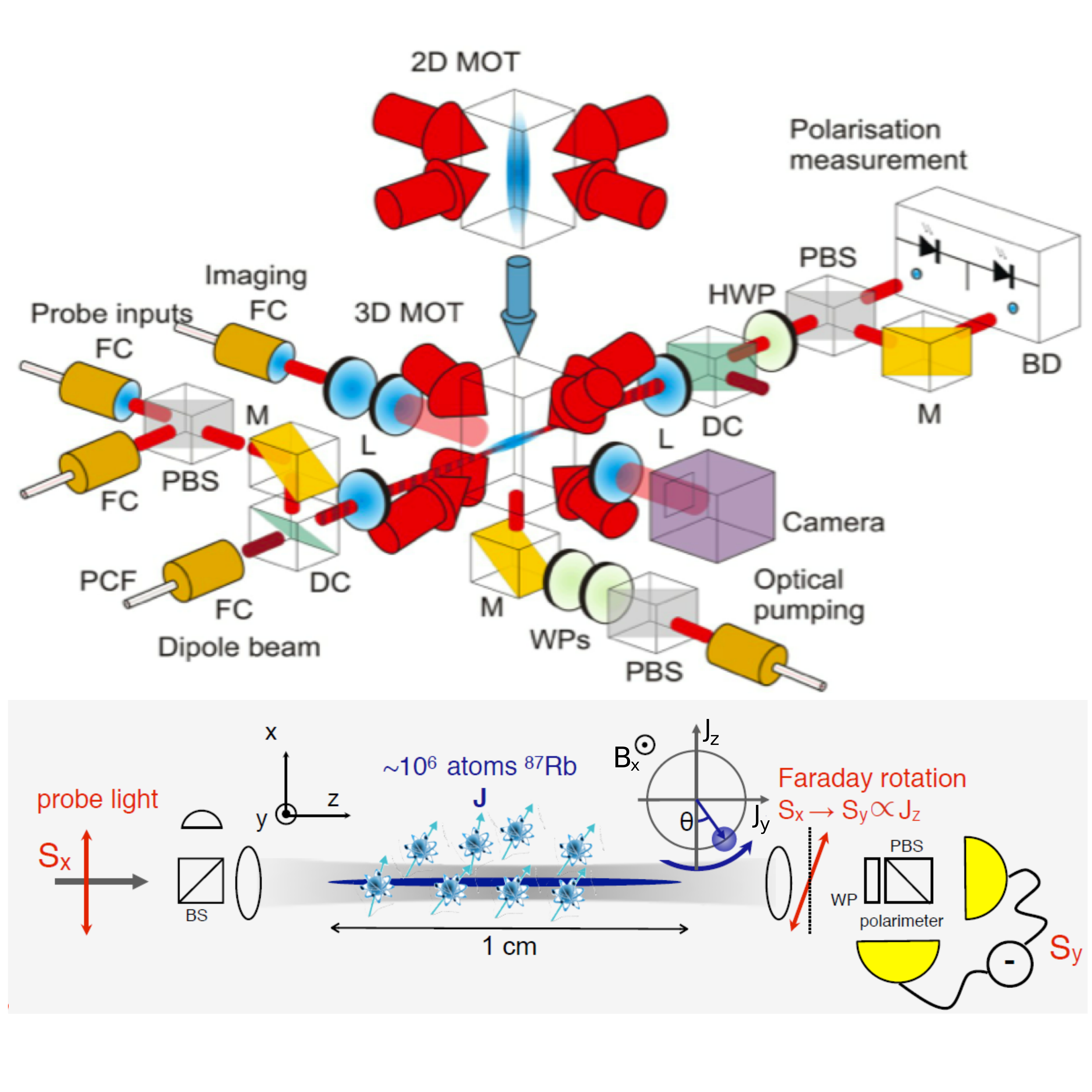}
\caption{A scheme of an experimental setup consisting of an ensemble of cold atoms probed with light pulses in a QND way. (top) A 3D scheme of the experimental apparatus. (bottom) A scheme of the probing for an LG test: $\hat x$-polarized light pulses interact with the total spin $\vec J$ of an ensemble of $\sim 10^6$ cold Rubidium atoms, independently rotating due to an external magnetic field pointing in the $\hat x$ direction. The light  polarization experiences a Faraday rotation and is then detected. In particular the output $S_y$ component of the light polarization is measured projectively. (Figure taken from G. Colangelo, experimental apparatus of the lab of M. W. Mitchell at ICFO, Barcelona.)}\label{fig:qndexp}
\end{figure*}

\begin{figure*}[h!]
\centering
\includegraphics[width=0.8\columnwidth]{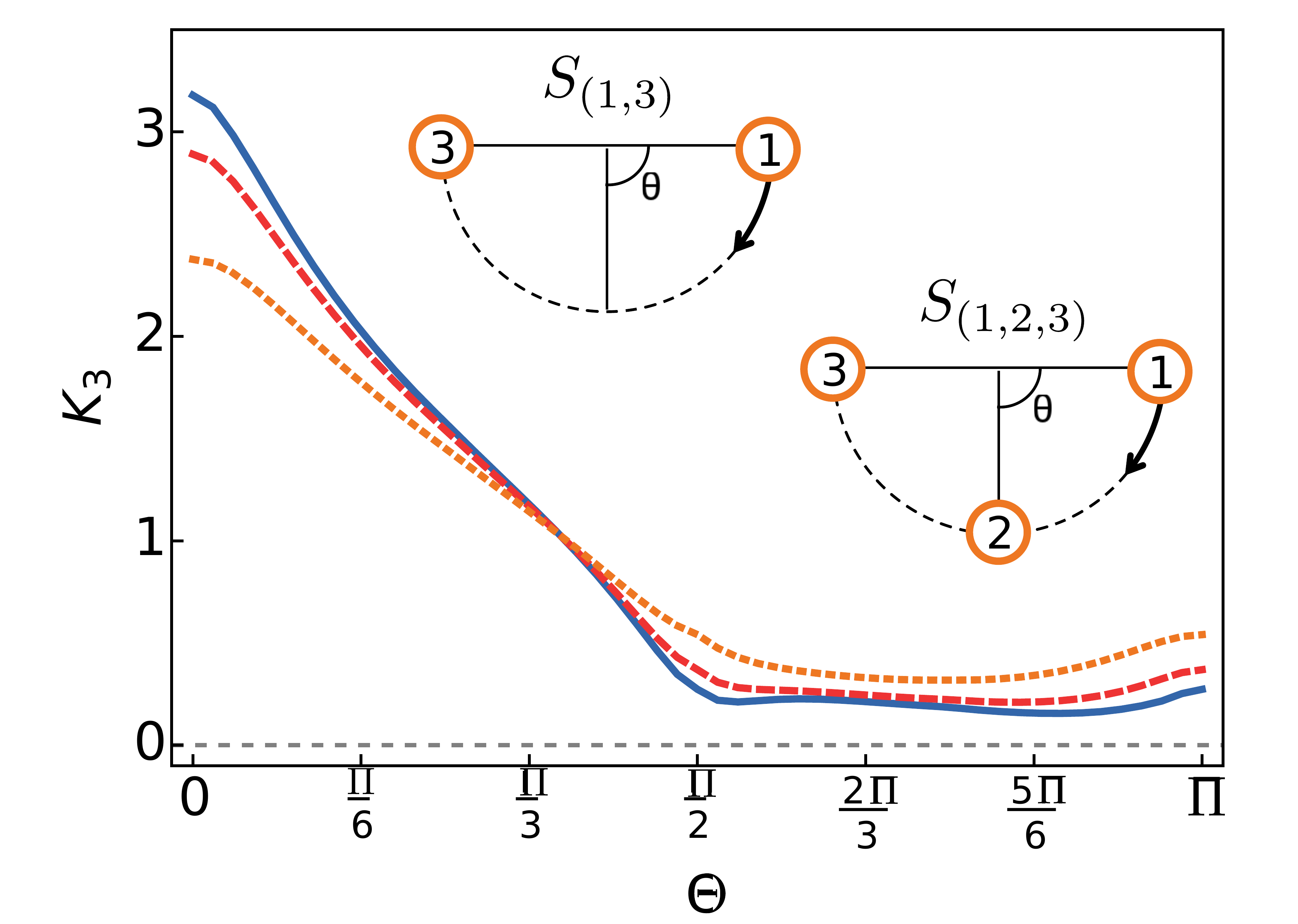}
\caption{Plot of $K_3$ as function
of the constant delay $\theta$ for $N_L=5\cdot 10^8$ probe
photons and truncations at $c=\sigma,\sigma/2,0$, respectively,
orange dotted, red dashed and blue solid line, with
$\sigma=\frac{N_L}{4} (1+2g^2N_AN_L)$. Such a scheme is not sufficient
to violate the LG inequality. Here, scattering and losses
are not taken into account.}\label{figK3cont1s}
\end{figure*}

\begin{figure*}[h!]
\centering
\includegraphics[width=0.8\textwidth]{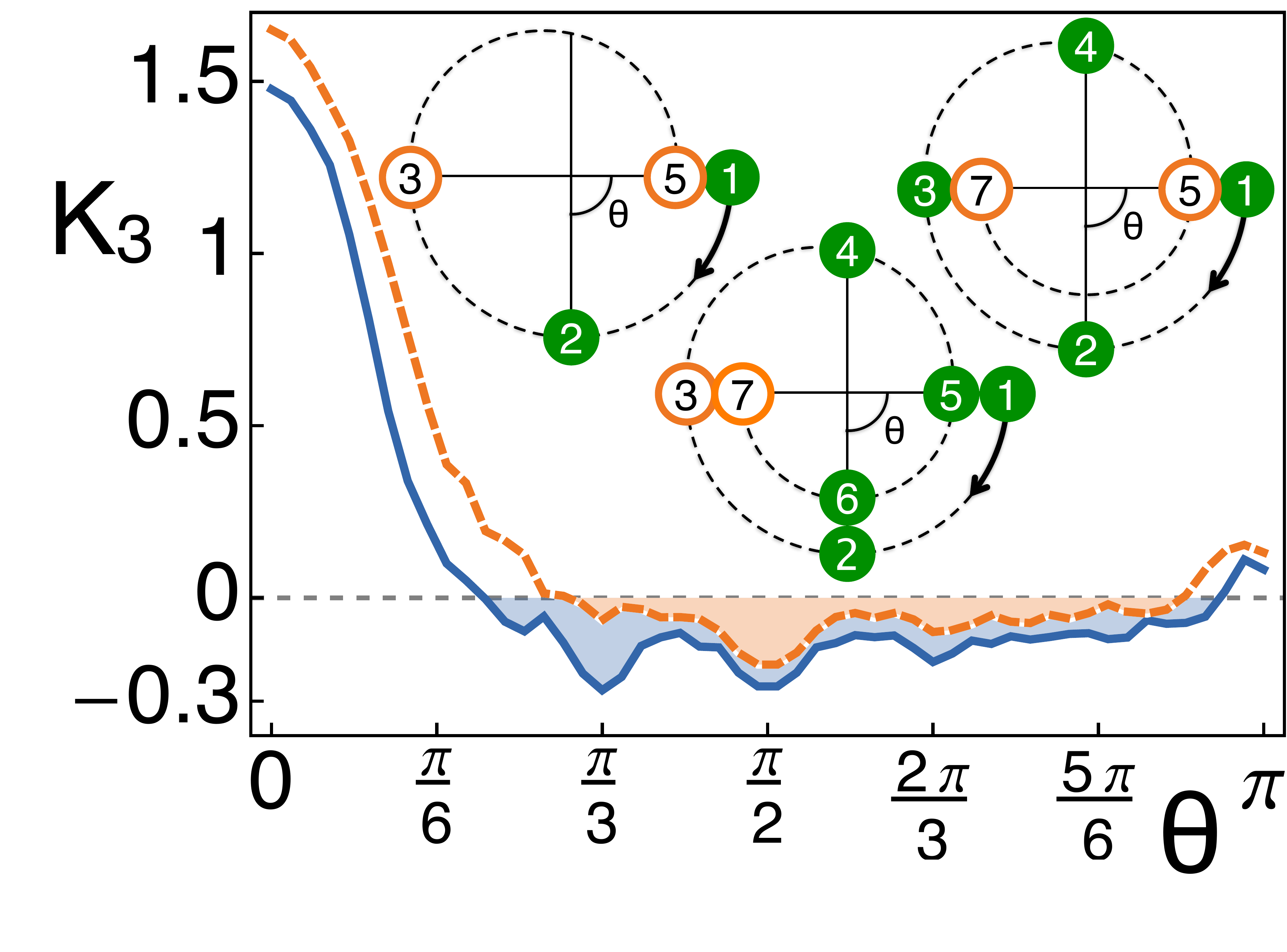}
\caption{Three-point LGI violations within longer measurement sequences. 
Upper and lower curves show $K_3$ versus $\theta$ for optimized $7$- and $9$-measurement sequences, respectively. Both plots are obtained for $N_L= 5\times 10^8$ including scattering and loss effects. In the inset there are depicted the three optimal sequences of a $7$-measurement scheme with delay $\theta=\frac{\pi} 2$.}\label{figK3from7}
\end{figure*}

We can show that in this system a violation of macrorealism is possible to reach with realistic parameters, even taking 
noise into account as in Eq.~(\ref{loss}). To do so we consider
sequences of measurements and evolutions (see the insets of Figs.~\ref{figK3cont1s},\ref{figK3from7}), and evaluate numerically the expected value of the time correlations $C_{ij}$. 
To compute a certain $C_{ij}$ we have to consider all the possible sequences that include measurements of $Q_i$ and $Q_j$ and take the optimal ones. 

In our system we expect the $n$ outcomes $\vec y=(y_1,\ldots,y_n)$ of $(Q_1,\ldots, Q_n)$ 
to be distributed according to a Gaussian function 
\begin{equation}
\Pr(\vec y)=G^{(n)}_{\Gamma}(\vec y)=\frac{\exp{\left(- (\vec y - \vec{\mu})^T \Gamma^{-1} (\vec y - \vec{\mu}) \right)}}{\sqrt{(2\pi)^n \det \Gamma}} \ ,
\end{equation}
with mean $\vec{\mu} \simeq \vec 0$ approximately zero and covariance matrix $\Gamma$. Then we consider the bivariate marginal relative to the pair of outcomes $\mathcal{Y}_{ij}=(y_i, y_j)$ of $Q_i$ and $Q_j$. This is also a Gaussian distribution
$G^{(2)}_{\Gamma^{\mathcal{Y}_{ij}}}(y_i,y_j)$ with mean value zero and covariance matrix $\Gamma^{\mathcal{Y}_{ij}}$, the submatrix of $\Gamma$ relative to $S_y^{(i)}$ and $S_y^{(j)}$. Here each covariance matrix has been first evolved with specific forms of Eqs.~(\ref{eq:gaussatorot},\ref{eq:gaussqnd},\ref{loss}), relative to the concrete measurement scheme.

Afterwards,
in order to evaluate correctly the LG expressions, we have to relabel the outcomes with a certain function $f(y)$ such that $-1\leq f(y) \leq 1$. 
A simple way to ensure the correct normalization of the outcomes is to truncate the probability distribution after a certain region $\mathcal R:=[-c,c] \times [-c,c]$ and then relabel the outcomes as
\begin{equation}\label{relabelfunc}
f(y)=
\left\{
\begin{array}{c}
\frac{y}{c} \qquad  |y| \leq c  \\
{\rm sgn}(y) \qquad |y| > c
\end{array}
\right. \ .
\end{equation}
In particular in the limit $c \rightarrow 0$ we obtain completely discretized outcomes $f(y)= {\rm sgn}(y)$, as in the original LG proposal. 

In this way the correlations assume the form
\begin{equation}\label{corrcsigma}
C_{ij}= P_{\mathcal R} \frac{C^{\rm tr}_{ij}}{c^2} + 4P_{++}-1
 \ ,
\end{equation}
where we called $P_{\mathcal R}=\Pr(y_i,y_j \in \mathcal R)$, 
\begin{equation}
C^{\rm tr}_{ij} = \frac 1 {P_{\mathcal R}} \int_{{\bf y} \in \mathcal R}  y_i y_j G^{(2)}_{\Gamma^{\mathcal{Y}_{ij}}}(y_i,y_j) {\rm d}y_i{\rm d}y_j \ ,
\end{equation}
and $P_{++}=\Pr(y_i>c,y_j >c)$. Since the probability distributions are Gaussian, Eq.~(\ref{corrcsigma}) can be evaluated from just the final covariance matrix
\begin{equation}
\Gamma^{\mathcal{Y}_{ij}}=
\left(
\begin{array}{cc}
 A &   B \\
 B &  C 
\end{array}
\right) \ ,
\end{equation}
and in the special case $c=0$ the analytical result is 
\begin{equation}\label{cijdiscr}
C_{ij}=\left(1-\frac{2\alpha} \pi \right) {\rm sgn}(B):=C_{ij}^{\rm disc} \ ,
\end{equation}
where $\alpha= \arctan ( \tfrac{\sqrt{AC-B^2}}{|B|} )$.
For simplicity we compute $K_3$ defined as in Eq.~(\ref{LGdef}) for equally delayed measurements. 

The result is plotted in Fig.~\ref{figK3cont1s}, where we can see that a simple $3$-measurement protocol is not sufficient to see a violation of macrorealism and that the discretized relabelling provides the best result. 

\begin{figure*}
\includegraphics[width=0.45 \textwidth]{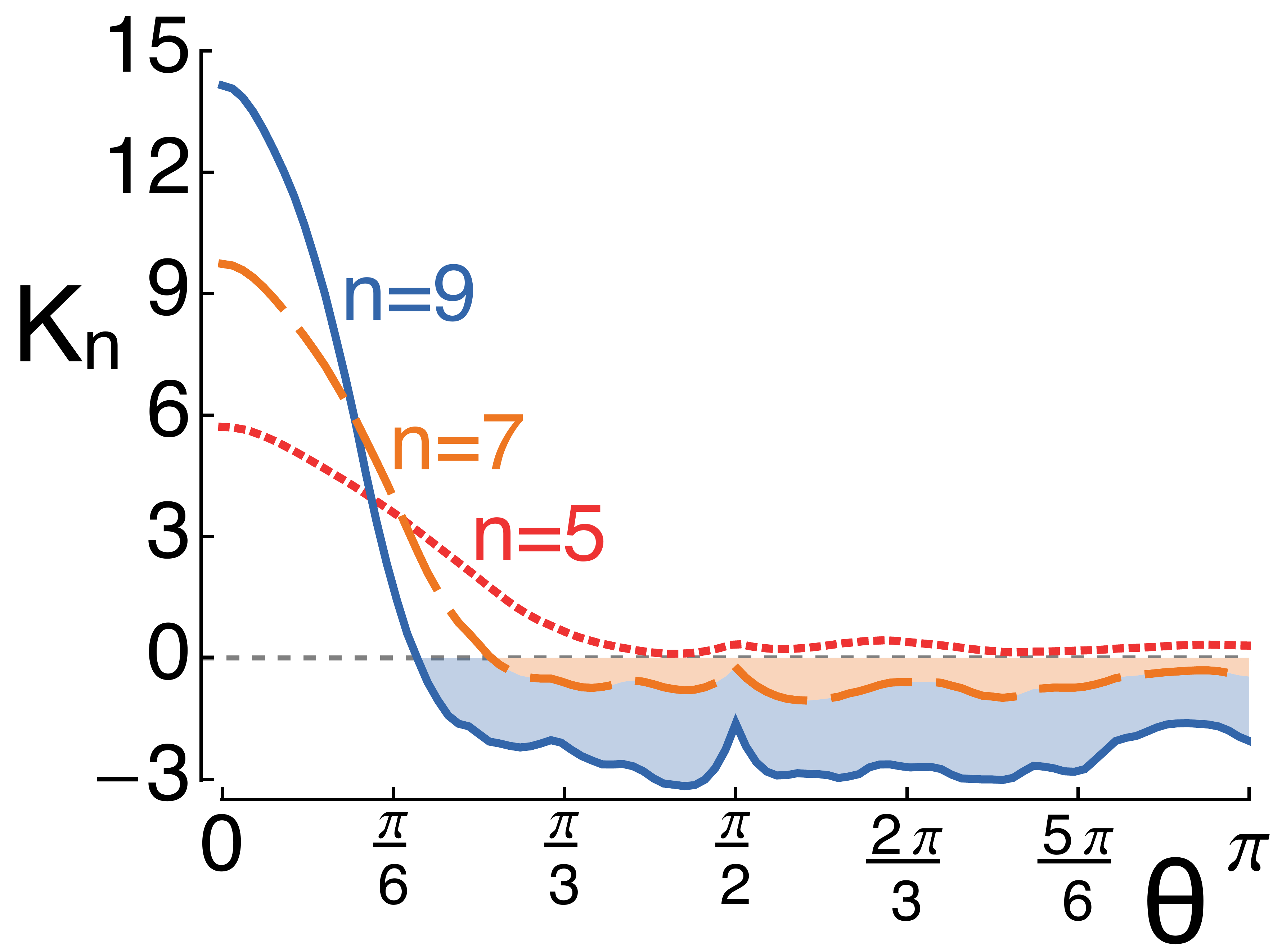}
\includegraphics[width=0.45 \textwidth]{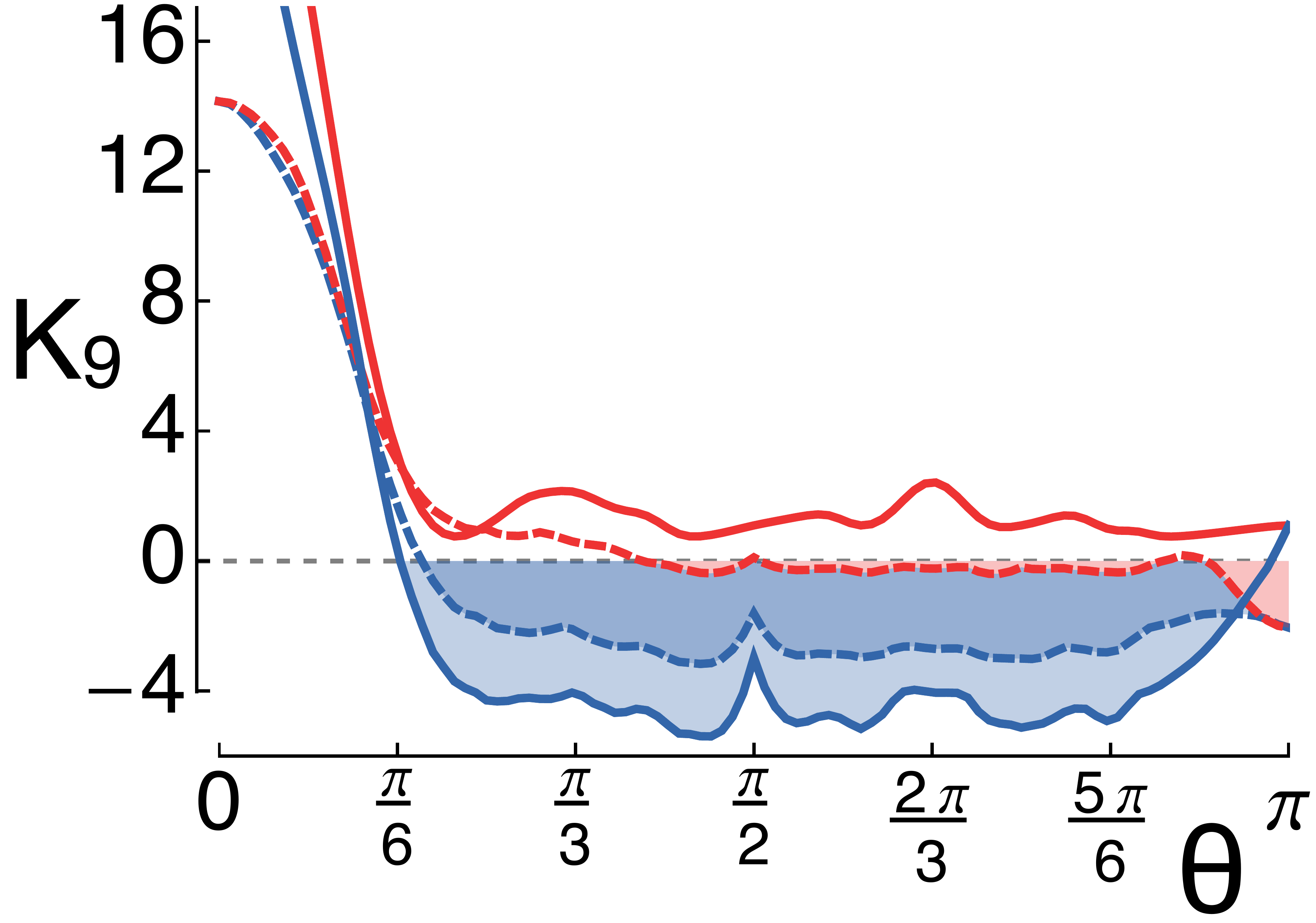}
\caption{Numerical evaluation of $n$-measurement LGIs as functions of $\theta$. (left)
$K_n$ for (from top to bottom) $n=5,7,9$ in the presence of scattering. (right) numerical evaluation of $K_{9}$, the two lower blue (upper red) curves are results with (without) back action. In solid (dashed) lines results with (without) scattering. 
All plots are obtained using the same parameters, taken from experiment (see text).}\label{plotKndiscsc}
\end{figure*}

Then, we considered a test with $n=5,7,9$ measurements with discretized outcomes, i.e., $c=0$, and evaluate $K_5,K_7,K_9$ of Eq.~(\ref{K_n}). We computed numerically the covariance matrix resulting from a certain sequence
and then optimize the value of $C_{i j}$ over all possible sequences. 
Realistic parameters were used: $g=10^{-7}$, $N_A=2\cdot 10^6$, $N_L=5\cdot 10^8$, and $\chi=\exp(-\eta N_L)$, where $\eta=0.5 \cdot 10^{-9}$~\cite{ColangeloNJP2013}. 

The results are plotted in Fig.~\ref{plotKndiscsc}, where  we take into account losses due to off-resonant scattering as in (\ref{loss}). Our analysis shows that with realistic parameters a violation of the tests based on $K_7$ and $K_9$ is achievable. 

Here in general the question arises, however, whether this witnesses a genuine quantum effect or it is due to a classical clumsiness of the measurements.
For non-ideal QND measurements there are two ingredients that contribute to the violation: a direct disturbance on the measured variable $J_z$, that must be attributed to {\it classical clumsiness}, and a {\it back action} that affects a conjugate variable $J_y$ and can be interpreted as a fundamental quantum effect.
For example, the violation around $\theta=\pi$ is correctly explained in terms of classical clumsiness: measurements at angles $k\pi$, which should be perfectly correlated or anticorrelated, are decorrelated due to scattering and losses of atoms.

Nevertheless, our formalism allows to separate and distinguish the two contributions by simulating a QND measurement where the quantum back action is ``turned off'', i.e., we set $J_y^{\rm (out)}=J_y^{\rm (in)}$ in the input/output relations (\ref{eq:gaussqnd}). 
The results in this last case are shown in the red upper curves of Fig.~\ref{plotKndiscsc}(right). These indicate that the violation genuinely comes from the effect of quantum back action in most cases. Scattering becomes important only at some specific phases, and is responsible for the violation only for $\theta$ approaching $\pi$.

The above tests involve a large number of correlation terms (e.g., $21$ for computing $K_7$).
We can considerably simplify the protocol and reduce the number of measurement sequences by considering just a triple $(\mathcal M_a, \mathcal M_b, \mathcal M_c)$, extracted out of the $n$-measurement scheme, and the corresponding correlators combined as in the definition of $K_3$, namely
\begin{equation}\label{k3abc}
K_3 = C_{ab} + C_{bc} + C_{ac} + 1 \geq 0 .
\end{equation}

We compute the best achievable $K_3$ optimizing over all possible triples and all possible sequences. 
The results are plotted in Fig.~\ref{figK3from7}, where it can be seen that a violation is obtainable even in these simplified protocols, especially around $\theta=\tfrac{\pi} 2$ and $\theta=\tfrac{\pi} 3$.

An explicit measurement scheme that comes from our analysis as a practically feasible test of macrorealism is depicted in the inset of Fig.~\ref{figK3from7}: $3$ sequences of $7$ measurements each are performed, and only $3$ couples of outcomes are registered (yellow-white filled circles in the figure), while the others are discarded (green circles). Each of the correlations $C_{ab}$ between the outcomes of $(Q_3, Q_5, Q_7)$ is then computed using the corresponding sequence. Ideally with a constant delay of $\theta=\tfrac{\pi} 2$ between the measurements and without  decoherence we should get $C_{35}=C_{57}=-1$, while keeping $C_{37}<1$ due to the various discarded measurements made between $\mathcal M_3$ and $\mathcal M_7$. In particular ideally the fact that $C_{37}<1$ is due to the {\it quantum back-action} of the QND measurements.

This scheme shows less violation as compared to extracting the full $K_7$ with the optimal $7$-measurement protocol, but requires fewer measurement sequences and involves calculation of a simpler combination of correlations, potentially making the test more robust in the presence of experimental uncertainties. 

\section{Quantifying the clumsiness of QND measurements}\label{sec:invasivityquant}

The next step is to try to formally address the (NIM) assumption separately in order to tighten the clumsiness loophole. Here we show in detail that indeed the fact that the QND measurement has a back action entirely on the conjugate variable allows to discriminate rigorously between clumsy effects, like unwanted scattering, and unavoidable quantum disturbance, i.e., the back action itself.
We do this by focusing our attention to study the invasivity of a QND measurement from a macrorealist perspective
and defining an invasivity parameter similar to Eq.~(\ref{invpar}) but tied to detect our clumsiness.

Let us recall that in abstract probability theories the assumption of macrorealism (MR) guarantees that, at any time, relative to a set of $n$ observables, {\it there exists a corresponding joint probability distribution} $\Pr_{\rm ideal}=\Pr(x_1,\dots, x_n)_{\rm (Q_1,\dots,Q_n)}$ for the whole set of potential outcomes, independently on any actual measurement $\mathcal M_i$. Then, the assumption of non-invasive measurability (NIM) implies that from $\Pr_{\rm ideal}$ there can be extracted bivariate marginals $\Pr(x_i, x_j)_{\rm ideal}$, so to compute the correlations $\mean{Q_i Q_j}$ as 
\begin{equation}
\mean{Q_i Q_j} = \int {\rm d} x_1 \dots {\rm d} x_n x_i x_j \Pr_{\rm ideal} \ .
\end{equation}
Note that (NIM) implies that the marginals can be extracted all from the same probability distribution $\Pr_{\rm ideal}$, independently of the sequence of measurements actually performed. Here we want to avoid this assumption. Thus, we allow the different 
$\Pr(x_i, x_j)_{\rm (Q_i,Q_j)}$ to be independent probability distributions, {\it not necessarily corresponding to marginals} of $\Pr_{\rm ideal}$.

Furthermore, the LGIs (\ref{K_n}) must hold even in a more ``pragmatic'' situation, in which there exists a certain probability distribution $\Pr(x_1,\dots, x_n)_{\rm (Q_1,\dots,Q_n)}$ for the whole set of outcomes from which there are extracted all the $\Pr(x_i,x_j)_{\rm (Q_i,Q_j)}$. This happens when, in an $n$ measurement protocol, we perform just a single sequence of $n$ measurements $\mathcal S_{\rm n-seq}=(\mathcal M_1\rightarrow \dots \rightarrow \mathcal M_n)$ and all the outcomes are registered. This provides {\it a posteriori} a probability distribution $\Pr(x_1,\dots, x_n)_{\rm n-seq}$, from which we can  
compute all the correlators $\aver{Q_i Q_j}$. 
In this case Eq.~(\ref{K_n}) must hold, as in the ideal case. 

Then, as we discussed concerning the proposal of Wilde and Mizel, in the simplest case of a $3$-measurement protocol it is needed just a quantifier of the clumsiness of the measurement $\mathcal M_2$, that estimates its potential contribution to the correlator $\aver{Q_3 Q_1}$.
Our approach to define such quantifier consists in separating the effect of a measurement as coming from two different contributions
\begin{equation}
\mathcal M(Q,P) = \mathcal M_Q(Q,P) \circ \mathcal M_P(Q,P) \ ,
\end{equation}
with $\mathcal M_Q(Q,P) = (Q^\prime,P)$ and $\mathcal M_P(Q,P) = (Q,P^\prime)$, i.e., a disturbance effect $\mathcal M_Q$ on the 
the measured variable $Q$ itself from a disturbance $\mathcal M_P$ on just its conjugate $P$.

More generally, assuming an initial state described by an outcome probability distribution $\Pr_{\rm in}(x,p)$, the effect of the 
measurement can be described by a mapping 
\begin{equation}
\Pr_{\rm in}(x,p)\mapsto \Pr_{\rm in}(\mathcal M(x,p)) = \Pr_{\rm out}(x,p)
\end{equation}
to a different probability distribution $\Pr_{\rm out}(x,p)$. Then, among the effects that a measurement can have on the whole probability distribution, 
we can quantify the disturbance directly visible into the variable $Q$ by taking a distance measure
\begin{equation}
\mathcal I (\mathcal M_Q) = \int |\Pr_{\rm in}(x)-\Pr_{\rm out}(x) | {\rm d} x \ ,
\end{equation}
between the marginals $\Pr_{\rm in/out}(x)=\int \Pr_{\rm in/out}(x,p) {\rm d}p$ relative to its outcomes. 

In a scheme consisting of $n$-time steps the initial state corresponds to a probability distribution $\Pr_{\rm in}(x_1,\dots,x_n,p_1,\dots,p_n)$ valued in a $2n$-dimensional phase space.
When a sequence $\mathcal S$ of measurements is performed, the state is mapped to some $\Pr_{\rm out}({\bf x},{\bf p})$ with ${\bf x}=(x_1,\dots,x_n)$ and ${\bf p}=(p_1,\dots,p_n)$. To quantify the disturbance not attributable to the Heisenberg principle, i.e., the clumsiness of such sequence we then define a parameter
\begin{equation}\label{eq:abstractpar}
\mathcal I (\mathcal S) = \int |\Pr_{\rm in}({\bf x})-\Pr_{\rm out}({\bf x}) | {\rm d} {\bf x} \ ,
\end{equation}
that is quantifying the effect of the sequence on just the variables ${\bf x}$ by considering the marginals
$\Pr_{\rm in/out}({\bf x}) = \int \Pr_{\rm in/out}({\bf x},{\bf p}) {\rm d}{\bf p}$.

Here the intuitive notion of ``clumsiness'' is formalized as some {\it uncontrollable}
disturbance $\mathcal M_Q$ that acts directly on the measured variable $Q$ only and is distinguished from an indirect disturbance
made on its conjugate $P$ that might act back on $Q$ because of their initial correlations, possibly enhanced by the time evolution.

\begin{figure*}[h!]
\centering
\includegraphics[width=0.8\textwidth,clip]{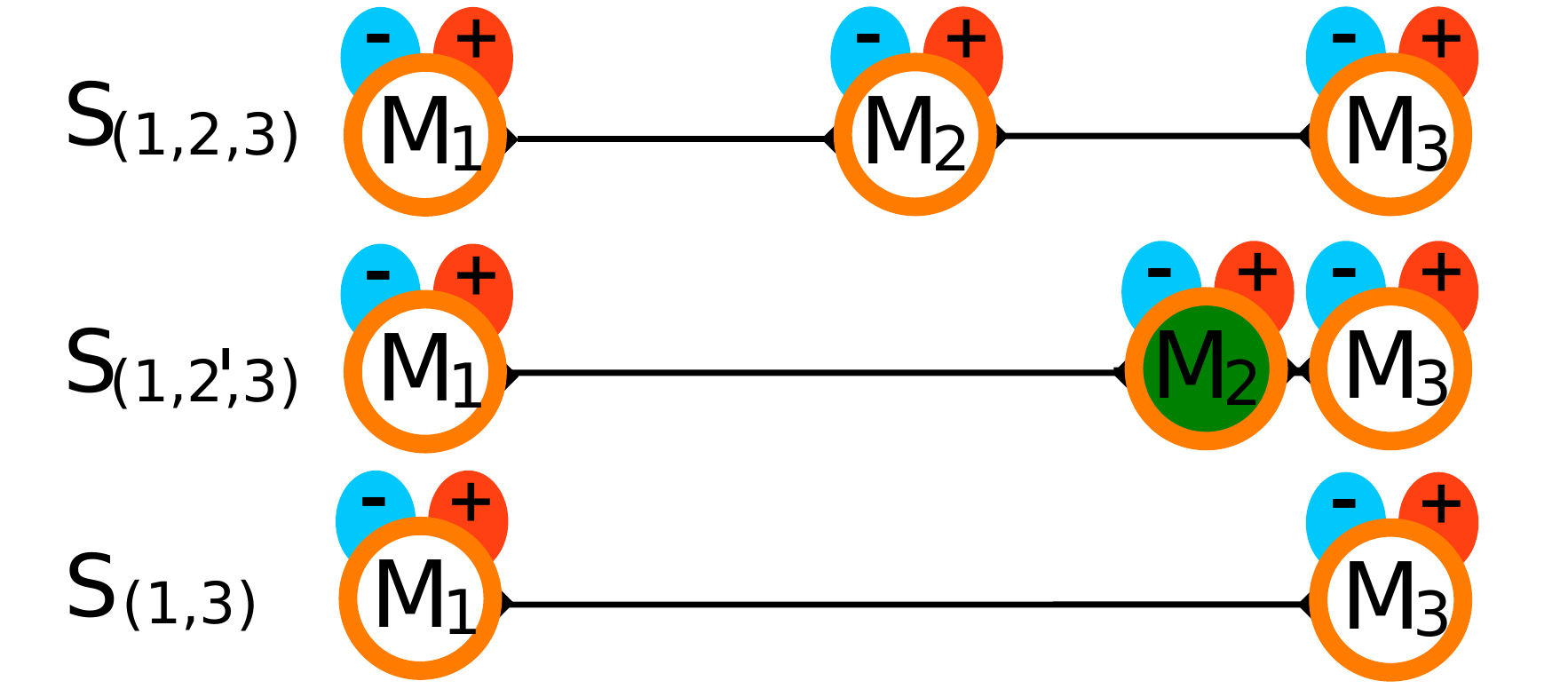}
\caption{Schematic representation of our proposal to quantify the contribution of clumsiness in the simplest $3$-measurement LG test. An auxiliary sequence $\mathcal S_{\rm (1,2^\prime,3)}$ is performed in which $\mathcal M_2$ 
is right before $\mathcal M_3$. In this way we quantify the disturbance made by the presence of $\mathcal M_2$ directly onto the measured variable $Q$ through Eq.~(\ref{invpar}).}\label{fig:ourproposalk3}
\end{figure*}

Thus, for a $3$-measurement LG test we can define a clumsiness parameter for the second measurement in the sequence $\mathcal{S}_{1,2,3}$ as 
\begin{equation}\label{invpar}
\mathcal{I}(\mathcal M_2)=\int \left|\Pr(x_1,x_3)_{\rm \mathcal S_{\rm (1,2^\prime,3)}}-\Pr (x_1,x_3)_{\rm \mathcal S_{\rm (1,3)}}\right| {\rm d} x_1 {\rm d} x_3  \ ,
\end{equation}
by switching off a particular evolution between $t_1$ and $t_3$ and performing the second measurement at any
instant $t_1\leq t_2^\prime\leq t_3$. Equivalently, keeping the evolution between $t_1$ and $t_3$, we can just perform the measurement $Q_2$ right before the third, i.e., with $t_2=t_3$, see Fig.~\ref{fig:ourproposalk3}. In this way only the contribution of the effect $(\mathcal M_2)_Q$ is taken into account as it contributes
to the correlators since the bound
\begin{equation}
|\mean{Q_3 Q_1}_{\rm (1,2^\prime,3)}-\mean{Q_3 Q_1}_{\rm (1,3)}|\leq \mathcal{I} (\mathcal M_2)
\end{equation}
holds.We can then define a modified LG inequality as 
\begin{equation}\label{lgthreeinv}
\mean{Q_2Q_1}+\mean{Q_3Q_2}+ \mean{Q_3Q_1}+ \mathcal{I}(\mathcal M_2) + 1 \geq 0 \ ,
\end{equation}
that takes into account possible unwanted clumsiness of $\mathcal M_2$. Thus, a violation of Eq.~(\ref{lgthreeinv}) witnesses a disturbance of $\mathcal M_2$ on a conjugate variable $P$ that acts back on $Q$ due to the evolution between $t_1$ and $t_3$. In particular in our protocol involving sequences of QND measurements we correctly exclude explanations in terms of classical disturbances, since these last can be correctly quantified through the parameter (\ref{invpar}), i.e., by performing two measurements in rapid succession (cf. Fig.~\ref{fig:ourproposalk3}).

Note that the invasivity parameter as defined in Eq.~(\ref{invpar}) is symmetric in the probability distributions, meaning that equivalently it quantifies the clumsiness as it contributes to the correlations $\mean{Q_3Q_1}$ due to the ``absence'' of measurement $\mathcal M_2$.
Moreover, in the case just discussed the measurements performed are the minimal possible needed to compute the $K_3$
and thus the outcomes are always registered. However, in order to enhance the violation, one is free to perform more complicated sequences with additional measurements, the outcomes of which are not registered, as we have seen in 
the $7$-measurement protocol that we proposed (cf. Fig.~\ref{figK3from7}). Then, one has to address also the clumsiness of all these additional measurements and consider their contribution in all the correlators $\mean{Q_iQ_j}$. 
This can be done by considering the sequence $\mathcal S_{ij}$ from which $C_{ij}$ is computed and measure its clumsiness with respect
to a reference sequence $\mathcal S_{\rm ref}$ that contains the minimal possible set of measurements, i.e., the $n$ measurements that give rise to $\Pr({\bf x})$ plus eventually the additional measurements contained in all 
sequences of the protocol (see Fig.~\ref{fig:seqMIall} as a practical example). These last can be in a sense thought as part of the time evolution. Thus we define
\begin{equation}\label{invparseq}
\mathcal{I}_{ij}=\int \left|\Pr(x_i,x_j)_{\rm \mathcal S^\prime_{ij}}-\Pr (x_i,x_j)_{\rm \mathcal S_{\rm ref}}\right| {\rm d} x_i {\rm d} x_j  \ ,
\end{equation}
where the auxiliary sequence $\mathcal S^\prime_{ij}$ is made such that the additional measurements not contained in $\mathcal S_{\rm ref}$, or vice versa, are
performed in rapid succession, i.e. without time evolution.

In the case in which $\mathcal S_{ij}$ contains less measurements than $\mathcal S_{\rm ref}$ the parameter (\ref{invparseq}) quantifies the contribution of the clumsiness due to the ``absence'' of these additional measurements, as in the previous case. 
The $n$-measurement LG-like tests resulting from this construction are then the following.
\bboo{(LGIs for clumsy QND measurements)}
The following set of inequalities
\begin{equation}\label{eq:LGIclumsyQND}
KI_n:=\sum_{1\leq j \leq i \leq n} \mean{Q_i Q_j} + \left\lfloor \frac{n}{2}\right \rfloor + \mathcal{I} \geq 0 \ ,
\end{equation}
where $\mathcal{I} = \sum_{i,j} \mathcal{I}_{ij}$ is defined according to Eq.~(\ref{invparseq}),
holds in every $n$-measurement scheme
under the assumption of {\it Macrorealism} even in the presence of hidden clumsiness that directly perturbs the measured variable $Q$. 
\eeoo

A violation of Eq.~(\ref{eq:LGIclumsyQND}) would prove that either (i) the system is not macrorealistic or (ii) the evolution
reveals a classical disturbance of the measurements that is otherwise completely hidden in a conjugate variable.

Applied to a QND measurement the parameter $\mathcal{I}$ correctly quantifies the disturbance introduced in a sequence with many measurements by effects such as off-resonant scattering, as we will also show numerically. 
However Eq.~(\ref{eq:LGIclumsyQND}) might be employed in other systems as well, allowing to account for disturbances of the measurements
that are not ascribable to the Heisenberg uncertainty principle.

More in detail our construction exploits the fact that ideally the outcomes of repeated QND measurements of the same variable $Q$ agree perfectly with each other. Thus, when the coherent evolution between two measurements is turned off, the back action only influences some conjugate variable $P$ and all visible difference between two consecutive outcomes of $Q$ is due entirely to clumsiness, as well as in the classical case. This is also the advantage of performing sequences of QND measurements in tests of macrorealism as compared to other, even indirect, measurements: QND measurements cause the minimal possible perturbation to the system implied by the Heisenberg principle, that in particular does not affect directly the measured variable. The same is not always true for other kinds of measurements. 

\subsection{Numerical results for a realistic potential test}

Here let us apply the previous construction to a practical example, i.e., the simplest measurement scheme providing a visible violation that we described previously (cf Fig.~\ref{figK3from7}). The results will show that indeed even a violation of Eq.~(\ref{eq:LGIclumsyQND}) can be attained, i.e., the clumsiness loophole can be actually tightened in a realistic protocol.

Let us consider the protocol depicted in Fig.~\ref{figK3from7}. 
It consists of a series of three sequences $(\mathcal S_{\rm (3,5)},\mathcal S_{\rm (5,7)},\mathcal S_{\rm (3,7)})$ of $7$ measurements delayed by $\theta=\frac{\pi} 2$ constantly between each other. From these sequences the inequality
\begin{equation}\label{eq:k357}
K_3=C_{35}+C_{57}+C_{37}+1 \geq 0 \ ,
\end{equation}
is checked as a test of macrorealism.

Then, to account for clumsiness, we can compute three parameters (\ref{invparseq}), one for each correlator, taking as reference
the following ``minimal'' sequence
\begin{equation}
\mathcal S_{\rm ref} =(\mathcal M_1 \stackrel{\frac{\pi} 2}{\rightarrow} \mathcal M_2 \stackrel{\frac{\pi} 2}{\rightarrow} \mathcal M_3
\stackrel{\pi}{\rightarrow} \mathcal M_5 \stackrel{\pi}{\rightarrow} \mathcal M_7) \ ,
\end{equation}
that consists of the $5$ measurements contained in all sequences. 

\begin{figure*}[h]
\centering
\includegraphics[width=0.8 \textwidth,clip]{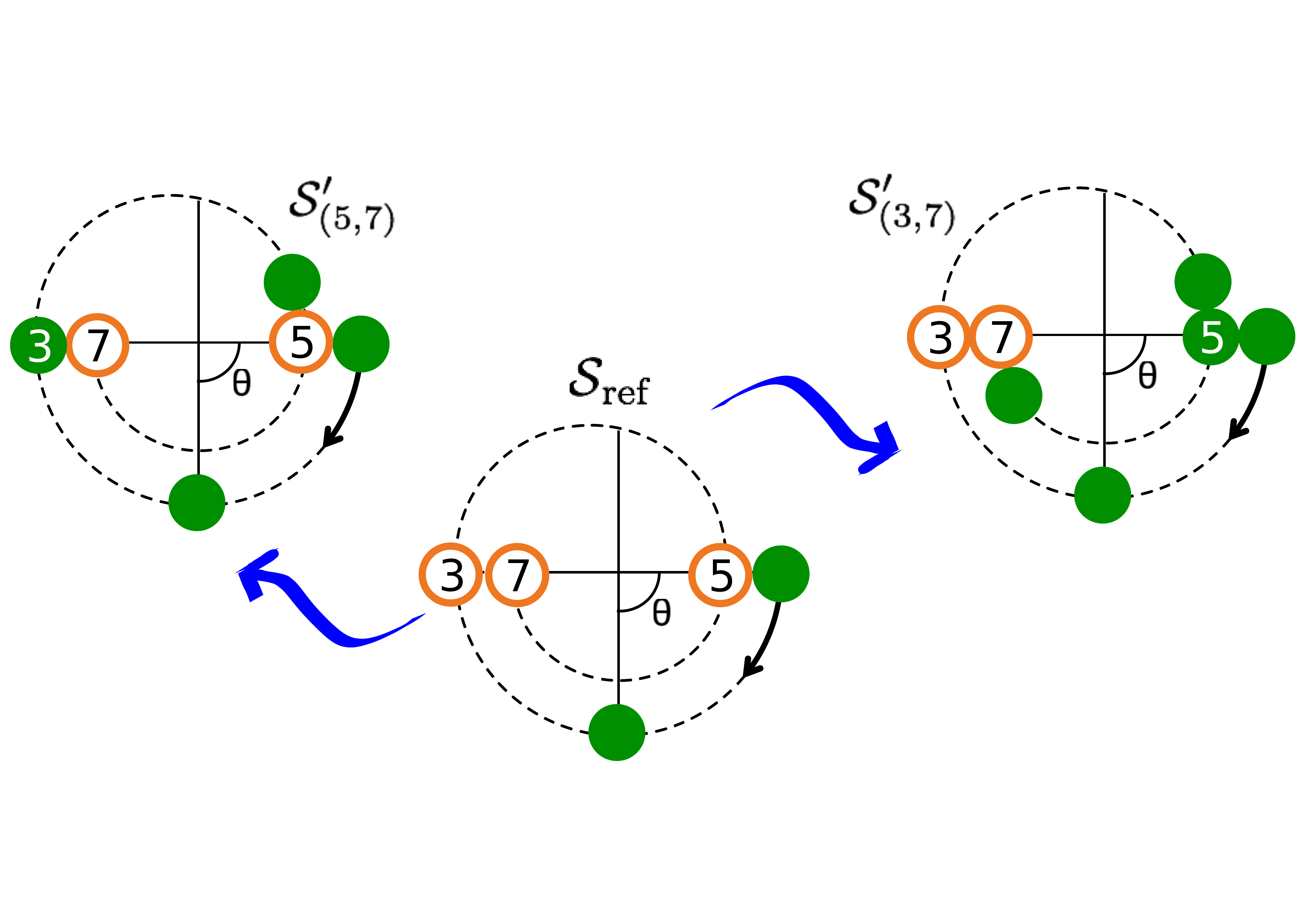}
\caption{Two auxiliary sequences of $7$ measurements $(\mathcal S_{\rm (3,7)}^\prime,\mathcal S_{\rm (5,7)}^\prime)$ are performed to compute
the parameter $\mathcal{I}=\mathcal{I}_{37}+\mathcal{I}_{57}$, as in Eq.~(\ref{discinvpar}). The reference sequence $\mathcal S_{\rm ref}$ contains the measurements performed in every one of the sequences $(\mathcal S_{\rm (3,5)},\mathcal S_{\rm (5,7)},\mathcal S_{\rm (3,7)})$ of the protocol.}
\label{fig:seqMIall}
\end{figure*}

Since $C_{35}$ is taken from $\mathcal S_{\rm (3,5)}$ that is completely equivalent to $\mathcal S_{\rm ref}$ we have $\mathcal{I}_{35}=0$ identically. Then we have to consider the auxiliary sequences (see Fig.~\ref{fig:seqMIall})
\begin{subequations}
\begin{align}
\mathcal S_{\rm (3,7)}^\prime&=(\mathcal M_1 \stackrel{\frac{\pi} 2}{\rightarrow} \mathcal M_2 \stackrel{\frac{\pi} 2}{\rightarrow} \mathcal M_3
\stackrel{\pi}{\rightarrow} \mathcal M_4 \stackrel{\pi}{\rightarrow} \mathcal M_5 \stackrel{0}{\rightarrow} \mathcal M_6 \stackrel{0}{\rightarrow} \mathcal M_7) \\
\mathcal S_{\rm (5,7)}^\prime&=(\mathcal M_1 \stackrel{\frac{\pi} 2}{\rightarrow} \mathcal M_2 \stackrel{\frac{\pi} 2}{\rightarrow} \mathcal M_3
\stackrel{\pi}{\rightarrow} \mathcal M_4 \stackrel{0}{\rightarrow} \mathcal M_5 \stackrel{\pi}{\rightarrow} \mathcal M_7) \ ,
\end{align}
\end{subequations}
and compute
\begin{align}\label{discinvpar}
\mathcal{I}_{37}=\sum_{y_3,y_7=\pm} \left|\Pr(y_3,y_7)_{\rm \mathcal S_{\rm (3,7)}^\prime}- \Pr(y_3,y_7)_{\rm \mathcal S_{\rm ref}}\right|,
\end{align}
and in total $\mathcal{I}=\mathcal{I}_{37}+\mathcal{I}_{57}$ with $\mathcal{I}_{57}$ analogous to Eq.~(\ref{discinvpar}). 
We can then check numerically the value of $\mathcal I$ for the protocol of Fig.~\ref{figK3from7}. We compute the covariance matrix resulting from the two sequences needed, i.e., $\Gamma_{\rm \mathcal S_{\rm (3,7)}^\prime}$ and $\Gamma_{\rm \mathcal S_{\rm (5,7)}^\prime}$ and the corresponding discretized probabilities, like e.g.,
\begin{equation}\label{discretpplu}
P(+,+)_{\rm \mathcal S_{\rm (3,7)}^\prime}=\int_{y_i\geq 0 , y_i\geq 0} {\rm d} y_i {\rm d} y_j G(y_i,y_j)_{\rm \mathcal S_{\rm (3,7)}^\prime}
\end{equation}
coming from an underlying continuous Gaussian distribution $G(y_i,y_j)_{\rm \mathcal S_{\rm (3,7)}^\prime}$ with $\vec{\mu}=(\mean{y_i},\mean{y_j})\simeq \vec 0$ as mean vector. 

We set the scattering parameter as $\chi=\exp(-\eta N_L)$ and evaluate numerically $\mathcal{I}(N_L)$ as a function of the tuneable parameter $N_L$ for three values of the number of atoms $N_A=\{0.5 \cdot 10^{5},0.2 \cdot 10^{6},0.5 \cdot 10^{6} \}$, leaving fixed the other important parameters to their experimental values as before, i.e., $\eta=0.5 \cdot 10^{-9}, g=10^{-7}$.

\begin{figure*}[h]
\includegraphics[width=0.5 \textwidth,clip]{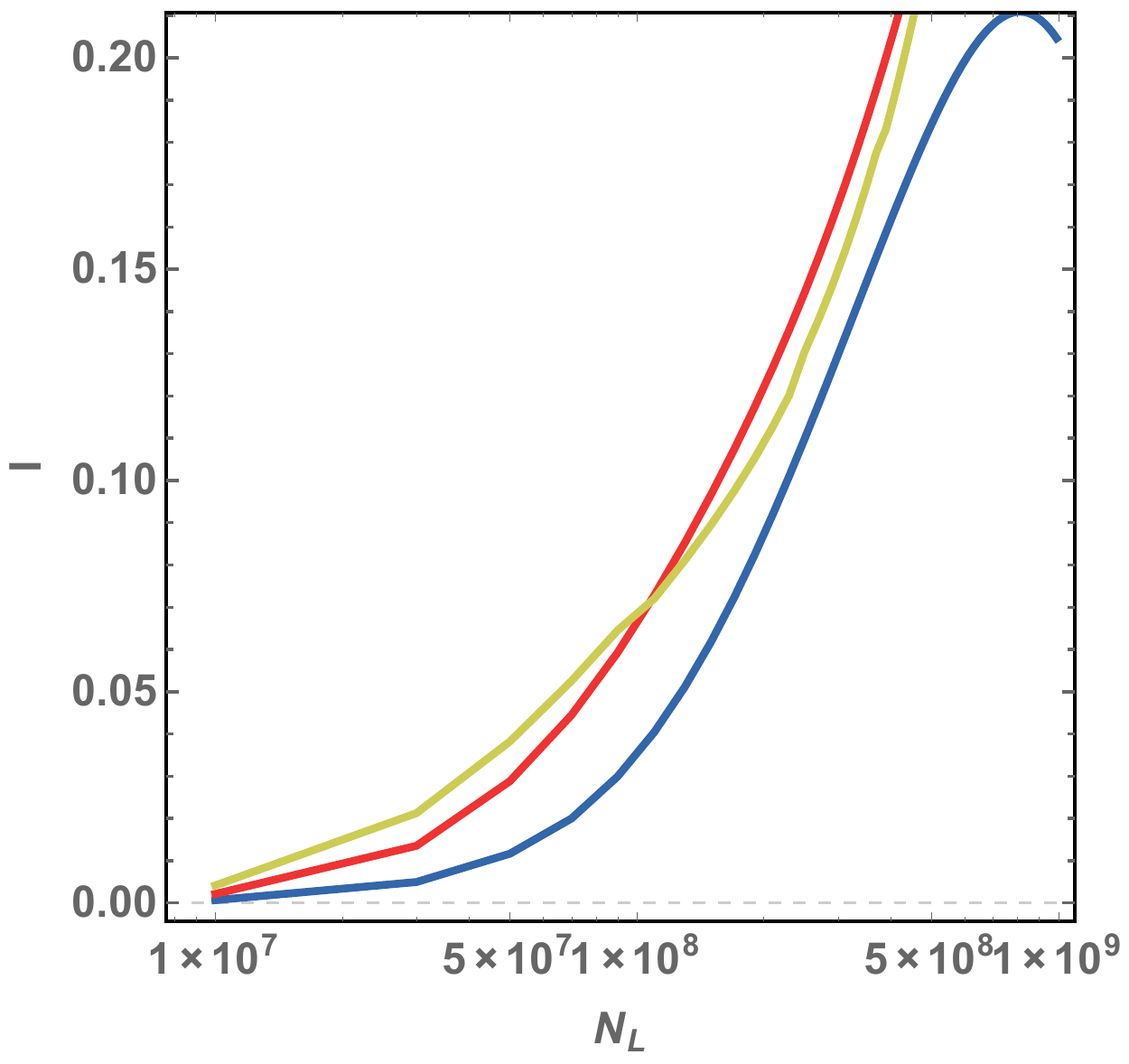}
\hspace{0.01\textwidth}
\includegraphics[width=0.5 \textwidth,clip]{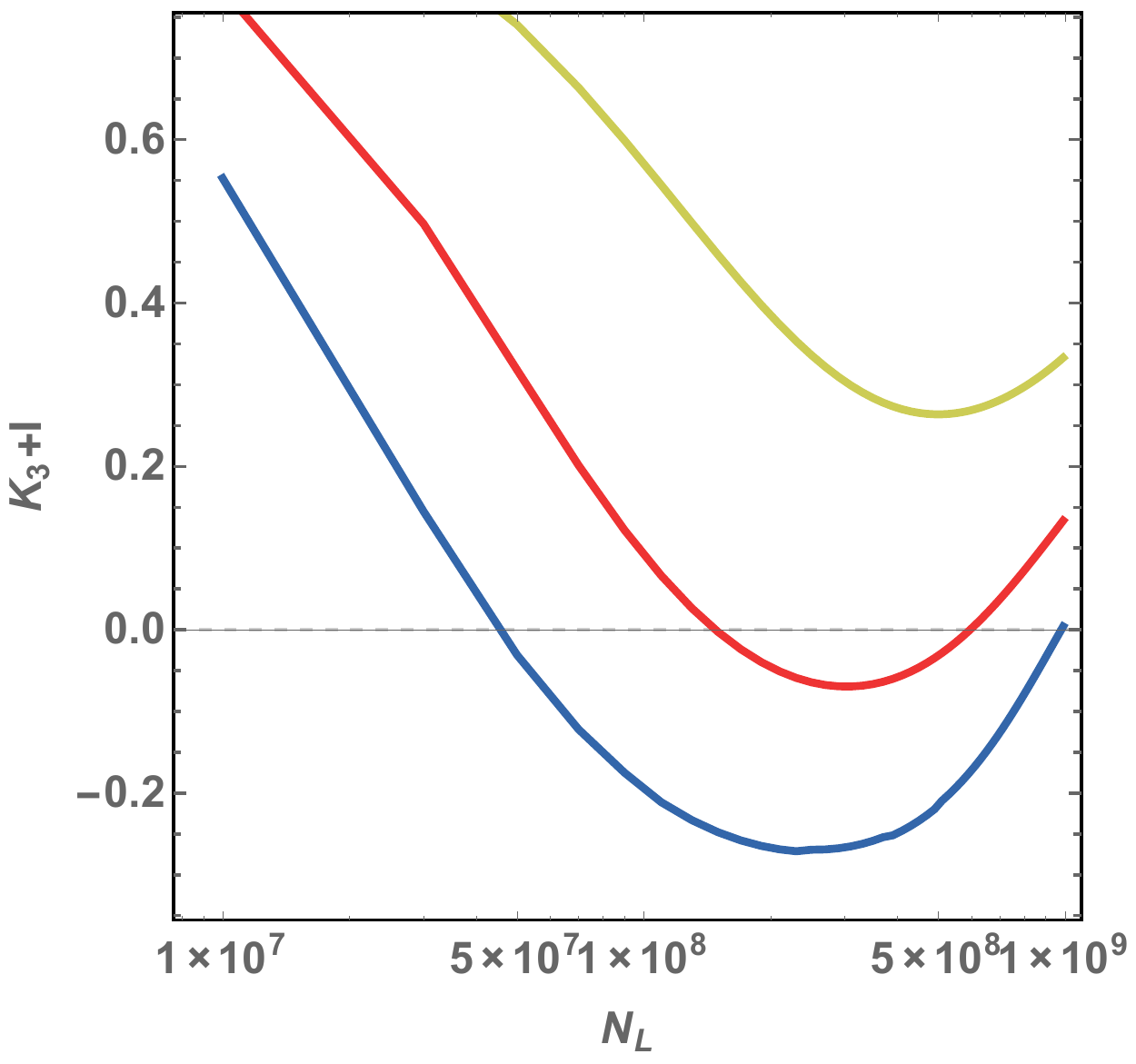}
\caption{(left) The invasivity parameter $\mathcal{I}$ as a function of $N_L$ for $N_A=0.5 \cdot 10^{6}$ (blue), $N_A=0.2 \cdot 10^{6}$ (red), $N_A=0.5 \cdot 10^{5}$ (yellow). (right) The expression $KI_3(N_L)$ in Eq.~(\ref{k3I7meas}) evaluated for the same numbers of atoms. The other parameters are $\eta=0.5\cdot 10^{-9}, g=10^{-7}$.}
\label{inv35}
\end{figure*}

The results are plotted in Fig.~\ref{inv35}(left) and also show as desirable that by increasing the macroscopicity of the system (i.e., the number of atoms $N_A$) the effect of clumsiness decreases, suggesting that the measurements become adroit in a classical sense when the size of the system increases. 

Afterwards we can also directly compute the expression $KI_3$ as in Eq.~(\ref{eq:LGIclumsyQND}), i.e.,
\begin{equation}\label{k3I7meas}
KI_3=C_{35}+C_{57}+C_{37}+1+\mathcal{I}_{37}+\mathcal{I}_{57}
\end{equation}
as a function of $N_L$ and for the same three values of the number of atoms. The results are plotted in Fig.~\ref{inv35}(right) and show that a violation of Eq.~(\ref{eq:LGIclumsyQND}) can be seen under realistic conditions. 

Thus, in this chapter we have shown how it is possible to violate a LGI in a macroscopic ensemble of $N\sim 10^6$ atoms probed with light in a non-demolition way. Then 
we have also shown how to suitably quantify the clumsiness of the QND measurements and how much it contributes to the LGI violation. Finally we have shown numerically that, even taking account of the classical clumsiness, a violation of macrorealism can be achieved in such macroscopic systems in a realistic setting.

This represents a big step forward as compared to most of the past proposals and experiments that have been done with microscopic systems \cite{GogginPNAS2011,xuSR2011,DresselPRL2011,SuzukiNJP2012,WaldherrPRL2011,GeorgePNAS2013,AthalyePRL2011,SouzaNJP2011,KneSimGau12,RobensPRX2015} and also 
with respect to the few tests performed to date with a macroscopic flux quibit of a SQUID \cite{PalaciosNatPhot2010,groen13}, particularly the experiment of Palacios-Laloy {\it et al} \cite{PalaciosNatPhot2010}. In this last case in fact, although the result was in very good agreement with the predictions and showed a significant violation of a LGI, no indication was provided that the obtained violation could not be due to unwanted or hidden clumsiness.
This was a fundamental weakness of the test that was unavoidable in that experimental setting, as also explained in \cite{WildeFP2012}. Here, by using a different system and exploiting the fact that the back action of QND measurements is initially hidden in the conjugate variable, we were able to tighten to some extent this loophole. This opens the possibility to use QND measurements 
to make clumsiness-free tests of macrorealism in a very wide variety of experimental settings \cite{BraginskyS1980}, 
going from atomic \cite{MitchellNJP2012,SewellNatPhot2013} to optical \cite{GrangierN1998} and opto-mechanical \cite{optomechanics14,ligo09} systems and potentially allowing for going up to an every-day life scale of macroscopicity \cite{ligo09}. 

To conclude let us write a few words about the definition itself of macroscopicity in this framework. Although an intuitive picture of a macroscopic object might be more or less clear, it is difficult to formalize it as a mathematical property of a system. Furthermore, in different contexts the definition can vary a lot \cite{leggett02,frowisdur12,Nimmrichter13}, depending on the properties that one is interested to observe. For example, one can define 
a macroscopically entangled state \cite{frowisdur12} as something that is close to a Schr\"odinger cat state, but this does not capture the notion of 
a classical macroscopic state, such as an every-day life object.

In the case of LG tests, the idea is that the state has to be macroscopic in a sense similar to our classical intuition and one aims to detect the impossibility of assigning a definite value to a macroscopic observable (e.g., quoting Einstein, the position of the moon). In particular no entanglement is needed, but just the state being a superposition of \emph{macroscopically distinct states}, or alternatively but equivalently that the measurement has a macroscopic effect on the observable's value. 
One possible definition in this sense, proposed by Leggett himself \cite{leggett02}, is called \emph{Extensive Difference} and is computed by taking the difference in expectation values of the measured (supposed macroscopic) quantity between two states and compare it to some ``reference'' value for the specific system.
Although quite naive, this is the only definition to date that goes in the direction of quantifying how macroscopically the measured observable is perturbed by the measurement. In the system that we are considering here, i.e., ensembles of cold atoms, the extensive difference can be roughly estimated to scale as $\sim \sqrt N$.

A satisfactory formal definition of macroscopicity, however, is still lacking and
it is an interesting open question to provide a quantitative parameter that allows to compare the macroscopicity of different experimental settings.

\chapter{Conclusions and future directions} 

\label{Conclusions} 


\section{Further developments on Spin Squeezing parameters}

The results that we have shown in Sec.~\ref{subs:completessij}, in particular Obs.~\ref{obs:completessispinj} can be easily generalized further in several directions. First of all, we said that the only condition needed to derive Eq.~(\ref{completessiqubitsj})
is a LUR constraint (namely Eq.~(\ref{3spinslur})). Then one can easily show that similar
entanglement conditions can be found for other sets of collective observables $\{ A_k \}$, provided that some LUR constraint
$\sum_k (\Delta a_k)^2 \geq {\rm const.}$ holds for the local components of $A_k$. 

\paragraph{$su(d)$ Squeezing Inequalities.}
For example, for a system of $d$-level particles the space of local operators is wider and spanned by an $su(d)$ basis. Then, 
a set of entanglement criteria based on an $su(d)$ collective basis $G_k=\sum_{n} g_k^{(n)}$ can be derived for ensembles of many qudits based on the same reasoning of Obs.~\ref{obs:completessispinj} (see Ref.~\cite{vitagliano11}).

\bboo{($su(d)$-squeezing inequalities).}
Consider a system of $N$ particles with $d$ levels and the $d^2-1$ collective $su(d)$ generators $\{ G_k=\sum_{n} g_k^{(n)}\}$, constructed from a basis $\{g_k\}$ of the $su(d)$ algebra $[g_k,g_l]=i f_{klm} g_m$.
The following set of inequalities
\begin{equation}\label{gensepcond}
(N-1) \sum_{k \in I} (\tilde{\Delta} G_{k})^{2} - \sum_{k \notin I}\aver{\tilde{G}_{k}^{2}} +2N(N-1) \frac{d-1} d \geq 0 ,
\end{equation}
holds for all separable states for all subsets $I$ of indices $k\in \{0,\dots,d^2-1\}$. 
In Eq.~(\ref{gensepcond}) we defined the
modified second moments as $\aver{\tilde{G}_{k}^{2}}=\aver{\tilde{G}_{k}^{2}}-\sum_n \aver{(g_k^{(n)})^2}$ and analogously the modified variance $(\tilde{\Delta} G_{k})^{2}$.
\eeoo

For brevity we omit the proof of the statement, that however is based on the existence of a LUR constraint $\sum_k (\Delta g_k)^2 \geq 2(d-1)$ for the single particle $su(d)$ generators \cite{vitagliano11}. 
Thus, every state $\rho$ that violates one inequality in the set Eq.~(\ref{gensepcond}) must be entangled.

Then, analogously as in Sec.~\ref{subs:completessij} one can derive
a single $su(d)$-squeezing parameter that embraces the whole set (\ref{gensepcond}) and show that it is invariant under
all the operations $U^{\otimes N}$ with unitary $U$. Even more, the $su(d)$ squeezing parameter can be shown to be a monotone under permutationally invariant separable operations and can be thus used as a measure of entanglement in permutationally invariant systems \cite{vitaglianototh2015}.

In fact such a set of inequalities is complete, i.e., no more inequalities can be added to the set in order to detect more entangled states, when only the second moments of collective quantities can be measured. So far all the spin squeezing inequalities are based on permutationally invariant quantities, like the variances of collective operators. A natural generalization would then be to give up this symmetry.

\paragraph{Translationally invariant inequalities.}

To go a step further then, one can consider sets of operators that are not permutationally invariant, but still {\it translationally invariant}. For example the following operators
\begin{equation}
(J_{\rm TI})_l(q) = \sum_n e^{iq n} j_l^{(n)} \ ,
\end{equation}
where $q\in \mathbb R$ is a real number, are a translationally, but not permutationally invariant generalizations of the collective spin components\footnote{Here we put a single real number $q$ in the definition, but in general for each component $J_l$ a different
scalar component $q_l$ of a vector $\vec q=(q_x,q_y,q_z)$ can be considered.}.
These operators are not Hermitean, but still one can define 
some modified second moments as
\begin{equation}\label{TI1modsec}
\aver{(\tilde{J}_{\rm TI})^2_l(q)} := \aver{(J_{\rm TI})_l^\dagger(q) (J_{\rm TI})_l(q)}-\sum_n \aver{(j_l^{(n)})^2} = \sum_{n \neq m} e^{iq(n-m)}\aver{j_l^{(n)} j_l^{(m)}} \ ,
\end{equation}
and the corresponding modified variances $(\tilde{\Delta} (J_{\rm TI})_{l})^{2}(q) := 
\aver{(\tilde{J}_{\rm TI})^2_l(q)} - \aver{(J_{\rm TI})_l^\dagger(q)} \aver{(J_{\rm TI})_l(q)}$.

Then one can prove the following 

\bboo{(Translationally invariant SSIs).}
Every inequality in the following set
\begin{equation}\label{TIcompletessi}
(N-1)\sum_{l \in I} (\tilde{\Delta} (J_{\rm TI})_{l})^{2}(q) - 
\sum_{l \notin I} \aver{(\tilde{J}_{\rm TI})^2_l(q)} + N(N-1)j^2\geq 0 \ ,
\end{equation}
must be satisfied by all separable states of $N$ spin-$j$ particles for all possible subsets $I$ of indices $k \in \{x,y,z\}$ and every
real number $q\in \mathbb R$.  
\eeoo

Again, Eq.~(\ref{TIcompletessi}) is derived with exactly the same reasoning as in Obs.~\ref{obs:completessispinj}. In fact, still 
in this case the only constraint needed is Eq.~(\ref{3spinslur}). In this case to find a suitable single parameter one should optimize over all subsets $I$ and all real numbers $q$. This would provide a characterization of full separability in systems in which only the translationally invariant quantities (\ref{TI1modsec}) can be measured. A specific example of such a situation is 
provided by solid state systems probed with neutron scattering, where (\ref{TI1modsec}) can be extracted from the corresponding scattering cross sections \cite{marty14}.

\paragraph{$k$-entanglement and $k$-range criteria.}

Finally, another natural extension of the spin squeezing inequalities, partly already made and discussed 
in Sec.~\ref{sec:extrdicke}, is to find criteria for the detection of the depth of entanglement, or analogously of the so called
{\it range} of entanglement. This last is usually considered in many body states that are thermal states of spin chain models and consists in the minimal distance $k$ between two entangled sites of a chain.
To define such extended framework we can make a sort of coarse graining of a multi-spin system by considering
e.g., collective spins $L_m=\sum_{n=1}^{k} j_m^{(n)}$ for groups of $k$ particles as a local basis from which to construct the collective operators
\begin{equation}
J_m=\sum_{n^\prime=1}^{N/k} L_m^{(n^\prime)} \ ,
\end{equation}
where the index $n^\prime$ labels the different groups of $k$ particles, $\frac N k$ being their total number. In this case the local variance would satisfy a Bloch vector length condition $\aver{L_m}^2 \leq (kj)^2$ and we can still use this constraint to derive SSIs.
In fact we are basically considering the collective operators $\{ J_m\}$ for systems of $\frac N k$ spin-$kj$ particles.

We can define the modified second moments as
\begin{equation}\label{eq:kmodsecmom}
\aver{\tilde{J}_{m,\rm k}^2}:= \aver{J_m^2} - \sum_{n^\prime=1}^{N/k} \aver{(L_m^{(n^\prime)})^2} = \aver{J_m^2} 
- \sum_n \aver{(j_m^{(n)})^2} - \sum_{n^\prime=1}^{N/k} \sum_{n\neq l=1}^k \aver{(j_m^{(n)} j_m^{(l)})^{(n^\prime)}} \ ,
\end{equation}
where the label ${\rm k}$ refers to a coarse graining in groups of $k$ particles and one can note that there is an additional subtracted term with respect to Eq.~(\ref{eq:modsecmomdef}), i.e.,
\begin{equation}
\aver{\tilde{J}_{m,\rm k}^2}=\aver{\tilde{J}_{m}^2} - \sum_{n^\prime=1}^{N/k} \sum_{n\neq l=1}^k \aver{(j_m^{(n)} j_m^{(l)})^{(n^\prime)}} \ ,
\end{equation}
and in particular if we {\it assume permutationally invariance}, we can simplify it to 
\begin{equation}
\aver{\tilde{J}_{m,\rm k}^2}=\aver{\tilde{J}_{m}^2} - N(k-1) \aver{j_m \otimes j_m} = N(N-k)\aver{j_m \otimes j_m}=
\frac{N-k}{N-1}\aver{\tilde{J}_{m}^2} \ ,
\end{equation}
where $\aver{j_m \otimes j_m}$ is the bipartite correlation, that is equal between all the couples of particles.
Then, one can derive spin squeezing inequalities based on $\aver{\tilde{J}_{m,\rm k}^2}$ that can distinguish the depth or the range of entanglement\footnote{Note that a state with entanglement depth equal to $k$ has also a range of entanglement of $k$ but not vice-versa.}. For example we can find the following (analytical) condition, that would detect the depth of entanglement of states close to Dicke states.

\bboo{(Linear SSI for the depth of entanglement).}
The following inequality
\begin{equation}\label{eq:lindickedepth}
\frac{N-k} k \left(\frac{N-k}{N-1}\aver{\tilde{J}_{z}^2} - \aver{J_z}^2 \right)
-\frac{N-k}{N-1} \left( \aver{\tilde{J}_{x}^2} + \aver{\tilde{J}_{y}^2}\right) + N(N-k)j^2\geq 0 \ ,
\end{equation}
where $\tilde{J}_{m}^2$ and $(\tilde{\Delta} J_{m})^2$ are defined as in Eq.~(\ref{eq:modsecmomdef}), must be satisfied by all permutationally invariant $k$ producible states of $N$ spin-$j$ particles. 
\eeoo

\bbpr
Let us consider the coarse grained local operators for a group of $k$ particles $L_m=\sum_{n=1}^{k} j_m^{(n)}$ and let us define the modified second moments $\aver{\tilde{J}_{m,\rm k}^2}$ as in Eq.~(\ref{eq:kmodsecmom}).
Following the same reasoning as in Obs.~\ref{obs:completessispinj}, but substituting $N\rightarrow \frac N k$, $\aver{\tilde{J}_{m}^2}\rightarrow \aver{\tilde{J}_{m,\rm k}^2}$ and exploiting the bound $\aver{L_m}^2 \leq (kj)^2$ for the local operators we find that the following set of inequalities
\begin{equation}\label{eq:kprodcondPI1}
\left(\frac N k -1\right) \sum_{m \in I} (\tilde{\Delta} J_{m,\rm k})^2 - \sum_{m \notin I} \aver{\tilde{J}_{m,\rm k}^2} 
+ \frac N k \left(\frac N k -1\right) k^2j^2 \geq 0 \ ,
\end{equation}
where $I$ is any subset of indices $m\in\{ x,y,z\}$ must hold for all states that are separable in groups of $k$ particles, i.e., all $k$ producible states. Let us choose $I=\{z\}$. We have that Eq.~(\ref{eq:kprodcondPI1}) reduces to
\begin{equation}
\left(\frac N k -1\right) (\tilde{\Delta} J_{z,\rm k})^2 - \left( \aver{\tilde{J}_{x,\rm k}^2} +\aver{\tilde{J}_{y,\rm k}^2}  \right)
+ \frac N k \left(\frac N k -1\right) k^2j^2 \geq 0 \ ,
\end{equation}
and assuming permutational invariance we have $\aver{\tilde{J}_{m,\rm k}^2}=\frac{N-k}{N-1}\aver{\tilde{J}_{m}^2}$, which proves the statement.
\eepr

As a comment, note that here we are using the same modified second moments as for Eq.~(\ref{completessiqubitsj})
and that the unpolarized Dicke state violates Eq.~(\ref{eq:lindickedepth}) for all $k<N$ and is detected as $N$ entangled as it should. This inequality has some advantages with respect to Eq.~(\ref{eq:oursormolcrit}) being {\it analytical and linear}. It can be also straightforwardly extended to system with fluctuating number of particles. The price that we pay is that we have to
compute the average local second moments $\sum_n \aver{(j_m^{(n)})^2}$ and assume that the state is {\it permutationally invariant}. Also, Eq.~(\ref{eq:oursormolcrit}) is an optimal inequality, while it is not clear whether Eq.~(\ref{eq:lindickedepth}) can detect the same set of states or a smaller one.

We conclude here by leaving open an additional question for further investigation.
We have already seen that in general we can give up the assumption of fixed number of particles and allow a fluctuating $\aver{N}$ and also that such generalization is not only possible, but also desirable for many experimental situations.
Another natural question is then: does it make sense to give up also the assumption of fixed local dimension $d=2j+1$ and allow a fluctuating 
number of levels $\aver{d}$ accessible to the particles (or analogously a fluctuating spin quantum number $\aver{j}$)?

\subsection{Spin Squeezing in spin models}

Another interesting question is to study entanglement in a thermal state and look at possible connections with its thermodynamical properties. 
As usual, a possibility is to study the simplest models available in the literature, for which they are known or they can be computed numerically the eigenenergies and some properties of the thermal states. 

In \cite{tothPRA09,vitagliano14} it has already been shown that indeed the optimal SSIs can detect as entangled thermal states of spin models. In particular very interestingly entangled states can been detected such that they are PPT for all possible bipartitions of the $N$-particle system. Furthermore, for some permutationally invariant models, spin squeezing was also connected to entanglement and quantum criticality \cite{vidal04,vidalpalacios04,wang10}.

Thus, those numerical results indicate that the SSIs can be very useful in detecting entanglement in spin models, especially mean field models, that have a permutationally invariant hamiltonian\footnote{In fact the SSIs are very efficient in detecting permutationally invariant states. See also the discussion in Appendix~\ref{ch:SSandPI}}, opening the possibility to use thermodinamical quantities, like energy \cite{tothpra05} and susceptibilities \cite{Wiesniak05,brukner06} as entanglement witnesses for condensed matter systems (see also the review article \cite{amico08} and references therein). 

Furthermore, it has become widely known that the behavior of entanglement in thermal states is influenced by the underlying quantum criticality of the Hamiltonian (see \cite{amico08} and references therein). The ground state becomes somehow highly entangled when approaching a quantum critical point. Moreover the scaling of bipartite entanglement with the size of the partition depends on the universality
class of the transition \cite{sachdev,vidallatorre03,calabresecardy04,calabresecardy06,cramereisertplenio07,calabresecardy09}. 
On the other hand, much less is known about entanglement in the thermal states even at very small temperatures above the quantum critical points. 
Therefore it is very interesting to study spin squeezing in thermal states of models that are quantum critical: on the one hand by extending the study of Refs.~\cite{vidal04,vidalpalacios04,wang10} and analyzing our generalized Spin Squeezing parameter (\ref{eq:genssparameter}) in permutationally invariant quantum critical models, such as the Lipkin-Meshkov-Glick \cite{Lipkin1965188,Meshkov1965199,Glick1965211} or the Dicke model \cite{Dicke1954Coherence}. On the other hand, it would be very interesting to compute 
Eq.~(\ref{eq:genssparameter}) in translationally invariant models, that are more suitable to describe actual condensed matter systems.

\paragraph{SS in thermal states of XY chains.}

As a toy example one can try to study entanglement of a thermal state of the XY model \cite{lieb61,baroucmcoy1,baroouchmcoy2,baroouchmcoy3,baroouchmcoy4}. As usually we choose these models because on the one hand they are so simple that entanglement of the thermal states might be studied also analytically \cite{itsjin05}. On the other hand they are complex enough to show non trivial and \emph{universal} phenomena, like quantum phase transitions (see also \cite{sachdev}). Thus let us consider the XY Hamiltonian
written as
\begin{equation}
H=\sum_{i=1}^{N} r\cos(\theta)\cos(\phi) \sigma_{i}^{x}\sigma_{i+1}^{x}+r\cos(\theta)\sin(\phi) \sigma_{i}^{y}\sigma_{i+1}^{y} + r\sin(\theta) \sigma_{i}^{z} .
\end{equation}
where we used a set of polar couplings $r, \cos(\theta), \cos(\phi)$
such that the characteristic energy scale of the system is the single parameter $r$. Afterwards, we will be mainly interested in the ratio between this energy scale and the temperature $\omega = \beta r$.

Another energy scale that is important in quantum phase transitions is the first gap $\Delta=E_1-E_0$. The energy levels of the model are
\begin{equation}
\epsilon_k = r\sqrt{(\sin \theta -\cos \theta \cos \varphi_k)^2+\delta^2\sin^2 \varphi_k} ,
\end{equation}
and in particular,
the first gap of respectively the XX and the Ising model assume the forms
\begin{gather}
\Delta_{\rm XX} = 0 \for | \sin \theta/\cos \theta | \leq 1 \ , \\
\Delta_{\rm Is} = 2r|\sin \theta - \cos \theta | ,
\end{gather}
thus, the XX model is quantum critical (gapless) in the whole $| \sin \theta/\cos \theta | \leq 1$ region, while the Ising model becomes critical when $\theta = \pi /2 + n \pi$.

In the recent literature there have been found 
connections between the criticality of the system and the scaling of bipartite entanglement between two blocks \cite{vidallatorre03,calabresecardy04,itsjin05,calabresecardy06,cramereisertplenio07,calabresecardy09,calabresecardytonni15}, 
or of their mutual information \cite{melko10,singh11}. 

Here we might follow a sort of complementary approach and try to connect the quantum criticality with the genuine multipartite entanglement of the whole system. 

Moreover there is also a region, $r^{2}=1$, in the XY phase plane in which the ground state is factorized, and thus not entangled. There it would be interesting to see whether for small temperatures the state becomes entangled or not. Some numerical studies and conjectures about this have already been made.

We might study our generalized Spin Squeezing parameter of Eq.~(\ref{eq:genssparameter}) 
\begin{equation}
\xi^2_{\rm G}(\rho,N):=\frac{\trace(\Gamma_\rho) - \sum_{k=1}^I \lambda^{>0}_k(\mathfrak Z_\rho)}{Nj} \ ,
\end{equation}
also as a function of $N$ and compare it with other entanglement measures. In this respect, we can consider first the $N=2$ particles parameter and compare it with e.g., the concurrence \cite{glaser03,yin11}, that for states that are invariant under translations, parity and reflections, has the following expression
\begin{equation}\label{Concexprpar}
\mathcal C := \frac 1 2 \max \{0, \mathcal C^{I}, \mathcal C^{II} \} \ ,
\end{equation}
where
\begin{equation}
\mathcal C^{II} = \sqrt{ (\aver{\sigma_{x} \otimes \sigma_{x}} - \aver{\sigma_{y} \otimes \sigma_{y}})^{2} } -\sqrt{(1-\aver{\sigma_{z} \otimes \sigma_{z}})^{2}} \ ,
\end{equation}
and
\begin{equation}
\mathcal C^{I} = \sqrt{ (\aver{\sigma_{x} \otimes \sigma_{x}} + \aver{\sigma_{y} \otimes \sigma_{y}})^{2} } -\sqrt{(1+\aver{\sigma_{z} \otimes \sigma_{z}})^{2}-4\aver{\sigma_{z} \otimes \id}^{2}} \ .
\end{equation}
The concurrence might seem quite unrelated to the Spin Squeezing parameter.
However, we can see that for states with certain symmetries
the two measures actually coincide. For example for states that have $\aver{\sigma_{z} \otimes \id}^{2}=0$ and all the three correlations are negative $\aver{\sigma_{k} \otimes \sigma_{k}}<0$. These states appear precisely as thermal states of certain XY spin chains and numerically it can be seen that in that case the two measures $\mathcal C(\rho)$ and $\xi^2_{\rm G}(\rho,2)$ do coincide \cite{vitaglianonotp} (see also \cite{glaser03,yin11}).

\paragraph{The scaling limit of SS parameter.}

In the critical regions a system becomes scale invariant and one can apply the renormalization group method to extract information about observable quantities. In the vicinity of critical points the system becomes also universal, in the sense that the microscopic details of the 
model are not important: any model in the critical point can be described by a quantum field theory that is the universal theory for the specific transition. For example, the universal theory describing the quantum XX critical line is the following free bosonic theory (in imaginary time) \cite{sachdev}
\begin{equation}
\mathcal L = \frac 1 {2\pi v_F} \left( (\partial_{\tau} \phi)^2 + v_F^2 (\nabla \phi)^2 \right) \ ,
\end{equation}
where $v_F:= 4 r a$ and $a$ is the lattice spacing. In the scaling limit the scale becomes the velocity $v_F$ in spite of the coupling strength $r$. 

Numerically it is easy to compute $\xi^2_{\rm G}(\rho,N)$ and study its scaling \cite{vitaglianonotp}, for example in the vicinity of quantum critical points and it reproduces some of the well known results obtained with Conformal Field Theories \cite{vidallatorre03,calabresecardy04,calabresecardy06,calabresecardy09}. 

On the other hand, we might also try to define analytically $\xi^2_{\rm G}(\rho,\infty)$ in a scaling limit
\begin{equation}
N\rightarrow \infty \ , \quad a \rightarrow 0 \ , \quad l:=aN = {\rm const.} \ ,
\end{equation}
where we left the generality of having a chain of finite length $l$. 
It could be used as a witness of quantum criticality and even measured experimentally, e.g., with neutron scattering, to detect a Quantum Phase Transition in a real system. 
Another interesting fact is that it just depends on the ratio between the temperature and the coupling strength $\omega=T/v_F$.  Therefore it might have a non trivial behaviour also for high temperatures $T \sim v_F$, provided that either the coupling $J$ or the external field $h$ are big enough.

There is a conceptual issue however, coming from the scaling limit of the correlation function $C_d=\aver{j^{(0)}_{k} j^{(d)}_{k}}$. We have to substitute the integer discrete distance $d$ with a real continuous variable $x:=d a$ and since $a\rightarrow 0$ consider the correlations at large distances $d \gg 1$ in order to compute $C(x)$ at finite $x$. Moreover it generally makes sense to study just the asymptotic behaviour $x\gg \zeta(\omega)$ of $C(x)$ with respect to some length scale $\zeta(\omega)$. 
On the contrary, $\xi^2_{\rm G}(\rho,N)$ contains the correlations {\it at all length scales}, i.e., for $1\leq d \leq N$. It is an expression that, being permutationally invariant takes account of average correlations between all the $N$ spins.

Thus there is a tension that we have to resolve in some way. An idea might be to e.g., set some {\it non-universal} constant like $\lim_{x\rightarrow 0}C(x):= \alpha$, that we compute in some specific models. Also, for translational invariant systems other possibilities might come from studying a translationally invariant extended parameter (see Eq.~(\ref{TIcompletessi}) and the related discussion).


\section{Applications of QND measurement based LG tests}

The other topic that we investigated in this thesis has been the foundational question of experimentally disproving 
the independence of observable values from measurements. In particular we have proposed to 
witness the disturbance implied by the Heisenberg uncertainty principle 
using QND measurements, thought to be the quantum measurements closest to the ideal non-invasive and adapt for
an LG-like test. 

We have shown that a test in a system consisting of millions of atoms is already feasible, and would provide an evidence of truly quantum effects
at a macroscopic scale independently from entanglement. Moreover we have provided a way to quantify the clumsiness of a QND measurement, that in principle allows to
make in the future clumsiness-free LG tests in different experimental settings \cite{BraginskyS1980,MitchellNJP2012,SewellNatPhot2013,GrangierN1998,optomechanics14,ligo09} and up to bigger and bigger scales.
Thus, the next natural step would be to try to increase the scale as much as possible so to push the ideal boundary of the quantum/classical divide
forward, up to an every-day life scale \cite{ligo09}

On the other hand, afterwards we can also try to extend further the idea and go beyond a close to original, clumsiness free, LG test. 
In fact, with the experiment that we propose we aim to show that the statistical correlations coming from quantum theory cannot be explained in terms of a more fundamental classical theory in a way complementary with respect to Bell tests.
We want to experimentally show to some extent that there is an ultimate limit on the trade off between precision in a measurement and disturbance on the measured system; a fact stated as the Heisenberg uncertainty principle.

This has been always one of the issues with quantum theory from a macrorealist viewpoint. Another fundamental one, actually a true problem of the theory, has been termed {\it the measurement problem}: what happens to the {\it actual} state of a system after a measurement is made? Does the collapse of the state physically happen? How is it compatible with the unitary evolution, then? 

\paragraph{Investigating the measurement problem}

Thus, on the one hand we can extend the foundational study itself and look more deeply at the measurement problem.
One idea could be to decompose the protocol that we proposed, by looking separately at all the ingredients that contribute
to the violation of a LGI. This means at first to decompose the QND measurement in all its passages: (i) the preparation of the system, (ii) the preparation of the probe, (iii) the system-meter interaction, (iv) the projective measurement of the probe.

This is a decomposition that follows the lines of the original formalization of a quantum measurement, made at first by von Neumann and later improved in several directions by many other authors. Then, we can see how every single element contributes to the violation of a LGI and how classical noise affects each of them.  

In particular, we have defined a quantifier for the clumsiness, with the idea that it comes mainly from a noisy system-meter interaction. We can then look for more advanced strategies that might reveal more in detail where the clumsiness actually comes from and how to limit it as much as possible. For example, an open question in this respect could be: does it help to have an initial state of the system that is spin squeezed? And a probe polarization-squeezed light? 
Is the spin squeezing parameter connected to the LGI violation? And eventually, which particular Spin Squeezing parameter quantifies this the better?

Also, we can study quantitatively the single effect of the final projective measurement of the probe and how it influences the LG test. In fact we know that actually the quantum back action comes all from the system-meter interaction and is later revealed by an independent free evolution of the system. Thus, a question arises from our study: does the projection influence the violation of an LGI at all? Can we avoid to account for this effect to explain the whole result of the test?

Later one can also study in more details what happens when the free evolution of the system is turned on. We know that this affects the LG test in revealing the quantum back action, otherwise hidden in a conjugate variable. A refinement of our study can then be to try to separately take into account an imperfect evolution and contrive even more the explanation of an LGI violation in terms of clumsy measurements. This would put more severe constraints on the presence of fundamental disturbances on the system, confining quantitatively the contribution coming from the Heisenberg principle.

Finally one can also ask the question: is there any reasonable classical model of evolution and invasive measurements that could explain the LGI violation as resulting from our protocol?

\paragraph{Extended figures of merit for QND measurements}

On the other hand we can try to complement the analysis of quantum measurements by relating our results, and in particular the invasivity parameter that we defined, with existing figures of merit for non-classicality of QND measurements.

It is in fact well known that to be genuinely QND a measurement should fulfill basically two criteria \cite{BraginskyS1980,GrangierN1998,MitchellNJP2012}: a good {\it quantum-state preparation}, that consists in the ability to generate correlations between the meter and the output signal variable and an {\it information-damage trade-off} that involves the ability to correlate the meter with the input system variable.

To show experimentally that a genuine QND measurement has been performed there are different figures of merit. In particular in the literature there are two generally accepted quantitative criteria that define a {\it Standard Quantum Limit} for the QND measurement, a threshold below which such measurements can be considered non-classical. 

The quantities needed are $\Delta X^2_M$, which describes the noise in the measurement referred to the input, $\Delta X^2_S$, i.e., the variance in the system added by the measurement and the post measurement conditional variance $\Delta X^2_{S|M}$.
All of them have to be normalized with respect to some Standard Quantum Limit defined independently.

Then, for the information-damage trade-off criteria of non-classicality one asks $\Delta X^2_M \Delta X^2_S <1$, while $\Delta X^2_{S|M}<1$ signs a non classical quantum-state preparation. 

To show genuine QND features in atomic ensembles a protocol discussed in \cite{MitchellNJP2012} was adopted that consists of three measurements where the phase $\phi=\frac{S_y^{\rm (out)}}{S_x^{\rm (in)}}$ is measured. With the variances of the first two measurements one quantifies the QND genuineness of the process. In particular one measures $\mathrm{var}(J_z|\phi_1)=\mathrm{var}(\phi_1-\chi \phi_2)-\mathrm{ var}(\phi_{\mathrm{RO}})$, where $\chi\equiv \mathrm{cov}(\phi_1, \phi_2) \mathrm{var}(\phi_1)$ and $\mathrm{var}(\phi_{\mathrm{RO}})$ is the variance of {\it read out}, i.e., the polarization variance of the input light pulse.

To quantify the damage due to the mesurement one needs to compare the correlations among all three measurements and consider the parameter $r_A\equiv \widetilde{\mathrm{cov} }(\phi_1,\phi_3)/\widetilde{\mathrm{cov} }(\phi_1,\phi_2)$ where $\widetilde{\mathrm{cov} }(X)\equiv \mathrm{cov} (X)-\mathrm{cov} (X_\mathrm{RO})$, so to define
the following figures of merit
\begin{gather}
\Delta X^2_M =\frac{\mathrm{ var}(\phi_1)}{J_0} \quad \ , \quad  \Delta X^2_S =\frac{\widetilde{\mathrm{var}}(\phi_2)-\widetilde{\mathrm{var}}(\phi_1)}{r_A J_0} \ , \\
\Delta X^2_{S|M}=\frac{\mathrm{ var}(J_z|\phi_1)}{r_A J_0} \ ,
\end{gather}
where $J_0=\frac{N_A} 4$ is the variance of an input completely $x$-polarized atomic state.
With these definitions one can then use the above mentioned non-classicality criteria $\Delta X^2_M \Delta X^2_S <1$ and 
$\Delta X^2_{S|M}<1$.

It is worth to emphasize that the protocol employed consists of a sequence of three consecutive measurements with very small delay, in some analogy with what we proposed in order to quantify the clumsiness in an LG test. An important difference is that 
the quantifier that we used for the clumsiness is taken from the whole output probability distributions, while these figures of merit for non-classicality are based on just variances and covariances.

Note also that the SQL can be defined with some degree of arbitrariness and therefore a non-classicality criterion could come from a violation of a LGI as well. We can then try to look at how our clumsiness parameter is related to these existing figures of merit for QND measurements and try to define an improved one that takes the LGI violation as standard, device-independent reference for non-classicality in that framework.


\section{Conclusions}

To end the thesis let us summarize and try to draw some general conclusions from this work.
Our study has been mainly oriented to theoretical foundational questions of quantum mechanics, to witness ``truly quantum'' effects, such as entanglement and the violation of realism, at macroscopic scales. 

A particular focus has been given to spin systems composed of very many particles and Gaussian states, i.e., roughly speaking, states that can be described in a collective Bloch sphere. In particular, Spin Squeezed States have been investigated with some depth in their connection to multipartite entanglement, as opposed to the near-to-classical Coherent Spin States. An optimal spin squeezing parameter has been defined that characterizes the full separability of Gaussian spin states of a large number of particles (i.e., of some macroscopic collective spin states) and other criteria have been developed that can detect very efficiently their depth of entanglement. Thus, in this framework the quantumness is detected in spin squeezed states as the presence of multipartite entanglement and this can be done efficiently in many actual experiments, also at macroscopic scales.

On the other hand, we have also shown that even CSS, when probed with QND measurements, can 
be used to witness non-classical features of quantum mechanics at macroscopic scales, namely the violation of macrorealism.
In that case entanglement is not needed to see the quantumness of the system, although it is intriguing to study possible connections
between violation of macrorealism in atomic ensembles and spin squeezing.

In the end, to summarize it, we studied what can be called the quantum/classical divide, especially focusing on macroscopic objects. Then, what did we learn from these studies about the difference between classical and quantum principles?

First of all, we learned that there is a very large ambiguity in the very definition of the quantum/classical divide.
We have seen already within the framework of spin squeezed states that several different and not-so-related definitions of Standard Quantum Limit
can be given. Many times ``quantumness'' is defined merely as the presence of entanglement, a phenomenon that violates classical principles. Certainly, in this sense entanglement
shows the quantumness of a state.
However, from this point of view there is an ambiguity in the definition: the same system can be considered as a single whole or as composed of parties,
according to a certain labelling. Then, unless there is some intrinsic way of distinguish the subsystems (such as e.g., different spatial locations) the labelling can be completely arbitrary and the state can be viewed equivalently as entangled or not.

Furthermore we have also seen that even ``classical'' states can provide evidence of macroscopic quantum effects, such as the violation of macrorealism. Thus, once more, the question remains largely open: where is the substantial difference between almost-classical and truly-quantum objects?

We have seen that the Heisenberg uncertainty relations are connected, though in different ways to both the detection of entanglement in many particle states and to the violation of macrorealism. Maybe then, a suggestion that comes out from our study is that the direction to follow
is to look at the possible incompatibility between observables and the minimal mutual uncertainty that one induces onto the other.

Ultimately, we might argue that the main incompatibility of quantum mechanics with classical principles is that it is impossible, according to QM,
to assign definite values to incompatible observables on the same state \emph{prior to and independent of} any measurement performed on the system. This is the lack of realism that troubles whoever wants to interpret the quantum wave function as the ontic state of the system. According to QM the outcomes of observables \emph{depend on the context} in which they are measured. A property that can be interpreted as the fact that 
measurements have always an effect on the whole system, even if it is composed by space-like distant parties.

Then, a suggestion is that the viewpoint can be switched from states to observables and try to resolve the ambiguity in defining the quantumness of a state by looking at the uncertainty principle to be satisfied by incompatible observables. In this respect, further developments in the study of Leggett-Garg-like tests might help, as being complementary to other analogous tests of quantum principles, like tests of non-contextuality and of non-locality. 

On the other hand, Coherent States are widely thought to be the nearest-to-classical states in many respects. Even in some cases one can define the quantumness of the state whenever it is somehow very different from a CS, for example by having a high degree of squeezing or a negative Wigner function. 
One of their characteristic is that they saturate Heisenberg uncertainty relations with two incompatible observables having the same uncertainty.
Thus, looking deeper at the difference between Squeezed/Coherent states might also help to understand the meaning of the fundamental unrealism
introduced by quantum mechanics as stated 
in the Heisenberg uncertainty principle and define somehow univocally the quantum/classical divide. This last, however, is widely thought not to be a very sharp bound (something similar to a universal Standard Quantum Limit) and it is still not clearly connected to the ``macroscopicity'' of the system, that is something not even properly defined yet.

\afterpage{\blankpage} 

\addtocontents{toc}{\vspace{2em}} 

\appendix 

\chapter{Gaussian states formalism and QND measurements}\label{sec:gaussstatqnd} 


\section{Definition of Gaussian states}

We have seen in Sec.~\ref{sec:geneses} that a $\hat z$-polarized Coherent Spin State can be described with a binomial probability distribution for the outcomes of $J_z$, as in Eq.~(\ref{eq:cssbinprob}). Then, in the limit of large number of particles $N\gg 1$ this distribution can be well approximated by a Gaussian with mean value $\aver{J_x}=0$ and variance $(\Delta J_x)^2=\frac N 4$.
This probability distribution is given by the coefficients of the state $|\Phi_{\rm z-CSS}\rangle = |J, J_z \rangle_{\rm z}$ with respect to the basis $|J, J_x \rangle_{\rm x}$, i.e.,
\begin{equation}
\Pr(J_x^{\rm (out)}=x)=\bigg| \langle J,J_z |J,x\rangle_{\rm x} \bigg|^2 \stackrel{N\rightarrow \infty}{\simeq}
\frac 1 {\sqrt{\pi N/2}} \exp\left( - \frac{2x^2}{N} \right) \ ,
\end{equation}
that is a Gaussian with mean $\mu_x=\aver{J_x^{\rm (out)}}=0$ and variance $\sigma_x^2=(\Delta J_x^{\rm (out)})^2=\frac N 4$ in the approximation of continuously distributed outcomes.
The same reasoning can be made for the outcomes of the conjugate spin component $J_y$, that are distributed according to
\begin{equation}
\Pr(J_y^{\rm (out)}=y)=\bigg| \langle J,J_z |J,y\rangle_{\rm y} \bigg|^2 \stackrel{N\rightarrow \infty}{\simeq}
\frac 1 {\sqrt{\pi N/2}} \exp\left( - \frac{2y^2}{N} \right) \ ,
\end{equation}
that is again a Gaussian with mean $\mu_y=\aver{J_y^{\rm (out)}}=0$ and variance $\sigma_y^2=(\Delta J_y^{\rm (out)})^2=\frac N 4$ in the same limit.
Note, however, that since $J_x$ and $J_y$ are not compatible with each other, there is not a joint probability distribution for the outcomes of the two observables.
A Spin Squeezed State can be also described very similarly with a Gaussian distribution function for the outcomes of $J_x$ (or $J_y$)
\begin{equation}
\Pr(J_x^{\rm (out)}=x)\stackrel{N\rightarrow \infty}{\simeq}\frac 1 {\sqrt{\pi \xi^2 N/2}} \exp\left( - \frac{2(x-\mu_x)^2}{\xi^2N} \right) \ ,
\end{equation}
with a mean value $\mu_x \neq 0$ in general (but that can be put back to zero with feedback schemes) and a squeezed variance $\sigma_x^2= \xi^2 \frac N 4$. Ideally, the probability distribution for the conjugate variable is such that $\sigma_y^2 = \frac{N}{4 \xi^2}$, but is still a Gaussian. All these are examples of states belonging to a class of so called {\it Gaussian states}. Basically these are states that can be described just in terms of the first two moments of the collective spin components $J_k$, since they have Gaussian probability distribution for their outcomes. For general mixed states we have the following definition
\bbdf{\bf (Gaussian states).} 
Let us consider a density matrix $\rho$ and a vector of operators $\vec V=(V_1,\dots,V_n)$ acting on the Hilbert space of states. 
Then the state $\rho$ is called {\it Gaussian} whenever the outcome probability distribution 
\begin{equation}
\Pr(V_k=x_k)=\trace(\rho | x_k\rangle_{\rm k}\langle x_k|) = G(\mu_k,\sigma_k)
\end{equation}
is a Gaussian function $G(\mu_k,\sigma^2_k)$ with a certain mean value $\mu_k=\aver{V_k}$ and variance 
$\sigma_k^2=(\Delta V_k)^2$ for all vector components $V_k$. 

In general there is not a joint probability distribution $\Pr(x_1,\dots,x_n)$ for the outcomes of all the components $(V_1, \dots , V_n)$ whenever $[V_k,V_l]\neq 0$ for some indices $(k,l)$. 
However we can still define a vector of mean values $\aver{\vec V}$ and a covariance matrix\footnote{Note that for collective spin components these are the same definitions that we used in the previous chapter, e.g., $\Gamma_{kl}=\frac 1 2 \aver{J_k J_l + J_l J_k}-\aver{J_k}\aver{J_l}$ as in Eq.~(\ref{eq:gammamatdef})} 
\begin{equation}
\Gamma_V:= \frac 1 2 \aver{\vec V \wedge \vec V+(\vec V \wedge \vec V)^T} -\aver{\vec V} \wedge \aver{\vec V}
\end{equation}
and extract all the information about a Gaussian state just from those quantities. 
\eedf 

Thus the mean vector $\aver{\vec J}$ and the covariance matrix $\Gamma_J$ are in a sense the defining quantities of a Gaussian spin state and 
the only constraints to the physicality of a certain state are given by some form of the Heisenberg uncertainty relations involving just those first two moments. To give a more intuitive picture, a Gaussian spin state can be imagined as a point (corresponding to the vector $\aver{\vec J}$) in a Bloch sphere, surrounded by an uncertainty region given by the covariance matrix $\Gamma_J$.

For states that are (almost) completely polarized in a certain direction $\hat z$ we can apply the Holstein-Primakoff mapping from the $su(2)$ to the Heisenberg-Weyl algebras, already introduced in Obs.~\ref{obs:holprimtransf}. In the limit $N\gg 1$ in which the $\hat z$ spin component can be considered constant $J_z \simeq Nj \id$ we can define the operators 
\begin{equation}
(x,p)=\left( \frac{J_x}{\sqrt{\aver{J_z}}}, \frac{J_y}{\sqrt{\aver{J_z}}}\right) \simeq \left( \frac{J_x}{\sqrt{Nj}}, \frac{J_y}{\sqrt{Nj}}\right) \ ,
\end{equation}
which satisfy the Heisenberg-Weyl commutation relations $[x,p]=2i \id$.
In this way the definition of Gaussian states is given in terms of the corresponding quadrature phase $(x,p)$ operators in analogy with the single mode bosonic {\it Gaussian states} 
\bbdf{\bf (Bosonic Gaussian states).} 
Let us consider a single bosonic mode with quadrature phase operators $\vec X=(x,p)$ obeying the Heisenberg-Weyl commutation relations $[x,p]=2i \id$. We define the {\it Wigner quasi-probability distribution} as 
\begin{equation}
W(x^{\rm out},p^{\rm out})= \frac 1 \pi \int \de a \langle x^{\rm out}-a | \rho |x^{\rm out}+a \rangle e^{-iap^{\rm out}} \ ,
\end{equation}
where $\rho$ is the density matrix describing the state of the mode and $(x^{\rm out},p^{\rm out})$ are the eigenvalues of the quadrature phase operators. Here the operator $D(a)=e^{iap}$ for $a\in \mathbb R$ is called {\it Weyl or displacement} operator and $\chi_\rho(a) = \trace(\rho D(a))=\langle x^{\rm out}-a | \rho |x^{\rm out}+a \rangle$ is called {\it characteristic function}. 

Then, a state is called {\it Gaussian} whenever $\chi_\rho(a)$ is a Gaussian function of $a$, or equivalently 
\begin{equation}
W(x^{\rm out},p^{\rm out})= \frac 1 {2\pi \sqrt{\det \Gamma}} \exp\left(-\frac 1 2 (\delta \vec X^{\rm out})^T \Gamma^{-1}\delta \vec X^{\rm out} \right) \ ,
\end{equation}
where we defined the vector $\delta \vec X^{\rm out} = (x^{\rm out}-\aver{x},p^{\rm out}-\aver{p})$ of fluctuations about the mean, is a Gaussian function of $(x^{\rm out},p^{\rm out})$. 

A single mode bosonic Gaussian state can be completely characterised by the mean vector $\aver{\vec X}=(\aver{x},\aver{p})$ and the covariance matrix $\Gamma_X=\frac 1 2 \aver{\vec X \wedge \vec X+(\vec X \wedge \vec X)^T} -\aver{\vec X} \wedge \aver{\vec X}$ and to be a physical state has to satisfy an Heisenberg uncertainty relation of the form 
\begin{equation}
\Gamma_X + i \Sigma_X \geq 0 \ ,
\end{equation}
where $2i(\Sigma_X)_{k l}= [X_k, X_l]$.  
\eedf 

Note that $W(x^{\rm out},p^{\rm out})$ need not to be a true probability distribution and in general can attain negative values. However for Gaussian states it is always positive. Geometrically, by identifying a completely polarized spin Gaussian state with a single mode bosonic Gaussian state we are mapping the Bloch sphere into the Heisenberg-Weyl plane. This mapping in fact holds in the asymptotic limit in which the radius of the sphere (i.e., the spin length $Nj$) goes to infinity 
$N\gg 1$. In this approximation, in particular, the spin components $J_k$ can be assumed to have a {\it continuous and unbounded spectrum} of outcomes $|J_k^{\rm (out)}|\leq Nj \rightarrow \infty$.

\section{Gaussian coherent rotations, QND measurements and noise}

Within the framework of Gaussian states one can define operations that map Gaussian states into Gaussian states. In total generality these can be described as linear operations acting on the mean vector $\aver{\vec V}$ and the covariance matrix $\Gamma_V$
\begin{subequations}
\begin{align}\label{eq:gaussopgen}
\aver{\vec V} &\mapsto M \aver{\vec V} + N_V \ , \\
\Gamma_V &\mapsto M \Gamma_V M^T + N_\Gamma \ , 
\end{align}
\end{subequations}
where $M$, $N_V$ and $N_\Gamma$ are real matrices. These, in order to be completely positive and trace non-increasing maps have to satisfy additional constraints, such that basically the Heisenberg uncertainty relations are preserved.

In the following let us focus on the particular system composed by 
an ensemble of $N_A$ atoms interacting with pulses of $N_L$ photons. The atomic state can be 
described by the collective spin vector of operators $\vec J=(J_x,J_y,J_z)$, with $J_k=\sum_{i=1}^{N_A} j_k^{(i)}$, while the light pulses are conveniently described by the Stokes vectors $\vec S=(S_x,S_y,S_z)$, with $S_k=\frac 1 2 (a_+, a_-)^\dagger \sigma_k (a_+, a_-)$ and $a_\pm$ being the annihilation operators for circular plus-minus polarizations\footnote{These are $su(2)$ operators obtained through a Schwinger representation. See Appendix~\ref{ch:SSandPI}}.

To describe a Gaussian state of the whole system we consider the joint vector 
\begin{equation}
\vec V= \vec J \oplus \vec S^{(1)} \oplus \dots \oplus \vec S^{(n_p)} \ ,
\end{equation}
where $n_p$ is the total number of light pulses, and correspondingly the mean vector $\aver{V}$ and the covariance matrix
$\Gamma_V$. Here we will describe, among all, basically three kinds of Gaussian operations: coherent rotations, QND measurements and decoherence due to classical noise. All of them can be described as linear maps at the level of the covariance matrix, as in Eqs.~(\ref{eq:gaussopgen}).

\paragraph{Atomic rotation driven by external field}

When the atomic ensembles are subject to an external magnetic field, they experience a spin rotation driven by an Hamiltonian like
\begin{equation}
H_B= \kappa \vec J \cdot \vec B \ ,
\end{equation}
where $\vec B$ is described as a classical external field. Accordingly, after an interaction time of $\Delta t$ the mean vector and the covariance matrix are modified just in their atomic part as
\begin{subequations}
\begin{align}\label{eq:gaussatorot}
\aver{\vec J} &\mapsto O_B \aver{\vec J} \ , \\
\Gamma_J &\mapsto O_B \Gamma_J O_B^T \ , 
\end{align}
\end{subequations}
where $O_B$ is an orthogonal matrix describing a rotation of a phase $\theta=\kappa B \Delta t$ about the direction of the field, e.g., 
\begin{equation}
O_B =  \left(\begin{array}{ccc}
    \cos(\kappa B \Delta t) & -\sin(\kappa B \Delta t) & 0 \\ 
    \sin(\kappa B \Delta t) & \cos(\kappa B \Delta t) & 0 \\ 
    0 & 0 & 1 
  \end{array}\right) \ ,
\end{equation}
when the field is oriented along the $\hat z$ axis. In the Bloch sphere this corresponds to a coherent rotation of $\aver{J}$ about the field axis $\hat B$ that preserves the shape of the uncertainty region. 

\paragraph{QND measurement of $J_z$}

The QND measurement is an indirect measurement, that can be decomposed into three steps: the preparation of the probe state, the interaction and the projective measurement of the probe. As an indirect measurement, it is performed with a system composed of the target $\hil_T$ and a meter $\hil_M$.
The additional requirement for such indirect measurement to be a true QND is that the observable $O_T$ to be measured is conserved during the evolution, while an observable of the meter $O_M$ is perturbed, acquiring information about the value of $O_T$. As a consequence, the easiest is to choose the interaction hamiltonian $H_I$ such that $[H_I,O_T]=0$ and
$[H_I,O_M]\neq 0$.

In the case of our interest, the QND measurement is performed in an atomic ensemble interacting with light pulses used as a meter. Moreover it can be  
achieved even within the framework of Gaussian states, i.e., as a Gaussian probabilistic operation. The observable to be measured is a component of the collective spin, say $J_z$, of an initially prepared Gaussian state.
The meter observable is the polarization component $S_y$ of a Gaussian state of the light. Ideally each of the probe pulse  
is prepared in the coherent state such that $\vec S^{(i)}=(\frac{N_L} 2, 0, 0)$ and 
$\Gamma_{S^{(i)}}=\diag(0,\frac{N_L} 4,\frac{N_L} 4)$.

The light-atoms interaction is ideally described by the QND hamiltonian
\begin{equation}
H_I= g J_z S_z \ ,
\end{equation}
that indeed is such that $[H_I,J_z]=0$ and $[H_I,S_y]\neq 0$. After a pulse $\vec S^{(i)}$ has passed through the atomic ensemble and interacted with it for a very small time $\tau$, the mean vector and covariance matrix of the system+meter is updated as
\begin{subequations}
\begin{align}\label{eq:gaussqnd}
\aver{\vec V} &\mapsto M^{(i)}_{\rm QND} \aver{\vec V} \ , \\
\Gamma_V &\mapsto M^{(i)}_{\rm QND} \Gamma_V (M_{\rm QND}^{(i)})^T \ , 
\end{align}
\end{subequations}
where at a first order approximation in $g\tau$ (it is thus assumed, just as a convenient approximation, that $g\tau \ll 1$), the linear transformation has the form $M^{(i)}_{\rm QND}=\left(
\begin{matrix}
M_A & B^{(i)}_{\rm at}\\
B^{(i)}_{\rm l}&M_L
\end{matrix}
\right)$, where the submatrices $B^{(i)}_{\rm at}$ and $B^{(i)}_{\rm l}$ represent the back action of the interaction on the atoms and light respectively.
In particular, $M^{(i)}_{\rm QND}$ is the identity apart from the elements $\Gamma_{J_y,S^{(i)}_z}$ and $\Gamma_{S^{(i)}_y,J_z}$ that are given by the following input/output relations
\begin{subequations}
\begin{align}\label{eq:gaussqnd}
S_y^{\rm (out)} &= S_y^{\rm (in)} + g J_z^{\rm (in)}S_x^{\rm (in)} \ , \\
J_y^{\rm (out)} &= J_y^{\rm (in)} + g J_x^{\rm (in)}S_z^{\rm (in)} \ .
\end{align}
\end{subequations}
Thus, in this step of the measurement there is a back action. The characterizing feature of the QND measurement is that such back action is entirely on a conjugate variable, $J_y$, that ideally is perturbed by the minimal amount allowed by the Heisenberg uncertainty principle.

The last step of the QND measurement is a projection of the meter observable $S_y$. In this case this formally corresponds, at the level of the Gaussian framework, to a {\it random} update of the mean vector, depending on the actual outcome, and a {\it deterministic}, projective transformation of the covariance matrix, namely
\begin{subequations}
\begin{align}\label{eq:gaussqndproj}
\Gamma_V &\mapsto \Gamma_V- \frac{\Gamma_V \Pi_{S^{(i)}_y} \Gamma_V^T}{\trace(\Gamma_V \Pi_{S^{(i)}_y})} \ , 
\end{align}
\end{subequations}
where $\Pi_{S^{(i)}_y}$ is the projector onto the $S^{(i)}_y$ element of the unit vector in the space of $\vec V$. 
This step corresponds to the acquisition of information and there is no further back action on the system. 
 
\paragraph{Uncoherent noise} Apart from a coherent evolution, the atomic ensemble can experience a loss of coherence and reduction of polarization due to various noise effects, such as, e.g., off-resonant scattering of the QND probe light. All these processes, can be described by Gaussian transformations of the form (\ref{eq:gaussopgen}) for suitable matrices $M$, $N_V$ and $N_\Gamma$ satisfying the constraint
\begin{equation}
N_\Gamma + i\Sigma^\prime - i M\Sigma M^T \geq 0 \ ,
\end{equation}
where $i(\Sigma)_{a b}=[V_a,V_b]$ and $\Sigma^\prime$ similarly defined are the matrices providing the commutation relations of the vector $\vec V$ elements respectively before and after the transformation. This is a constraint for the physicality of the output state that comes from the Heisenberg uncertainty relations.

\afterpage{\blankpage}

\chapter{Spin Squeezing and permutational symmetry}\label{ch:SSandPI} 

All the spin squeezing inequalities are based on collective observables $J_k=\sum_n j_k^{(n)}$, that are $N$-particle permutationally invariant operators, namely $P_j^\dagger J_k P_j=J_k$ for each of the $N!$ permutations $P_j$ of the particles. All the SSIs, then, are particularly tied for detecting permutationally invariant states, i.e., states $\rho$ such that $P_j^\dagger \rho P_j=\rho$.  

In fact, for example, all the states that maximally violate each of the SSIs (\ref{completessiqubitsj}) are permutationally invariant. In general
to each physical value of a spin squeezing parameter corresponds a permutationally invariant state. 
This means that the value of a spin squeezing parameter $\xi_X^2(\rho)$ of a state $\rho$ remains the same if one substitutes $\rho$ with its
permutationally invariant component $\rho_{\rm PI}:=\frac 1 {N!} \sum_j P_j^\dagger \rho P_j$, i.e., $\xi_X^2(\rho)=\xi_X^2(\rho_{\rm PI})$ for all states $\rho$.

This can be also seen by reformulating the inequalities in terms of average $n$-body correlations. In particular since Eq.~(\ref{completessiqubitsj}) contain just first and modified second moments of such collective quantities, they can be reformulated in terms of
average $2$-body correlations only.

For that, we define the average two-particle density matrix as
\begin{equation}
\label{avertwopar}
\rho_{{\rm av2}}:= \tfrac 1 {N(N-1)} \sum_{m\neq n} \rho_{mn} ,
\end{equation}
where $\rho_{mn}$ is the two-particle reduced density matrix for the $m^{\rm th}$ and $n^{\rm th}$ particles. This is a $2$-particle permutationally invariant state. Then, we formulate our entanglement conditions (\ref{completessiqubitsj}) with the density matrix $\rho_{{\rm av2}}.$

\bboo{(Optimal SSIs in terms of $\rho_{{\rm av2}}$).} The optimal Spin Squeezing Inequalities Eq.~(\ref{completessiqubitsj}) can be given in terms of the average two-body density matrix as 
\begin{gather}\label{tq}
N \sum_{l\in I} \left( 
\aver{j_{l} \otimes j_{l}}_{{\rm av2}} - \aver{j_{l}\otimes\id}_{{\rm av2}}^{2}  \right) \geq \Sigma- j^2,
\end{gather}
where we have defined the expression $\Sigma$ as the sum of all the two-particle correlations of the local spin operators
\begin{equation}
\Sigma := \sum_{l=x,y,z} \aver{j_{l} \otimes j_{l}}_{{\rm av2}}.
\end{equation}
The right-hand side of Eq.~(\ref{tq}) is nonpositive.  For the $j=\frac{1}{2}$ case, the right-hand side of Eq.~(\ref{tq}) is zero for all symmetric states,
i.e., states $|\varphi\rangle$, such that $P_j |\varphi\rangle=|\varphi\rangle$ for all permutations $P_j$.
For $j>\frac{1}{2}$ the right-hand side of Eq.~(\ref{tq}) is zero only for some symmetric states. 
\eeoo

\bbpr
Equation~(\ref{completessiqubitsj}) can be transformed into
\begin{equation}\label{symmexp1}
(N-1)\sum_{l=x,y,z} (\tilde{\Delta} J_l)^2 +N(N-1)j^2 \geq (N-1)\sum_{l \notin I} (\tilde{\Delta} J_l)^2 + \sum_{l \notin I}\aver{\tilde{J}_l^2} \ ,
\end{equation}

Let us now turn to the reformulation of Eq.~(\ref{symmexp1})  in terms of the two-body reduced density matrix.
The modified second moments and variances can be expressed with the average two-particle density matrix as
\begin{eqnarray}
\aver{\tilde{J}^{2}_l} &=& \sum_{m\neq n} \aver{j_{l}^{(n)} j_{l}^{(m)} } = 
N(N-1) \aver{j_{l} \otimes j_{l}}_{{\rm av2}} , \nonumber \\
(\tilde{\Delta}J_l)^{2}  
&=& - N^{2} \aver{j_{l}\otimes\id}_{{\rm av2}}^{2} + 
N(N-1) \aver{j_{l}\otimes j_{l}}_{{\rm av2}} .\nonumber \\\label{twopart}
\end{eqnarray}
Substituting Eq.~(\ref{twopart}) into Eq.~(\ref{symmexp1}), we obtain Eq.~(\ref{tq}).
As well as in Eq.~(\ref{symmexp1}), the right-hand side of Eq.~(\ref{tq}) is zero for symmetric states of spin-$\frac{1}{2}$ particles.
\eepr

Note that, as in the spin-$\frac{1}{2}$ case, there are states detected as entangled that have a separable two-particle density matrix \cite{tothPRL07,tothPRA09,vitagliano14}. 
Such states are, for example, permutationally invariant states with certain symmetries for which the reduced single-particle density matrix is completely mixed.
For large $N,$ due to permutational invariance and the symmetries mentioned above, the two-particle density matrices are very close to the identity matrix as well and hence they are separable. Still, some of such states can be detected as entangled by our optimal Spin Squeezing inequalities. Examples of such states are precisely the permutationally invariant singlet states discussed in Sec.~\ref{ssec:completessi12}.

Our inequalities are entanglement conditions.
We can thus compare them to the most useful entanglement condition known so far, the condition based on the positivity of the partial transpose (PPT), introduced in Th.~\ref{theo:PPT}. Staying within the framework of permutationally invariant states,
we can consider the special case of symmetric states, for which the PPT criterion can be stated in other equivalent forms, more similar to the SSIs \cite{tothguhne09}. We find that in this special case, the PPT condition 
applied to the reduced two-body density matrix detects all states detected by the spin-squeezing inequalities.
 
\bboo{(SSIs are equivalent to PPT for symmetric states).}The PPT criterion for the average two-particle density matrix defined in Eq.~(\ref{avertwopar})  
detects all symmetric entangled states that the optimal SSIs detect for $j>\frac{1}{2}.$ The two conditions are equivalent for symmetric states of particles with  $j=\frac{1}{2}.$ 
\eeoo

\bbpr
We will connect the violation of Eq.~(\ref{tq}) to the violation 
of the PPT criterion by the reduced two-particle density matrix $\varrho_{\rm av2}.$ 
If a quantum state is symmetric, its
reduced state $\varrho_{\rm av2}$ is also symmetric. For such states, 
the PPT condition is equivalent to \cite{tothguhne09}
\begin{equation}\label{PPT}
\aver{A \otimes A}_{\rm av2} - \aver{A\otimes \id}_{\rm av2}^{2} \geq 0
\end{equation}
holding for all Hermitian operators $A.$ Based on Observation 3, it can be seen by straightforward comparison of Eqs.~(\ref{tq}) and (\ref{PPT}) that, for  $j=\frac{1}{2},$  Eq.~(\ref{tq})  holds for all possible choices of $I$ and for all possible choices of 
coordinate axes, i.e., all possible $j_l,$ if and only if
Eq.~(\ref{PPT})  holds  for all Hermitian operators $A.$ For  $j>\frac{1}{2}$ there is no equivalence between the two statements. Only from the latter follows the former. 
\eepr

In the derivation of the SSIs it has been implicitly assumed that the particles are distinguishable. 
This is also the situation that one can encounter in many experimental systems, such as e.g., trapped ions, in which the particles are distinguished by their location, i.e., a definite site in a lattice.
However many spin squeezing experiments are done with Bose-Einstein condensates. In this situation the particles (bosons) cannot be always considered distinguishable, since they must satisfy the requirement that their collective state
must be in the symmetric subspace, and thus it must be ideally either fully separable or truly $N$-partite entangled. In practice any depth of entanglement smaller than $N$ would mean that some noise or some other way to distinguish the particles was present. 

Still, even for indistinguishable bosons the spin squeezing conditions signal entanglement and e.g., the relation between shot-noise limit and separable states holds formally as well.
This fact also makes bosonic systems like BEC an ideal setup to produce states with an high entanglement depth.
In such systems, the most natural formalism is the Fock operator description of multi particle systems, and in particular the so called {\it Schwinger representation} of the collective spin operators.
From two modes $(a_H,a_H^\dagger),(a_V,a_V^\dagger)$ we can construct a spin-$j=\frac 1 2$ representation as 
\begin{equation}
J_k=\frac 1 2 (a_H^\dagger,a_V^\dagger) \sigma_k (a_H,a_V) \ ,
\end{equation}
where note that these are already collective operators and in particular $J_z=\frac 1 2 (N_H-N_V)=\frac N 2 - N_L$ is the difference between the number operators of modes $H$ and $V$. These modes can be e.g., two photon polarisations (Horizontal and Vertical), or two atomic levels $(|H\rangle, |V\rangle)$. These operators are also sometimes called {\it pseudo-spin}, and act only on the permutationally symmetric subspace of states. Their definition can be used to conveniently define the collective spin operators in any system of many bosons, e.g., also for atoms in a Bose-Einstein condensate. They are what is actually measured in many experimental systems (i.e., population difference between two levels), including photons and Bose-Einstein condensates. 

\addtocontents{toc}{\vspace{2em}} 


\begin{thebibliography}{100}

\bibitem{aspect822}
J.~D. A.~Aspect and G.~Roger.
 Experimental test of bell's inequalities using time-varying
  analyzers.
 {\em Phys. Rev. Lett.}, (49):1804, 1982.

\bibitem{aspect81}
P.~G. A.~Aspect and G.~Roger.
 Experimental tests of realistic local theories via bell's theorem.
 {\em Phys. Rev. Lett.}, (47):460, 1981.

\bibitem{aspect821}
P.~G. A.~Aspect and G.~Roger.
 Experimental realization of einstein--podolsky--rosen--bohm
  gedankenexperiment: A new violation of bell's inequalities.
 {\em Phys. Rev. Lett.}, (49):91, 1982.

\bibitem{amico08}
L.~Amico, R.~Fazio, A.~Osterloh, and V.~Vedral.
 Entanglement in many-body systems.
 {\em Rev. Mod. Phys.}, 80:517--576, May 2008.

\bibitem{appel09}
J.~Appel, P.~J. Windpassinger, D.~Oblak, U.~B. Hoff, N.~Kj{\ae}rgaard, and
  E.~S. Polzik.
 Mesoscopic atomic entanglement for precision measurements beyond the
  standard quantum limit.
 {\em Proceedings of the National Academy of Sciences},
  106(27):10960--10965, 2009.

\bibitem{optomechanics14}
M.~Aspelmeyer, T.~J. Kippenberg, and F.~Marquardt.
 Cavity optomechanics.
 {\em Rev. Mod. Phys.}, 86:1391--1452, Dec 2014.

\bibitem{AthalyePRL2011}
V.~Athalye, S.~S. Roy, and T.~S. Mahesh.
 Investigation of the leggett-garg inequality for precessing nuclear
  spins.
 {\em Phys. Rev. Lett.}, 107:130402, Sep 2011.

\bibitem{parisi_qm}
G.~Auletta, M.~Fortunato, and G.~Parisi.
 {\em Quantum Mechanics}.
 Cambridge University Press, 1 edition, 4 2009.

\bibitem{AvisPRA2010}
D.~Avis, P.~Hayden, and M.~M. Wilde.
 Leggett-garg inequalities and the geometry of the cut polytope.
 {\em Phys. Rev. A}, 82:030102, Sep 2010.

\bibitem{ballentine87}
L.~E. Ballentine.
 Realism and quantum flux tunneling.
 {\em Phys. Rev. Lett.}, 59:1493--1495, Oct 1987.

\bibitem{baroouchmcoy3}
E.~Barouch and B.~M. McCoy.
 Statistical mechanics of the $\mathrm{XY}$ model. iii.
 {\em Phys. Rev. A}, 3:2137--2140, Jun 1971.

\bibitem{baroouchmcoy2}
E.~Barouch and B.~M. McCoy.
 Statistical mechanics of the $xy$ model. ii. spin-correlation
  functions.
 {\em Phys. Rev. A}, 3:786--804, Feb 1971.

\bibitem{baroucmcoy1}
E.~Barouch, B.~M. McCoy, and M.~Dresden.
 Statistical mechanics of the $\mathrm{XY}$ model. i.
 {\em Phys. Rev. A}, 2:1075--1092, Sep 1970.

\bibitem{BellP1964}
J.~S. Bell.
 On the {E}instein-{P}odolsky-{R}osen paradox.
 {\em Physics}, 1(3):195--200, 1964.

\bibitem{bell66}
J.~S. Bell.
 On the problem of hidden variables in quantum mechanics.
 {\em Rev. Mod. Phys.}, 38:447, 1966.

\bibitem{bell87}
J.~S. Bell.
 {\em Speakable and Unspeakable in Quantum Mechanics}.
 Cambridge University Press, 1987.

\bibitem{benatti95}
F.~Benatti, G.~Ghiraldi, and R.~Geassi.
 Testing macroscopic quantum coherence.
 {\em Il Nuovo Cimento B}, (110):593, 1995.

\bibitem{bengtsson06}
I.~Bengtsson and K.~\.Zyczkowski.
 {\em Geometry of quantum states}.
 Cambridge University Press, 2006.

\bibitem{bennett93}
C.~H. Bennett, G.~Brassard, C.~Cr\'epeau, R.~Jozsa, A.~Peres, and W.~K.
  Wootters.
 Teleporting an unknown quantum state via dual classical and
  einstein-podolsky-rosen channels.
 {\em Phys. Rev. Lett.}, 70:1895--1899, Mar 1993.

\bibitem{bohm51}
D.~Bohm.
 {\em Quantum Theory}.
 Dover Publications, 1951.

\bibitem{bohrEPR}
N.~Bohr.
 Can quantum-mechanical description of physical reality be considered
  complete?
 {\em Phys. Rev.}, 48:696--702, Oct 1935.

\bibitem{bohreinstdeb}
N.~Bohr.
 {\em Discussion with Einstein on epistemological problems in atomic
  physics}.
 Cambridge University Press, 1949.

\bibitem{braginsky}
V.~B. Braginsky and F.~Y. Khalili.
 {\em Quantum Measurement}.
 Cambridge University Press, 1992.

\bibitem{BraginskyS1980}
V.~B. Braginsky, Y.~I. Vorontsov, and K.~S. Thorne.
 Quantum nondemolition measurements.
 {\em Science}, 209(4456):547--557, 1980.

\bibitem{budronith}
C.~Budroni.
 {\em Temporal quantum correlations and hidden variable models}.
 PhD thesis, University of Siegen, 2014.

\bibitem{BudroniPRL2014}
C.~Budroni and C.~Emary.
 Temporal quantum correlations and leggett-garg inequalities in
  multilevel systems.
 {\em Phys. Rev. Lett.}, 113:050401, Jul 2014.

\bibitem{budronivitagliano}
C.~Budroni, G.~Vitagliano, G.~Colangelo, R.~J. Sewell, O.~G\"uhne, G.~T\'oth,
  and M.~W. Mitchell.
 Quantum nondemolition measurement enables macroscopic leggett-garg
  tests.
 {\em Phys. Rev. Lett.}, 115:200403, Nov 2015.

\bibitem{Busch2007155}
P.~Busch, T.~Heinonen, and P.~Lahti.
 Heisenberg's uncertainty principle.
 {\em Physics Reports}, 452(6):155 -- 176, 2007.

\bibitem{Busch13}
P.~Busch, P.~Lahti, and R.~F. Werner.
 Proof of heisenberg's error-disturbance relation.
 {\em Phys. Rev. Lett.}, 111:160405, Oct 2013.

\bibitem{Busch14}
P.~Busch, P.~Lahti, and R.~F. Werner.
 Measurement uncertainty relations.
 {\em Journal of Mathematical Physics}, 55(4):--, 2014.

\bibitem{brukner06}
V.~V. C.~Brukner and A.~Zeilinger.
 Crucial role of quantum entanglement in bulk properties of solids.
 {\em Phys. Rev. A}, 73:012110, Jan 2006.

\bibitem{calabresecardy04}
P.~Calabrese and J.~Cardy.
 Entanglement entropy and quantum field theory.
 {\em Journal of Statistical Mechanics: Theory and Experiment},
  2004(06):P06002, 2004.

\bibitem{calabresecardy06}
P.~Calabrese and J.~Cardy.
 Entanglement entropy and quantum field theory: a non-technical
  introduction.
 {\em Int.J.Quant.Inf.}, (4):429, 2006.

\bibitem{calabresecardy09}
P.~Calabrese and J.~Cardy.
 Entanglement entropy and conformal field theory.
 {\em Journal of Physics A: Mathematical and Theoretical},
  42(50):504005, 2009.

\bibitem{calabresecardytonni15}
P.~Calabrese, J.~Cardy, and E.~Tonni.
 Finite temperature entanglement negativity in conformal field theory.
 {\em Journal of Physics A: Mathematical and Theoretical},
  48(1):015006, 2015.

\bibitem{chen03}
K.~Chen and L.-A. Wu.
 A matrix realignment method for recognizing entanglement.
 {\em Quantum Inf. Comput.}, (3):193, 2003.

\bibitem{choi75}
M.-D. Choi.
 Completely positive linear maps on complex matrices.
 {\em Linear Algebra Appl.}, (10):285, 1975.

\bibitem{choi82}
M.-D. Choi.
 Positive linear maps.
 {\em Proc. Symp. Pure Math.}, (38):583, 1982.

\bibitem{clarisse06}
L.~Clarisse and P.~Wocjan.
 On independent permutation separability criteria.
 {\em Quantum Inf. Comput.}, (6):277, 2006.

\bibitem{chsh69}
J.~F. Clauser, M.~A. Horne, A.~Shimony, and R.~A. Holt.
 Proposed experiment to test local hidden-variable theories.
 {\em Phys. Rev. Lett.}, 23:880--884, Oct 1969.

\bibitem{coffman00}
V.~Coffman, J.~Kundu, and W.~K. Wootters.
 Distributed entanglement.
 {\em Phys. Rev. A}, 61:052306, Apr 2000.

\bibitem{ColangeloNJP2013}
G.~Colangelo, R.~J. Sewell, N.~Behbood, F.~M. Ciurana, G.~Triginer, and M.~W.
  Mitchell.
 Quantum atom--light interfaces in the gaussian description for spin-1
  systems.
 {\em New J. Phys.}, 15(10):103007, 2013.

\bibitem{ligo09}
T.~L. collaboration.
 Observation of a kilogram-scale oscillator near its quantum ground
  state.
 {\em New Journal of Physics}, 11(7):073032, 2009.

\bibitem{cramereisertplenio07}
M.~Cramer, J.~Eisert, and M.~B. Plenio.
 Statistics dependence of the entanglement entropy.
 {\em Phys. Rev. Lett.}, 98:220603, May 2007.

\bibitem{depillis67}
J.~de~Pillis.
 Linear transformations which preserve hermitean and positive
  semidefinite.
 {\em Pacific J. Math.}, (23):129, 1967.

\bibitem{devicente05}
J.~I. de~Vicente and J.~S\'anchez-Ruiz.
 Separability conditions from the landau-pollak uncertainty relation.
 {\em Phys. Rev. A}, 71:052325, May 2005.

\bibitem{Dicke1954Coherence}
R.~H. Dicke.
 Coherence in spontaneous radiation processes.
 {\em Phys. Rev.}, 93:99--110, Jan 1954.

\bibitem{dirac30}
P.~M. Dirac.
 {\em The Principles of Quantum Mechanics}.
 Oxford University Press, 1930.

\bibitem{DresselPRL2011}
J.~Dressel, C.~J. Broadbent, J.~C. Howell, and A.~N. Jordan.
 Experimental violation of two-party leggett-garg inequalities with
  semiweak measurements.
 {\em Phys. Rev. Lett.}, 106:040402, Jan 2011.

\bibitem{Duan2011Entanglement}
L.-M. Duan.
 Entanglement detection in the vicinity of arbitrary dicke states.
 {\em Phys. Rev. Lett.}, 107:180502, Oct 2011.

\bibitem{dur00}
W.~D\"ur, G.~Vidal, and J.~I. Cirac.
 Three qubits can be entangled in two inequivalent ways.
 {\em Phys. Rev. A}, 62:062314, Nov 2000.

\bibitem{EPR1935}
A.~Einstein, B.~Podolsky, and N.~Rosen.
 Can quantum-mechanical description of physical reality be considered
  complete?
 {\em Phys. Rev.}, 47:777--780, May 1935.

\bibitem{eckert91}
A.~K. Ekert.
 Quantum cryptography based on bell's theorem.
 {\em Phys. Rev. Lett.}, 67:661--663, Aug 1991.

\bibitem{Elby199217}
A.~Elby and S.~Foster.
 Why squid experiments can rule out non-invasive measurability.
 {\em Physics Letters A}, 166(1):17 -- 23, 1992.

\bibitem{lieb61}
T.~S. Elliot~Lieb and D.~Mattis.
 Two soluble models of an antiferromagnetic chain.
 {\em Ann. Phys.}, (16):407--466, 1961.

\bibitem{EmariNoriRPR2014}
C.~Emary, N.~Lambert, and F.~Nori.
 Leggett--garg inequalities.
 {\em Reports on Progress in Physics}, 77(1):016001, 2014.

\bibitem{Esteve2008Squeezing}
J.~Esteve, C.~Gross, A.~Weller, S.~Giovanazzi, and M.~Oberthaler.
 Squeezing and entanglement in a bose--einstein condensate.
 {\em Nature}, 455(7217):1216--1219, 2008.

\bibitem{Fernholz2008Spin}
T.~Fernholz, H.~Krauter, K.~Jensen, J.~F. Sherson, A.~S. S\o{}rensen, and E.~S.
  Polzik.
 Spin squeezing of atomic ensembles via nuclear-electronic spin
  entanglement.
 {\em Phys. Rev. Lett.}, 101:073601, Aug 2008.

\bibitem{feynman42}
R.~Feynman.
 {\em The Principle of Least Action in Quantum Mechanics}.
 PhD thesis, Princeton University, 1942.

\bibitem{feynman82}
R.~P. Feynman.
 Simulating physics with computers.
 {\em International Journal of Theoretical Physics}, 21(6-7):467--488,
  1982.

\bibitem{freedmanclauser}
S.~J. Freedman and J.~F. Clauser.
 Experimental test of local hidden-variable theories.
 {\em Phys. Rev. Lett.}, 28:938--941, Apr 1972.

\bibitem{frowisdur12}
F.~Fr{\"o}wis and W.~D{\"u}r.
 Measures of macroscopicity for quantum spin systems.
 {\em New Journal of Physics}, 14(9):093039, 2012.

\bibitem{Genovese2005319}
M.~Genovese.
 Research on hidden variable theories: A review of recent progresses.
 {\em Physics Reports}, 413(6):319 -- 396, 2005.

\bibitem{GeorgePNAS2013}
R.~E. George, L.~M. Robledo, O.~J.~E. Maroney, M.~S. Blok, H.~Bernien, M.~L.
  Markham, D.~J. Twitchen, J.~J.~L. Morton, G.~A.~D. Briggs, and R.~Hanson.
 Opening up three quantum boxes causes classically undetectable
  wavefunction collapse.
 {\em Proceedings of the National Academy of Sciences},
  110(10):3777--3781, 2013.

\bibitem{ghiraldistanf}
G.~Ghirardi.
 Collapse theories.
 In {\em The Stanford Encyclopedia of Philosophy}. The Metaphysics
  Research Lab, Center for the Study of Language and Information, Stanford
  University, 2011.

\bibitem{giovannetti04}
V.~Giovannetti.
 Separability conditions from entropic uncertainty relations.
 {\em Phys. Rev. A}, 70:012102, Jul 2004.

\bibitem{Giovannetti2011Advances}
V.~Giovannetti, S.~Lloyd, and L.~Maccone.
 Advances in quantum metrology.
 {\em Nature Photon.}, 5(4):222--229, 2011.

\bibitem{gittsovich08}
O.~Gittsovich, O.~G\"uhne, P.~Hyllus, and J.~Eisert.
 Unifying several separability conditions using the covariance matrix
  criterion.
 {\em Phys. Rev. A}, 78:052319, Nov 2008.

\bibitem{GiustinaN2013}
M.~Giustina, A.~Mech, S.~Ramelow, B.~Wittmann, J.~Kofler, J.~Beyer, A.~Lita,
  B.~Calkins, T.~Gerrits, S.~W. Nam, R.~Ursin, and A.~Zeilinger.
 Bell violation using entangled photons without the fair-sampling
  assumption.
 {\em Nature}, 497(7448):227--230, 05 2013.

\bibitem{glaser03}
U.~Glaser, H.~B\"uttner, and H.~Fehske.
 Entanglement and correlation in anisotropic quantum spin systems.
 {\em Phys. Rev. A}, 68:032318, Sep 2003.

\bibitem{glauber63}
R.~J. Glauber.
 Coherent and incoherent states of the radiation field.
 {\em Phys. Rev.}, 131:2766--2788, Sep 1963.

\bibitem{gleason57}
A.~Gleason.
 Measures on the closed subspaces of a hilbert space.
 {\em Indiana Univ. Math. J.}, 6:885--893, 1957.

\bibitem{Glick1965211}
A.~Glick, H.~Lipkin, and N.~Meshkov.
 Validity of many-body approximation methods for a solvable model:
  (iii). diagram summations.
 {\em Nuclear Physics}, 62(2):211 -- 224, 1965.

\bibitem{GogginPNAS2011}
M.~E. Goggin, M.~P. Almeida, M.~Barbieri, B.~P. Lanyon, J.~L. O'Brien, A.~G.
  White, and G.~J. Pryde.
 Violation of the leggett--garg inequality with weak measurements of
  photons.
 {\em Proceedings of the National Academy of Sciences},
  108(4):1256--1261, 2011.

\bibitem{GrangierN1998}
P.~Grangier, J.~A. Levenson, and J.-P. Poizat.
 Quantum non-demolition measurements in optics.
 {\em Nature}, 396(6711):537--542, 12 1998.

\bibitem{ghz}
D.~M. {Greenberger}, M.~A. {Horne}, and A.~{Zeilinger}.
 {Going Beyond Bell's Theorem}.
 {\em ArXiv e-prints}, Dec. 2007.

\bibitem{groen13}
J.~P. Groen, D.~Rist\`e, L.~Tornberg, J.~Cramer, P.~C. de~Groot, T.~Picot,
  G.~Johansson, and L.~DiCarlo.
 Partial-measurement backaction and nonclassical weak values in a
  superconducting circuit.
 {\em Phys. Rev. Lett.}, 111:090506, Aug 2013.

\bibitem{Gross:2010aa}
C.~Gross, T.~Zibold, E.~Nicklas, J.~Est{\`e}ve, and M.~K. Oberthaler.
 Nonlinear atom interferometer surpasses classical precision limit.
 {\em Nature}, 464(7292):1165--1169, 04 2010.

\bibitem{Guhne2004Characterizing}
O.~G\"uhne.
 Characterizing entanglement via uncertainty relations.
 {\em Phys. Rev. Lett.}, 92:117903, Mar 2004.

\bibitem{guhnecova}
O.~G\"uhne, P.~Hyllus, O.~Gittsovich, and J.~Eisert.
 Covariance matrices and the separability problem.
 {\em Phys. Rev. Lett.}, 99:130504, Sep 2007.

\bibitem{guhnelew}
O.~G\"uhne and M.~Lewenstein.
 Entropic uncertainty relations and entanglement.
 {\em Phys. Rev. A}, 70:022316, Aug 2004.

\bibitem{guhnemechler}
O.~G\"uhne, M.~Mechler, G.~T\'oth, and P.~Adam.
 Entanglement criteria based on local uncertainty relations are
  strictly stronger than the computable cross norm criterion.
 {\em Phys. Rev. A}, 74:010301, Jul 2006.

\bibitem{Guhne2009Entanglement}
O.~G{\"u}hne and G.~T{\'o}th.
 Entanglement detection.
 {\em Phys. Rep.}, 474(1):1--75, 2009.

\bibitem{He2011Planar}
Q.~Y. He, S.-G. Peng, P.~D. Drummond, and M.~D. Reid.
 Planar quantum squeezing and atom interferometry.
 {\em Phys. Rev. A}, 84:022107, Aug 2011.

\bibitem{stanfkochen}
C.~Held.
 The kochen-specker theorem.
 In {\em The Stanford Encyclopedia of Philosophy}. The Metaphysics
  Research Lab, Center for the Study of Language and Information, Stanford
  University, 2014.

\bibitem{wootters97}
S.~Hill and W.~K. Wootters.
 Entanglement of a pair of quantum bits.
 {\em Phys. Rev. Lett.}, 78:5022--5025, Jun 1997.

\bibitem{hofman03}
H.~F. Hofmann and S.~Takeuchi.
 Violation of local uncertainty relations as a signature of
  entanglement.
 {\em Phys. Rev. A}, 68:032103, Sep 2003.

\bibitem{holstein40}
T.~Holstein and H.~Primakoff.
 Field dependence of the intrinsic domain magnetization of a
  ferromagnet.
 {\em Phys. Rev.}, 58:1098--1113, Dec 1940.

\bibitem{Horodecki19961}
M.~Horodecki, P.~Horodecki, and R.~Horodecki.
 Separability of mixed states: necessary and sufficient conditions.
 {\em Physics Letters A}, 223(1--2):1 -- 8, 1996.

\bibitem{horodecki06}
M.~Horodecki, P.~Horodecki, and R.~Horodecki.
 Separability of mixed quantum states: Linear contractions approach.
 {\em Open Syst. Inf. Dyn.}, (13):103, 2006.

\bibitem{horodeckirev}
R.~Horodecki, P.~Horodecki, M.~Horodecki, and K.~Horodecki.
 Quantum entanglement.
 {\em Rev. Mod. Phys.}, 81:865--942, Jun 2009.

\bibitem{hyllus12}
P.~Hyllus, L.~Pezz\'e, A.~Smerzi, and G.~T\'oth.
 Entanglement and extreme spin squeezing for a fluctuating number of
  indistinguishable particles.
 {\em Phys. Rev. A}, 86:012337, Jul 2012.

\bibitem{itsjin05}
A.~R. Its, B.-Q. Jin, and V.~E. Korepin.
 Entanglement in the xy spin chain.
 {\em Journal of Physics A: Mathematical and General}, 38(13):2975,
  2005.

\bibitem{jamiolkowski72}
A.~Jamio\l{}kowski.
 Linear transformation which preserve trace and positive
  semidefiniteness of operators.
 {\em Rep. Math. Phys.}, (3):275, 1972.

\bibitem{jo07}
G.-B. Jo, Y.~Shin, S.~Will, T.~A. Pasquini, M.~Saba, W.~Ketterle, D.~E.
  Pritchard, M.~Vengalattore, and M.~Prentiss.
 Long phase coherence time and number squeezing of two bose-einstein
  condensates on an atom chip.
 {\em Phys. Rev. Lett.}, 98:030407, Jan 2007.

\bibitem{Julsgaard:2001aa}
B.~Julsgaard, A.~Kozhekin, and E.~S. Polzik.
 Experimental long-lived entanglement of two macroscopic objects.
 {\em Nature}, 413(6854):400--403, 09 2001.

\bibitem{keyl02}
M.~Keyl.
 Fundamentals of quantum information theory.
 {\em Physics Reports}, (369):431--548, 2002.

\bibitem{Kitagawa1993Squeezed}
M.~Kitagawa and M.~Ueda.
 Squeezed spin states.
 {\em Phys. Rev. A}, 47:5138--5143, Jun 1993.

\bibitem{Klyachko03}
A.~A. Klyachko, M.~A. Can, S.~Binicio\ifmmode~\breve{g}\else \u{g}\fi{}lu, and
  A.~S. Shumovsky.
 Simple test for hidden variables in spin-1 systems.
 {\em Phys. Rev. Lett.}, 101:020403, Jul 2008.

\bibitem{KneSimGau12}
G.~C. Knee, S.~Simmons, E.~M. Gauger, J.~J.~L. Morton, H.~Riemann, N.~V.
  Abrosimov, P.~Becker, H.-J. Pohl, K.~M. Itoh, M.~L.~W. Thewalt, Briggs, and
  S.~C. Benjamin.
 {Violation of a Leggett--Garg inequality with ideal non-invasive
  measurements}.
 {\em Nature Communications}, 3:606, Jan. 2012.

\bibitem{kochen67}
S.~Kochen and E.~Specker.
 The problem of hidden variables in quantum mechanics.
 {\em Journal of Mathematics and Mechanics}, (17):59--87, 1967.

\bibitem{KoschorreckPhD}
M.~Koschorreck.
 {\em Generation of Spin Squeezing in an Ensemble of Cold Rubidium
  87}.
 PhD thesis, Institute of Photonic Sciences (ICFO), 2010.

\bibitem{KoschorreckPRL2010b}
M.~Koschorreck, M.~Napolitano, B.~Dubost, and M.~W. Mitchell.
 Quantum nondemolition measurement of large-spin ensembles by
  dynamical decoupling.
 {\em Phys. Rev. Lett.}, 105:093602, Aug 2010.

\bibitem{Koschorreck10}
M.~Koschorreck, M.~Napolitano, B.~Dubost, and M.~W. Mitchell.
 Sub-projection-noise sensitivity in broadband atomic magnetometry.
 {\em Phys. Rev. Lett.}, 104:093602, Mar 2010.

\bibitem{stanfmeas}
H.~Krips.
 Measurement in quantum theory.
 In E.~N. Zalta, editor, {\em The Stanford Encyclopedia of
  Philosophy}. The Metaphysics Research Lab, Center for the Study of Language
  and Information, Stanford University, 2013.

\bibitem{Kuzmich1998Atomic}
A.~Kuzmich, L.~Mandel, and N.~P. Bigelow.
 Generation of spin squeezing via continuous quantum nondemolition
  measurement.
 {\em Phys. Rev. Lett.}, 85:1594--1597, Aug 2000.

\bibitem{law01}
C.~K. Law, H.~T. Ng, and P.~T. Leung.
 Coherent control of spin squeezing.
 {\em Phys. Rev. A}, 63:055601, Apr 2001.

\bibitem{leggett02}
A.~J. Leggett.
 Testing the limits of quantum mechanics: motivation, state of play,
  prospects.
 {\em Journal of Physics: Condensed Matter}, 14(15):R415, 2002.

\bibitem{Leggett08}
A.~J. Leggett.
 Realism and the physical world.
 {\em Reports on Progress in Physics}, 71(2):022001, 2008.

\bibitem{LG85}
A.~J. Leggett and A.~Garg.
 Quantum mechanics versus macroscopic realism: Is the flux there when
  nobody looks?
 {\em Phys. Rev. Lett.}, 54:857--860, Mar 1985.

\bibitem{LGcomment87}
A.~J. Leggett and A.~Garg.
 Comment on ``realism and quantum flux tunneling''.
 {\em Phys. Rev. Lett.}, 59:1621--1621, Oct 1987.

\bibitem{LGcomment89}
A.~J. Leggett and A.~Garg.
 Comment on ``quantum limitations on measurement of magnetic flux''.
 {\em Phys. Rev. Lett.}, 63:2159--2159, Nov 1989.

\bibitem{li07}
W.~Li, A.~K. Tuchman, H.-C. Chien, and M.~A. Kasevich.
 Extended coherence time with atom-number squeezed states.
 {\em Phys. Rev. Lett.}, 98:040402, Jan 2007.

\bibitem{Lipkin1965188}
H.~Lipkin, N.~Meshkov, and A.~Glick.
 Validity of many-body approximation methods for a solvable model:
  (i). exact solutions and perturbation theory.
 {\em Nuclear Physics}, 62(2):188 -- 198, 1965.

\bibitem{LouchetChauvet10}
A.~Louchet-Chauvet, J.~Appel, J.~J. Renema, D.~Oblak, N.~Kjaergaard, and E.~S.
  Polzik.
 Entanglement-assisted atomic clock beyond the projection noise limit.
 {\em New Journal of Physics}, 12(6):065032, 2010.

\bibitem{Lucke2014Detecting}
B.~L\"ucke, J.~Peise, G.~Vitagliano, J.~Arlt, L.~Santos, G.~T\'oth, and
  C.~Klempt.
 Detecting multiparticle entanglement of dicke states.
 {\em Phys. Rev. Lett.}, 112:155304, Apr 2014.

\bibitem{Lucke2011Twin}
B.~L{\"u}cke, M.~Scherer, J.~Kruse, L.~Pezz\'e, F.~Deuretzbacher, P.~Hyllus,
  J.~Peise, W.~Ertmer, J.~Arlt, L.~Santos, A.~Smerzi, and C.~Klempt.
 Twin matter waves for interferometry beyond the classical limit.
 {\em Science}, 334(6057):773--776, 2011.

\bibitem{luders51}
G.~L\"uders.
 {\em Ann. Phys. (Leipzig)}, (8):322, 1951.

\bibitem{Ma2011Quantum}
J.~{Ma}, X.~{Wang}, C.~P. {Sun}, and F.~{Nori}.
 {Quantum spin squeezing}.
 {\em Phys. Rep.}, 509:89--165, Dec. 2011.

\bibitem{marty14}
O.~Marty, M.~Epping, H.~Kampermann, D.~Bru\ss{}, M.~B. Plenio, and M.~Cramer.
 Quantifying entanglement with scattering experiments.
 {\em Phys. Rev. B}, 89:125117, Mar 2014.

\bibitem{baroouchmcoy4}
B.~M. McCoy, E.~Barouch, and D.~B. Abraham.
 Statistical mechanics of the $\mathrm{XY}$ model. iv. time-dependent
  spin-correlation functions.
 {\em Phys. Rev. A}, 4:2331--2341, Dec 1971.

\bibitem{melko10}
R.~G. Melko, A.~B. Kallin, and M.~B. Hastings.
 Finite-size scaling of mutual information in monte carlo simulations:
  Application to the spin-$\frac{1}{2}$ $xxz$ model.
 {\em Phys. Rev. B}, 82:100409, Sep 2010.

\bibitem{mermin90}
N.~Mermin.
 Simple unified form for the major no-hidden-variables theorems.
 {\em Phys. Rev. Lett.}, (65):3373, 1990.

\bibitem{Meshkov1965199}
N.~Meshkov, A.~Glick, and H.~Lipkin.
 Validity of many-body approximation methods for a solvable model:
  (ii). linearization procedures.
 {\em Nuclear Physics}, 62(2):199 -- 210, 1965.

\bibitem{MitchellNJP2012}
M.~W. Mitchell, M.~Koschorreck, M.~Kubasik, M.~Napolitano, and R.~J. Sewell.
 Certified quantum non-demolition measurement of material systems.
 {\em New Journal of Physics}, 14(8):085021, 2012.

\bibitem{Nimmrichter13}
S.~Nimmrichter and K.~Hornberger.
 Macroscopicity of mechanical quantum superposition states.
 {\em Phys. Rev. Lett.}, 110:160403, Apr 2013.

\bibitem{orzel01}
C.~Orzel, A.~K. Tuchman, M.~L. Fenselau, M.~Yasuda, and M.~A. Kasevich.
 Squeezed states in a bose-einstein condensate.
 {\em Science}, 291(5512):2386--2389, 2001.

\bibitem{PalaciosNatPhot2010}
A.~Palacios-Laloy, F.~Mallet, F.~Nguyen, P.~Bertet, D.~Vion, D.~Esteve, and
  A.~N. Korotkov.
 {Experimental violation of a Bell's inequality in time with weak
  measurement}.
 {\em Nature Physics}, 6(6):442--447, Apr. 2010.

\bibitem{perelomov86}
A.~M. Perelomov.
 {\em Generalized coherent states and their applications}.
 Springer-Verlag Berlin, 1986.

\bibitem{peresLG88}
A.~Peres.
 Quantum limitations on measurement of magnetic flux.
 {\em Phys. Rev. Lett.}, 61:2019--2021, Oct 1988.

\bibitem{Peres1990107}
A.~Peres.
 Incompatible results of quantum measurements.
 {\em Physics Letters A}, 151(3--4):107 -- 108, 1990.

\bibitem{peres91}
A.~Peres.
 Two simple proofs of the kochen--specker theorem.
 {\em J. Phys. A: Math. Gen.}, (24):L175--L178, 1991.

\bibitem{peres_qm}
A.~Peres.
 {\em Quantum Theory: Concepts and Methods (Fundamental Theories of
  Physics)}.
 Springer, 1993 edition, 9 1995.

\bibitem{peresPPT}
A.~Peres.
 Separability criterion for density matrices.
 {\em Phys. Rev. Lett.}, 77:1413--1415, Aug 1996.

\bibitem{Pitowsky89}
I.~Pitowsky.
 {\em Quantum Probability - Quantum Logic}.
 Springer-Verlag Berlin and Heidelberg GmbH \& Co. K, Dec. 1989.

\bibitem{pleniovirmani}
M.~B. Plenio and S.~Virmani.
 An introduction to entanglement measures.
 {\em Quantum Inf. Comput.}, (7):1, 2007.

\bibitem{puentes13}
G.~Puentes, G.~Colangelo, R.~J. Sewell, and M.~W. Mitchell.
 Planar squeezing by quantum non-demolition measurement in cold atomic
  ensembles.
 {\em New Journal of Physics}, 15(10):103031, 2013.

\bibitem{Riedel2010Atom-chip-based}
M.~F. Riedel, P.~B{\"o}hi, Y.~Li, T.~W. H{\"a}nsch, A.~Sinatra, and
  P.~Treutlein.
 Atom-chip-based generation of entanglement for quantum metrology.
 {\em Nature}, 464(7292):1170--1173, 2010.

\bibitem{RobensPRX2015}
C.~Robens, W.~Alt, D.~Meschede, C.~Emary, and A.~Alberti.
 Ideal negative measurements in quantum walks disprove theories based
  on classical trajectories.
 {\em Phys. Rev. X}, 5:011003, Jan 2015.

\bibitem{rovelli_qg}
C.~Rovelli.
 {\em Quantum Gravity (Cambridge Monographs on Mathematical Physics)}.
 Cambridge University Press, 12 2007.

\bibitem{RoweN2001}
M.~A. Rowe, D.~Kielpinski, V.~Meyer, C.~A. Sackett, W.~M. Itano, C.~Monroe, and
  D.~J. Wineland.
 Experimental violation of a bell's inequality with efficient
  detection.
 {\em Nature}, 409(6822):791--794, 02 2001.

\bibitem{rudolph00}
O.~Rudolph.
 A separability criterion for density operators.
 {\em J. Phys. A: Math. Gen.}, (33):3951, 2000.

\bibitem{sachdev}
S.~Sachdev.
 {\em Quantum Phase Transitions}.
 Cambridge University Press, 2011.

\bibitem{SchleierSmith10}
M.~H. Schleier-Smith, I.~D. Leroux, and V.~Vuleti\ifmmode~\acute{c}\else
  \'{c}\fi{}.
 States of an ensemble of two-level atoms with reduced quantum
  uncertainty.
 {\em Phys. Rev. Lett.}, 104:073604, Feb 2010.

\bibitem{schrodinger26}
E.~Schr\"odinger.
 Der stetige \"ubergang von der mikrozur makromechanik.
 {\em Naturwissenschaften}, (14):664--666, 1926.

\bibitem{schrod35}
E.~Schr\"odinger.
 Die gegenw{\"a}rtige situation in der quantenmechanik i-iii [the
  present situation in quantum mechanics].
 {\em Naturwissenschaften}, 23(48-50):807--12, 823-- 28, 844--49,
  1935.

\bibitem{SewellNatPhot2013}
R.~J. Sewell, M.~Napolitano, N.~Behbood, G.~Colangelo, and M.~W. Mitchell.
 Certified quantum non-demolition measurement of a macroscopic
  material system.
 {\em Nat. Photon.}, 7(7):517--520, 07 2013.

\bibitem{stanfbell}
A.~Shimony.
 Bell's theorem.
 In {\em The Stanford Encyclopedia of Philosophy}. The Metaphysics
  Research Lab, Center for the Study of Language and Information, Stanford
  University,
  {\em http://plato.stanford.edu/archives/win2013/entries/bell-theorem/}, 2013.

\bibitem{shor97}
P.~W. Shor.
 Polynomial-time algorithms for prime factorization and discrete
  logarithms on a quantum computer.
 {\em SIAM Journal on Computing}, 26(5):1484--1509, 1997.

\bibitem{singh11}
R.~R.~P. Singh, M.~B. Hastings, A.~B. Kallin, and R.~G. Melko.
 Finite-temperature critical behavior of mutual information.
 {\em Phys. Rev. Lett.}, 106:135701, Mar 2011.

\bibitem{Sorensen2001Many-particle}
A.~S{\o}rensen, L.-M. Duan, J.~Cirac, and P.~Zoller.
 Many-particle entanglement with bose--einstein condensates.
 {\em Nature}, 409(6816):63--66, 2001.

\bibitem{Sorensen2001Entanglement}
A.~S. S\o{}rensen and K.~M\o{}lmer.
 Entanglement and extreme spin squeezing.
 {\em Phys. Rev. Lett.}, 86:4431--4434, May 2001.

\bibitem{SouzaNJP2011}
A.~M. Souza, I.~S. Oliveira, and R.~S. Sarthour.
 A scattering quantum circuit for measuring bell's time inequality: a
  nuclear magnetic resonance demonstration using maximally mixed states.
 {\em New Journal of Physics}, 13(5):053023, 2011.

\bibitem{SuzukiNJP2012}
Y.~Suzuki, M.~Iinuma, and H.~F. Hofmann.
 Violation of leggett--garg inequalities in quantum measurements with
  variable resolution and back-action.
 {\em New Journal of Physics}, 14(10):103022, 2012.

\bibitem{thooft07}
G.~'t~Hooft.
 The free-will postulate in quantum mechanics.
 {\em {\em arXiv:quant-ph/0701097}}, 2007.

\bibitem{tothpra04}
G.~T\'oth.
 Entanglement detection in optical lattices of bosonic atoms with
  collective measurements.
 {\em Phys. Rev. A}, 69:052327, May 2004.

\bibitem{tothpra05}
G.~T\'oth.
 Entanglement witnesses in spin models.
 {\em Phys. Rev. A}, 71:010301, Jan 2005.

\bibitem{tothapellaniz14}
G.~T{\'o}th and I.~Apellaniz.
 Quantum metrology from a quantum information science perspective.
 {\em Journal of Physics A: Mathematical and Theoretical},
  47(42):424006, 2014.

\bibitem{tothguhne09}
G.~T\'oth and O.~G\"uhne.
 Entanglement and permutational symmetry.
 {\em Phys. Rev. Lett.}, 102:170503, May 2009.

\bibitem{tothPRL07}
G.~T\'oth, C.~Knapp, O.~G\"uhne, and H.~J. Briegel.
 Optimal spin squeezing inequalities detect bound entanglement in spin
  models.
 {\em Phys. Rev. Lett.}, 99:250405, Dec 2007.

\bibitem{tothPRA09}
G.~T\'oth, C.~Knapp, O.~G\"uhne, and H.~J. Briegel.
 Spin squeezing and entanglement.
 {\em Phys. Rev. A}, 79:042334, Apr 2009.

\bibitem{tothmitchell10}
G.~T{\'o}th and M.~W. Mitchell.
 Generation of macroscopic singlet states in atomic ensembles.
 {\em New Journal of Physics}, 12(5):053007, 2010.

\bibitem{tothpra13}
G.~T\'oth and D.~Petz.
 Extremal properties of the variance and the quantum fisher
  information.
 {\em Phys. Rev. A}, 87:032324, Mar 2013.

\bibitem{urizar13}
I.~Urizar-Lanz, P.~Hyllus, I.~L. Egusquiza, M.~W. Mitchell, and G.~T\'oth.
 Macroscopic singlet states for gradient magnetometry.
 {\em Phys. Rev. A}, 88:013626, Jul 2013.

\bibitem{vengalatorre}
M.~Vengalattore, J.~M. Higbie, S.~R. Leslie, J.~Guzman, L.~E. Sadler, and D.~M.
  Stamper-Kurn.
 High-resolution magnetometry with a spinor bose-einstein condensate.
 {\em Phys. Rev. Lett.}, 98:200801, May 2007.

\bibitem{vidal00}
G.~Vidal.
 Entanglement monotones.
 {\em J. Mod. Opt.}, (47):355, 2000.

\bibitem{vidallatorre03}
G.~Vidal, J.~I. Latorre, E.~Rico, and A.~Kitaev.
 Entanglement in quantum critical phenomena.
 {\em Phys. Rev. Lett.}, 90:227902, Jun 2003.

\bibitem{vidal04}
J.~Vidal, G.~Palacios, and C.~Aslangul.
 Entanglement dynamics in the lipkin-meshkov-glick model.
 {\em Phys. Rev. A}, 70:062304, Dec 2004.

\bibitem{vidalpalacios04}
J.~Vidal, G.~Palacios, and R.~Mosseri.
 Entanglement in a second-order quantum phase transition.
 {\em Phys. Rev. A}, 69:022107, Feb 2004.

\bibitem{vitaglianonotp}
G.~Vitagliano.
 not published.

\bibitem{vitagliano14}
G.~Vitagliano, I.~Apellaniz, I.~L. Egusquiza, and G.~T\'oth.
 Spin squeezing and entanglement for an arbitrary spin.
 {\em Phys. Rev. A}, 89:032307, Mar 2014.

\bibitem{vitaglianobudroni15}
G.~Vitagliano, C.~Budroni, G.~Colangelo, and M.~Mitchell.
 in preparation.

\bibitem{vitagliano11}
G.~Vitagliano, P.~Hyllus, I.~L. Egusquiza, and G.~T\'oth.
 Spin squeezing inequalities for arbitrary spin.
 {\em Phys. Rev. Lett.}, 107:240502, Dec 2011.

\bibitem{vitaglianototh2015}
G.~Vitagliano and G.~T\'oth.
 to be published.

\bibitem{vonneumann32}
J.~von Neumann.
 {\em Mathematische Grundlagen der Quantenmechanik [Mathematical
  Foundations of Quantum Mechanics]}.
 Springer, Berlin, 1932.

\bibitem{WaldherrPRL2011}
G.~Waldherr, P.~Neumann, S.~F. Huelga, F.~Jelezko, and J.~Wrachtrup.
 Violation of a temporal bell inequality for single spins in a diamond
  defect center.
 {\em Phys. Rev. Lett.}, 107:090401, Aug 2011.

\bibitem{wang10}
X.~Wang, J.~Ma, L.~Song, X.~Zhang, and X.~Wang.
 Spin squeezing, negative correlations, and concurrence in the quantum
  kicked top model.
 {\em Phys. Rev. E}, 82:056205, Nov 2010.

\bibitem{wei03}
T.-C. Wei and P.~M. Goldbart.
 Geometric measure of entanglement and applications to bipartite and
  multipartite quantum states.
 {\em Phys. Rev. A}, 68:042307, Oct 2003.

\bibitem{Werner1989Quantum}
R.~F. Werner.
 Quantum states with einstein-podolsky-rosen correlations admitting a
  hidden-variable model.
 {\em Phys. Rev. A}, 40:4277--4281, Oct 1989.

\bibitem{Werner14}
R.~F. Werner.
 Steering, or maybe why einstein did not go all the way to bell's
  argument.
 {\em Journal of Physics A: Mathematical and Theoretical},
  47(42):424008, 2014.

\bibitem{Wiesniak05}
M.~Wie{\'s}niak, V.~Vedral, and {\v C}.~Brukner.
 Magnetic susceptibility as a macroscopic entanglement witness.
 {\em New Journal of Physics}, 7(1):258, 2005.

\bibitem{WildeFP2012}
M.~M. Wilde and A.~Mizel.
 Addressing the clumsiness loophole in a {Leggett-Garg} test of
  macrorealism.
 {\em Foundations of Physics}, 42(2):256--265, 2012.

\bibitem{Wineland1994Squeezed}
D.~J. Wineland, J.~J. Bollinger, W.~M. Itano, and D.~J. Heinzen.
 Squeezed atomic states and projection noise in spectroscopy.
 {\em Phys. Rev. A}, 50:67--88, Jul 1994.

\bibitem{Wiseman07}
H.~M. Wiseman, S.~J. Jones, and A.~C. Doherty.
 Steering, entanglement, nonlocality, and the einstein-podolsky-rosen
  paradox.
 {\em Phys. Rev. Lett.}, 98:140402, Apr 2007.

\bibitem{wocjan05}
P.~Wocjan and M.~Horodecki.
 Characterization of combinatorially independent permutation
  separability criteria.
 {\em Open Syst. Inf. Dyn.}, (12):331, 2005.

\bibitem{wolf12}
M.~M. Wolf.
 {\em Quantum Channels and Operations Guided Tour}.
 2012.

\bibitem{xuSR2011}
J.-S. Xu, C.-F. Li, X.-B. Zou, and G.-C. Guo.
 Experimental violation of the leggett-garg inequality under
  decoherence.
 {\em Sci. Rep.}, 1, 09 2011.

\bibitem{yin11}
X.~Yin, X.~Wang, J.~Ma, and X.~Wang.
 Spin squeezing and concurrence.
 {\em Journal of Physics B: Atomic, Molecular and Optical Physics},
  44(1):015501, 2011.

\bibitem{zhang90}
W.-M. Zhang, D.~H. Feng, and R.~Gilmore.
 Coherent states: Theory and some applications.
 {\em Rev. Mod. Phys.}, 62:867--927, Oct 1990.

\end{thebibliography}
\end{document}